\documentclass[twocolumn,showpacs,amsmath,amssymb,prd]{revtex4}
\usepackage{graphicx}
\usepackage{graphicx}
\usepackage{epsf}
\def\agt{\mathrel{\raise.3ex\hbox{$>$}\mkern-14mu\lower0.6ex\hbox{$\sim$}}}
\def\alt{\mathrel{\raise.3ex\hbox{$<$}\mkern-14mu\lower0.6ex\hbox{$\sim$}}}

\newcommand{\beq}{\begin{equation}}
\newcommand{\eeq}{\end{equation}}
\newcommand{\beqn}{\begin{eqnarray}}
\newcommand{\eeqn}{\end{eqnarray}}
\newcommand{\pa}{\partial}

\newcommand{\varep}{\varepsilon}

\begin{document}

\title{Evolution of magnetized, differentially rotating neutron stars:  Simulations in full 
general relativity}

\author{Matthew D. Duez$^1$}
\altaffiliation{Current address:  Center for Radiophysics and Space Research,
Cornell University, Ithaca, NY 14853}

\author{Yuk Tung Liu$^1$}

\author{Stuart L. Shapiro$^1$}
\altaffiliation{Also at the Department of Astronomy and NCSA, University
of Illinois at Urbana-Champaign, Urbana, IL 61801}

\author{Masaru Shibata$^2$}

\author{Branson C. Stephens$^1$}
\affiliation{$^1$Department of Physics, University of Illinois at 
Urbana-Champaign, Urbana, IL 61801, USA \\
$^2$Graduate School of Arts and Sciences, 
University of Tokyo, Komaba, Meguro, Tokyo 153-8902, Japan}

\begin{abstract}
We study the effects of magnetic fields on the evolution of differentially
rotating neutron stars, which can be formed in stellar core collapse or
binary neutron star coalescence.  Magnetic braking and the magnetorotational 
instability (MRI)
both act on differentially rotating stars to redistribute angular momentum. 
Simulations of these stars are carried out in axisymmetry using our
recently developed codes which integrate the coupled Einstein-Maxwell-MHD
equations.  We consider stars with two different equations of state (EOS), 
a gamma-law EOS with $\Gamma=2$, and a more realistic hybrid EOS, and we
evolve them adiabatically.
Our simulations show that the fate of the star depends on its mass and spin.
For initial data, we consider three categories of differentially rotating, 
equilibrium configurations, which we label normal, hypermassive and 
ultraspinning. Normal configurations have rest masses below the maximum 
achievable with uniform rotation, and angular momentum below the 
maximum for uniform rotation at the same rest mass. Hypermassive 
stars have rest masses exceeding the mass limit for uniform rotation. 
Ultraspinning stars are not hypermassive, but have angular momentum 
exceeding the maximum for uniform rotation at the same rest mass. 
We show that a normal star will evolve to a uniformly rotating 
equilibrium configuration. An ultraspinning star evolves to an 
equilibrium state consisting of a nearly uniformly rotating central 
core, surrounded by a differentially rotating torus with constant 
angular velocity along magnetic field lines, so that differential 
rotation ceases to wind the magnetic field. In addition, the final 
state is stable against the MRI, although it has differential 
rotation.
For a hypermassive neutron star, the MHD-driven angular momentum transport
leads to catastrophic collapse of the core.  The resulting rotating black hole 
is surrounded by a hot, massive, magnetized torus undergoing 
quasistationary accretion, and a magnetic field collimated along 
the spin axis---a promising candidate
for the central engine of a short gamma-ray burst.  
\end{abstract}

\maketitle

\section{Introduction}

Differentially rotating neutron stars can form from the collapse of
massive stellar cores, which likely acquire rapid differential
rotation during collapse even if they are spinning uniformly at the
outset~\cite{zm97,diffrot} (see also~\cite{liu01}). Differential rotation
can also arise from the mergers of binary neutron
stars~\cite{RS99,SU00,FR00}. In these new-born, dynamically stable, 
neutron stars, magnetic
fields and/or viscosity will transport angular momentum and cause a
substantial change in the configurations on a secular timescale. 

Some newly-formed differentially rotating neutron stars may be {\em
hypermassive}. Specifically, the mass limits for non-rotating stars
[the Oppenheimer-Volkoff limit] and for rigidly rotating stars (the
{\it supramassive} limit, which is only about 20\% larger) can be
significantly exceeded by the presence of {\it differential}
rotation~\cite{BSS,MBS04}.  Mergers of binary neutron stars could lead
to the formation of such hypermassive neutron stars (HMNSs) as
remnants.  This possibility was foreshadowed in Newtonian~\cite{RS99},
post-Newtonian~\cite{FR00}, and in full general relativistic
simulations \cite{SU00}. The latest binary neutron star merger
simulations in full general relativity~\cite{STUa,STUb,STUc} have
confirmed that HMNS formation is indeed a possible outcome. HMNSs
could also result from core collapse of massive stars.

Differentially rotating stars tend to approach rigid rotation when 
acted upon by processes which transport angular momentum. HMNSs,
however, cannot settle down to rigidly rotating neutron stars since 
their masses exceed the maximum allowed by rigid rotation. Thus, 
delayed collapse to a black hole and, possibly, mass loss may result 
after sufficient transport of angular momentum from the inner to 
the outer regions.  Several processes can act to transport angular 
momentum and drive the HMNS to collapse.  Previous calculations 
in full general relativity 
have modeled the evolution of HMNS driven by viscous angular momentum 
transport~\cite{DLSS} and by angular momentum loss due to gravitational 
radiation~\cite{STUb}.  In both cases, the core of the HMNS eventually 
collapses to a black hole.  However, in the case of viscosity-driven 
evolution, a large accretion torus is found to develop around the 
newly-formed black hole, while for gravitational wave-driven evolution, 
the disk present after collapse is very small.  The size of the disk is
of crucial importance, because if a large disk is produced, the 
post-collapse system may produce a short-duration gamma-ray burst (GRB).  

The merger of binary neutron stars has been proposed
for many years as an explanation of short-hard GRBs~\cite{GRB,GRB-BNS}. 
Possible associations between short GRBs and elliptical galaxies
reported recently~\cite{short} make it unlikely that short GRBs are
related to supernova stellar core collapse.  The merger of compact-object
binaries (neutron star-neutron star or black hole-neutron star) is now
the favored hypothesis for explaining short GRBs.  According to this 
scenario, after the merger, a stellar-mass black hole is formed,
surrounded by hot accretion torus containing $\sim 1$--$10\%$ of 
the total mass of the system.  Energy extracted 
from this system, either by MHD processes or neutrino-radiation, powers 
the fireball for the GRB.  The viability of this model depends on the presence 
of a significantly massive accretion disk after the collapse of the remnant 
core, which in turn depends on the mechanism driving the collapse.

Though magnetic fields likely play a significant role in the evolution 
of HMNSs, the numerical tools needed to study this problem have not been
available until recently.  In particular, the evolution of magnetized HMNSs
can only be determined by solving the coupled Einstein-Maxwell-MHD equations
self-consistently in full general relativity.  
Recently, Duez {\it et al.}~\cite{DLSS2} and Shibata
and Sekiguchi~\cite{SS} independently developed codes designed 
to do such calculations 
for the first time (see also~\cite{valencia:fn}).  
The first simulations of magnetized 
hypermassive neutron star collapse (assuming both axial and equatorial 
symmetry) were reported in~\cite{DLSSS1}, and the implications of these results 
for short GRBs were presented in~\cite{GRB2}.  These simulations proved that 
the amplification of small seed magnetic fields by a combination of magnetic 
winding and the magnetorotational instability (MRI) is sufficient to trigger 
collapse in hypermassive stars on the Alfv\'en timescale, 
confirming earlier predictions~\cite{BSS,Shapiro}.  In the present work, we 
describe these collapse calculations in more detail. 

We also compare the results 
for hypermassive stars with the evolution of two differentially rotating 
models below the supramassive limit in order to highlight the qualitatively 
different physical effects which arise in the evolution. Given a fixed equation 
of state (EOS), the sequence of 
uniformly rotating stars with a given rest mass has a maximum angular momentum 
$J_{\rm max}$. A nonhypermassive star having angular momentum $J>J_{\rm max}$ 
is referred to as an ``ultraspinning'' star. We perform simulations 
on the MHD evolution of two nonhypermassive stars -- one is ultraspinning 
and the other is not; we refer to the later as ``normal.''  
Instead of collapsing, they settle down to a 
new equilibrium state after several Alfv\'en times. The normal
star settles down to a uniformly rotating configuration. In contrast, the 
ultraspinning star settles down to 
a nearly uniformly rotating central core, surrounded by 
a differentially rotating torus. In this new equilibrium, the system has 
adjusted to a state where the angular velocity is constant along the magnetic 
field lines, which means that the residual differential rotation ceases
further magnetic winding. In addition, we find that the final state is 
also stable against the MRI, although it has differential rotation.

The key subtlety in all of these simulations is that the wavelengths 
of the MRI modes 
must be well resolved on the computational grid.  Since this wavelength is 
proportional to the magnetic field strength, it becomes very difficult to 
resolve for small seed fields.  However, the simulations reported here succeed 
in resolving the MRI. 

The structure of this paper is as follows. We give an overview of the 
MHD effects acting on differentially rotating stars in 
Sec.~\ref{overview}. The initial models are briefly discussed 
in Sec.~\ref{models}. In Sec.~\ref{sec:eqs} we summarize the 
set of coupled Einstein-Maxwell-MHD equations which are solved during 
the simulations. We outline our numerical methods for the evolution 
in Sec.~\ref{sec:methods}, and present our simulation 
results in Sec.~\ref{sec:results}. Finally, we summarize and discuss 
our main results in Sec.~\ref{sec:conclusions}. 
In what follows, we assume geometrized units such that $G=c=1$.

\section{Overview of MHD effects}
\label{overview}
Two distinct processes which are known to transport angular momentum 
in differentially rotating magnetized fluids are magnetic 
braking~\cite{spitzer,spruit,BSS,Shapiro} and the MRI~\cite{MRI0,MRI}.
Magnetic braking transports angular momentum on the Alfv\'en time
scale~\cite{BSS,Shapiro}:
\begin{equation}
  t_A \sim \frac{R}{v_A} \sim 10^2~{\rm s} \left( \frac{B}{10^{12}~{\rm G}}
\right)^{-1} 
\left( \frac{R}{15~{\rm km}}\right)^{-1/2}\left( \frac{M}{3M_{\odot}}
\right)^{1/2} \ ,
\end{equation}
where $R$ is the radius of the HMNS and $v_A$ is the Alfv\'en speed.  

At early times, the effects of magnetic braking 
grow linearly with time.  This can be seen by considering the magnetic 
induction equation in a perfectly conducting (MHD) plasma 
[see Eq.~(\ref{eqn:induction}) below]:  
\beq
\partial_t \tilde{B}^i + \partial_j (v^j \tilde{B}^i - v^i 
\tilde{B}^j)=0 \ ,
\eeq
where 
\beq
\tilde B^i \equiv \sqrt{\gamma}B^i = \sqrt{\gamma} n_{\nu}F^{*\nu i} \ .
\eeq
In the above formulae, $\gamma$ is the determinant of the spatial metric, 
$n_{\mu} = (-\alpha,0,0,0)$ is the normal to the spatial hypersurface, 
$\alpha$ is the lapse, 
$F^{*\mu\nu}$ is the dual of the Faraday tensor, and $v^i = u^i/u^0$ is 
the 3-velocity of the fluid.  Now, if the magnetic field is weak
and has a negligible backreaction on the fluid, the velocities will remain 
constant with time.  In cylindrical coordinates, we have (assuming 
axisymmetry)
\beqn
\partial_t \tilde{B}^i &\approx& 0 \ , \ \ \ \ \ \ \ \ \ \ \ \ \ \ \ \ \ 
(i=\varpi,z) \\
\partial_t \tilde{B}^{\varphi} &=& -\partial_i (v^i \tilde{B}^{\varphi} - 
v^{\varphi}\tilde{B}^i) \nonumber \\ 
& \approx  & \partial_i(v^{\varphi}\tilde{B}^i) \  ,
\ \ \ \ \ \ \ (i=\varpi,z)
\eeqn
where $\varpi$ is the cylindrical radius, and where 
we have used the fact that $v^{\varpi}=v^z = 0$ at $t=0$ and remains so under 
these assumptions.  Then, since $v^{\varphi} = \Omega$ (the angular velocity),
\beq
\partial_t \tilde{B}^{\varphi} \approx \tilde{B}^i\partial_i\Omega + 
\Omega\partial_i\tilde{B}^i \ \ \ \ \ (i=\varpi,z) .
\label{eqn:dtBphi_int}
\eeq
The second term vanishes by Maxwell's equations (the no-monopole constraint, 
$\partial_i \tilde{B}^i = 0$)
and the assumption of axisymmetry ($\partial_{\varphi} = 0$).  
At early times, Eq.~(\ref{eqn:dtBphi_int}) indicates that the toroidal
component of the field $B^T \equiv \varpi B^{\varphi}$ grows linearly 
according to
\beq
B^T(t;\varpi,z) \approx t\varpi B^i(0;\varpi,z)\partial_i\Omega(0;\varpi,z) 
\ \ \ \ \ (i=\varpi,z) \ . 
\label{eqn:dtBphi}
\eeq
The growth of $B^T$ is expected to deviate from this linear relation 
when the tension due to the winding up of 
magnetic field lines begins to change the angular velocity profile of 
the fluid.

The MRI is present in a weakly magnetized, rotating fluid wherever 
$\partial_{\varpi} \Omega < 0$~\cite{MRI,MRIrev}.  When 
the instability reaches the nonlinear regime, the distortions in the 
magnetic field lines and velocity field lead to turbulence.  To 
estimate the growth timescale $t_{\rm MRI}$ and the 
wavelength of the fastest growing mode $\lambda_{\rm max}$, 
we make use of a simple Newtonian linear analysis given in~\cite{MRIrev}
(see also~\cite{gammie04}).  Linearizing the MHD equations for a local patch 
of a rotating fluid and imposing $e^{i({\bf k}\cdot{\bf x} - \omega t)}$ 
dependence on the perturbations leads to the dispersion relation 
given in Eq.~(125) of~\cite{MRIrev}.  Specializing this equation for a 
constant entropy star and considering only modes in the vertical direction, 
this reduces to
\beq
\omega^4 - [2({\bf k}\cdot {\bf v}_A)^2 + \kappa^2]\omega^2 +
({\bf k}\cdot {\bf v}_A)^2[({\bf k}\cdot {\bf v}_A)^2 + \kappa^2 - 
4\Omega^2] = 0 \ ,
\label{disp}
\eeq
where ${\bf v}_A = {\bf B}/\sqrt{4\pi\rho}$ is the (Newtonian)
Alfv\'en velocity, $\rho$ is the mass density and 
$\kappa$ is the ``epicyclic frequency'' of Newtonian theory:
\beq
\kappa^2 \equiv \frac{1}{r^3}\frac{\partial(r^4\Omega^2)}{\partial r} =
4\Omega^2 + 2\Omega \frac{\partial \Omega}{\partial \ln r} \ .
\eeq
We consider vertical modes (${\bf k} = k{\bf e}_z$) since we are only 
looking for an estimate of $\lambda_{\rm max}$, and since these  
are likely to be the dominant modes.

Since the dispersion relation in Eq.~(\ref{disp}) is quadratic in
$\omega^2$, it can be easily solved for $\omega^2$ and then minimized to
find the frequency of the fastest-growing mode, $\omega_{\rm max}$:
\beq
 -\omega^2_{\rm max} = \frac{s^4}{4(s^2 + \kappa^2)}  = 
\frac{1}{4}\left(\frac{\partial \Omega}{\partial \ln r}\right)^2 \ , 
\eeq 
where $s^2 = 4\Omega^2 - \kappa^2 = -2\Omega \partial \Omega/\partial \ln r$.  
This maximum growth rate corresponds to 
\beq
({\bf k}\cdot {\bf v}_A)^2_{\max} = \frac{s^2}{4}
\left(\frac{s^2 + 2\kappa^2}{s^2 + \kappa^2}\right) \ .
\eeq
For the growth time and wavelength of the fastest growing mode, we then have
\beqn
t_{\rm MRI} & = & 2 \left(\partial \Omega/\partial \ln \varpi \right)^{-1} 
\ ,  \label{tmri} \\
\lambda_{\rm max} & = & \frac{4\pi v_A^z}{s}\sqrt{\frac{s^2 + \kappa^2}{s^2+2\kappa^2}} \ .
\eeqn
In order of magnitude, 
\beqn
\lambda_{\rm max} & \sim &  2\pi v_A^z/\Omega  \nonumber \\
& \sim & 3~{\rm cm} \ 
\left(\frac{\Omega}{4000 \ {\rm rad} \ {\rm s}^-1}\right)^{-1} 
\left(\frac{B}{10^{12} \ {\rm G}}\right)  \\ 
t_{\rm MRI} & \sim & 1/\Omega \sim 0.25~{\rm ms} \ 
\left(\frac{\Omega}{4000 \ {\rm rad} \ {\rm s}^-1}\right)^{-1}  \ . 
\eeqn 
Here $\Omega=4000~{\rm rad}~{\rm s}^{-1}$ corresponds to a rotation 
period $P=1.57$~ms.
For realistic HMNS magnetic fields, $\lambda_{\rm max}$ will be much smaller 
than $R$. We note that, since $\lambda_{\rm max} \propto v_A$, 
larger magnetic fields will result in longer MRI wavelengths.  When 
$\lambda_{\rm max} \agt R$, where
$R$ is the equatorial radius of the star, the MRI will be suppressed since the 
unstable perturbations will no longer fit inside the star.  This is why the
MRI is regarded as a weak-field instability.  Typically,
we set magnetic field amplitudes so that 
$\lambda_{\rm max} \sim R/10$ for the models we consider here.
We note that $t_{\rm MRI}$ is independent of the strength of the seed 
magnetic field. The MRI always grows on a dynamical timescale for a 
sufficiently differentially rotating configuration. Hence, the MRI 
is likely to be very important during the early evolution for realistic HMNSs. 
However, the resulting angular momentum transport is governed by the turbulence 
and is thus expected to 
occur on a timescale longer than $t_{\rm MRI}$.

Magnetic fields and turbulence tend to transport specific angular momentum 
from the rapidly rotating inner region of a differentially rotating star to the 
more slowly rotating outer layers.  This causes the inner part to contract 
and the 
outer layers to expand.  Since hypermassive stars depend on their strong
differential rotation for stability, this angular momentum transport process 
likely leads to collapse.  However, in Section~\ref{starB1}, we show that very
different behavior can result for rapidly rotating nonhypermassive models.  
In the example we explore, the star readjusts 
to a new equilibrium state consisting of a nearly rigidly rotating core 
surrounded by a differentially rotating torus in 
which the magnetic field lines are everywhere orthogonal to the gradient of the 
angular velocity (i.e., $B^j \partial_j \Omega=0$). Hence, magnetic winding 
shuts down even though the configuration is still differentially 
rotating. This possibility has been discussed previously by 
Spruit~\cite{spruit} in the context of Newtonian theory.

\section{Initial Models}
\label{models}
In order to study the effects of rotation and EOS,
we evolve four representative differentially rotating stars, 
which we call ``A'', ``B1'', ``B2'' and ``C''. 
Their properties are listed in Table~\ref{startable}. 
Stars~A and C are hypermassive; stars~B1 and B2 are not. 
These configurations are all dynamically stable.

\begin{table*}
\caption{Initial Models}
\begin{center}
\begin{tabular}{c c c c c c c c c c c}
\hline 
\hline
\vspace{0.03in}
Case & \small{EOS} & \small{$M_0/M_{0,\rm{TOV}}$${}^{\rm a}$} & 
\small{$M_0/M_{0,\rm{sup}}$${}^{\rm b}$}  & 
$M/M_{\rm sup}$${}^{\rm c}$ & $R_{\rm eq}/M$${}^{\rm d}$ & $J/M^2$${\ }^{\rm e}$ 
& $T_{\rm rot}/|W|$${}^{\rm f}$ & $\Omega_{\rm eq}/\Omega_c$${}^{\rm g}$ &
$P_c/M$${}^{\rm h}$ \\
\hline\hline
  A  & $\Gamma = 2$ & 1.69 & 1.46 & 1.49            & 4.48  &  1.0  & 0.249 & 0.33  & 38.4 \\
\hline
  B1  & $\Gamma = 2$ & 0.99 & 0.86 & 0.89            & 8.12  &  1.0  & 0.181 & 0.40  & 103 \\
\hline
  B2 & $\Gamma = 2$ & 0.98 & 0.85 & 0.86            & 4.84  &  0.38  & 0.040 & 0.34  & 105 \\
\hline
  C  & hybrid       & 1.28 & 1.14 & 1.17 & 2.75 &  0.82 & 0.241 & 0.185 & 15.5 \\
\hline \hline
\end{tabular}
\end{center}
\vskip 12pt
\begin{minipage}{16cm}
\raggedright
${}^{\rm a}$ {The ratio of the rest mass $M_0$ to the TOV rest mass limit for the given EOS.}
\\
${}^{\rm b}$ {The ratio of the rest mass $M_0$ to the rest mass limit for uniformly rotating
stars of the given EOS (the supramassive limit).  If this ratio is greater than unity, 
the star is hypermassive.}
\\
${}^{\rm c}$ {The ratio of the ADM mass $M$ to the gravitational mass limit for uniformly 
rotating stars of the given EOS.}
\\
${}^{\rm d}$ {The equatorial coordinate radius $R_{\rm eq}$ normalized by 
the ADM mass.}
\\
${}^{\rm e}$ {The ratio of the angular momentum $J$ to $M^2$ (the angular momentum 
parameter).}
\\
${}^{\rm f}$ {The ratio of the rotational kinetic energy to the gravitational 
binding energy [see Eqs.~(\ref{eq:Trot}) and (\ref{eq:Wdef})].}
\\
${}^{\rm g}$ {The ratio of the angular velocity at the equator to the 
central angular velocity.}
\\
${}^{\rm h}$ {The initial central rotation period $P_c$ normalized by the ADM mass.}
\end{minipage}
\label{startable}
\end{table*}

Stars~A, B1 and B2 are constructed using a $\Gamma=2$ polytropic EOS, 
$P=K\rho_0^{\Gamma}$, where $P$, $K$, and $\rho_0$ are the pressure, 
polytropic constant, and rest-mass density, respectively.  (In~\cite{DLSS}, 
which considered evolution with shear viscosity, star A was
referred to as ``star I'' and star~B1 was referred to as ``star V''). 
The rest mass of star~A 
exceeds the supramassive limit by 46\%, while the rest masses of stars~B1 
and B2 are below the supramassive limit. The angular momentum of star~B1 
exceeds the maximum angular momentum ($J_{\rm max}$) for a rigidly 
rotating star with the same rest mass and EOS, whereas star~B2 has 
angular momentum $J<J_{\rm max}$. Thus, star~B1 is ``ultraspinning,''
while star~B2 is ``normal.''  Stars~A, B1 and B2 may be scaled to any desired 
physical mass by adjusting the value of $K$~\cite{CST}.  In general, 
$M \propto K^{n/2}$, where $n$ is the polytropic index ($\Gamma = 1 + 1/n$, 
here $n=1$).  For example, choosing 
$K=2.42\times 10^{5} {\rm g}^{-1} {\rm cm}^5 {\rm s}^{-2}$ gives 
the maximum ADM mass (rest mass) of $2.12 M_{\odot}$ ($2.32 M_{\odot}$) 
for spherical neutron stars and $2.42 M_{\odot}$ ($2.67M_{\odot}$) for 
rigidly rotating neutron stars.

In order to consider the effects of a more realistic neutron star
equation of state, star~C is constructed from a cold hybrid
EOS~\cite{zm97,SS} defined as follows:
\beq
P = P_{\rm cold} = \left\{ \begin{array}{ll}
                           K_1\rho_0^{\Gamma_1} & \mbox{for $\rho_0 \leq \rho_{\rm nuc}$} \\    
                           K_2\rho_0^{\Gamma_2} & \mbox{for $\rho_0 \geq \rho_{\rm nuc}$} 
			   \end{array} \right. \ .
\eeq
We set $\Gamma_1=1.3$, $\Gamma_2=2.75$, $K_1=5.16 \times 10^{14}$~cgs,
$K_2=K_1\rho_{\rm nuc}^{\Gamma_1-\Gamma_2}$, and $\rho_{\rm nuc}=1.8
\times 10^{14}~{\rm g/cm^3}$. With this EOS, the maximum ADM
mass (rest mass) is $2.01M_{\odot}~(2.32M_{\odot})$ for spherical
neutron stars and $2.27M_{\odot}~(2.60M_{\odot})$ for rigidly rotating
neutron stars, which are similar values to those in realistic stiff
EOSs~\cite{EOS}.  Star~C exceeds the supramassive limit by 14\%.  The
various parameters of star~C are chosen in order to more closely mimic the 
HMNSs formed through binary neutron star mergers with realistic equations of 
state in~\cite{STUb}.

Following previous papers (e.g, \cite{CST,BSS,SBS,DLSS}), we choose the 
initial rotation law $u^0 u_{\varphi}=A^2(\Omega_c-\Omega)$, where $u^{\mu}$ is 
the four-velocity, $\Omega_c$ is the angular velocity along the rotational axis,
and $\Omega \equiv u^{\varphi}/u^0$ is the angular velocity. In the Newtonian
limit, this rotation law becomes
\beq
  \Omega = \frac{\Omega_c}{1 + {\frac{\varpi^2}{A^2}}}\ .
\eeq
The constant $A$ has units of length and determines the steepness of the
differential rotation. In this paper, $A$ is set equal to the
coordinate equatorial radius $R_{\rm eq}$ for stars~A, B1, and B2, while
$A=0.8R_{\rm eq}$ for star~C. The corresponding values of 
$\Omega_{\rm eq}/\Omega_c$ are shown in Table~\ref{startable} 
(where $\Omega_{\rm eq}$ is the angular velocity at the equatorial 
surface).  The magnitude of the angular momentum
is seen from the Kerr parameter $a/M = q \equiv J/M^2$.  Stars~A, B1, and
C have $q =$ 1.0, 1.0, and 0.82, respectively.  These stars rotate
very rapidly and are highly flattened due to centrifugal force.  Star~B2, 
on the other hand, has a comparatively low angular momentum parameter: 
$q=0.38$.

We must also specify initial conditions for the magnetic field.  We choose
to add a weak poloidal magnetic field to the equilibrium model 
by introducing a vector potential of the following form
$A_{\varphi}= \varpi^2 {\rm max}[A_b (P-P_{\rm cut}), 0]$, 
where the cutoff $P_{\rm cut}$ is 4\% of the maximum pressure, and $A_b$ 
is a constant which determines the initial strength of
the magnetic field. We characterize the strength of the initial magnetic 
field by $C\equiv {\rm max}(b^2/P)$, i.e.\ the maximum value on the grid 
of the ratio of the magnetic energy density to the pressure.  We choose
$A_b$ such that $C\sim 10^{-3}$--$10^{-2}$. We have verified 
that such small initial magnetic fields introduce negligible violations
of the Hamiltonian and momentum constraints in the initial data.

\section{Basic Equations}
\label{sec:eqs}
\subsection{Evolution of the gravitational fields}

We evolve the 3-metric $\gamma_{ij}$ and extrinsic curvature $K_{ij}$
using the Baumgarte-Shapiro-Shibata-Nakamura (BSSN) 
formulation~\cite{BSSN}.  The fundamental variables for
BSSN evolution are
\begin{eqnarray}
  \phi &\equiv& {1\over 12}\ln[\det(\gamma_{ij})]\ , \\
  \tilde\gamma_{ij} &\equiv& e^{-4\phi}\gamma_{ij}\ , \\
  K &\equiv& \gamma^{ij}K_{ij}\ , \\
  \tilde A_{ij} &\equiv& e^{-4\phi}(K_{ij} - {1\over 3}\gamma_{ij}K)\ , \\
  \tilde\Gamma^i &\equiv& -\tilde\gamma^{ij}{}_{,j} \ \ \ 
  ({\rm or}~~F_i \equiv \delta^{jk} \tilde \gamma_{ij,k}) \ . 
\label{eq:F_i}
\end{eqnarray}
The evolution equations for these  variables are as follows: 
\begin{eqnarray}
\label{evolve_gamma}
(\partial_t - {\mathcal{L}}_{\beta})\tilde\gamma_{ij}
		&=& -2\alpha\tilde A_{ij} \\
\label{evolve_phi}
(\partial_t - {\mathcal{L}}_{\beta})\phi
                &=& -{1\over 6}\alpha K \\
\label{evolve_K}
(\partial_t - {\mathcal{L}}_{\beta})K
                &=& -\gamma^{ij}D_jD_i\alpha + {1\over 3}\alpha K^2 \\
                & & + \alpha \tilde A_{ij}\tilde A^{ij}
                    + 4\pi\alpha (\rho+ S) \nonumber \\
\label{evolve_A}
(\partial_t - {\mathcal{L}}_{\beta})\tilde A_{ij}
                &=& e^{-4\phi}(-D_iD_j\alpha
                    + \alpha(R_{ij}-8\pi S_{ij}))^{TF} \nonumber \\
		& & + \alpha(K\tilde A_{ij} - 2\tilde A_{il}\tilde A^l{}_j) \ ,
\end{eqnarray}
where $\alpha$ is the lapse function, $\beta^i$ is the shift,
${\mathcal{L}}_{\beta}$ is the Lie derivative along the shift, and
$D_i$ is the covariant derivative with respect to the spatial 3-metric.
The Ricci tensor $R_{ij}$ can be written as the sum
\begin{equation}
R_{ij} = \tilde R_{ij} + R_{ij}^{\phi}.
\end{equation}
Here $R_{ij}^{\phi}$ is
\begin{eqnarray}
R^{\phi}_{ij} & = & - 2 \tilde D_i \tilde D_j \phi -
        2 \tilde \gamma_{ij} \tilde D^l \tilde D_l \phi \nonumber \\[1mm]
        & & + 4 (\tilde D_i \phi)(\tilde D_j \phi)
        - 4 \tilde \gamma_{ij} (\tilde D^l \phi) (\tilde D_l \phi),
\end{eqnarray}
where $\tilde D^i = \tilde \gamma^{ij} \tilde D_j$.
The ``tilde'' Ricci tensor $\tilde R_{ij}$ is the Ricci tensor
associated with $\tilde \gamma_{ij}$, and is computed by 
\begin{eqnarray}
\tilde R_{ij} & = & - \frac{1}{2} \tilde \gamma^{lm}
        \tilde \gamma_{ij,lm}
        + \tilde \gamma_{k(i} \partial_{j)} \tilde \Gamma^k
        + \tilde \Gamma^k \tilde \Gamma_{(ij)k}  + \nonumber \\[1mm]
        & & \tilde \gamma^{lm} \left( 2 \tilde \Gamma^k_{l(i}
        \tilde \Gamma_{j)km} + \tilde \Gamma^k_{im} \tilde \Gamma_{klj}
        \right) \ ,
\end{eqnarray}
where 
\beq
  \tilde \Gamma_{ijk} = \frac{1}{2}(\tilde \gamma_{ij,k} + 
\tilde \gamma_{ik,j} - \tilde \gamma_{jk,i}) \ .
\eeq
The evolution of $\tilde\Gamma^i$ is given by 
\begin{eqnarray}
\label{evolve_Gamma}
\partial_t\tilde\Gamma^i &=& \partial_j(2\alpha\tilde A^{ij} 
		+ {\mathcal{L}}_{\beta}\tilde\gamma^{ij}) \nonumber \\
	&=& \tilde\gamma^{jk}\beta^i{}_{,jk} 
		+ {1\over 3}\tilde\gamma^{ij}\beta^k{}_{,kj}
 - \tilde\Gamma^j\beta^i{}_{,j} \\
 & &+{2\over 3}\tilde\Gamma^i\beta^j{}_{,j} 
    + \beta^j\tilde\Gamma^i{}_{,j} - 2\tilde A^{ij}\partial_j\alpha 
	\nonumber \\
 & &- 2\alpha\left({2\over 3}\tilde\gamma^{ij}K_{,j} - 6\tilde A^{ij}\phi_{,j} 
    - \tilde\Gamma^i{}_{jk}\tilde A^{jk} + 8\pi\tilde\gamma^{ij}S_j
	\right) \ .
	\nonumber
\end{eqnarray}
The matter source terms $\rho$, $S_{ij}$, $S_i$ and $S$ are related to the 
stress-energy tensor $T^{\mu \nu}$ as follows: 
\begin{eqnarray}
  \rho &=& n_{\alpha}n_{\beta}T^{\alpha\beta}\ , \nonumber \\
  S_i  &=& -\gamma_{i\alpha}n_{\beta}T^{\alpha\beta}\ , \nonumber \\
  S_{ij} &=& \gamma_{i\alpha}\gamma_{j\beta}T^{\alpha\beta}\ ,\nonumber \\
  S &=& S^i{}_i \ .
\end{eqnarray}

In the code of Duez {\it et al.}~\cite{DLSS2}, additional constraint 
damping terms are included in the BSSN evolution system, as described
in~\cite{ybs02,excision} (see Eqs.~(45) and (46) of~\cite{ybs02}, 
Eqs.~(6)--(8), (11), (13) and (15) of~\cite{excision}).  
The gauge conditions used with this code are 
the following hyperbolic driver conditions~\cite{abpst01,excision}:
\begin{eqnarray}
\label{hb_lapse_nok3}
\nonumber
\partial_t \alpha &=& \alpha {\cal A} \ , \\
\partial_t {\cal A} &=& -a_1(\alpha\partial_tK 
    + a_2\partial_t\alpha + a_3 e^{-4\phi}\alpha K)\ , \\
\label{hb_shift}
\partial^2_t\beta^i &=& b_1(\alpha\partial_t\tilde\Gamma^i
 - b_2\partial_t\beta^i)\ ,
\end{eqnarray}
where $a_1$, $a_2$, $a_3$, $b_1$, and $b_2$ are freely specifiable
constants.  We usually choose $a_1 = 0.75$, $b_1 = 0.15$, 
$a_2$ and $b_2$ between $0.34/M$ and $0.56/M$, and $a_3$ between
$0.17/M$ and $0.28/M$, where $M$ is the ADM mass of the star. For runs 
with excision, we use $a_1 = b_1 = 0.75$, $a_3 = 0$,  
$0.34/M \leq a_2 \leq 0.56/M$ and $b_2=a_2$.

In the code of Shibata and Sekiguchi~\cite{SS}
the following dynamical gauge conditions are used:
\beqn
\pa_t \alpha & = & -\alpha K,\\
\pa_t \beta^i  & = & \tilde \gamma^{ij}(F_j + \Delta t \pa_t F_j),
\eeqn
where $\Delta t$ is the timestep, and $F_i$ is the function defined 
in Eq.~(\ref{eq:F_i}). For the evolution, constraint damping
terms are added to the equations for $\phi$ and $\tilde A_{ij}$ to
suppress high-frequency noise and maintain the accuracy of the
Hamiltonian constraint and tr$(\tilde A_{ij})=0$.

\subsection{Evolution of the electromagnetic fields}
The evolution equation for the magnetic field in a perfectly 
conducting MHD fluid ($F^{\mu \nu} u_{\nu}=0$) can be obtained in
conservative form by taking the dual of Maxwell's equation
$F_{[\mu \nu,\lambda]}=0$.  One finds
\beq
  \nabla_{\nu} F^{* \mu \nu} = \frac{1}{\sqrt{-g}} \partial_{\nu} 
(\sqrt{-g}\, F^{*\mu \nu}) = 0 \ , \label{Maxwell}
\eeq
where $\sqrt{-g} = \alpha \sqrt{\gamma}$, $F^{\alpha\beta}$ is the Faraday
tensor, and $F^{*\alpha\beta}$ is its dual.  Using the fact that the magnetic 
field as measured by a normal observer $n^a$ is given by 
$B^i = n_{\mu}F^{*\mu i}$, 
the time component of Eq.~(\ref{Maxwell}) gives the no-monopole constraint 
$\partial_j \tilde{B}^j =0$, where $\tilde{B}^j = \sqrt{\gamma}\, B^j$.
The spatial components of Eq.~(\ref{Maxwell}) give the magnetic induction 
equation, which can be written as 
\beq
  \partial_t \tilde{B}^i + \partial_j (v^j \tilde{B}^i - v^i 
\tilde{B}^j)=0 \ .
\label{eqn:induction}
\eeq 

\subsection{Evolution of the hydrodynamics fields}
The evolution equations for the fluid are as follows~\cite{DLSS2,SS}:
\beqn
  \label{rho_star_eqn}
  \partial_t \rho_* + \partial_j (\rho_* v^j) & = & 0 \ , \\
  \partial_t \tilde{S}_i 
 + \partial_j (\alpha \sqrt{\gamma}\, T^j{}_i) & = & \frac{1}{2} \alpha \sqrt{\gamma} 
\, T^{\alpha \beta} g_{\alpha \beta,i} \ , \\
  \partial_t \tilde{\tau} + \partial_i ( \alpha^2 \sqrt{\gamma}\, T^{0i} 
-\rho_* v^i) & = & s \ , \label{fonts_eng_eq}
\eeqn
where the density variable is $\rho_* = \alpha \sqrt{\gamma} \rho_0 u^0$, 
the momentum-density variable is $\tilde{S}_i = \alpha \sqrt{\gamma} T^0_i$, 
the energy-density variable as adopted by Duez {\it et al.}~\cite{DLSS2} 
is $\tilde{\tau} = \alpha^2 \sqrt{\gamma}\, T^{00} - \rho_*$, 
and the source term $s$ is
\beqn
  s &=& -\alpha \sqrt{\gamma}\, T^{\mu \nu} \nabla_{\nu} n_{\mu}  \cr
   &=& \alpha \sqrt{\gamma}\, [ (T^{00}\beta^i \beta^j + 2 T^{0i} \beta^j 
+ T^{ij}) K_{ij} \cr
 & & - (T^{00} \beta^i + T^{0i}) \partial_i \alpha ]\ .
\eeqn
The MHD stress-energy tensor is given by 
\beq
T^{\mu\nu} = (\rho_0 h + b^2)u^{\mu}u^{\nu} + (P+b^2/2)g^{\mu\nu} - b^{\mu}b^{\nu} \ , 
\eeq
where $b^{\mu} \equiv u_{\nu}F^{*\nu\mu}/\sqrt{4\pi}$, the specific enthalpy is 
given by $h \equiv 1 + \epsilon + P/\rho_0$, $\epsilon$ is the specific internal 
energy, and $b^2 = b_{\mu}b^{\mu}$.  

In the code of Shibata and Sekiguchi~\cite{SS}, 
the energy evolution variable is chosen to be 
$\sqrt{\gamma}\, n_{\mu} n_{\nu} T^{\mu \nu} = \tilde{\tau}+\rho_*$, and
the evolution equation may be obtained by adding Eq.~(\ref{rho_star_eqn}) to 
Eq.~(\ref{fonts_eng_eq}).  

The MHD system of equations is completed by a choice of EOS 
for the evolution.  For stars~A, B2, and B2, we adopt a 
$\Gamma$-law EOS $P = (\Gamma-1)\rho_0\epsilon$,
with $\Gamma = 2$.  For star~C, we adopt the following hybrid EOS:
\beq 
P=P_{\rm cold}+(\Gamma_{\rm th}-1)\rho_0(\varepsilon-\varepsilon_{\rm cold}) . 
\eeq
Here, $P_{\rm cold}$ and $\varepsilon_{\rm cold}$ denote the cold component 
of $P$ and $\varep$~\cite{SS}. The conversion efficiency of kinetic energy 
to thermal energy at shocks is determined by $\Gamma_{\rm th}$, which we 
set to 1.3 to conservatively account for shock heating. 

\subsection{Diagnostics}
\label{sec:diagnostics}

We monitor several global conserved quantities to check the accuracy of our 
simulations. The ADM mass $M$ and angular momentum $J$ are defined 
as integrals over surfaces at infinity as follows~\cite{integrals}:
\beqn
\label{Mdef}
M &=& {1\over 16\pi}\int_{r=\infty}\sqrt{\gamma}\gamma^{im}\gamma^{jn}
                                 (\gamma_{mn,j} - \gamma_{jn,m}) d^2S_i \ , \\ 
\label{Jdef}
 J_i &=& {{1}\over{8\pi}}\varepsilon_{ij}{}^k \int_{r=\infty}
                         x^j K_k^m d^2S_m \ .
\eeqn
In cases for which no singularity is present on the grid, these 
surface integrals can be converted to volume integrals using Gauss's 
theorem (see Appendix~A of~\cite{ybs02}):
\begin{eqnarray}
\label{Mfin}
M   &=& \int_V \left[ e^{5\phi}(\rho_0 + {1\over 16\pi}\tilde A_{ij}\tilde A^{ij}
                             - {1\over 24\pi}K^2) \right. \\
    & &\left.  \quad - {1\over 16\pi}\tilde\Gamma^{ijk}\tilde\Gamma_{jik}
			     + {1-e^{\phi}\over 16\pi}\tilde R \right] d^3x
			     \nonumber \\
\label{Jfin}
J_i &=& \varepsilon_{ij}{}^k\int_V \Bigl({1\over 8\pi}\tilde A^j_k
        + x^j S_k \\
\nonumber
    & & \quad +  {1\over 12\pi}x^j K_{,k}
        - {1\over 16\pi}x^j\tilde\gamma^{lm}{}_{,k}\tilde A_{lm}\Bigr) 
        e^{6\phi} d^3x \ .
\end{eqnarray}
These integrals should be exactly conserved.  However, using finite grids, 
we are unable to perform this integral out to infinity, and
we expect to see mass and angular momentum losses due to outflows
(of fluid, electromagnetic fields, and/or gravitational waves) through the boundaries. 
These fluxes can be measured, however, and are found to be quite small.

In axisymmetry, the volume integral for the angular momentum (which 
is entirely in the $z$-direction) simplifies considerably~\cite{Wald}:  
\beq
J = \int_V \tilde{S}_{\varphi} d^3x\ . \label{Jaxi}
\eeq
An additional conserved quantity is the total rest mass $M_0$: 
\begin{equation}
M_0 = \int_V \rho_{\star} d^3x\ . \label{M0int}
\end{equation}
In axisymmetry, gravitational radiation carries no angular momentum, 
and in this case our GRMHD codes are finite differenced such that $M_0$ and $J$ 
are identically conserved in the absence of flux through the boundaries.  
Hence, $M_0$ and $J$ are not useful diagnostics when volume 
integrals~(\ref{Jaxi}) and~(\ref{M0int}) are applicable.

For runs with black hole excision, a volume integral must be replaced with 
an integral over an inner surface surrounding the black hole plus a volume integral
extending over the rest of the grid (see~\cite{ybs02} for details).  The integral
for $J$ is then no longer identically conserved by our numerical scheme, and the 
total angular momentum is only constant to the extent that the excision evolution
is accurate.  During excision evolutions, we separately track the rest mass and angular momentum 
of matter outside the hole by carrying out the integrals in Eqs.~(\ref{Jaxi}) and 
(\ref{M0int}) over the region outside the apparent horizon.  
Though no longer exact, these integrals allow us to 
estimate the rest mass and angular momentum of the accretion torus.

When a black hole is present, we detect it by using an apparent horizon
finder (see~\cite{bcsst96} for details).  As the system approaches
stationarity, the apparent horizon will approach the event horizon. 
From the surface area of the apparent horizon $\mathcal{A}_{\rm AH}$, we compute the
approximate irreducible mass $M_{\rm irr}$ by
\begin{equation}
M_{\rm irr} \approx \sqrt{{\mathcal{A}_{\rm AH}}/16\pi^2} \ .
\end{equation}

In order to check the accuracy of our simulations, we monitor the L2
norms of the violation in the constraint equations.  In terms of the BSSN 
variables, the constraint equations become,
respectively,
\begin{eqnarray}
\label{Hamiltonian_BSSN}
  0 = \mathcal{H} &=& 
                \tilde\gamma^{ij}\tilde D_i\tilde D_j e^{\phi}
                - {e^{\phi} \over 8}\tilde R  \\
		& & + {e^{5\phi}\over 8}\tilde A_{ij}\tilde A^{ij}
		    - {e^{5\phi}\over 12}K^2 + 2\pi e^{5\phi}\rho, 
	\nonumber \\
\label{momentum_BSSN}
  0 = {\mathcal{M}}^i &=&
  \tilde D_j(e^{6\phi}\tilde A^{ji})- {2\over 3}e^{6\phi}\tilde D^i K
  - 8\pi e^{6\phi}S^i \ .
\end{eqnarray}
We normalize $\mathcal{H}$ and ${\mathcal{M}}^i$ and compute the L2 norms 
on the grid as described in~\cite{dmsb03}.

In order to understand the evolution of the magnetic field, it is useful
to compute field lines.  Below, we plot field lines corresponding to the
poloidal magnetic field.  In axisymmetry, these field lines correspond 
to the level
surfaces of $A_{\varphi}$ (see Appendix~\ref{app:field_lines}), which 
is computed from $B^{\varpi}$ and $B^z$.
To visualize the toroidal field, we also plot the 3D field lines 
projected onto the equatorial plane (see Appendix~\ref{app:field_lines} 
for details of the method).  

We measure several invariant energy integral diagnostics during the evolution.
We define the adiabatic internal energy $E_{\rm int,ad}$, the internal
energy from heat, $E_{\rm heat}$, rotational kinetic energy $T_{\rm rot}$, the 
electromagnetic energy $E_{\rm EM}$, 
and gravitational potential energy $W$, as follows: 
\begin{eqnarray}
E_{\rm int,ad} &=& \int_V (\rho_0\epsilon_{\rm cold}) d{\cal V}\ , \\
E_{\rm heat}   &=& \int_V (\rho_0\epsilon_{\rm heat}) d{\cal V}\ , \\
T_{\rm rot}    &=& \int_V {1\over 2}\Omega T_{\rm fluid}^0{}_{\varphi} 
d{\cal V}/u^0 \ ,
\label{eq:Trot} \\
E_{\rm EM}     &=& \int_V n_{\mu} n_{\nu} T_{\rm EM}^{\mu \nu} d{\cal V}/
(\alpha u^0) \ , \\
W              &=&  M - M_0 - E_{\rm int,ad} - E_{\rm heat} - T_{\rm rot} - E_{\rm EM}  \ ,  \ \ \ \ \ \ \  \label{eq:Wdef}
\end{eqnarray}
where $d{\cal V} = \alpha u^0 \sqrt{\gamma}\, d^3x$ is the proper 3-volume
element, $T_{\rm fluid}^{\mu \nu}=\rho_0 h u^{\mu} u^{\nu} + Pg^{\mu \nu}$ 
is the perfect fluid stress-energy tensor, 
$T^{\mu \nu}_{\rm EM} = b^2 u^{\mu} u^{\nu} + b^2 g^{\mu \nu}/2 
- b^{\mu} b^{\nu}$ is the stress-energy tensor associated with the 
electromagnetic field,
$\epsilon_{\rm cold}$ refers to a the cold initial
polytrope or hybrid EOS internal energy, and $\epsilon_{\rm heat}$ 
is the energy due to shock heating 
$\epsilon_{\rm heat} = \epsilon - \epsilon_{\rm cold}$.

\section{Numerical Methods}
\label{sec:methods}
Duez {\it et al.}~\cite{DLSS2} and Shibata and Sekiguchi~\cite{SS} have
independently developed new codes to evolve magnetized fluids in 
dynamical spacetimes
by solving the Einstein-Maxwell-MHD system of equations self-consistently.
Both codes evolve the Einstein field equations without
approximation, and both use high-resolution shock capturing techniques
to track the MHD fluid.  Several tests have been performed
with these codes, including MHD shocks, nonlinear MHD wave propagation, 
magnetized Bondi
accretion, MHD waves induced by linear gravitational waves, and magnetized
accretion onto a neutron star.  Details of our techniques for evolving the
Einstein-Maxwell-MHD system as well as tests can be found in~\cite{DLSS2,SS}. 
In this paper, we have performed several simulations for identical 
initial data using both codes and 
found that the results are essentially the same.

The simulations presented in this paper assume axial and equatorial 
symmetry. We evolve only the $x$-$z$ plane [a (2+1) dimensional problem]. 
We adopt 
the cartoon method~\cite{cartoon} for evolving the BSSN equations, and use 
cylindrical coordinates for evolving the induction and MHD equations. In this 
scheme, the coordinate $x$ is identified with the cylindrical 
radius $\varpi$, and the $y$-direction corresponds to the azimuthal 
direction. For example, for any vector $V^i$, $V^x \equiv 
V^{\varpi}$, and $V^y \equiv \varpi V^{\varphi}$.

When black holes appear in our simulations, we avoid the singularity by
using black hole excision.  This technique involves removing from the
grid a region inside the event horizon which contains the spacetime
singularity.  Rather than evolving inside this region, boundary
conditions are placed on the fields immediately outside.  For details on
our excision techniques, see~\cite{alcubierre,ybs02,excision}.

As in many hydrodynamic simulations, we add a tenuous, uniform-density 
``atmosphere'' to cover the computational grid outside the star. For 
stars A, B1, and B2, the rest-mass density in the atmosphere is set to 
$\rho_a = 10^{-7} \rho_{\rm max}(0)$, where $\rho_{\rm max}(0)$ is 
the initial maximum rest-mass density.  The initial pressure in the 
atmosphere is set to the cold polytropic value ($P = K\rho_a^{\Gamma}$).
If the density in a given grid cell drops below $\rho_a$ after an 
evolution step, we simply set $\rho = \rho_a$.  We also impose limits 
on the pressure in order to prevent negative values of the internal 
energy and to prevent spurious heating of the atmosphere.  In particular,
if the pressure drops below $P_{\rm min} = 0.5 K \rho^{\Gamma}$, we set 
$P=P_{\rm min}$; similarly, if $P$ rises above 
$P_{\rm max} = 10 K\rho^{\Gamma}$, we set $P=P_{\rm max}$. 
Our main results are not sensitive to the adopted (small) value 
of $\rho_a$; similarly for $P_{\rm min}$ and $P_{\rm max}$.

Due to the hybrid EOS, we found that a different atmosphere scheme is 
appropriate when evolving star C.  For this case, we choose 
$\rho_a=10^9~{\rm g/cm^3} \approx 10^{-6} \rho_{\rm max}(0)$. 
The specific internal energy $\varepsilon$ of 
the atmosphere is set to be 
$K_1(100\rho_a)^{\Gamma_1-1}/(\Gamma_1-1)\equiv \varepsilon_{\rm
min}$.  If the value of $\varepsilon$ becomes smaller than this value,
we artificially set $\varepsilon=\varepsilon_{\rm min}$. We also limit
the maximum value of $\varepsilon$ as $30\varepsilon_{\rm cold}$; if
the value of $\varepsilon$ exceeds this value, we artificially set
$\varepsilon=30\varepsilon_{\rm cold}$.

\section{Numerical Results}
\label{sec:results}
\subsection{Star~A}
\label{starA}

We have performed simulations on star~A with fixed initial field 
strength ($C=2.5 \times 10^{-3}$). We use a uniform grid with size 
$(N,N)$ in cylindrical coordinates $(\varpi,z)$, which covers the region 
$[0,L]$ in each direction. We have performed simulations with 
$L=4R_{\rm eq}$ and $5R_{\rm eq}$ and found that the results 
depend only weakly on $L$.  In the following, we present results 
with $L=4.5R_{\rm eq}$. For star~A, $R_{\rm eq}=4.5M=18.6~{\rm km} 
(M/2.8M_{\odot})$. To check the convergence of our numerical results, 
we perform simulations with four different grid resolutions: 
$N=$ 250, 300, 400 and 500. Unless otherwise stated, 
all results presented in the following subsections are from  
the simulation data with resolution $N=500$.  We will first describe
the general features of the evolution and then discuss the effects 
of resolution, the behavior of the various components of the
energy, and the excision evolution.

\subsubsection{General features of the evolution}

\begin{figure*}
\begin{center}
\epsfxsize=1.8in
\leavevmode
\hspace{-0.7cm}\epsffile{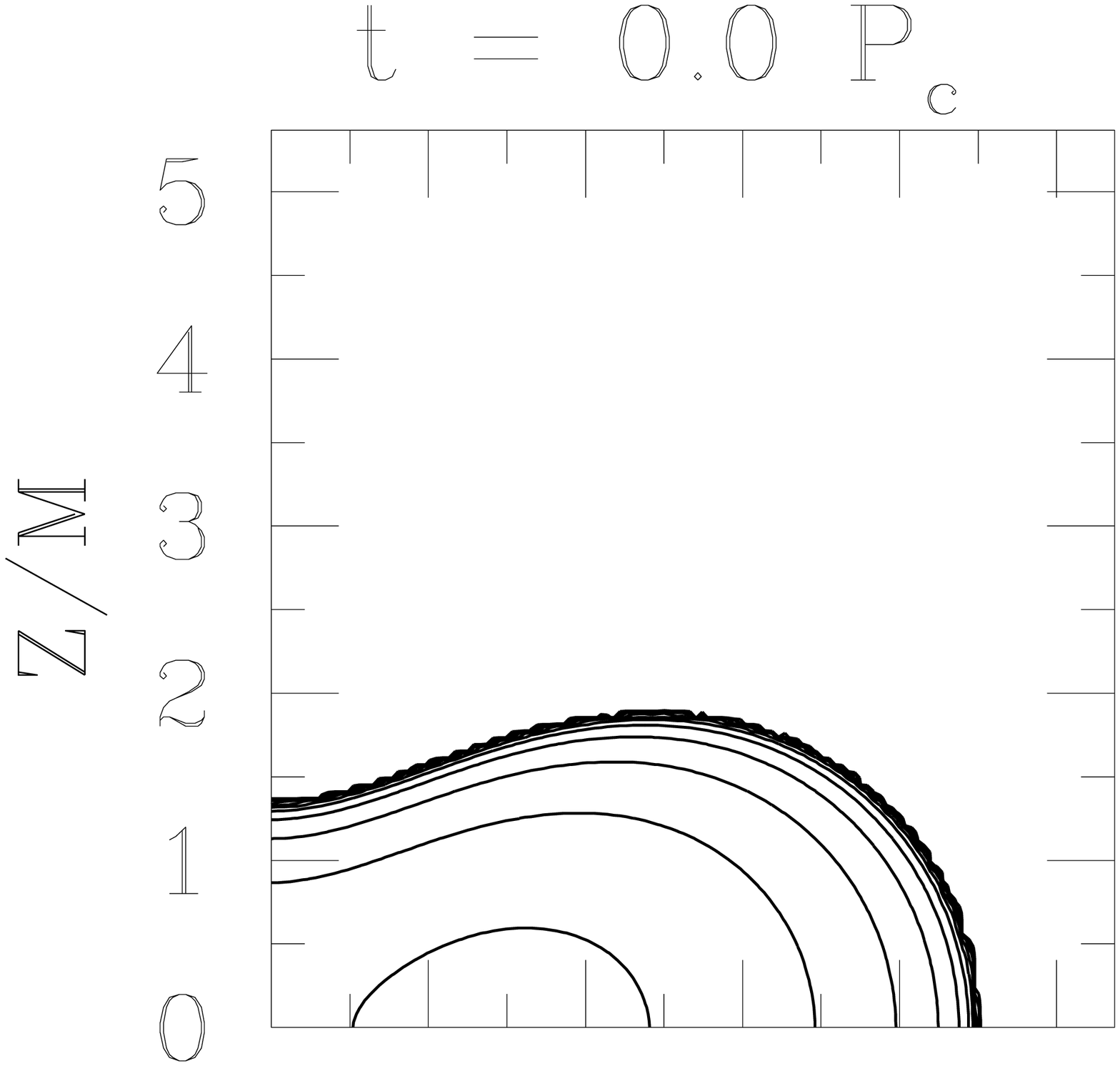}
\epsfxsize=1.8in
\leavevmode
\hspace{-0.5cm}\epsffile{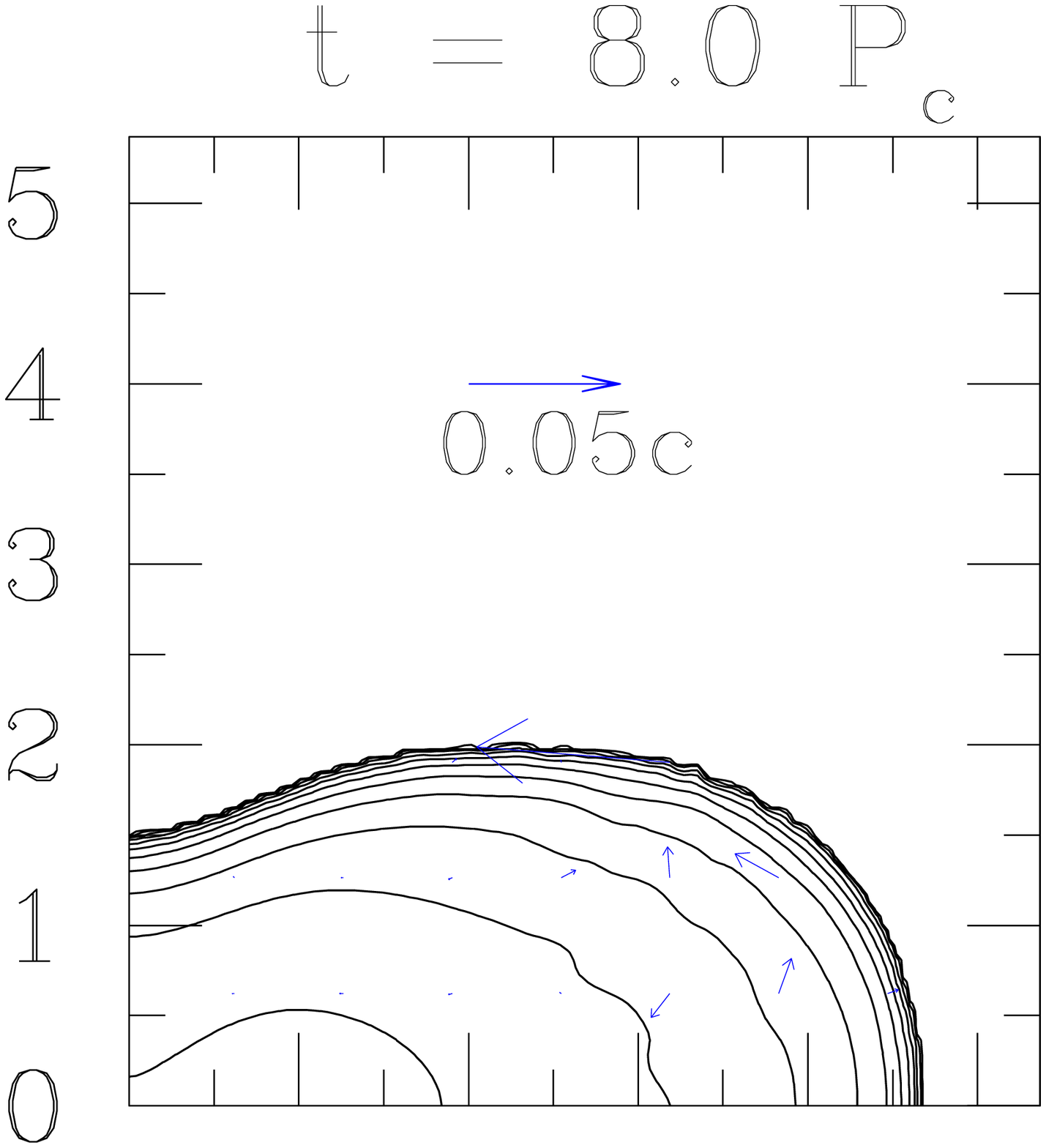}
\epsfxsize=1.8in
\leavevmode
\hspace{-0.5cm}\epsffile{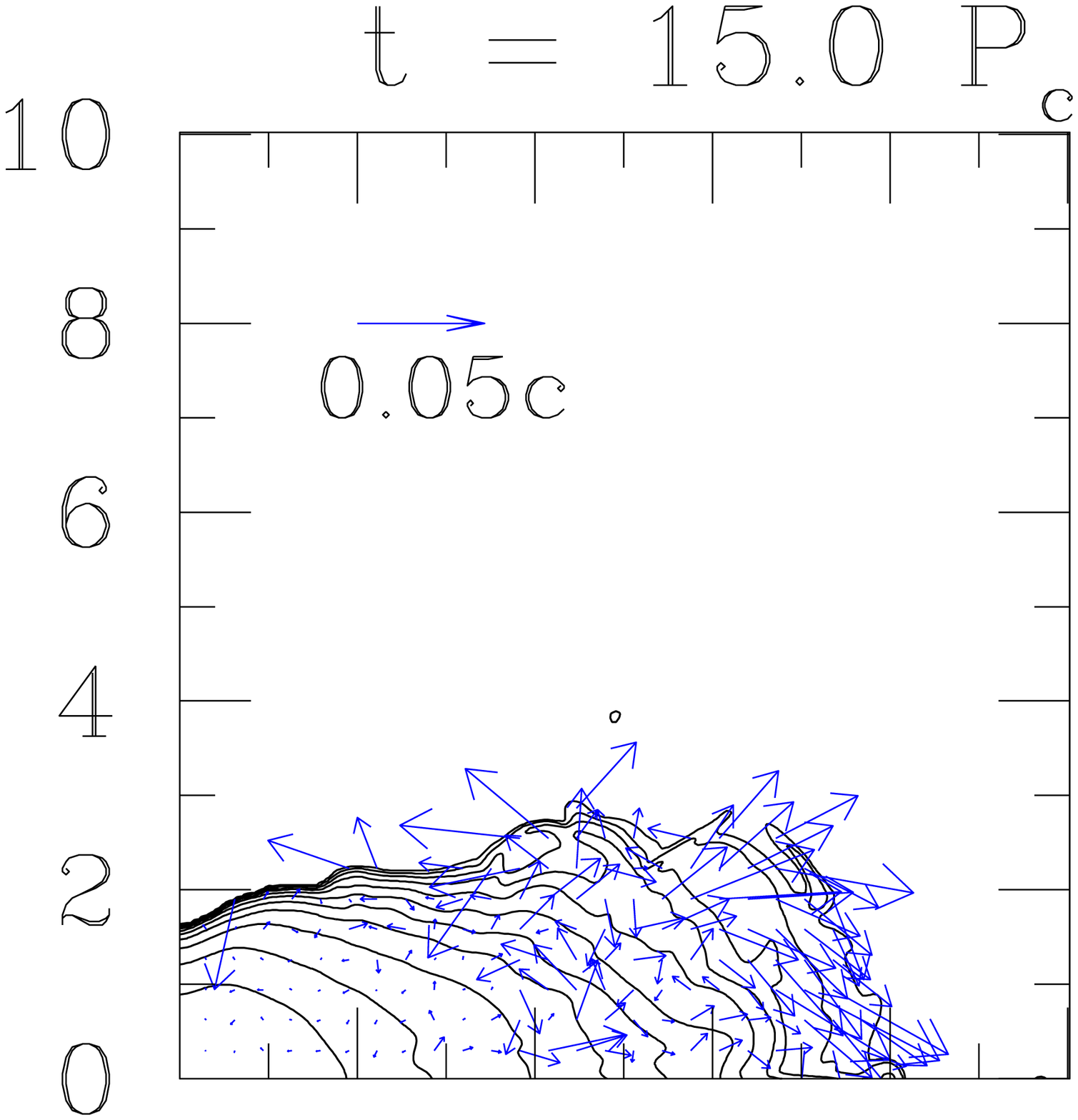}
\epsfxsize=1.8in
\leavevmode
\hspace{-0.5cm}\epsffile{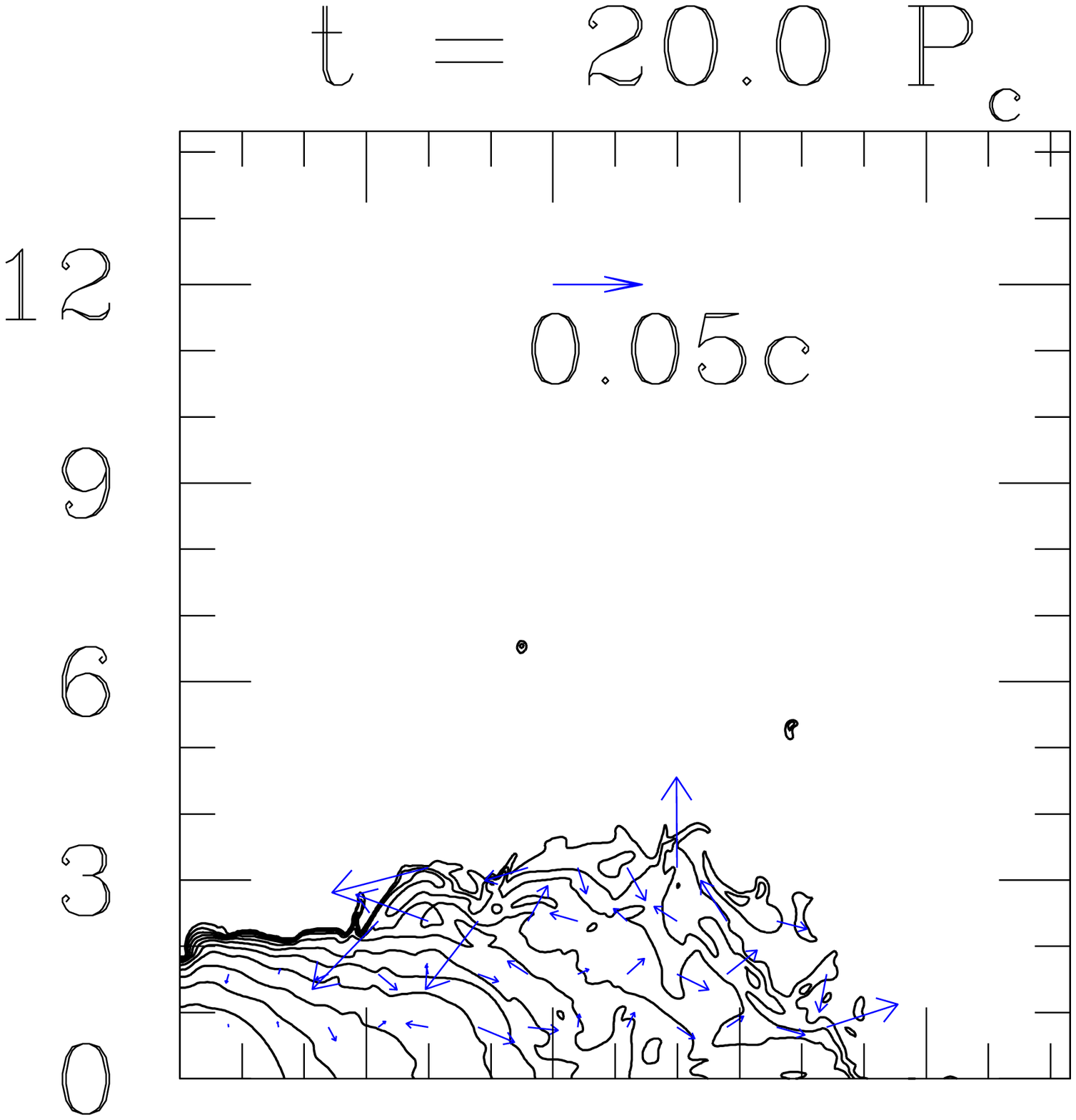} \\
\vspace{-0.5cm}
\epsfxsize=1.8in
\leavevmode
\hspace{-0.7cm}\epsffile{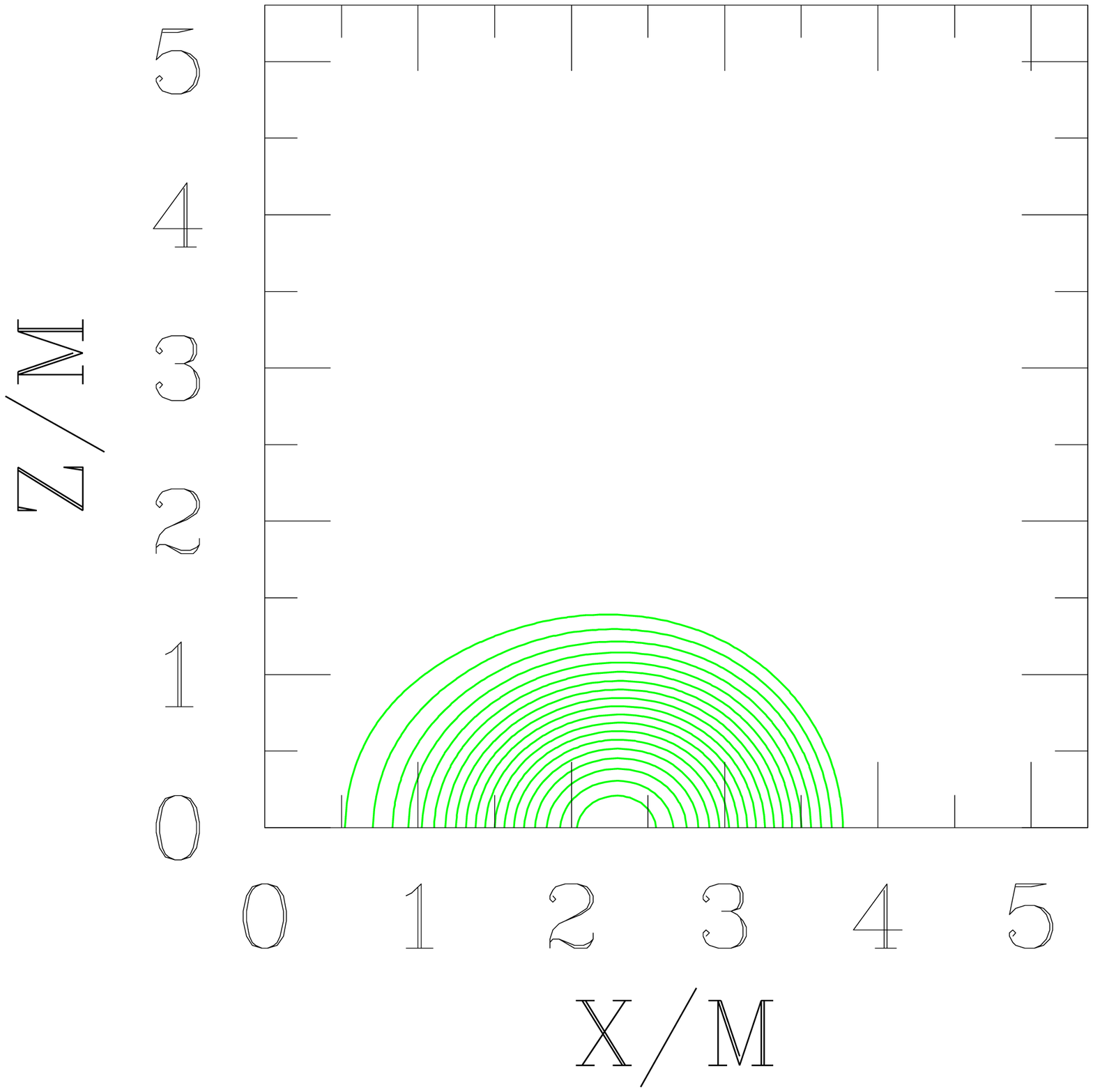}
\epsfxsize=1.8in
\leavevmode
\hspace{-0.5cm}\epsffile{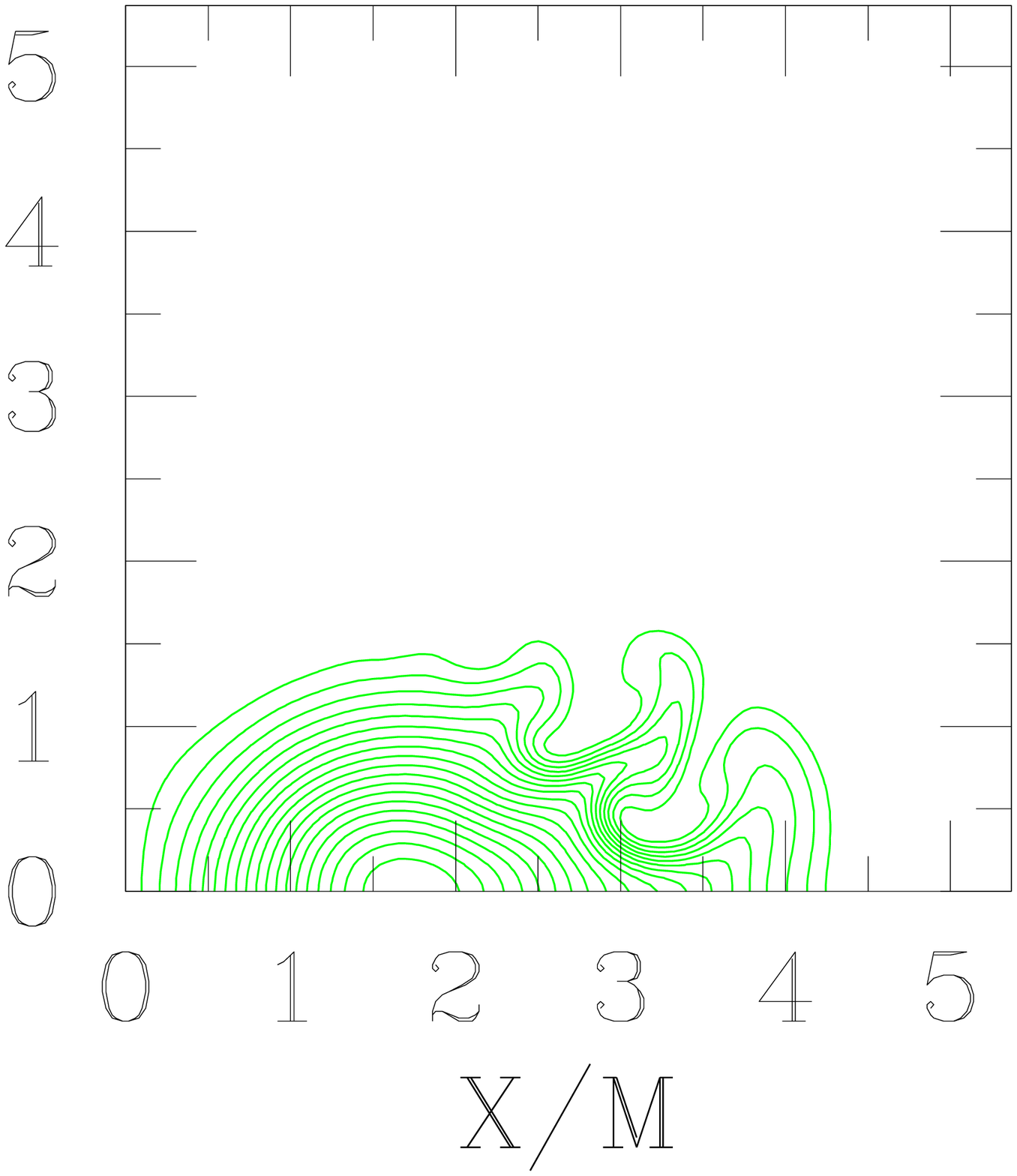}
\epsfxsize=1.8in
\leavevmode
\hspace{-0.5cm}\epsffile{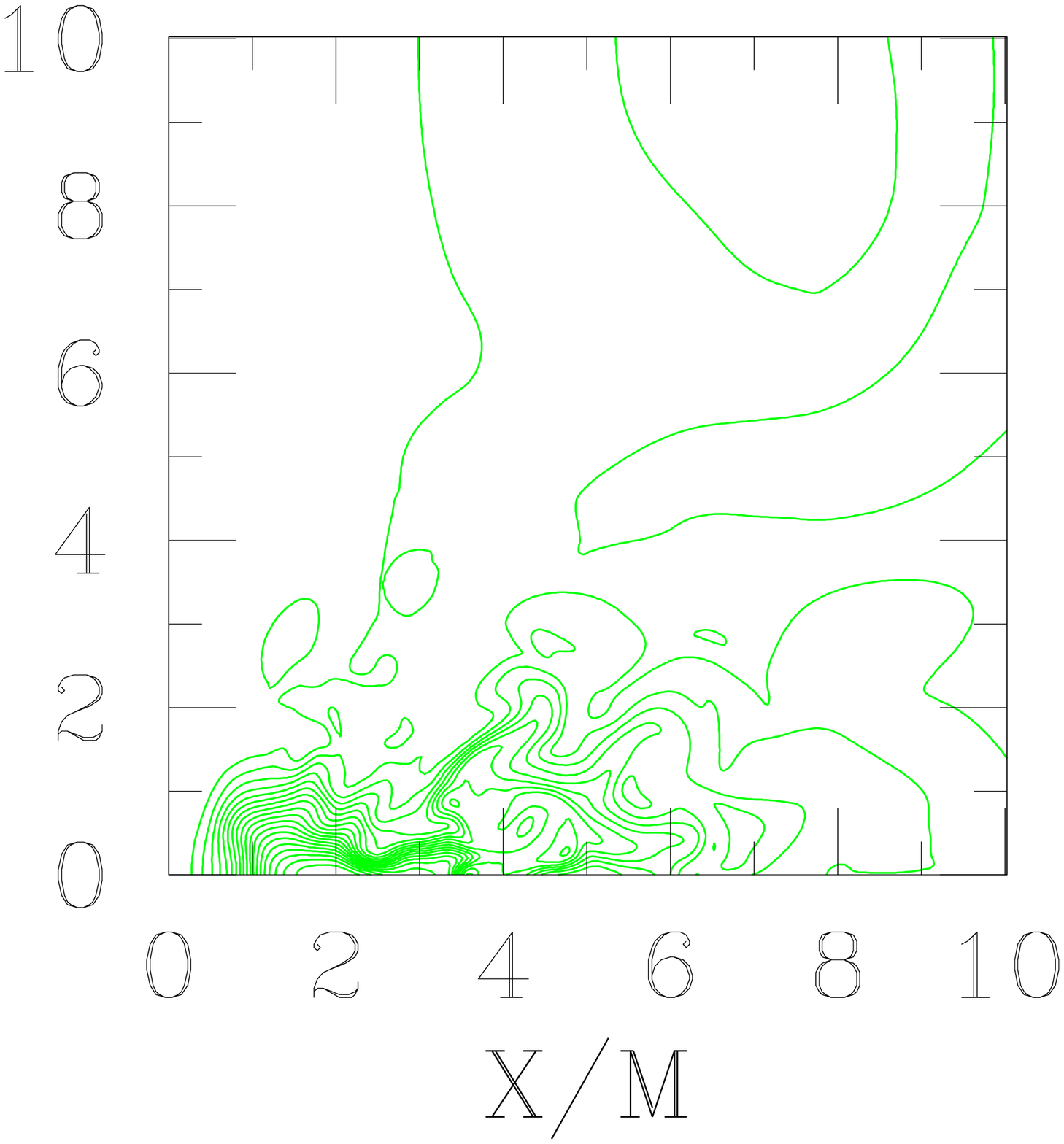}
\epsfxsize=1.8in
\leavevmode
\hspace{-0.5cm}\epsffile{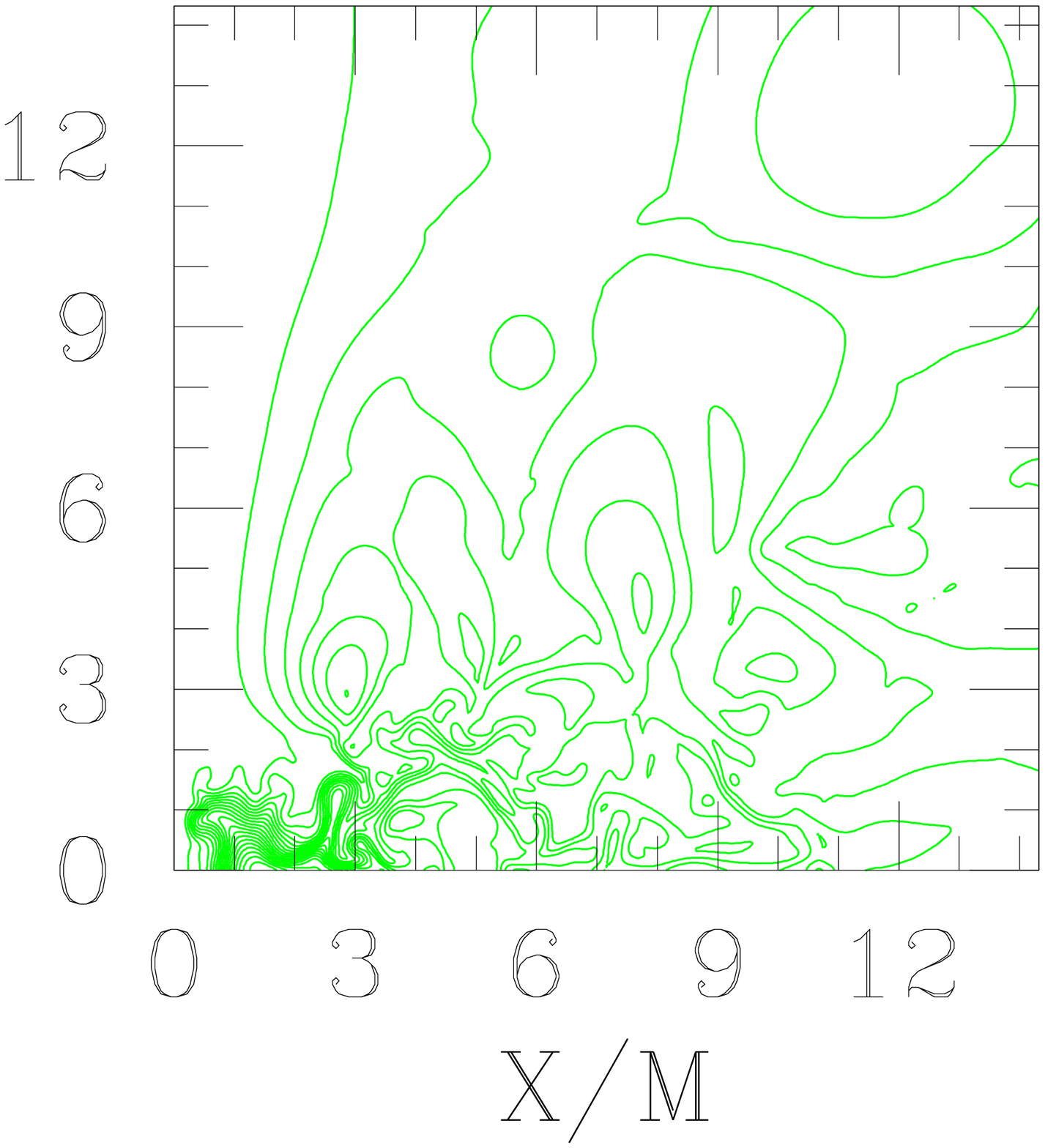}
\end{center}

\begin{center}
\epsfxsize=1.8in
\leavevmode
\hspace{-0.7cm}\epsffile{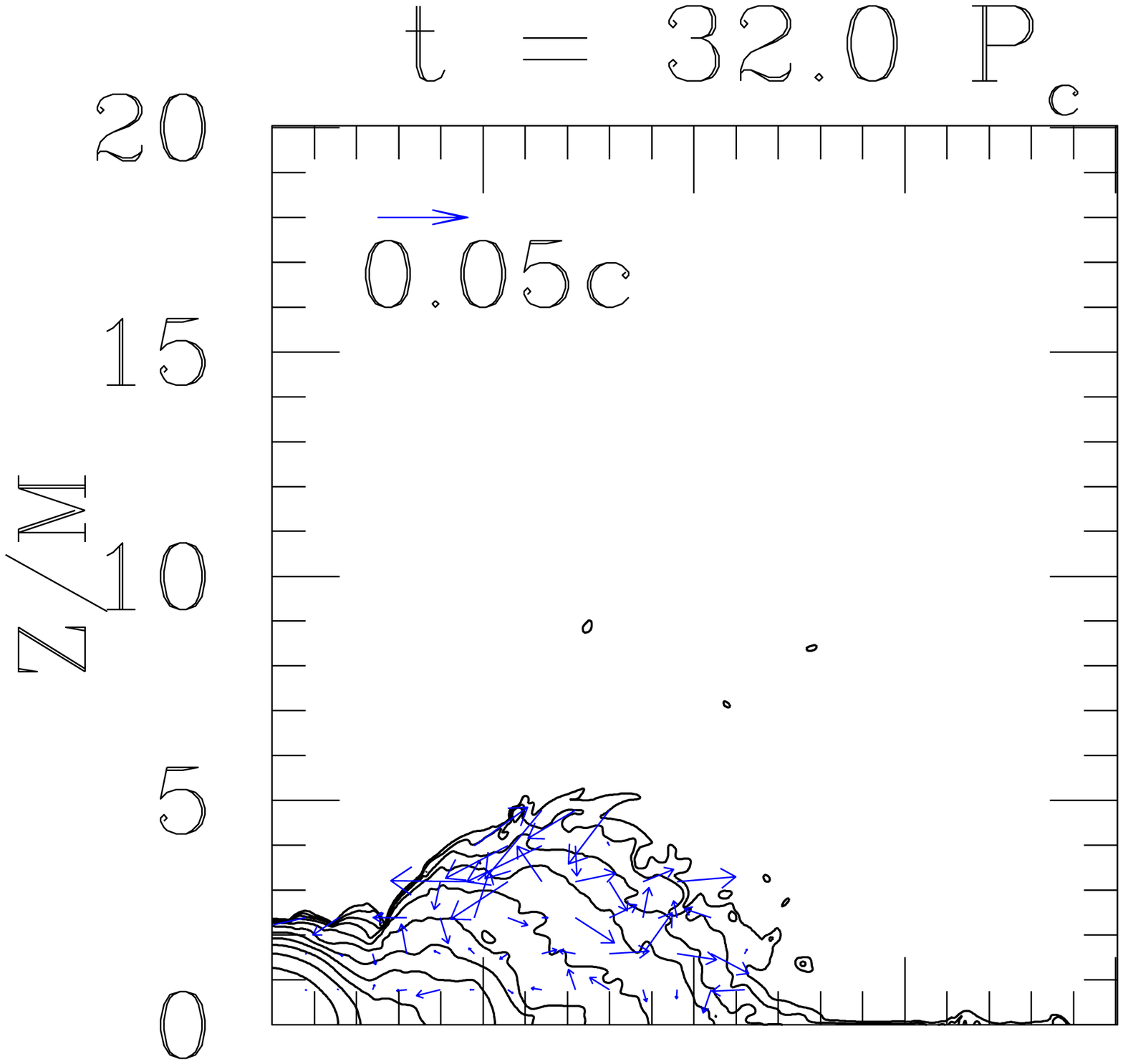}
\epsfxsize=1.8in
\leavevmode
\hspace{-0.5cm}\epsffile{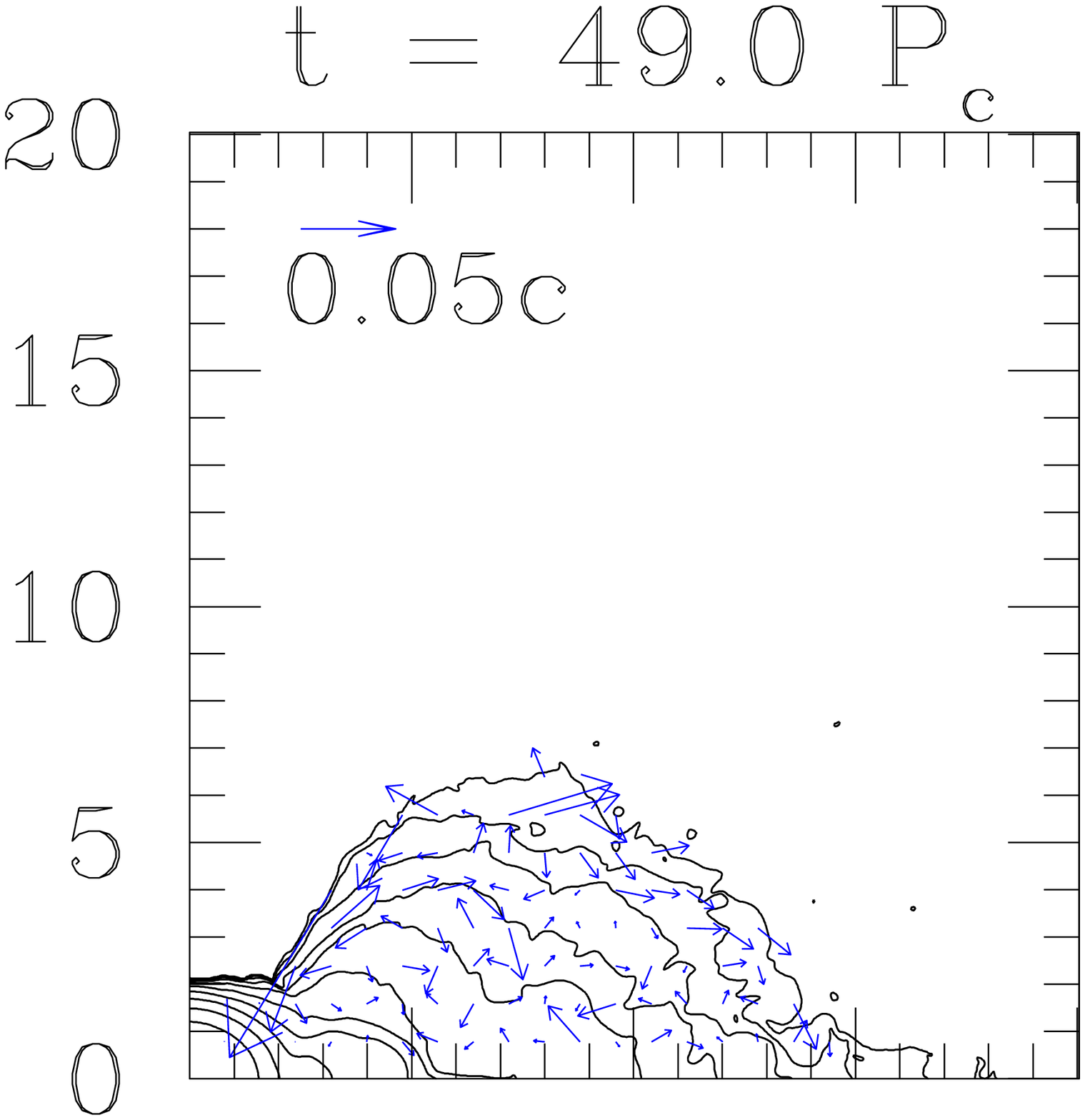}
\epsfxsize=1.8in
\leavevmode
\hspace{-0.5cm}\epsffile{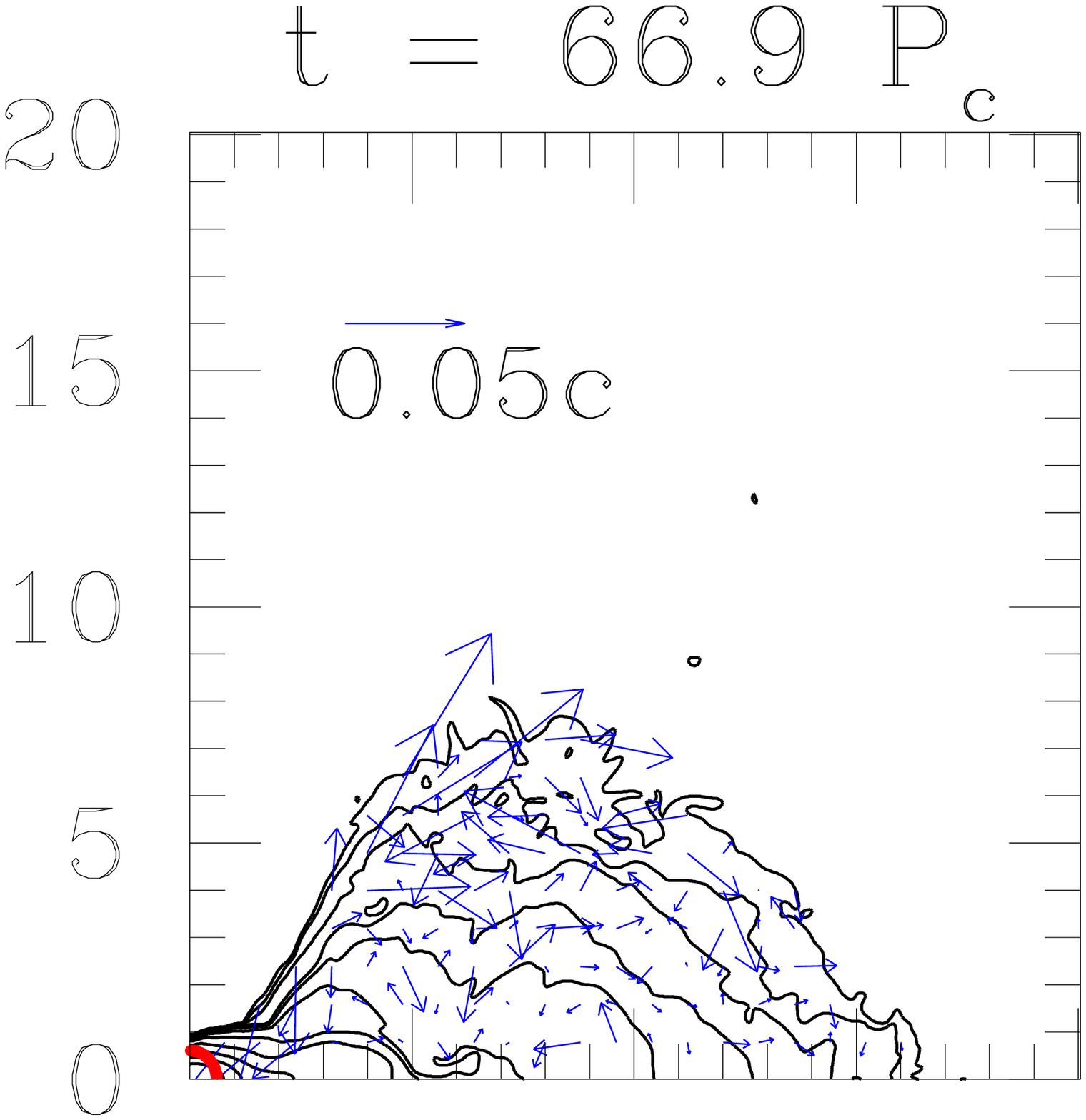}
\epsfxsize=1.8in
\leavevmode
\hspace{-0.5cm}\epsffile{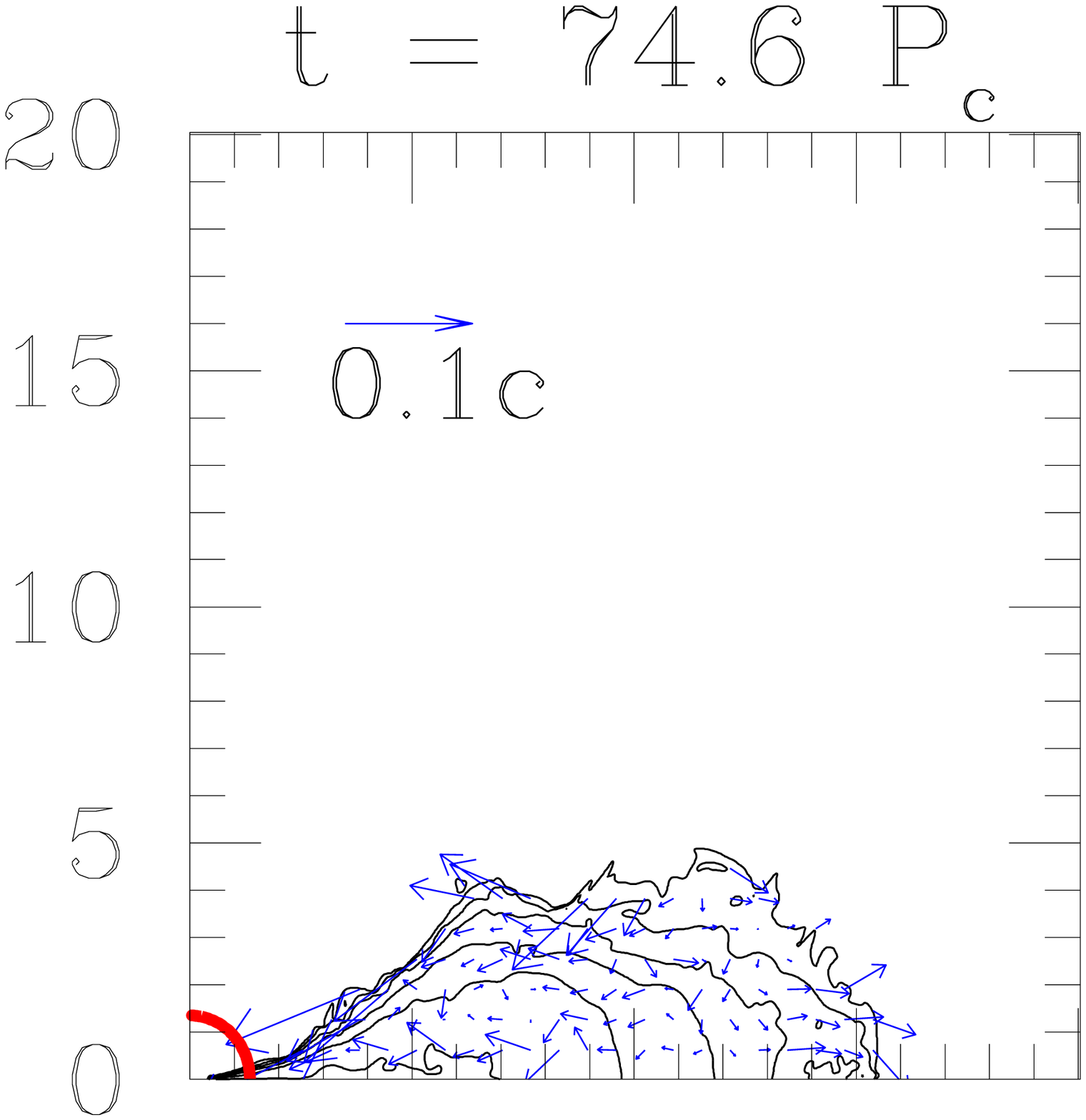} \\
\vspace{-0.5cm}
\epsfxsize=1.8in
\leavevmode
\hspace{-0.7cm}\epsffile{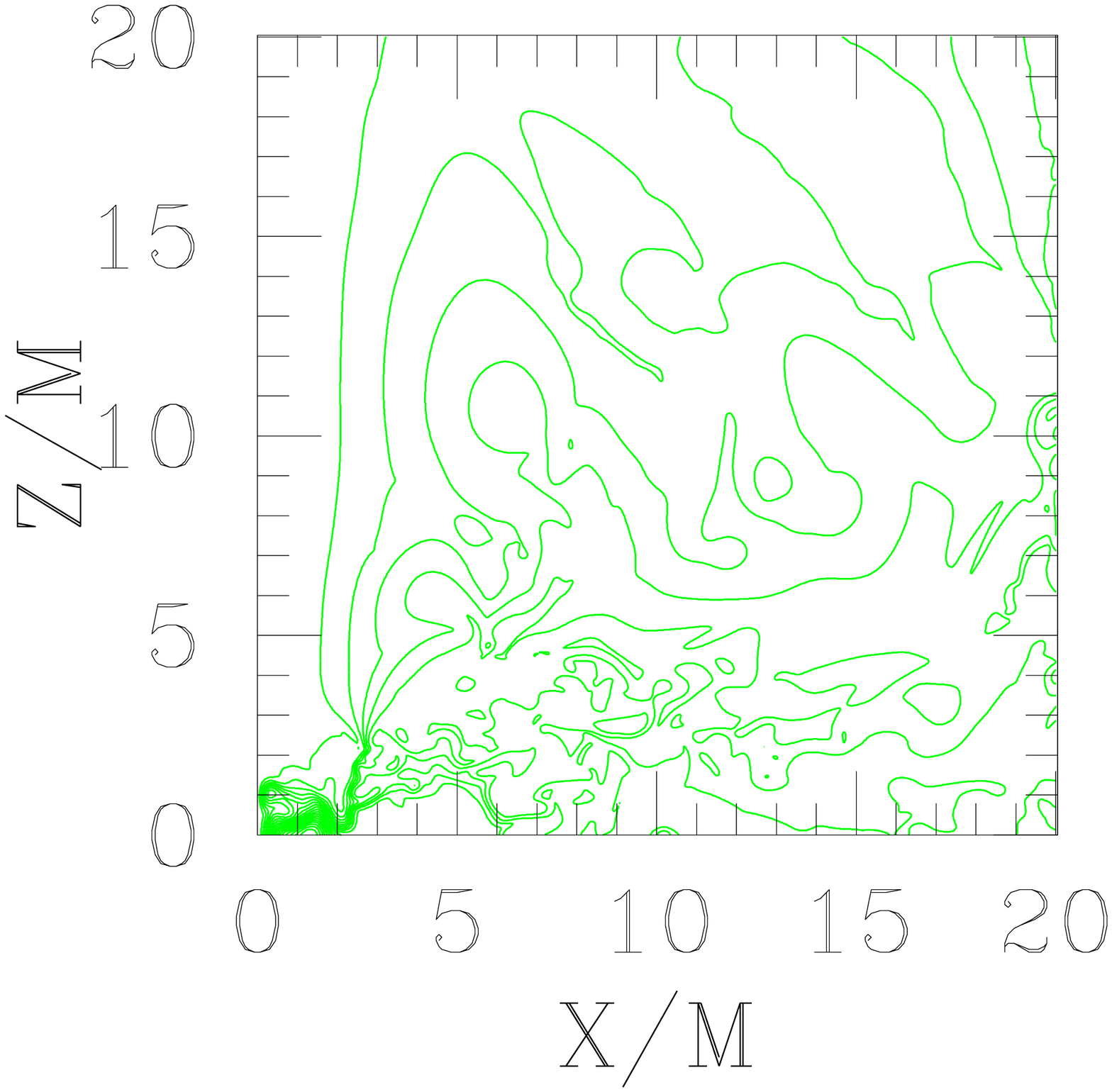}
\epsfxsize=1.8in
\leavevmode
\hspace{-0.5cm}\epsffile{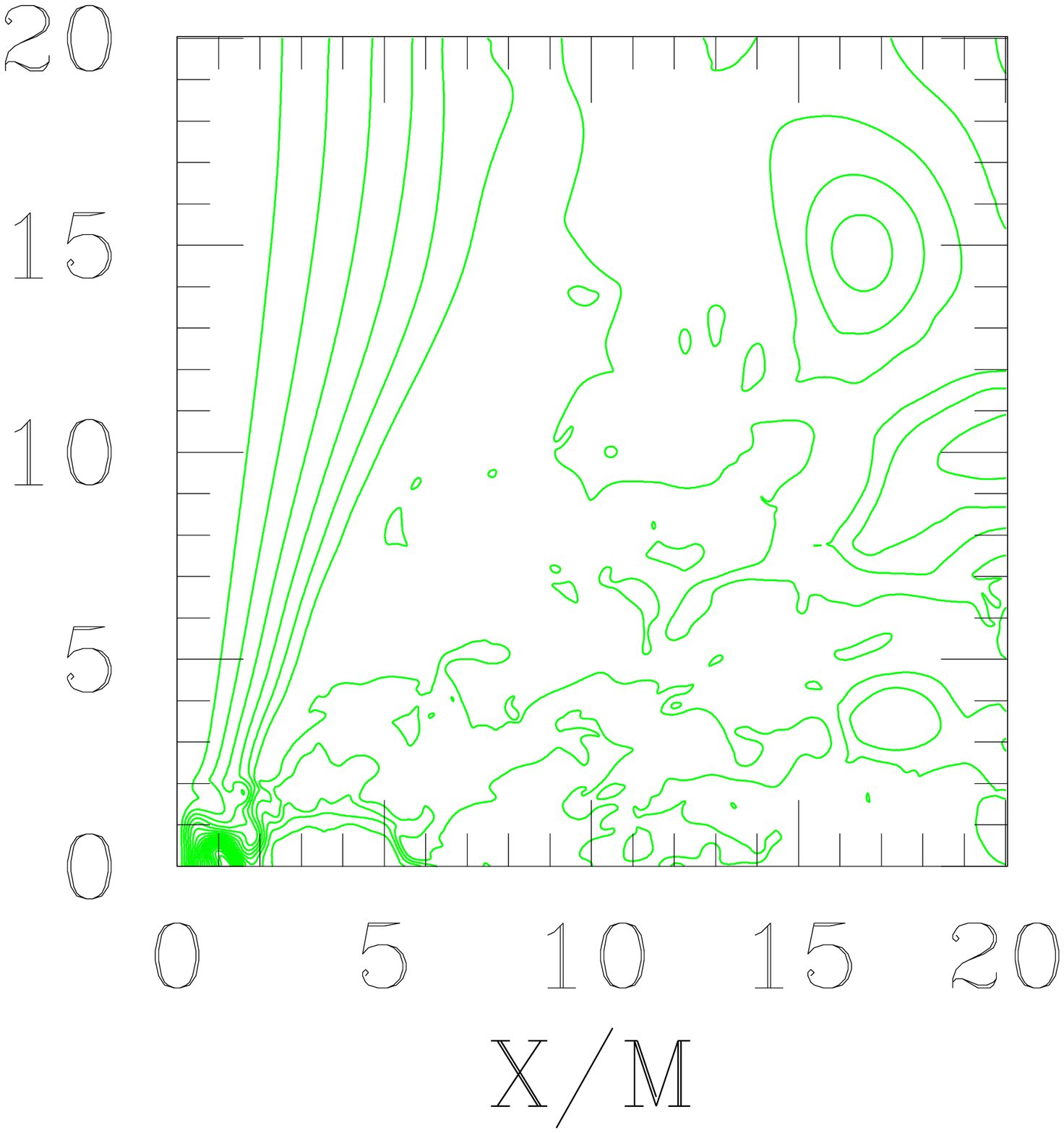}
\epsfxsize=1.8in
\leavevmode
\hspace{-0.5cm}\epsffile{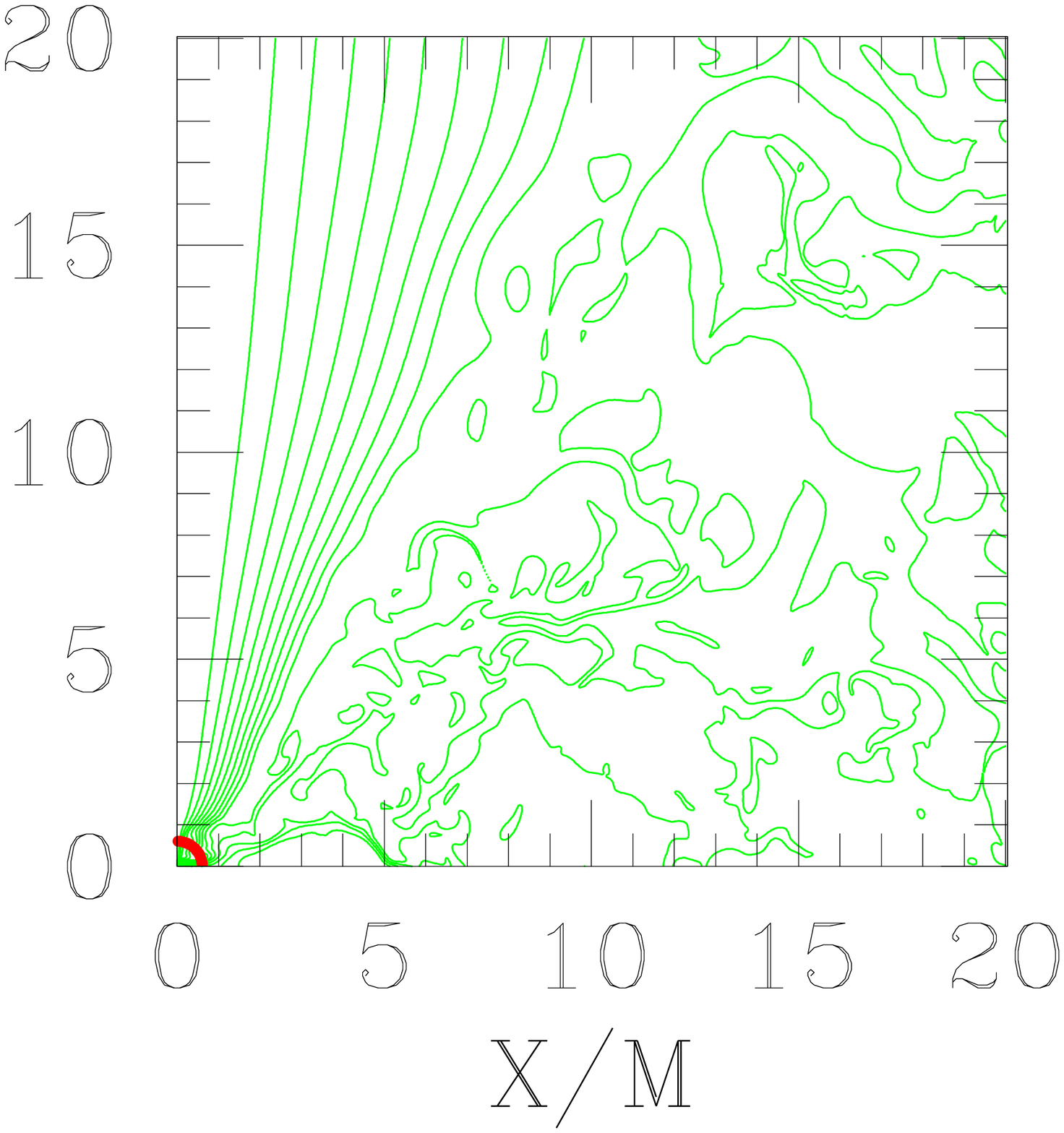}
\epsfxsize=1.8in
\leavevmode
\hspace{-0.5cm}\epsffile{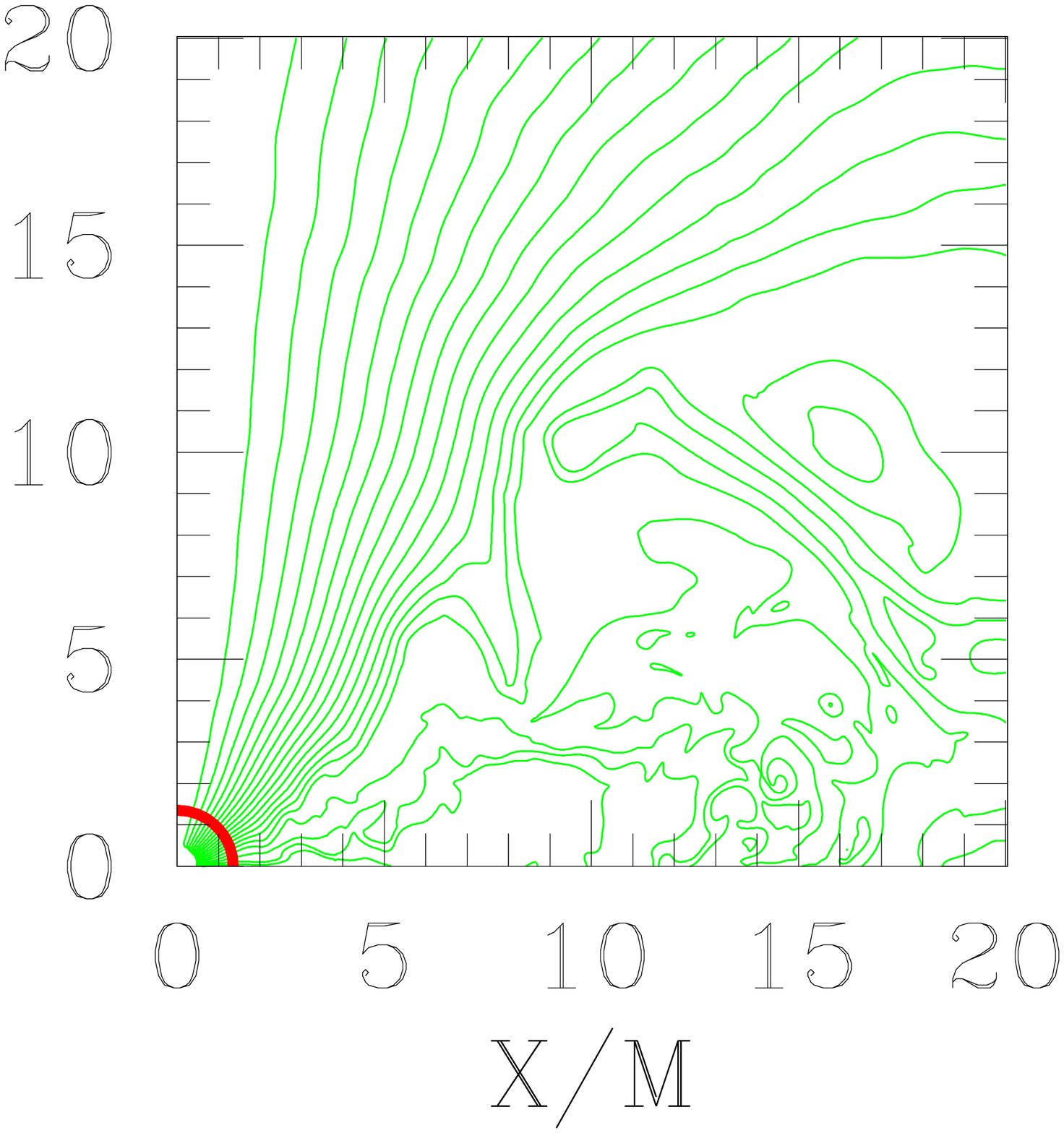}
\caption{Snapshots of  rest-mass density contours and poloidal 
magnetic field lines for star~A at selected times. 
The first and third rows show snapshots of the rest-mass density
contours and velocity vectors on the meridional plane. The second and fourth
rows show the corresponding field lines (lines of constant $A_{\varphi}$)
for the poloidal magnetic field at the same times.
The density contours are drawn for $\rho_0/\rho_{\rm max}(0)=
10^{-0.36 i - 0.09}~(i=0$--10), where $\rho_{\rm max}(0)$ is the 
maximum rest-mass density at $t=0$. 
The field lines are drawn for $A_{\varphi} = A_{\varphi,\rm min}
+ (A_{\varphi,\rm max} - A_{\varphi,\rm min}) i/20~(i=1$--19),
where $A_{\varphi,\rm max}$ and $A_{\varphi,\rm min}$ are the maximum
and minimum values of $A_{\varphi}$, respectively, at the given time.
The thick solid (red) curves denote the apparent horizon. 
In the last panel, the field lines are terminated inside the black hole
at the excision boundary.
\label{fig:StarA_contours}}
\end{center}
\end{figure*}

\begin{figure*}

\begin{center}
\epsfxsize=1.9in
\leavevmode
\epsffile{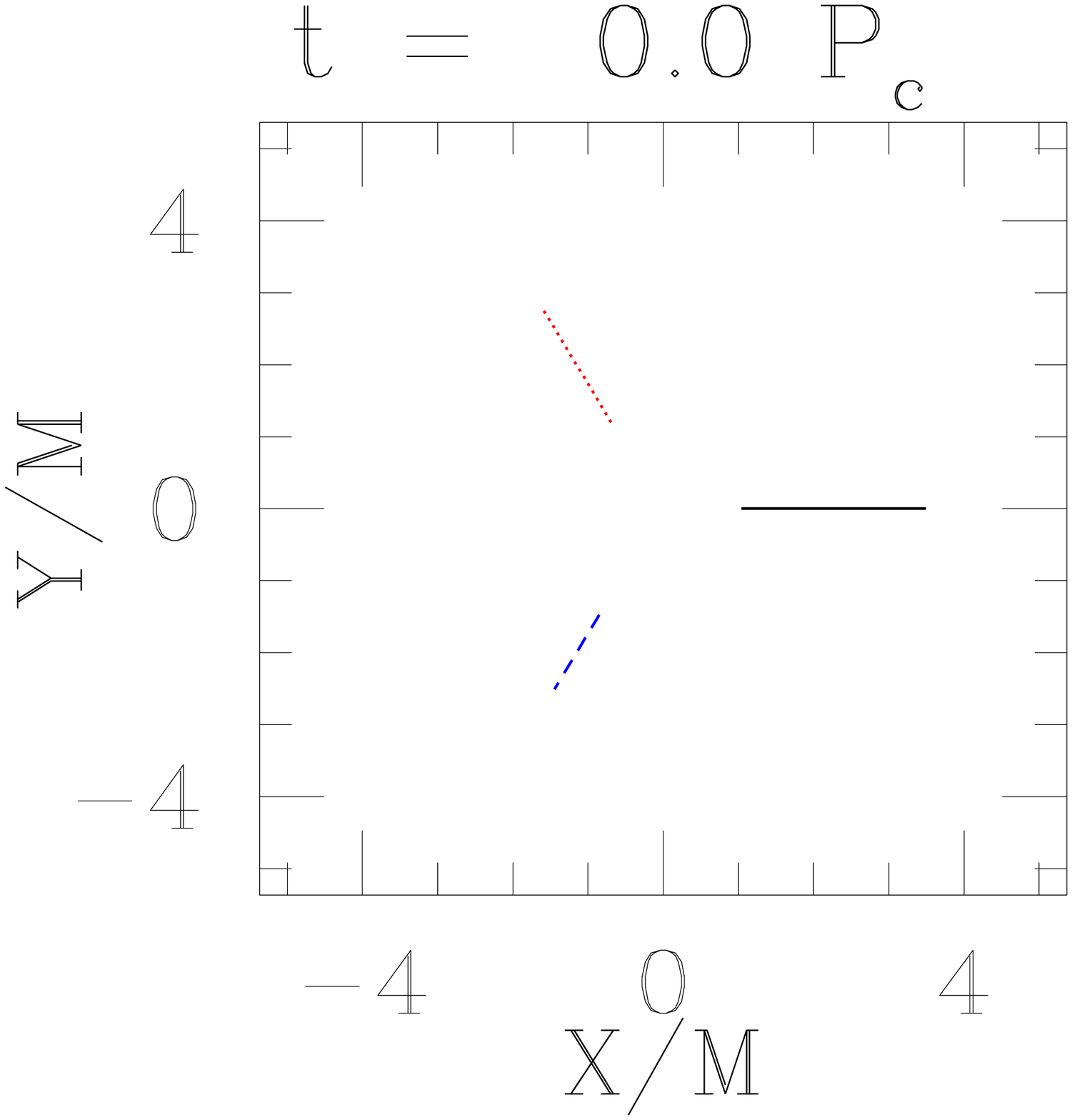}
\epsfxsize=1.9in
\leavevmode
\epsffile{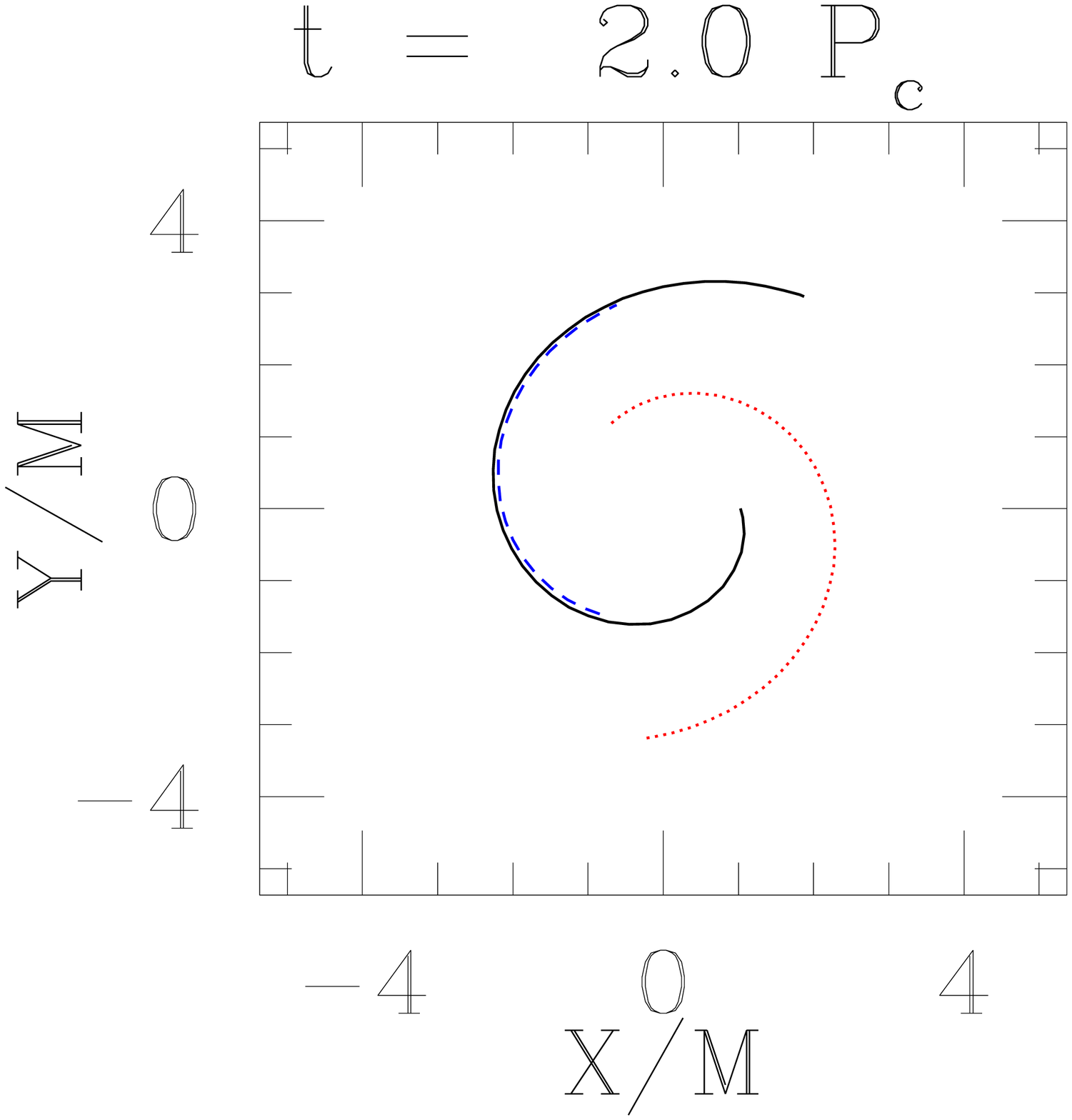}
\epsfxsize=1.9in
\leavevmode
\epsffile{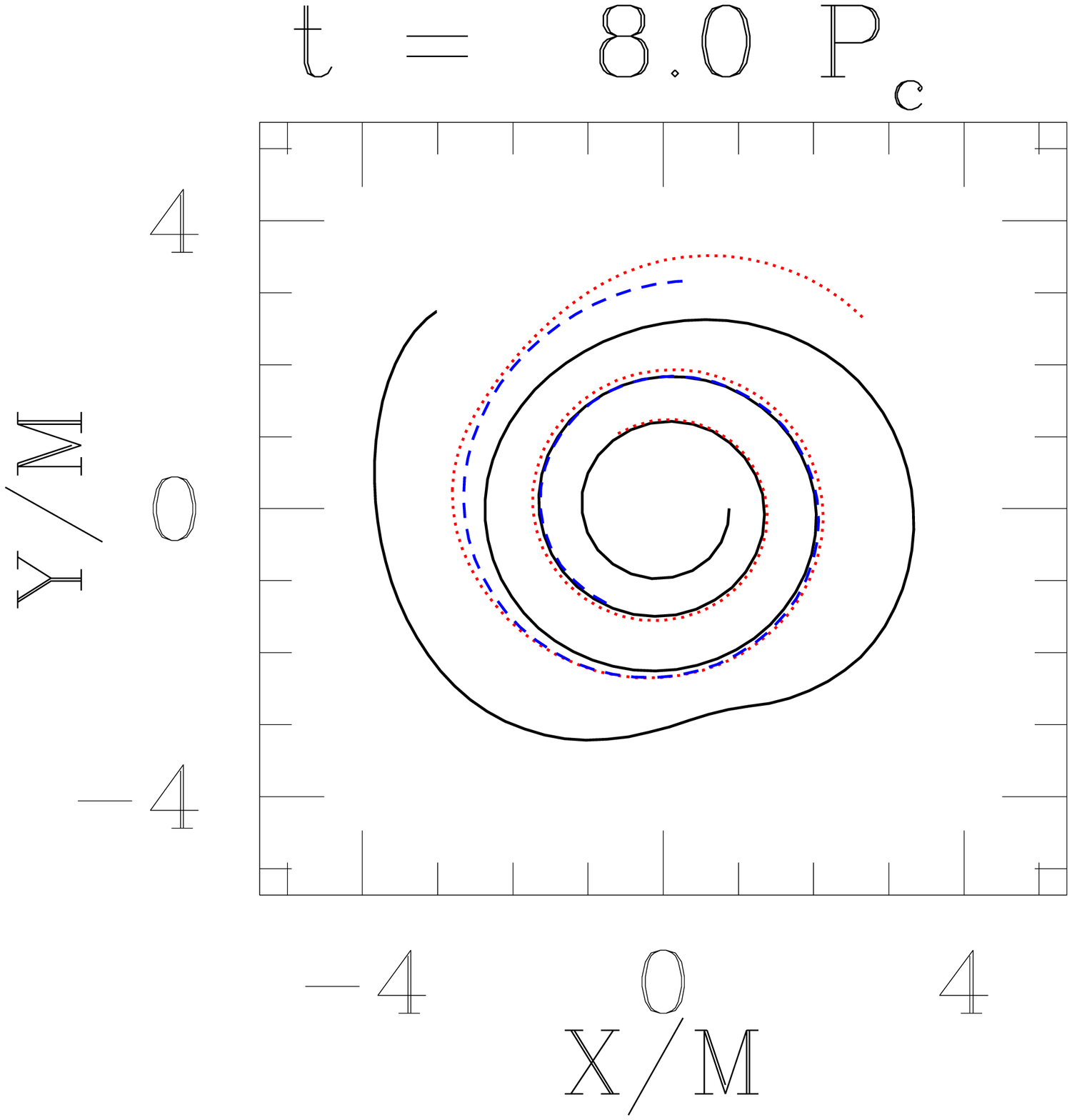}
\end{center}

\begin{center}
\epsfxsize=1.9in
\leavevmode
\epsffile{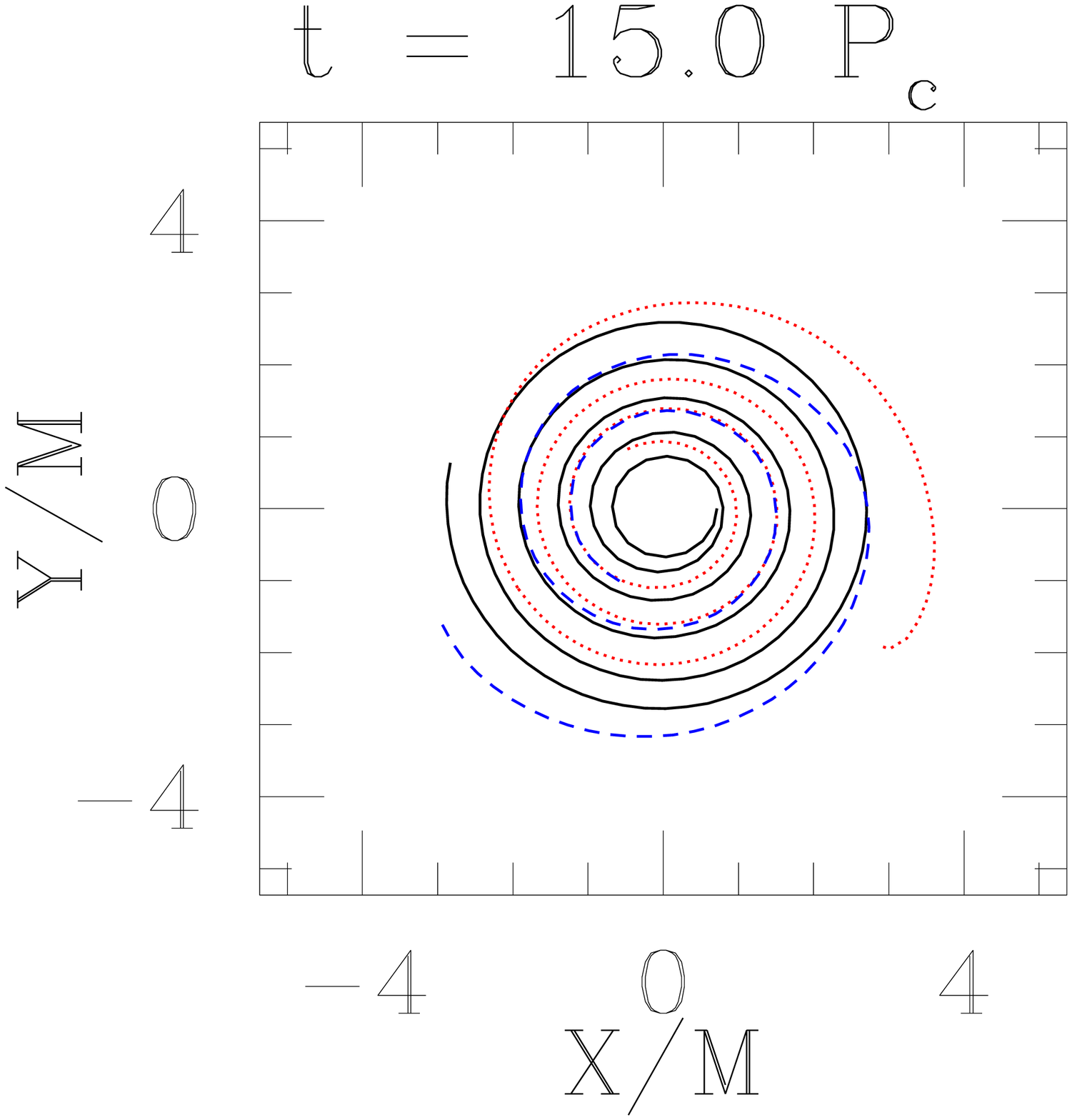}
\epsfxsize=1.9in
\leavevmode
\epsffile{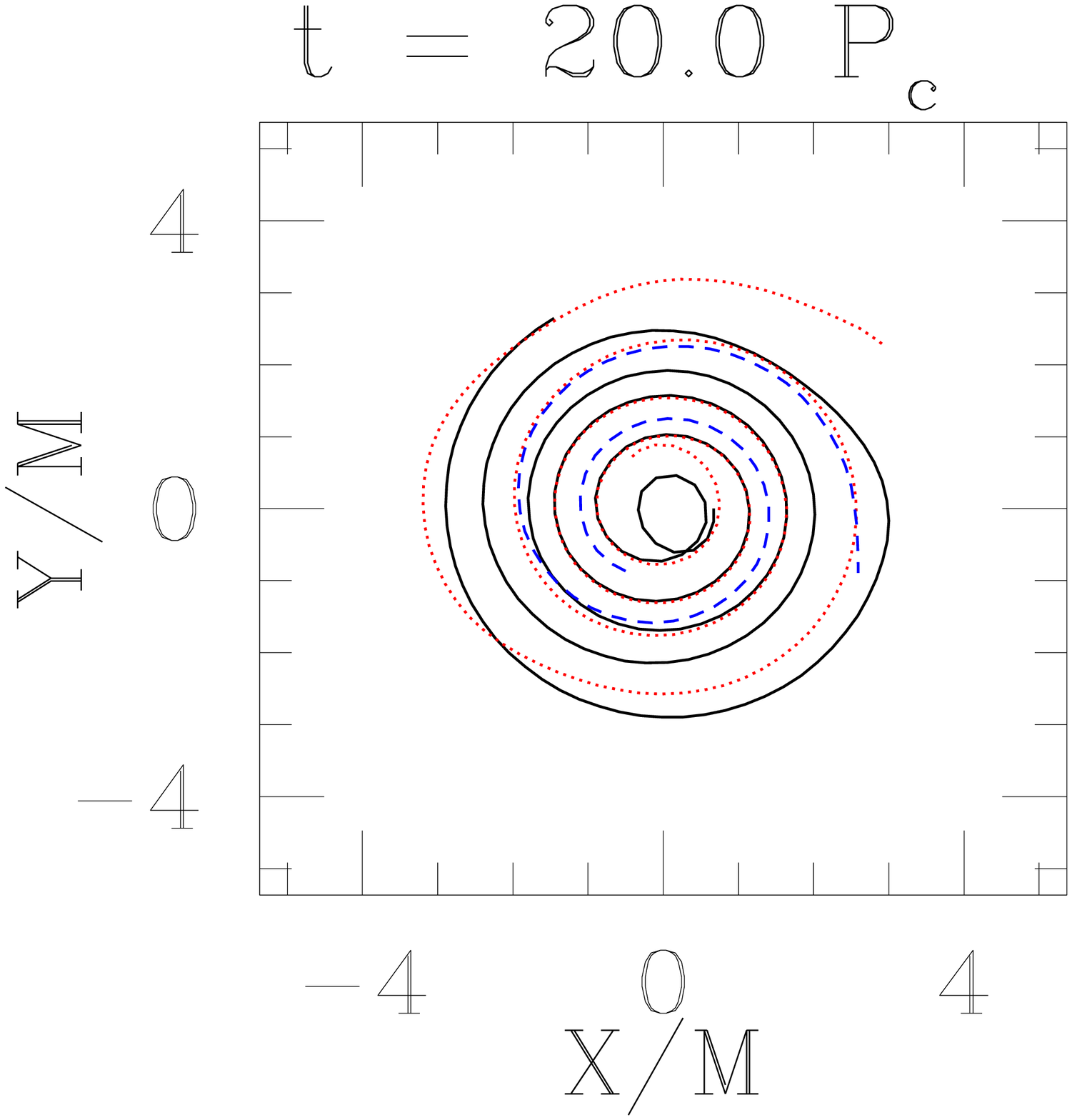}
\epsfxsize=1.9in
\leavevmode
\epsffile{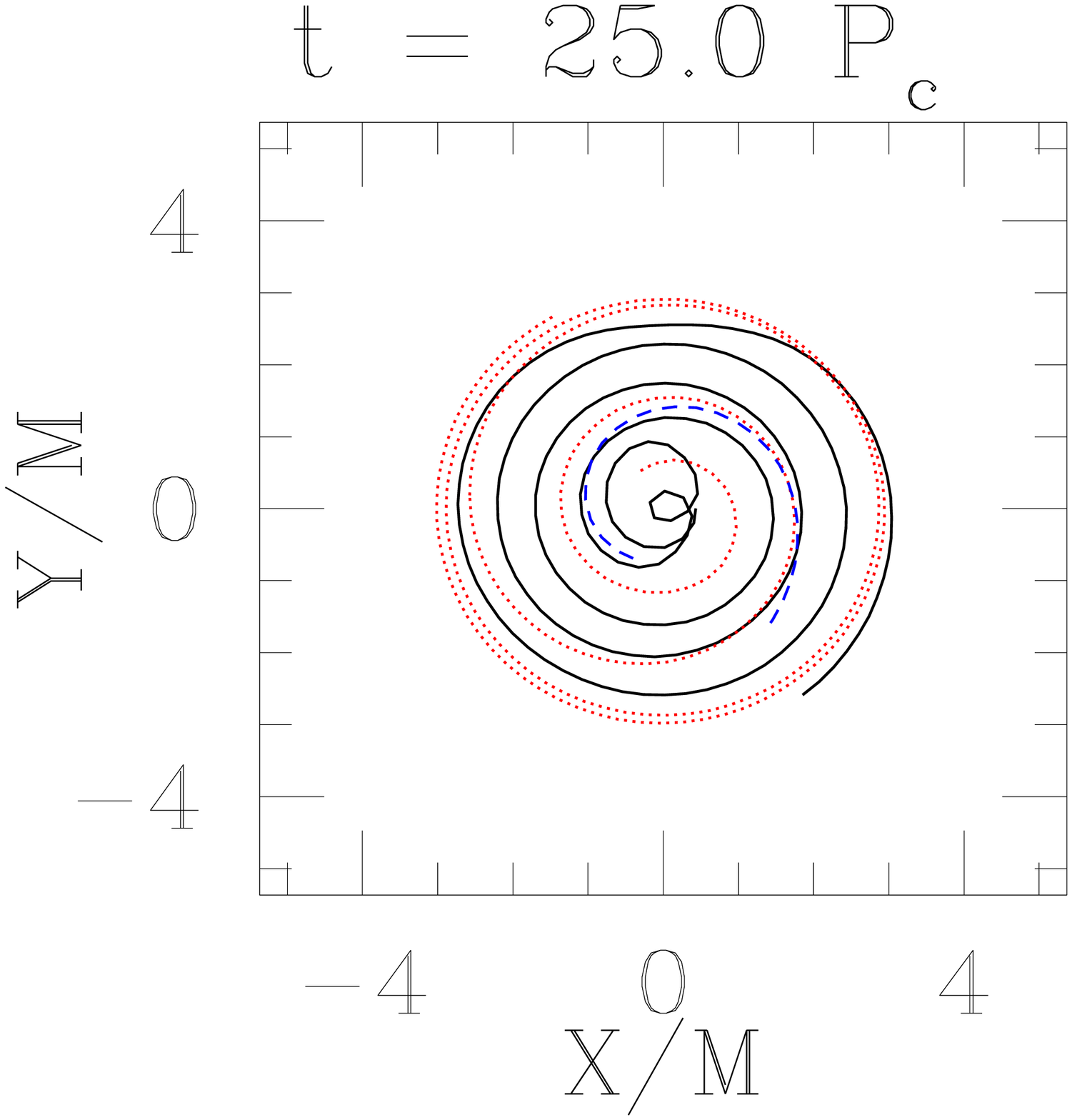}
\end{center}

\begin{center}
\epsfxsize=1.9in
\leavevmode
\epsffile{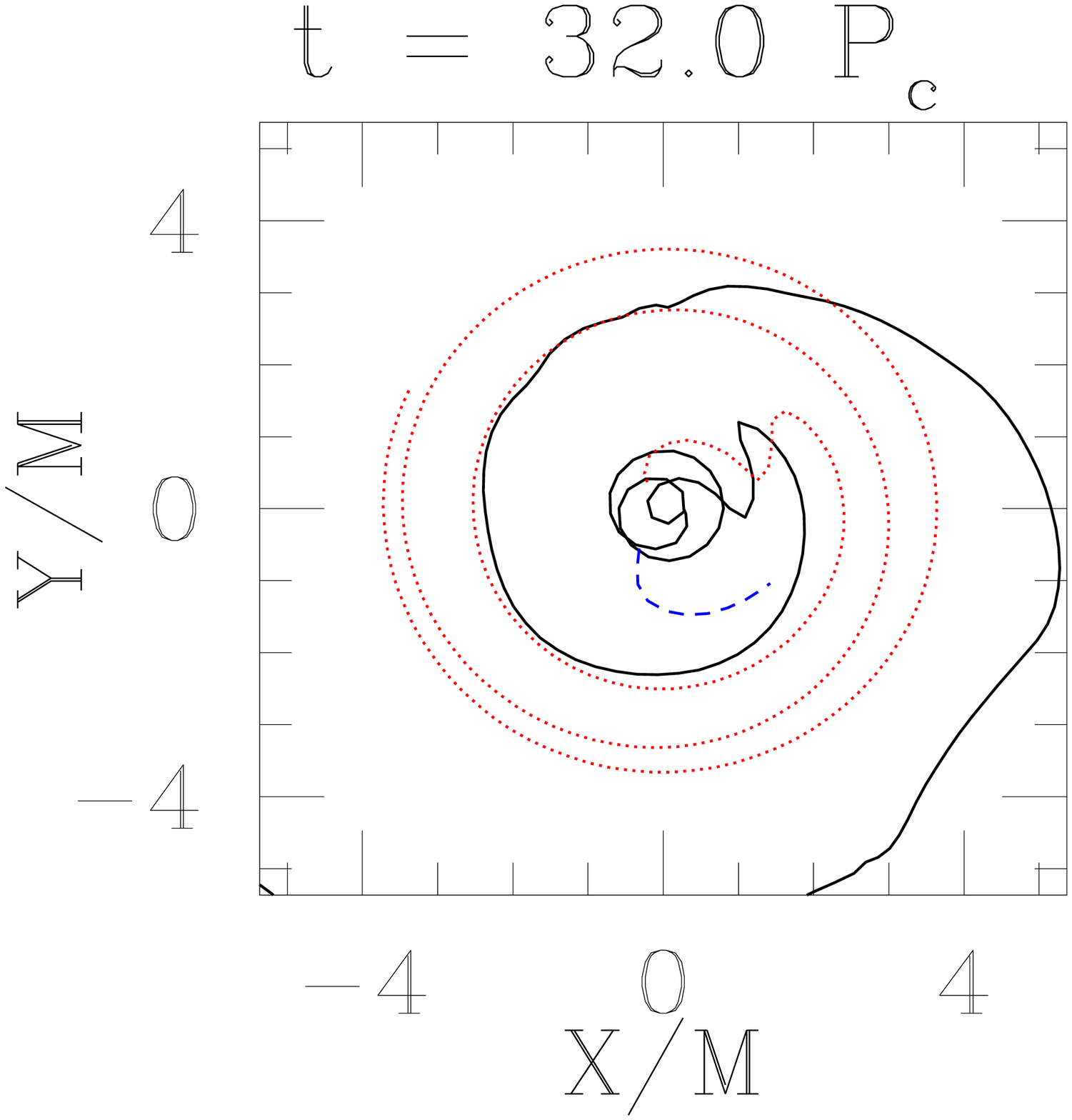}
\epsfxsize=1.9in
\leavevmode
\epsffile{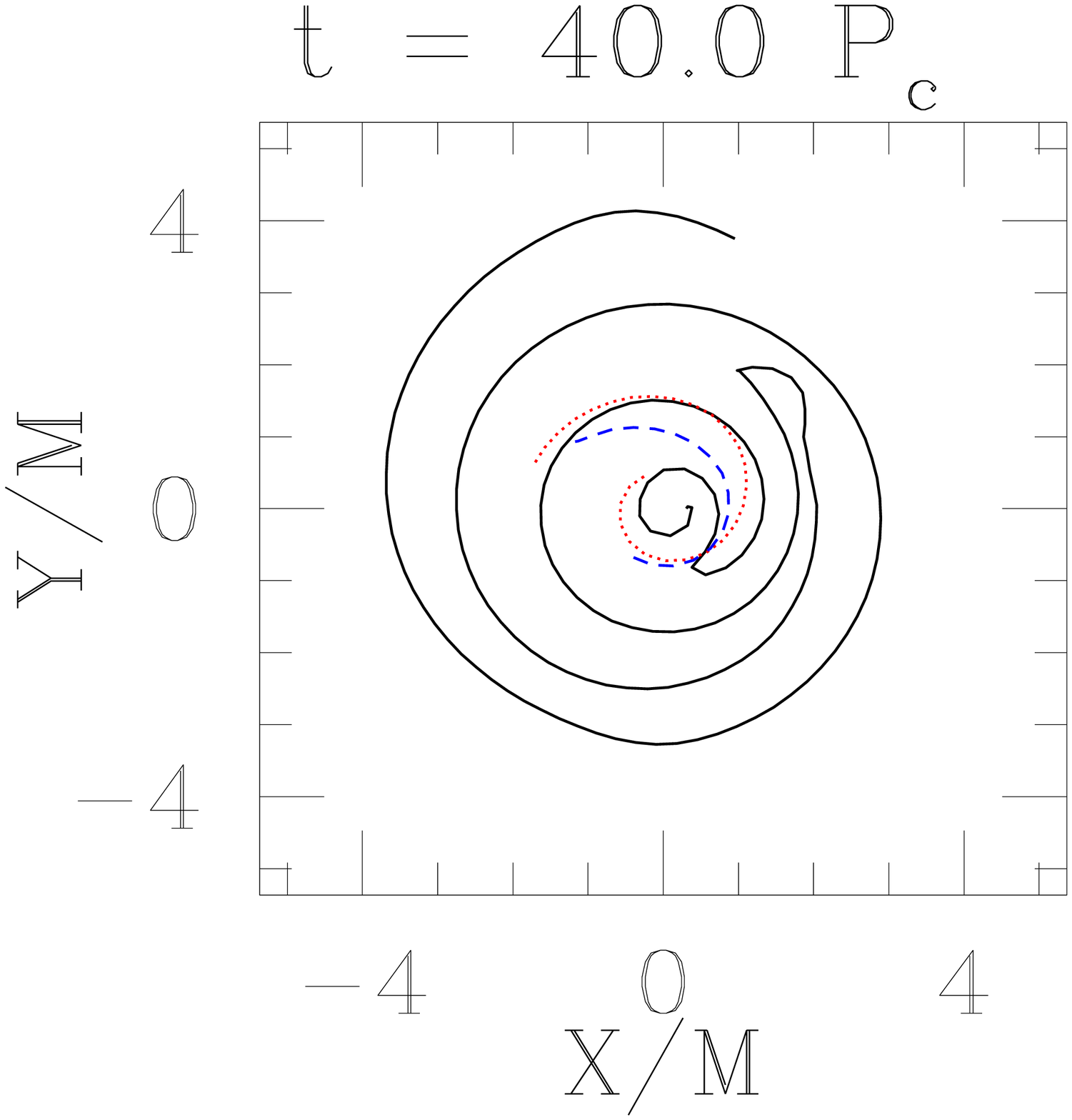}
\epsfxsize=1.9in
\leavevmode
\epsffile{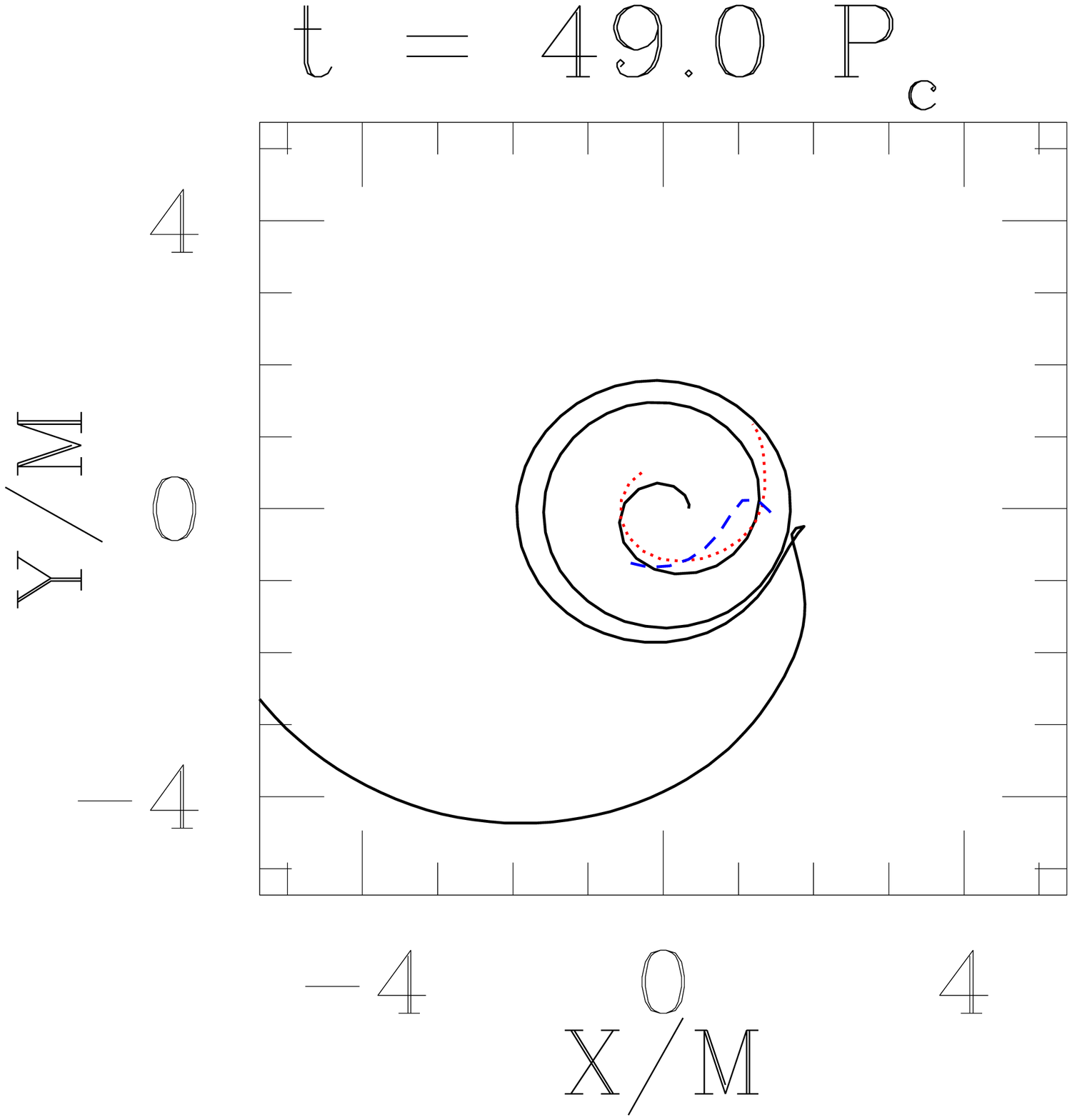}
\end{center}

\begin{center}
\epsfxsize=1.9in
\leavevmode
\epsffile{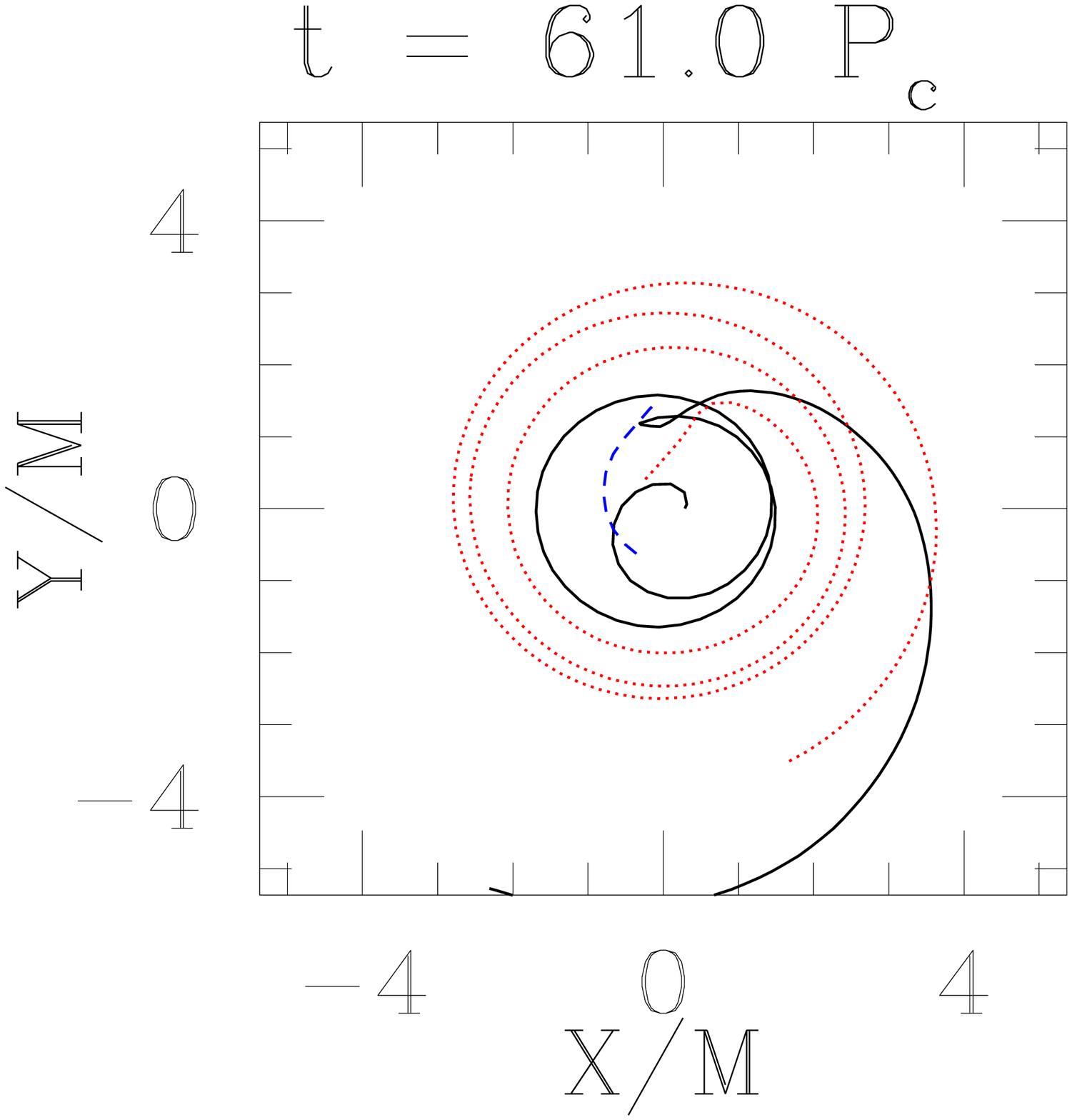}
\epsfxsize=1.9in
\leavevmode
\epsffile{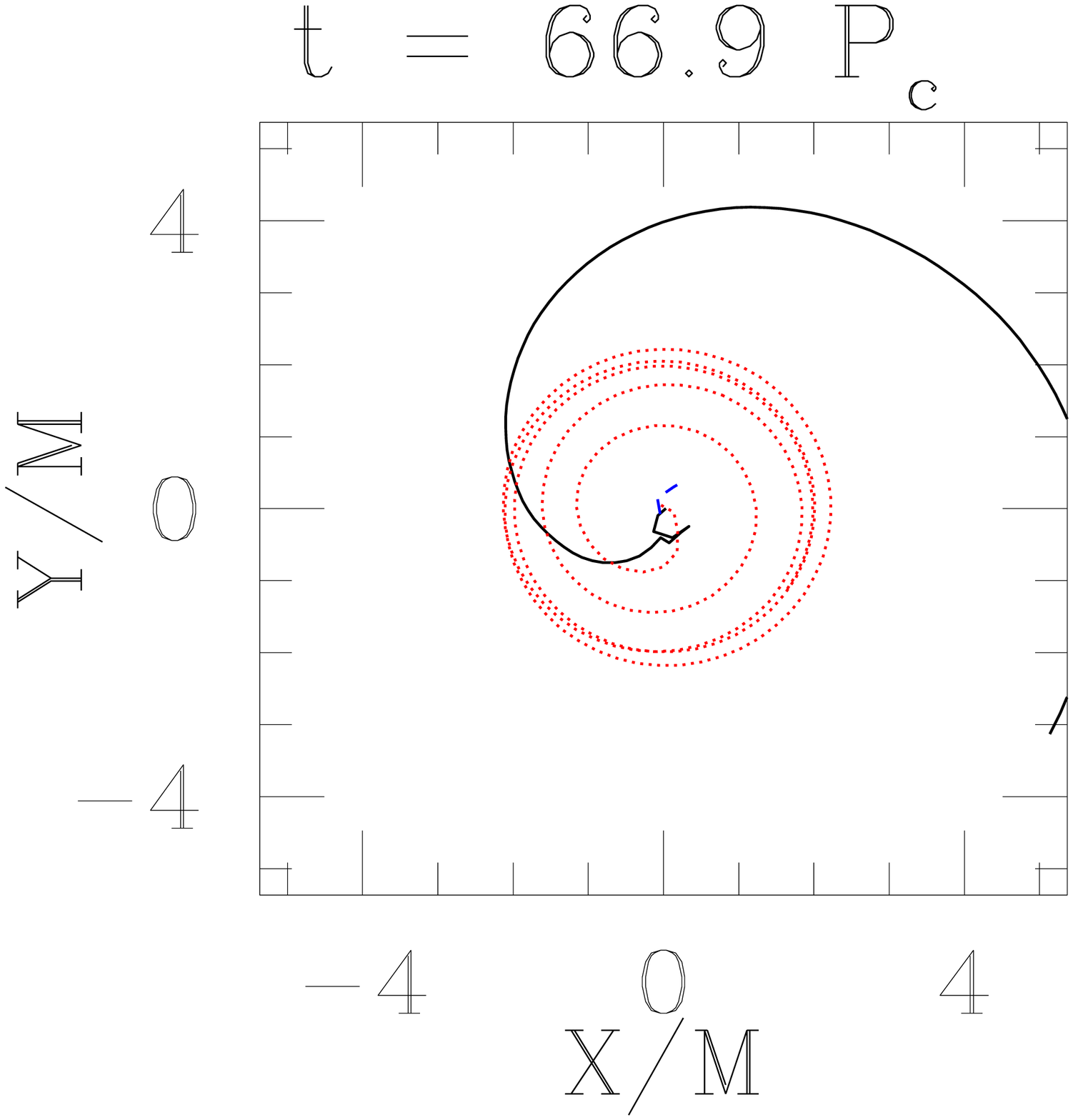}
\epsfxsize=1.9in
\leavevmode
\epsffile{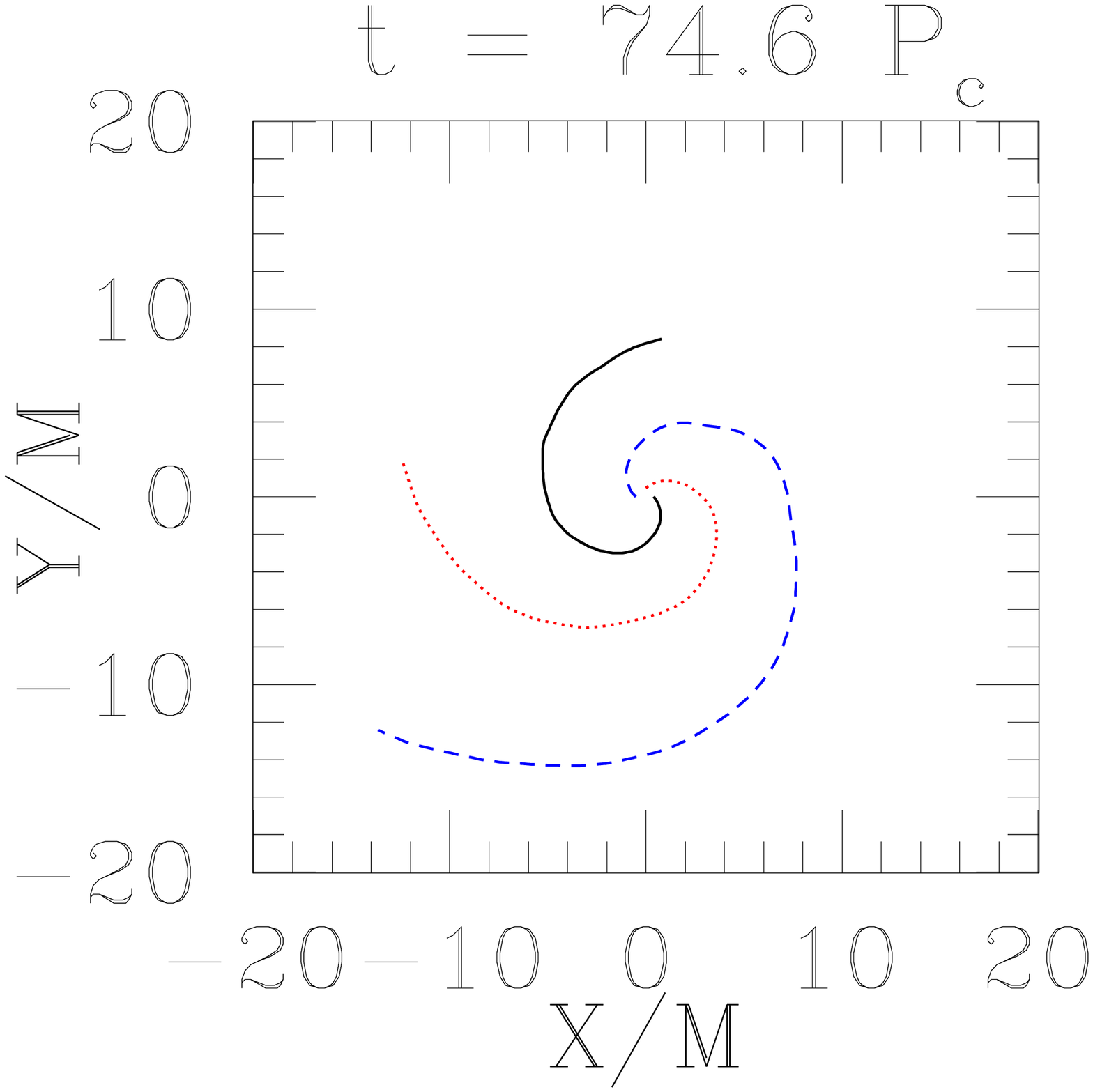}
\end{center}

\caption{Snapshots of the projected 3D magnetic field lines for star~A (see  
Appendix~\ref{app:field_lines} for details) at selected times. Only 
three lines are drawn 
in each panel to prevent overcrowding of field lines.
\label{fig:StarA_Blines}}
\end{figure*}

\begin{figure}
\vspace{-4mm}
\begin{center}
\epsfxsize=3.in
\leavevmode
\epsffile{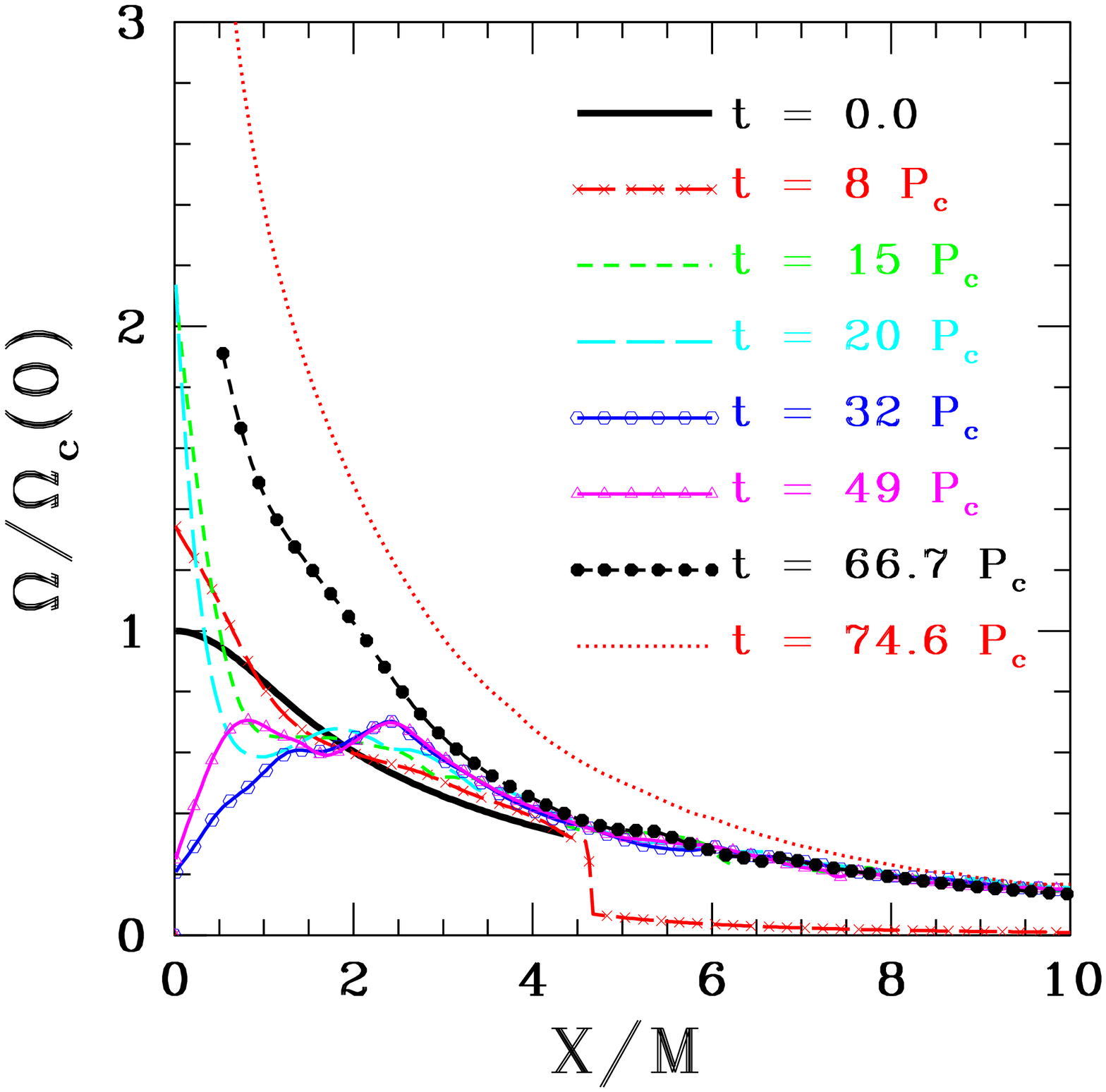}
\caption{Angular velocity profiles for star~A at selected times (corresponding
to the times in Fig.~\ref{fig:StarA_contours}).  The last two profiles
correspond to the moment
of excision and a late time in the excision run.
\label{omprof}}
\end{center}
\end{figure}

Figure~\ref{fig:StarA_contours} shows the snapshots of density contours 
and poloidal magnetic field lines (lines of constant $A_{\varphi}$) in 
the meridional plane. Figure~\ref{fig:StarA_Blines} shows the snapshots 
of three-dimensional (3D) magnetic field lines projected onto the equatorial 
plane. 

In the early phase of the evolution, the frozen-in poloidal magnetic fields 
lines are wound up by the differentially rotating matter, creating 
a toroidal field which grows linearly in 
time (see Fig.~\ref{fig:StarA_Blines} and Fig.~\ref{rescomp}d) 
with the growth rate predicted by Eq.~(\ref{eqn:dtBphi}). 
When the magnetic field becomes sufficiently strong, magnetic stresses 
act back on 
the fluid, causing a redistribution of angular momentum. The core 
of the star contracts while the outer layers expand. At $t \gtrsim 
6P_c$, the effect of the MRI becomes evident, as shown in 
Figs.~\ref{rescomp}c and ~\ref{mrirate}, 
where we see that the maximum value of $|B^x|$ ($\equiv |B^{\varpi}|$) 
suddenly increases, 
growing exponentially for a short period (about one $e$-folding). 
We find that the MRI first occurs in the outer layers of the star near 
the equatorial plane. This is consistent with the linear analysis, as
Eq.~(\ref{tmri}) together with star~A's angular velocity profile gives 
a shorter $t_{\rm MRI}$ near the outer part of the star. The effect of 
the MRI can be seen in Fig.~\ref{fig:StarA_contours}, 
where we see that the poloidal field lines are distorted. The growth 
of the central density slows down once $|B^x|_{\rm max}$ and 
$|B^y|_{\rm max}$ ($\equiv |\varpi B^{\varphi}|_{\rm max} = |B^T|_{\rm max}$) 
saturate at $t \sim 20 P_c$.  This may be 
caused by MRI-induced turbulence redistributing some of the angular 
momentum to slow down the contraction of the core. The amplitude of 
the toroidal field begins to decrease after $t \gtrsim 20P_c \sim t_A$ 
(see Figs.~\ref{fig:StarA_Blines} and~\ref{rescomp}d) and the core 
of the star becomes less differentially rotating (Fig.~\ref{omprof}).
This is consistent with the results of~\cite{Shapiro}, which predict that
the magnetic field growth by magnetic winding should saturate after
an Alfv\'en time, the magnetic energy having grown to an appreciable 
fraction of the initial rotational kinetic energy. 

The combined effects of magnetic braking and MRI eventually trigger 
gravitational collapse to a black hole at $t \approx 66P_c \approx 
36 (M/2.8M_{\odot})~{\rm ms}$ when an apparent horizon forms. A 
collimated magnetic field forms near the polar 
region at this time (see Fig.~\ref{fig:StarA_contours}). However, 
a substantial amount of toroidal field is still present 
(see Fig.~\ref{fig:StarA_Blines}). Without black hole excision, the 
simulation becomes inaccurate soon after the formation of the apparent 
horizon because of grid stretching. To follow the subsequent 
evolution, a simple excision technique is 
employed~\cite{alcubierre,excision}. 
We are able to track the evolution for another $300M \approx 8 P_c$. 
We find that not all the matter promptly falls into the black hole. 
The system settles down to a quasiequilibrium state consisting of a 
black hole surrounded by a hot torus and a collimated magnetic 
field near the polar region (see the panels corresponding to time 
$t=74.6P_c$ in Fig.~\ref{fig:StarA_contours}). The irreducible 
mass of the black hole is about $0.9M$ and the rest-mass of the 
torus is about $0.1M$ (Fig.~\ref{excision}). We estimate that 
$J/M^2 \sim 0.8$ for the final black hole.  This system is a 
promising central engine for the short-hard gamma-ray bursts 
(see Sec.~\ref{StarC} and~\cite{GRB2}). 

\subsubsection{Resolution study}

\begin{figure}
\epsfxsize=3.in
\epsffile{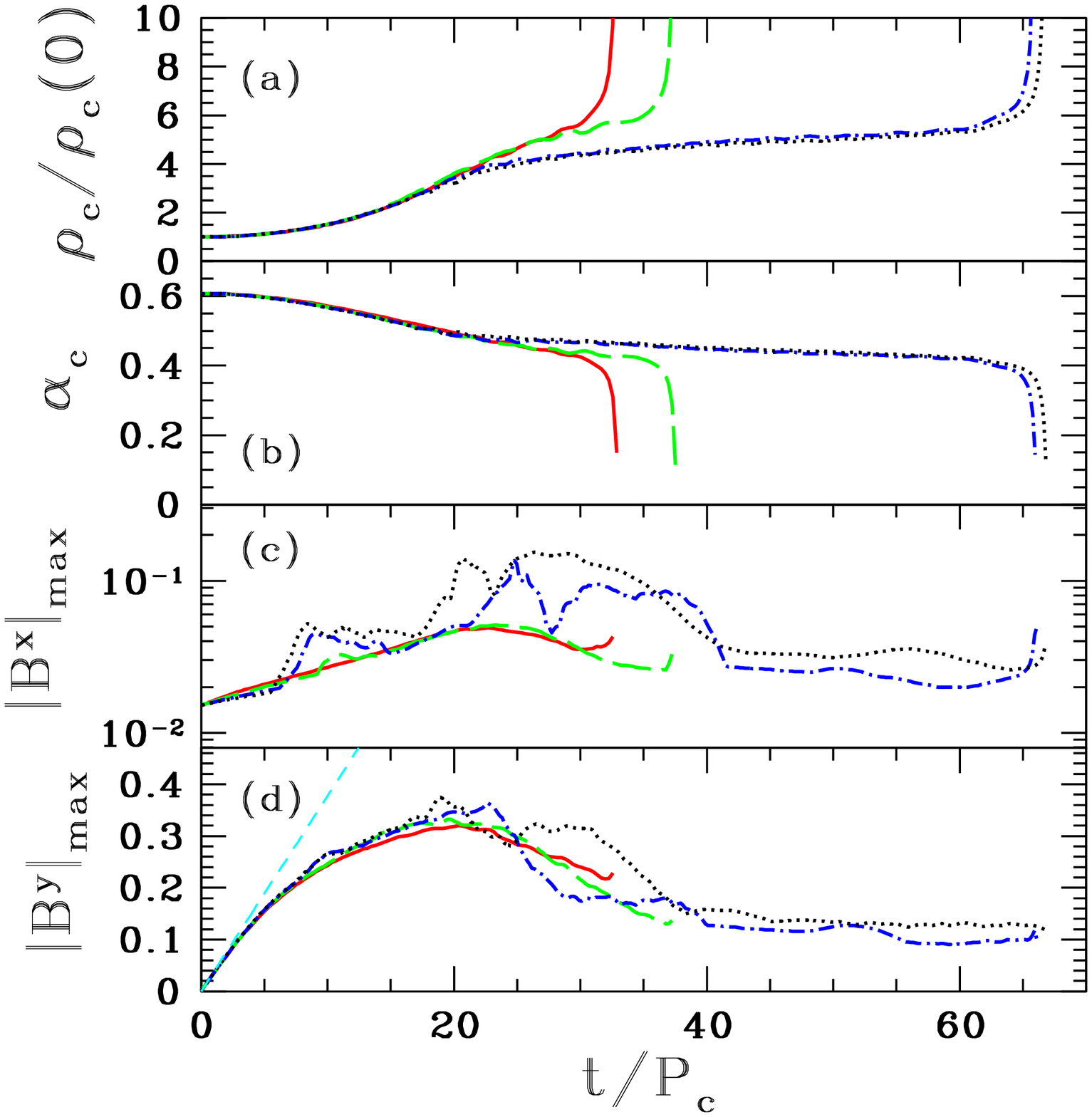}
\caption{Evolution of the central rest-mass density $\rho_c$, central lapse
$\alpha_c$, and
maximum values of $|B^x|$ and $|B^y|$.
$|B^x|_{\rm max}$ and $|B^y|_{\rm max}$ are plotted in units of 
$\sqrt{\rho_{\max}(0)}$. 
The solid (red), long-dashed (green), dot-dash (blue), 
and dotted (black) curves denote
the results with $N$=250, 300, 400, and 500 respectively.
The dashed (cyan) line in (d) represents the predicted
linear growth of $|B^y|_{\rm max}$ at early times from Eq.~(\ref{eqn:dtBphi}).
\label{rescomp}}
\end{figure}

\begin{figure}
\epsfxsize=3.in
\epsffile{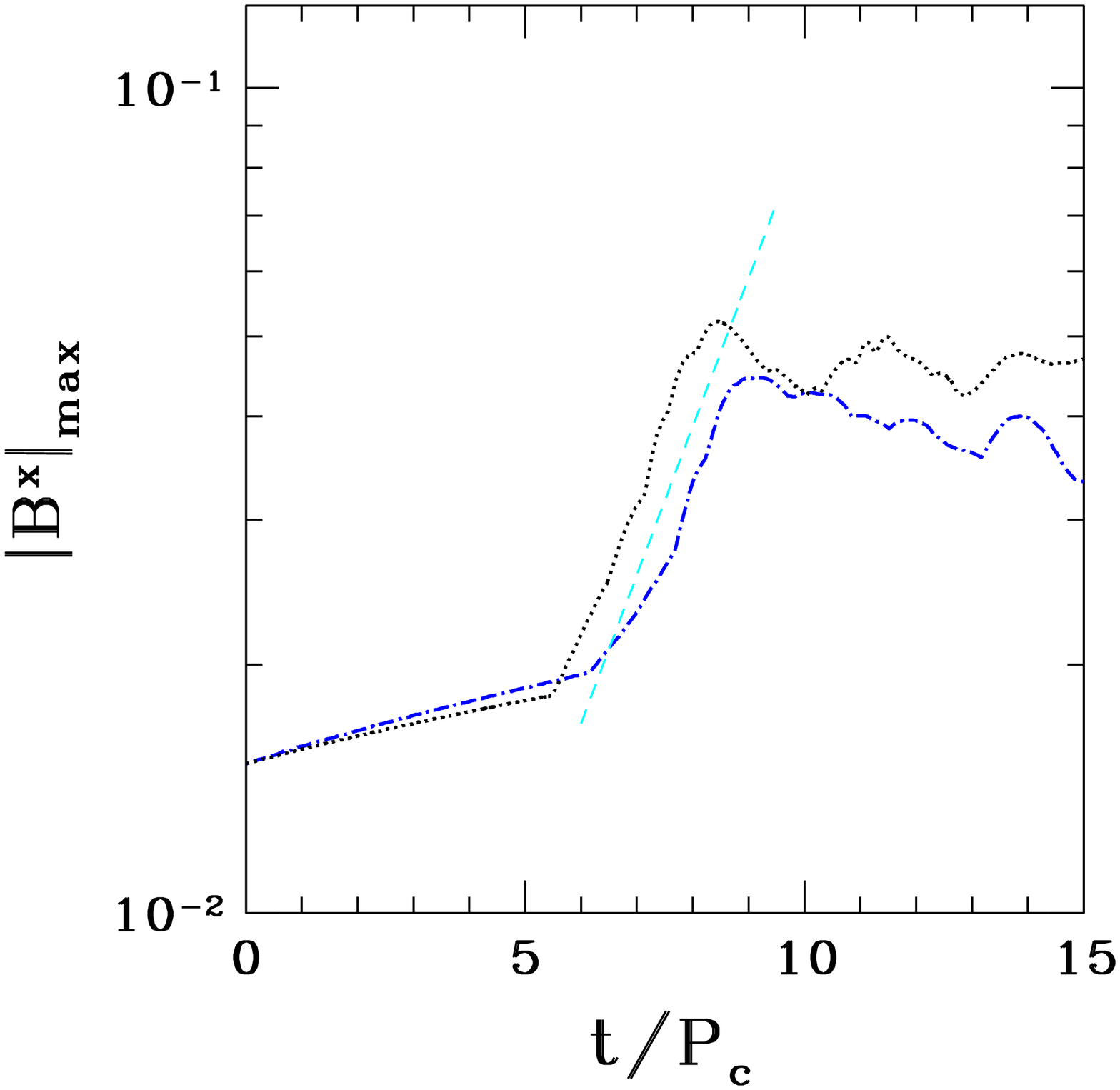}
\caption{Evolution of $|B^x|_{\rm max}$ plotted in the same units 
as in Fig.~\ref{rescomp} for the two highest resolution runs of
star~A.  The dot-dashed (blue), and dotted (black) curves 
denote the results with $N$=400 and 500, respectively.  The dashed (cyan) 
line represents an approximate slope $\omega=0.18/P_c$ for the 
exponential growth rate of the MRI, $\delta B^x \propto e^{\omega t}$. 
\label{mrirate}}
\end{figure}

Four simulations were performed with different resolutions 
(see Fig.~\ref{rescomp}): $N=$250, 300, 400 and 500. 
We find that the results converge approximately when $N\gtrsim 400$. 
On the other hand, results are far from convergent for $N \lesssim 300$. 
For example, $|B^x|_{\rm max}$ is much smaller at lower  
resolutions than for runs with 
higher resolutions, and the growth rate of $|B^x|_{\rm max}$ is
underestimated. Hence, the effect of MRI, which is responsible for the 
growth of $|B^x|_{\rm max}$, is not computed accurately for low 
resolutions. This is 
because the wavelength of the fastest growing MRI mode is not well-resolved 
for low resolutions. We find that we need a resolution 
$\Delta/\lambda_{\rm max} \lesssim 0.14$ ($N \gtrsim 400$) in order to 
resolve the MRI modes.  The straight dashed line in Fig.~\ref{rescomp}d 
corresponds to the linear growth rate predicted by Eq.~(\ref{eqn:dtBphi}).  
This slope agrees with the actual growth of $|B^y|_{\rm max}$ in the early 
(magnetic winding) phase of the simulation, but as back-reaction 
(magnetic braking) becomes important, the toroidal field begins to saturate.

Figure~\ref{mrirate} shows the onset of the MRI in more detail for 
the two highest resolutions.  Also shown is an approximate fit to 
the growth rate (the short dashed line).  This line shows that the perturbation 
grows approximately as $\delta B^x \propto e^{\omega t}$, where
$\omega \approx 0.18/P_c$.  This is a somewhat lower rate than that predicted
from linear theory, which gives $\omega_{\rm max} \sim 1/P_c$, where
$\omega_{\rm max}$ corresponds to the fastest growing MRI mode.  This 
discrepancy could be due to the fact that the linear analysis is inaccurate by 
a significant factor.  One drawback of the linear analysis is the 
assumption of Newtonian gravity, but star~A is highly relativistic. 
In addition, the linear analysis treats the MRI as a purely local 
phenomenon, assuming a uniform background state over length scales much longer 
than the wavelengths of the perturbations.  However, since the expected 
$\lambda_{\rm max}$ is only one order of magnitude smaller than the initial 
equatorial radius, these assumptions may lead to significant discrepancies 
between the predicted and actual properties of the MRI.

\subsubsection{Evolution of the energies vs.\ time}

\begin{figure}
\vspace{-4mm}
\begin{center}
\epsfxsize=3.in
\leavevmode
\epsffile{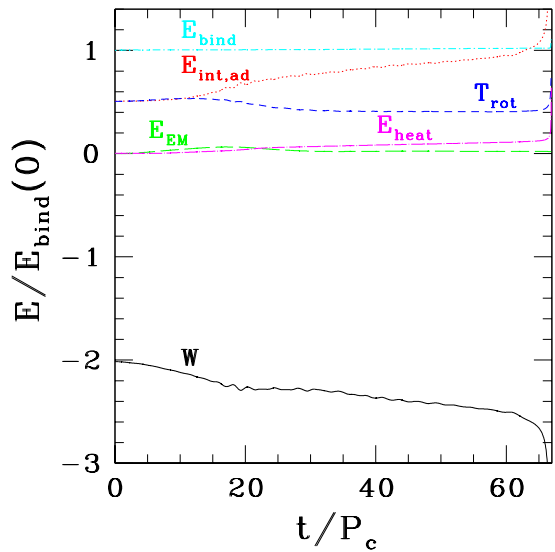}
\caption{Components of the energy vs.\ time for star~A.  
All energies are normalized to the binding energy at
$t=0$, where the binding energy is defined as $E_{\rm bind} \equiv M_0 - M$.
In the evolution, $E_{\rm bind}$ should be nearly conserved.
\label{energies}}
\end{center}
\end{figure}

Figure~\ref{energies} shows the evolution of various energies defined 
in Sec.~\ref{sec:diagnostics}. We see that the magnetic energy $E_{\rm EM}$ 
remains small throughout the entire evolution, even though the magnetic field 
drives the secular evolution. The gravitational potential energy $W$ and 
the adiabatic 
part of the internal energy $E_{\rm int,ad}$ change the most, which results
from the drastic change in the configuration of the star. 
The rotational kinetic energy decreases substantially before the core
collapses, presumably because the bulk of the mass of the star rotates slower
than at $t=0$. A substantial amount of heat ($E_{\rm heat}$) 
is also generated by shocks.

\subsubsection{Evolution with excision}
\label{sec:SA-excision}

\begin{figure}
\vspace{-4mm}
\begin{center}
\epsfxsize=3.in
\leavevmode
\epsffile{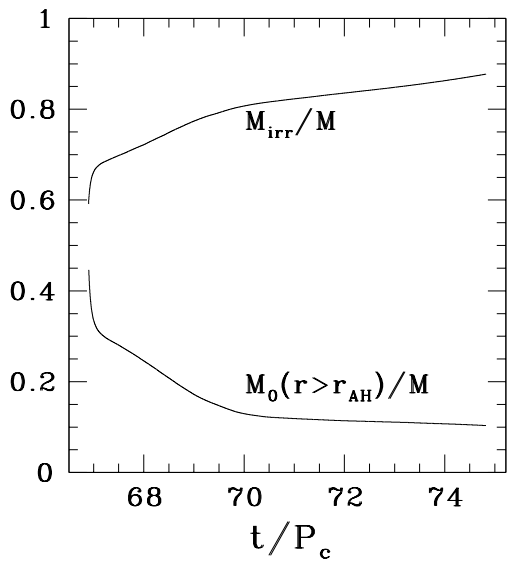}
\caption{Evolution of the irreducible mass 
and the total rest-mass outside the apparent horizon.  (Here,
$r_{\rm AH}$ is the local coordinate radius of the apparent horizon.)
\label{excision}}
\end{center}
\end{figure}

Soon after the formation of the apparent horizon, the simulation
becomes inaccurate due to grid stretching and an excision technique is 
required to follow the subsequent evolution.  
During the excision
evolution, we track the irreducible mass of the black hole by computing
the area of the apparent horizon $\mathcal{A}_{\rm AH}$ and using  
$M_{\rm irr} \approx \sqrt{\mathcal{A}_{\rm AH}/16\pi}$.  The irreducible 
mass and the total rest mass outside the apparent horizon are 
shown in Fig.~\ref{excision}.  The total ADM mass of the final state 
system, consisting of a BH surrounded by a massive accretion torus, is 
well defined.  In contrast, there is no rigorous definition for the mass 
of the black hole itself.  To obtain a rough estimate, we proceed as 
follows.  First, the angular momentum of the black hole is computed from
\beq
J_{\rm hole}=J-J_{\rm matter}(r>r_{AH})
\eeq
where the angular momentum of the matter outside the horizon is given by
\beq
J_{\rm matter}(r>r_{AH}) = \int_{V,r>r_{AH}} \tilde{S}_{\varphi} d^3x\ , 
\eeq
as in Eq.~(\ref{Jaxi}).  Then to estimate the black hole mass, we use
\beq
M_{\rm hole}\approx \sqrt{M_{\rm irr}^2 + (J_{\rm hole}/2M_{\rm irr})^2} \ , 
\eeq
which is an approximate relation for the spacetime of our numerical
simulation, but would be exact for a Kerr spacetime. We thus find 
$M_{\rm hole } \sim 0.9 M$, where $M$ is the total ADM mass of the 
system, and $J_{\rm hole}/M_{\rm hole}^2 \sim 0.8$.

The black hole grows at an initially rapid rate following its formation. 
However, the accretion rate $\dot M_0$ gradually 
decreases and the black hole settles down to a quasi-equilibrium state. 
By the end of the simulation, $\dot M_0$ has decreased
to a steady rate of $\approx 0.01 M_0/P_c$, giving an accretion timescale of 
$\sim 10$--$20P_c \approx 5$--$10~{\rm ms} (M/2.8M_{\odot})$.  
Also, we find that the specific internal thermal energy in the torus 
near the surface is substantial because of shock heating.  The possibility
that this sort of system could give rise to a GRB is discussed 
in Section~\ref{StarC} and~\cite{GRB2}.

\subsubsection{Constraint violations}

\begin{figure}
\epsfxsize=3.in
\epsffile{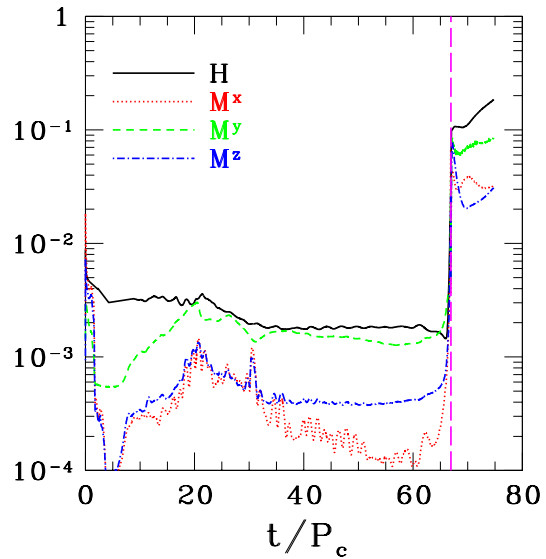}
\caption{L2 norms of the errors in the Hamiltonian ($\mathcal{H}$) and
momentum constraints ($\mathcal{M}^i$) for star~A.  The long-dashed,
vertical (magenta) line represents the initial time for the excision run.
\label{constraints}}
\end{figure}

We monitor the violation of Hamiltonian and momentum constraints
during the evolution. Figure~\ref{constraints} shows the L2 norm
of the constraints. We see that in the pre-excision phase, the violation
of all constraints are a few~$\times 10^{-3}$. Prior to excision,
the constraints are satisfied to better than 1\%. This
indicates that our numerical evolution data accurately satisfy the constraint
equations. After excision, the constraint errors jump to $\sim 10$\%, but they 
remain constant for $\gtrsim 300M \approx 8P_c$. We thus can track 
the evolution reliably for $\gtrsim 2800M$ in total, which is a 
nontrivial feat for highly relativistic, nonvacuum, and dynamical 
spacetime simulations.

\begin{figure}[thb]
\begin{center}
\epsfxsize=3.in
\leavevmode
\epsffile{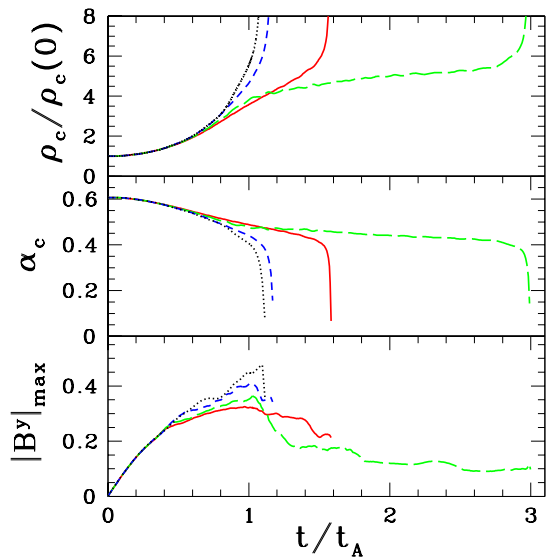}
\caption{Selected parameters plotted against scaled time ($t/t_{\rm A}$)
for evolutions of star~A with four different magnetic field strengths:
$C = 1.25 \times 10^{-3}$ (solid red lines), $C = 2.5 \times 10^{-3}$ 
(green long-dashed lines), $C = 5.0 \times 10^{-3}$ (blue short-dashed
lines), and $C = 10^{-2}$ (black dotted lines).  All runs were performed
with the same resolution ($400^2$ zones with outer boundaries at $20 M$).
When plotted against scaled time, the curves line up at early times 
($t \alt 0.5 t_{\rm A} = 11 P_c$) when the evolution is dominated by 
magnetic winding.
\label{scaling}}
\end{center}
\end{figure}

\begin{figure}[thb]
\begin{center}
\epsfxsize=3.in
\leavevmode
\epsffile{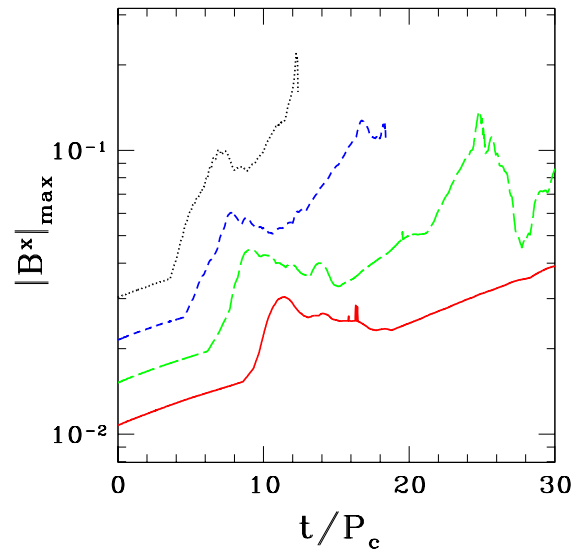}
\caption{Maximum value of $|B^x|$ plotted vs.\ $t/P_c$
for evolutions of star~A with four different magnetic field strengths.  
The line styles correspond to the same values of $C$ as in 
Fig.~\ref{scaling}.  The behavior
of $|B^x|_{\rm max}$ is dominated by the effects of the MRI and thus does 
not scale with the Alfv\'en time. The curves corresponding to the two highest 
values of $C$ (dotted and dashed) terminate at the time when the star 
collapses.
\label{bxmax}}
\end{center}
\end{figure}

\subsection{Star~A, comparison of different values of $C$}
\label{scaling_sec}
In order to test the scaling of our results for different values of the
initial magnetic field strength, we have examined three other values of $C$ in
addition to the value of $2.5\times 10^{-3}$ chosen for the results of 
Section~\ref{starA}.  
Namely, we consider $C = \left\{1.25, 2.5, 5.0,  10\right\} \times 10^{-3}$, 
and the results are shown in Figs.~\ref{scaling} and \ref{bxmax}.  For the 
portion of the simulations in which magnetic winding dominates, 
the behavior is expected to scale with the Alfv\'en time~\cite{Shapiro}.  
In other words, 
the same profiles should be seen for the same value of $t/t_A$.  (The 
Alv\'en time is inversely proportional to the magnetic field strength and hence 
proportional to $C^{-1/2}$.)  From Fig.~\ref{scaling}, it is evident that
this scaling holds very well for the toroidal field and for the central 
density and lapse, while $t \alt 0.4 t_A$.  After the toroidal field saturates,
the evolution is driven mainly by the MRI, which does not scale with the
Alfv\'en time.  The scaling also does not hold during the collapse phase, 
when the evolution is no longer quasi-stationary.  Though the scaling
breaks down at late times in these simulations, the qualitative outcome
is the same in all cases.  

The behavior of $|B^x|_{\rm max}$ for these four different values of $C$ 
is shown in Fig.~\ref{bxmax}.  The sudden sharp rise of $|B^x|_{\rm max}$ 
signals the onset of the MRI, and the approximate agreement of the slopes 
for different values of $C$
indicates that the exponential growth rate of the MRI does not depend on the 
initial magnetic field strength (as expected from the linear analysis).  
In cases with a very weak 
initial magnetic field, turbulence induced by the MRI may become important 
much earlier than the effects of magnetic braking, since the timescale for 
the growth of the MRI does not depend on the initial magnetic field strength.  
In this case, the scaling with $t_A$ would not hold during any phase of the 
evolution.  However, since both the MRI and magnetic braking lead to 
similar angular momentum transfer, the qualitative outcome may again be 
the same.    

\subsection{Star B1}
\label{starB1}

Here, we present results for the evolution of star~B1 with 
$C = 2.5 \times 10^{-3}$.  This run was performed with resolution $400^2$ 
and outer boundaries located at $4.5 R$ ($36.4M$).  Since this model is 
not hypermassive, the redistribution of angular momentum through MHD 
effects will not lead to collapse.  However, since this star is ultraspinning 
and angular momentum is conserved in axisymmetric spacetimes, 
it cannot relax to a uniform rotation state everywhere unless a 
significant amount of angular momentum can be dumped to the magnetic 
field.  We find that this model simply 
seeks out a magnetized equilibrium state which consists 
of a fairly uniformly rotating core surrounded by a differentially rotating 
torus. This is similar to the final state we found in~\cite{DLSS} for the 
same model when evolved with shear viscosity. 

\begin{figure}
\begin{center}
\epsfxsize=3.in
\leavevmode
\epsffile{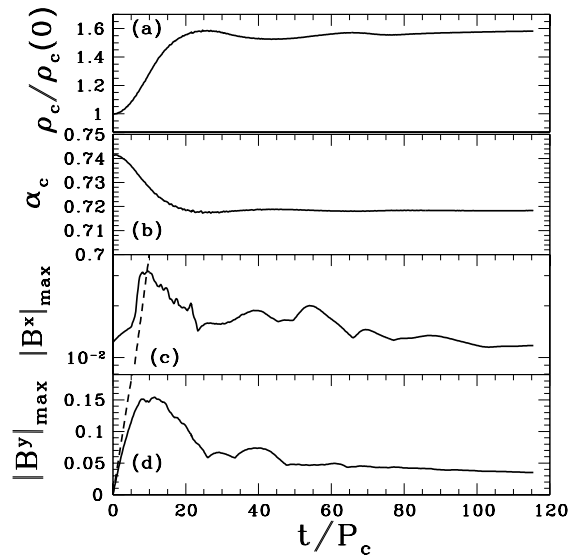}
\caption{Evolution of central rest-mass density $\rho_c$, central lapse 
$\alpha_c$, maximum values of $|B^x|$ and $|B^y|$ for star~B1. The magnetic
fields $|B^x|_{\rm max}$ and $|B^y|_{\rm max}$ are plotted in units of
$\sqrt{\rho_{\rm max}(0)}$.
Note that the lines become fairly horizontal
at late times, indicating that an equilibrium has been reached.
The dashed line in (c) represents an approximate slope of 
$\omega=(0.37/P_c)$ for the exponential growth rate of the MRI, 
$\delta B^x \propto e^{\omega t}$. The dashed line in (d)
represents the predicted linear growth of $|B^y|_{\rm max}$ computed
from Eq.~(\ref{eqn:dtBphi}).
\label{sBfigs}}
\end{center}
\end{figure}

\begin{figure}
\begin{center}
\epsfxsize=3.in
\leavevmode
\epsffile{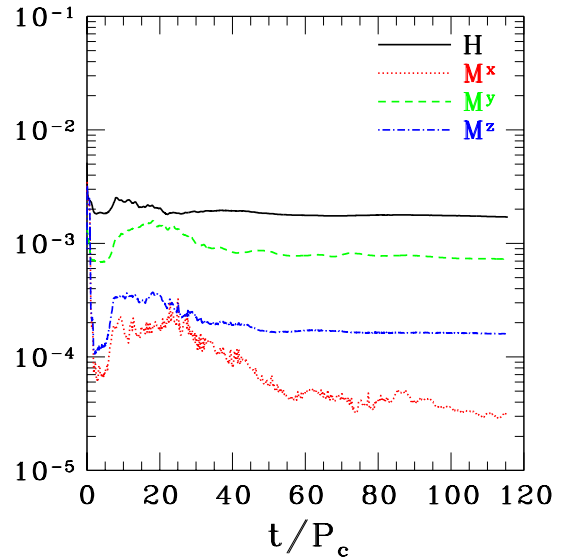}
\caption{L2 norms of the errors in Hamiltonian and momentum constraints 
during the evolution of star~B1.
\label{consB}}
\end{center}
\end{figure}

Figure~\ref{sBfigs} presents the evolution of some relevant 
quantities for this case.  From the central density and lapse, it is evident
that the star has settled into a more compact equilibrium configuration.  This
is consistent with the expectation that magnetic braking should transfer 
angular momentum from the core to the outer layers.  A brief episode of 
poloidal magnetic field growth due to the MRI is indicated by the plot of 
$|B^x|_{\rm max}$ in Figure~\ref{sBfigs}.  The instability saturates and 
quickly dies away~\cite{fn3}, leaving the strength of the poloidal field 
largely unchanged.
Early in the evolution, the maximum value of the toroidal component $|B^y|$ 
rises due to magnetic winding.  This growth saturates at 
$t \sim 10 P_c \sim 0.5 t_A$ .  We note, however, that the toroidal magnetic 
field is non-zero in the final equilibrium state, though it is no longer 
growing due to magnetic winding.  The accuracy of the spacetime evolution is 
demonstrated by Fig.~\ref{consB}, which shows that the Hamiltonian and momentum 
constraint errors remain very small throughout the simulation.  

\begin{figure*}
\begin{center}
\epsfxsize=1.8in
\leavevmode
\hspace{-0.7cm}\epsffile{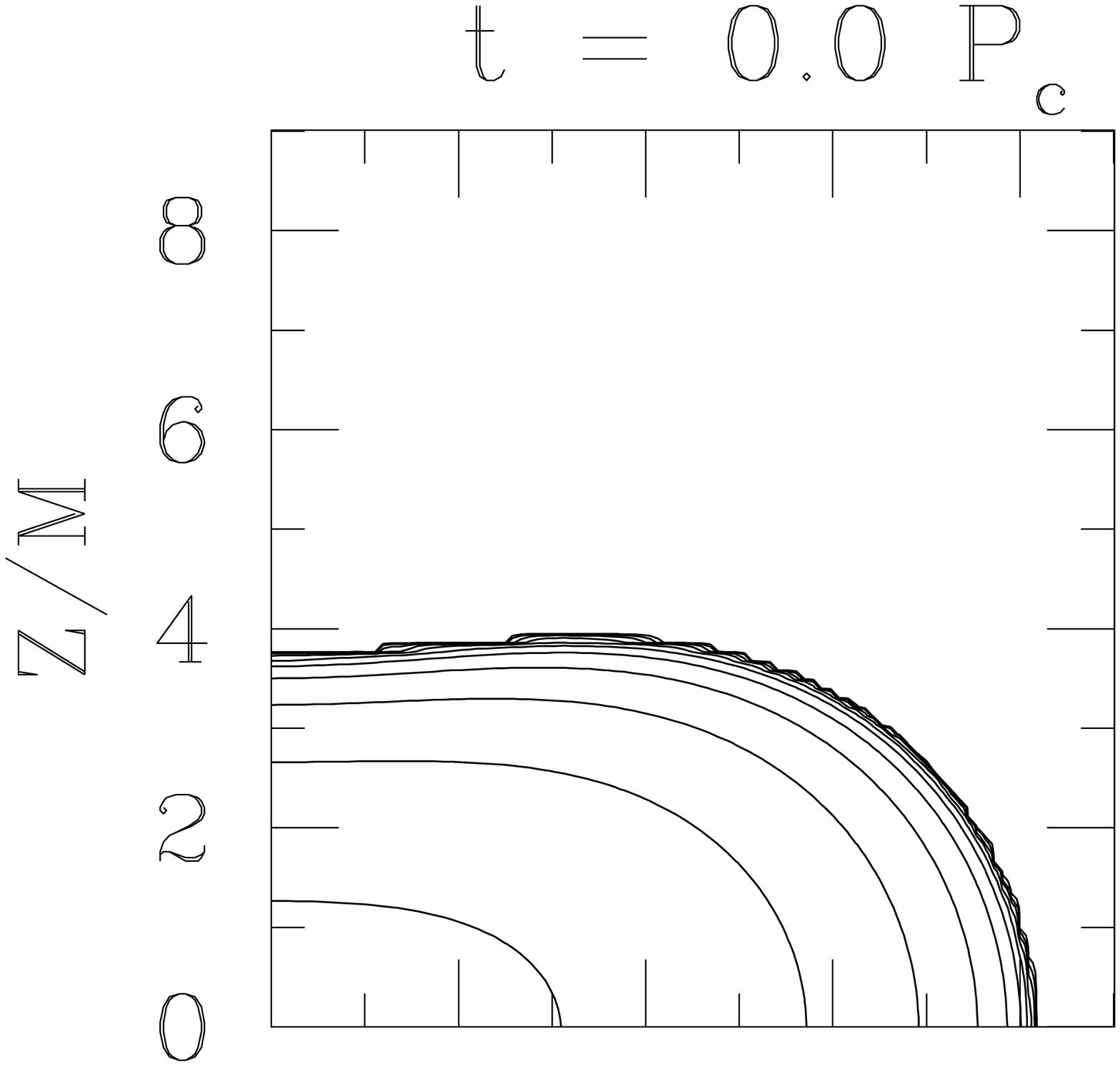}
\epsfxsize=1.8in
\leavevmode
\hspace{-0.5cm}\epsffile{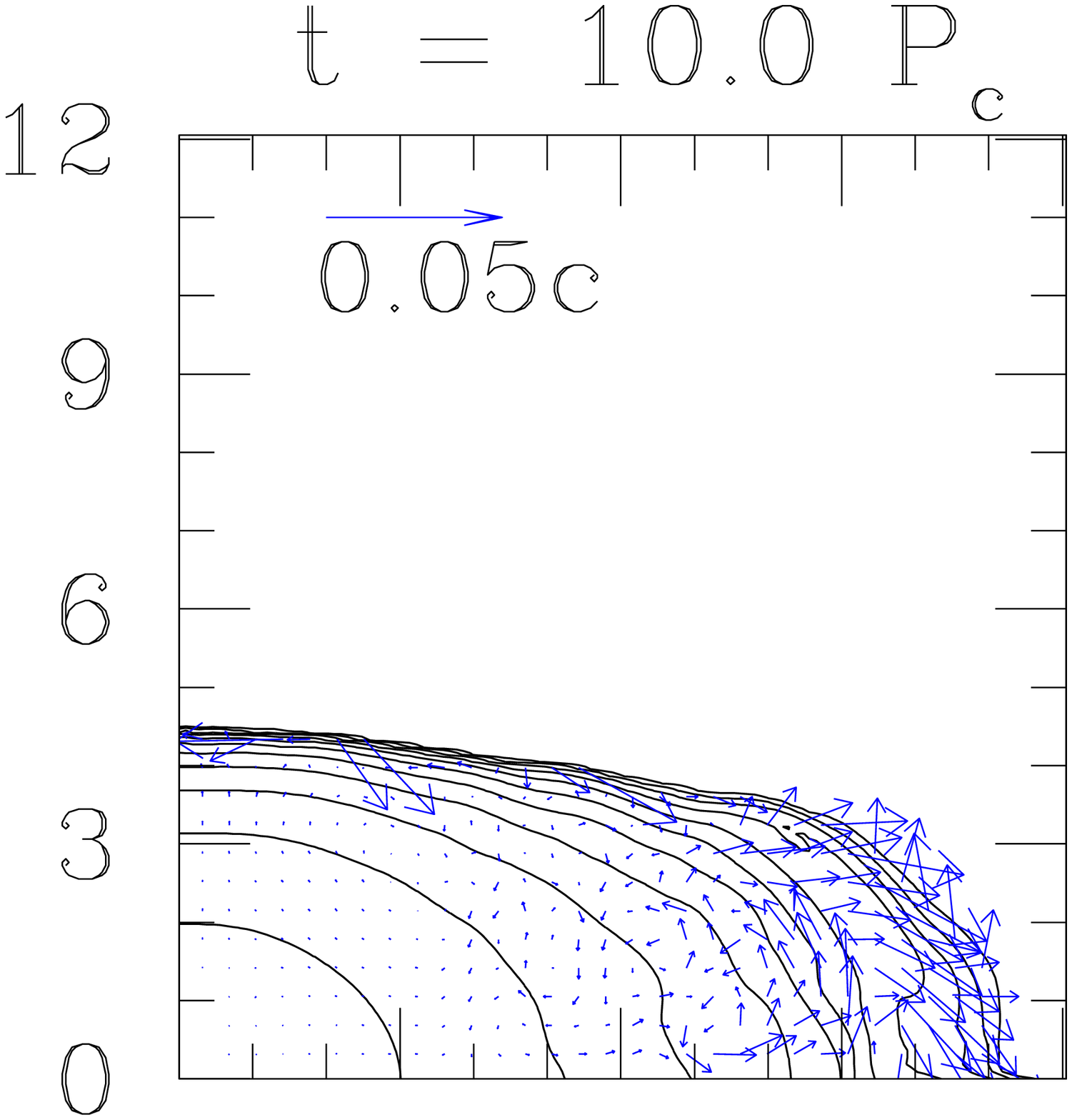}
\epsfxsize=1.8in
\leavevmode
\hspace{-0.5cm}\epsffile{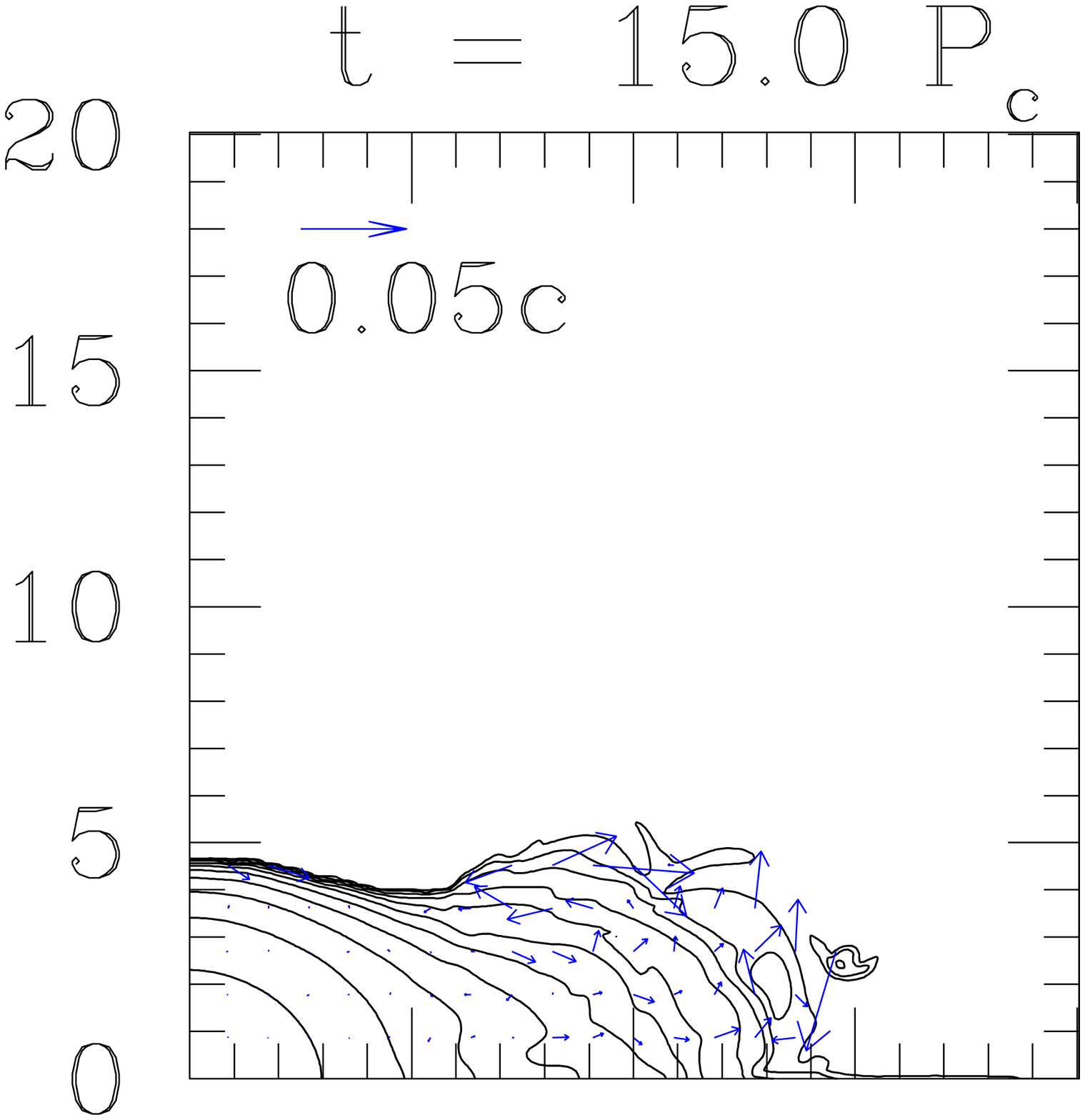}
\epsfxsize=1.8in
\leavevmode
\hspace{-0.5cm}\epsffile{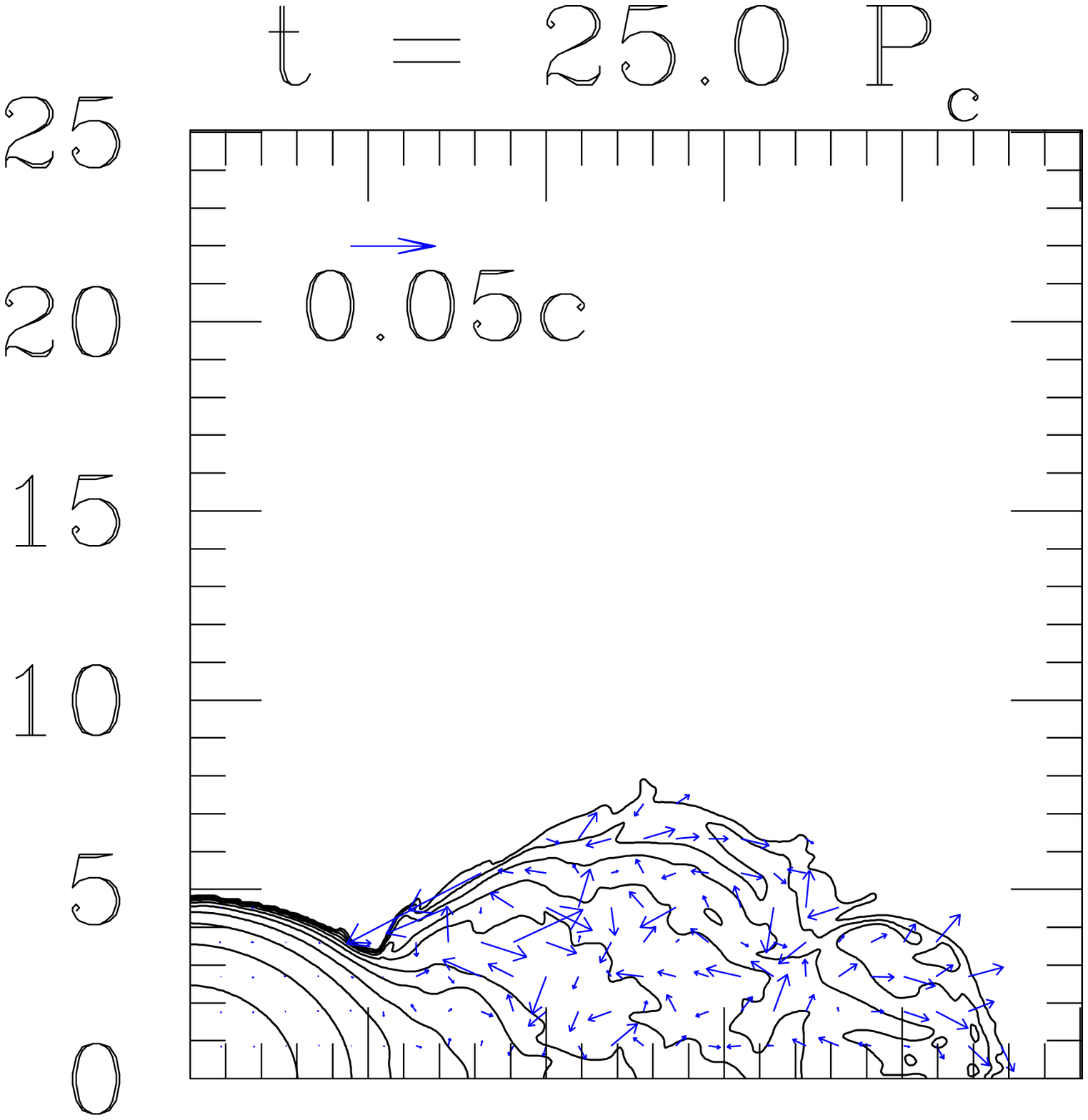} \\
\vspace{-0.5cm}
\epsfxsize=1.8in
\leavevmode
\hspace{-0.7cm}\epsffile{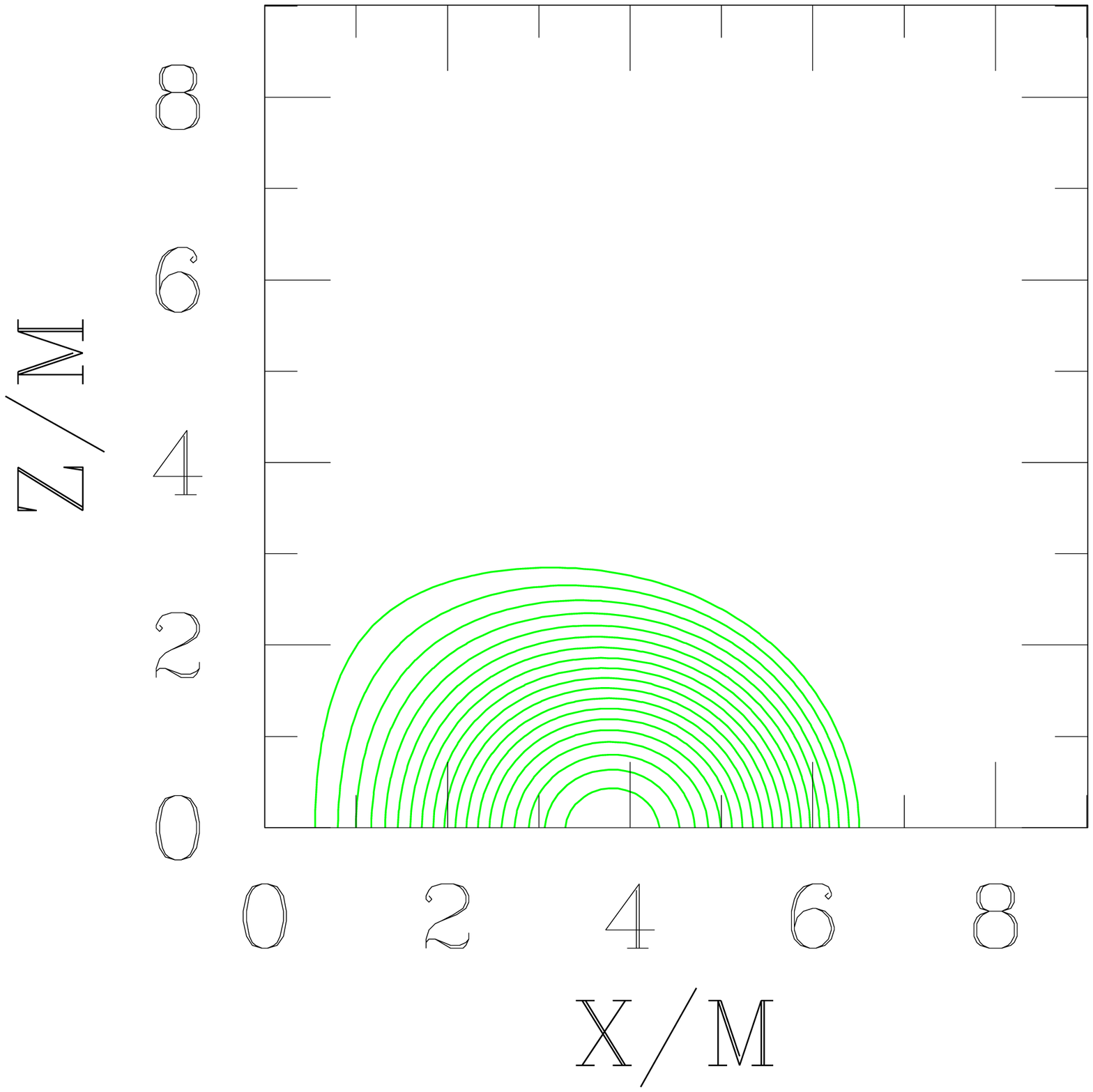}
\epsfxsize=1.8in
\leavevmode
\hspace{-0.5cm}\epsffile{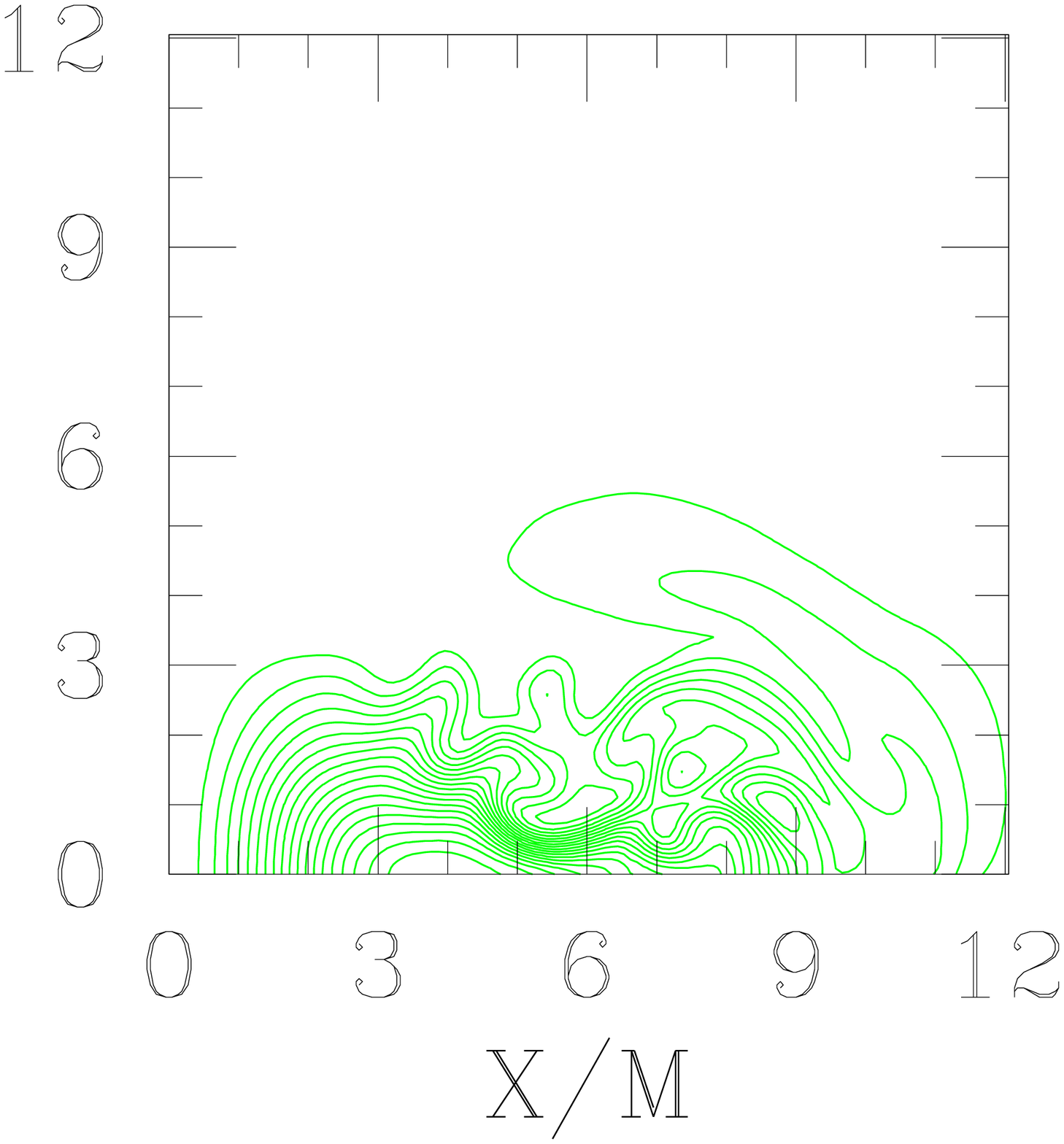}
\epsfxsize=1.8in
\leavevmode
\hspace{-0.5cm}\epsffile{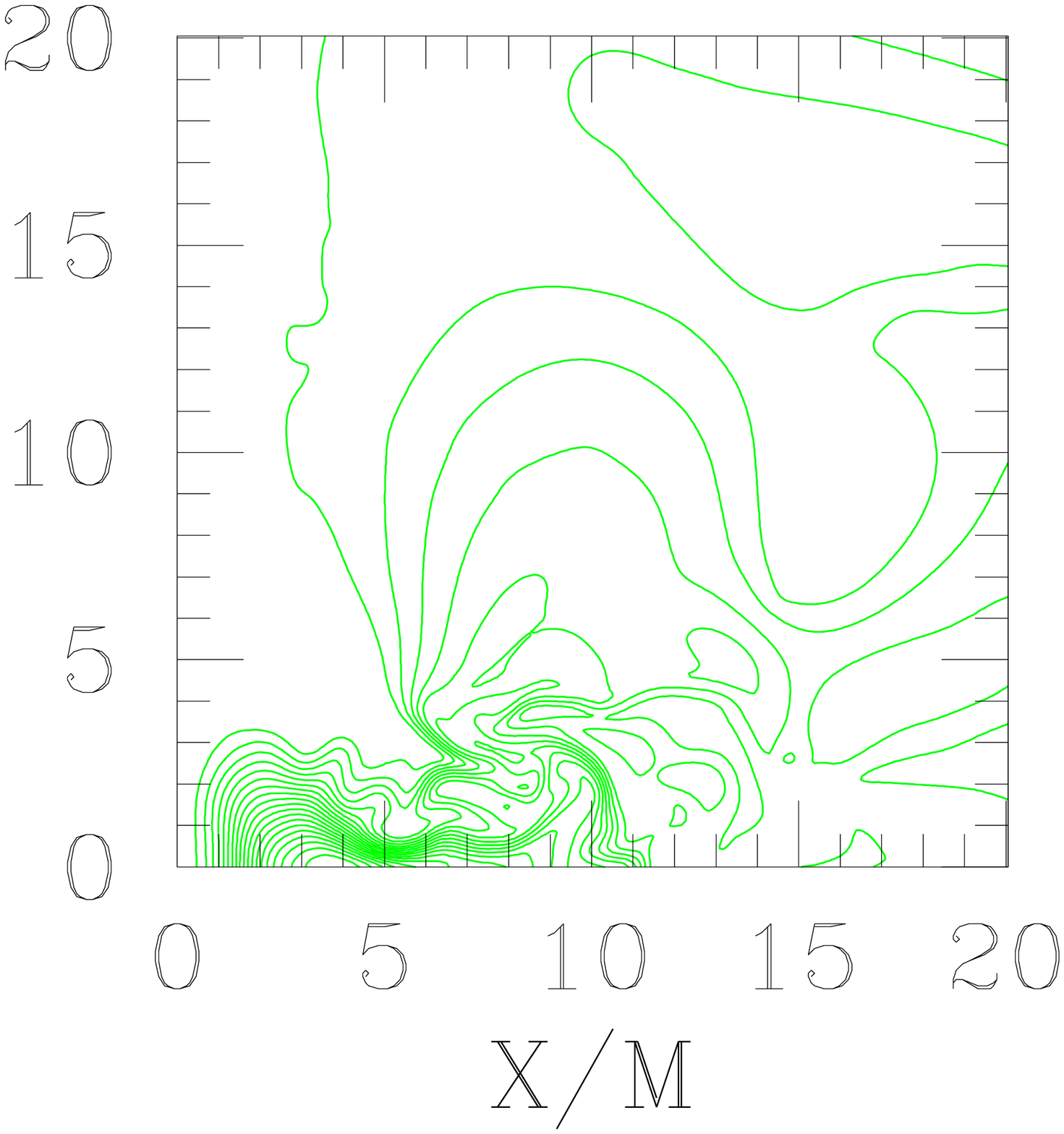}
\epsfxsize=1.8in
\leavevmode
\hspace{-0.5cm}\epsffile{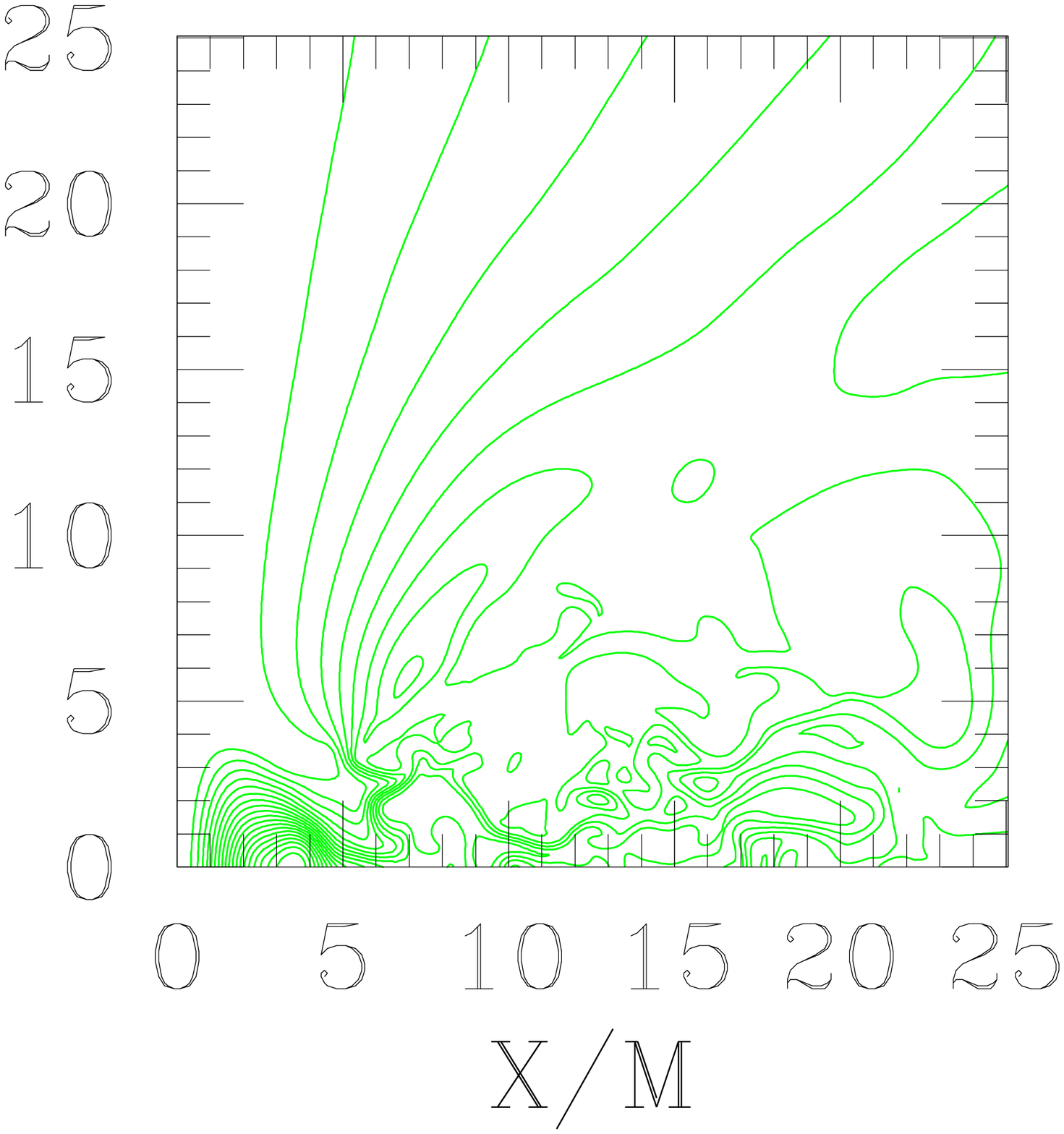}
\end{center}

\begin{center}
\epsfxsize=1.8in
\leavevmode
\hspace{-0.7cm}\epsffile{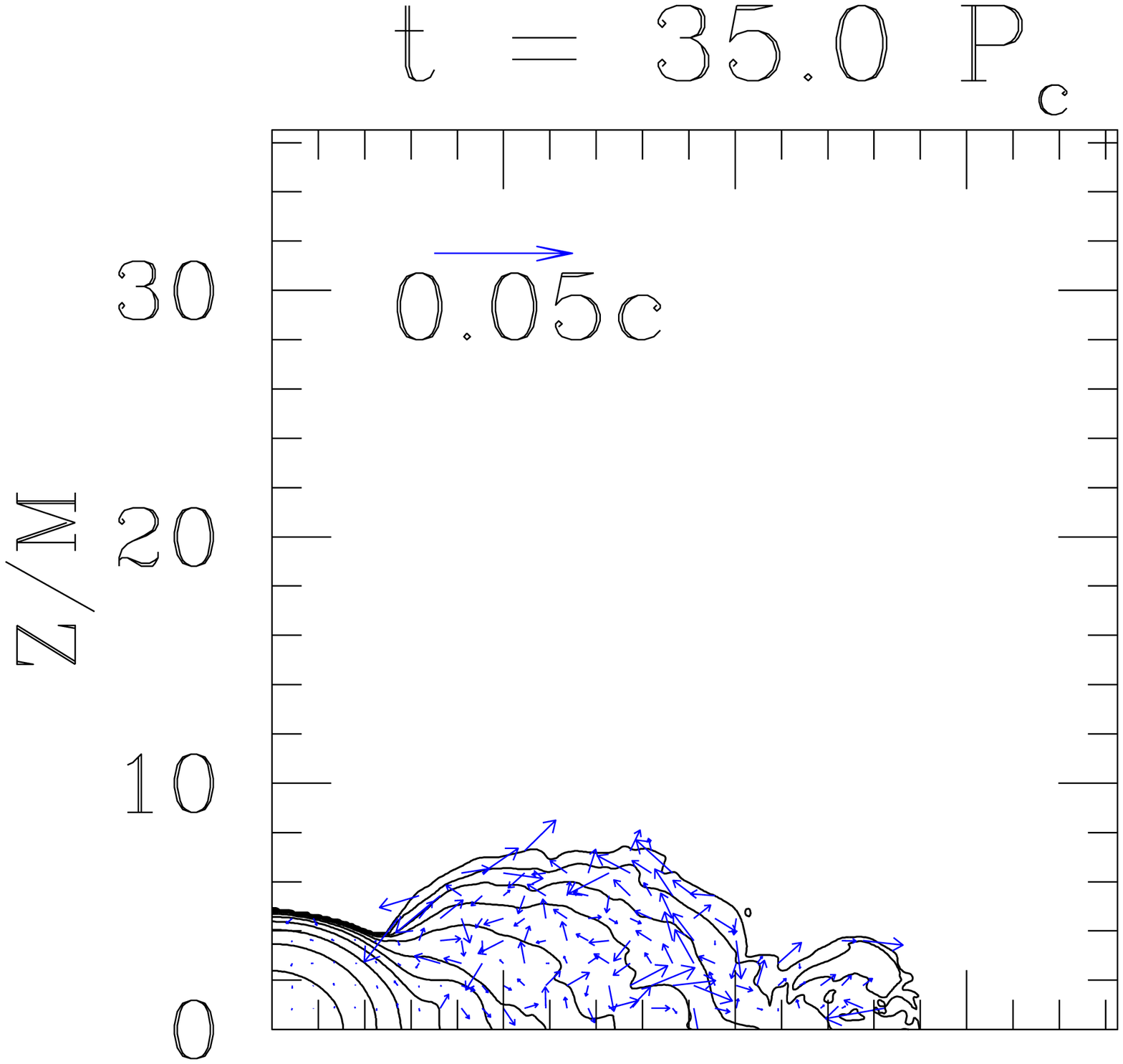}
\epsfxsize=1.8in
\leavevmode
\hspace{-0.5cm}\epsffile{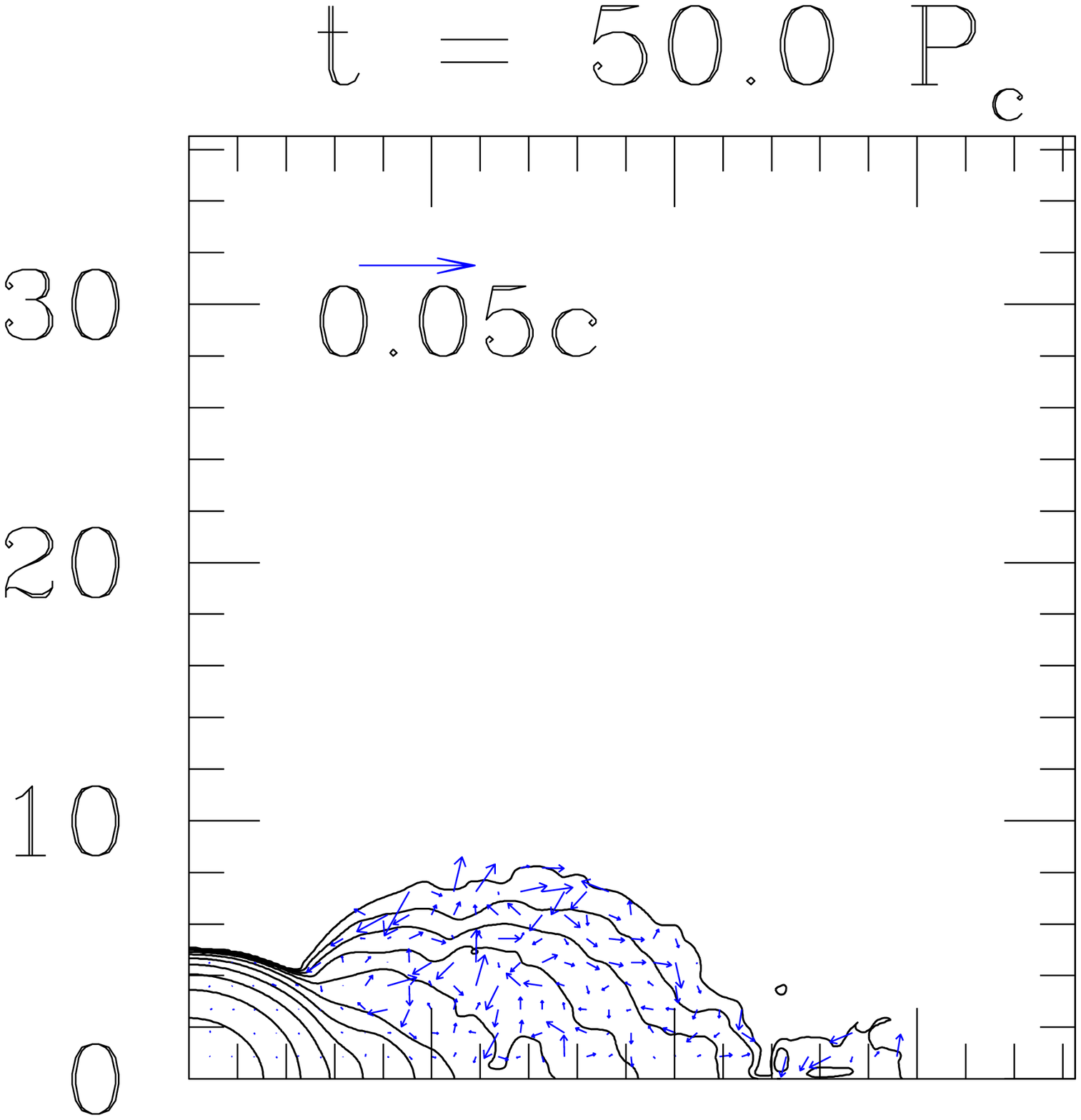}
\epsfxsize=1.8in
\leavevmode
\hspace{-0.5cm}\epsffile{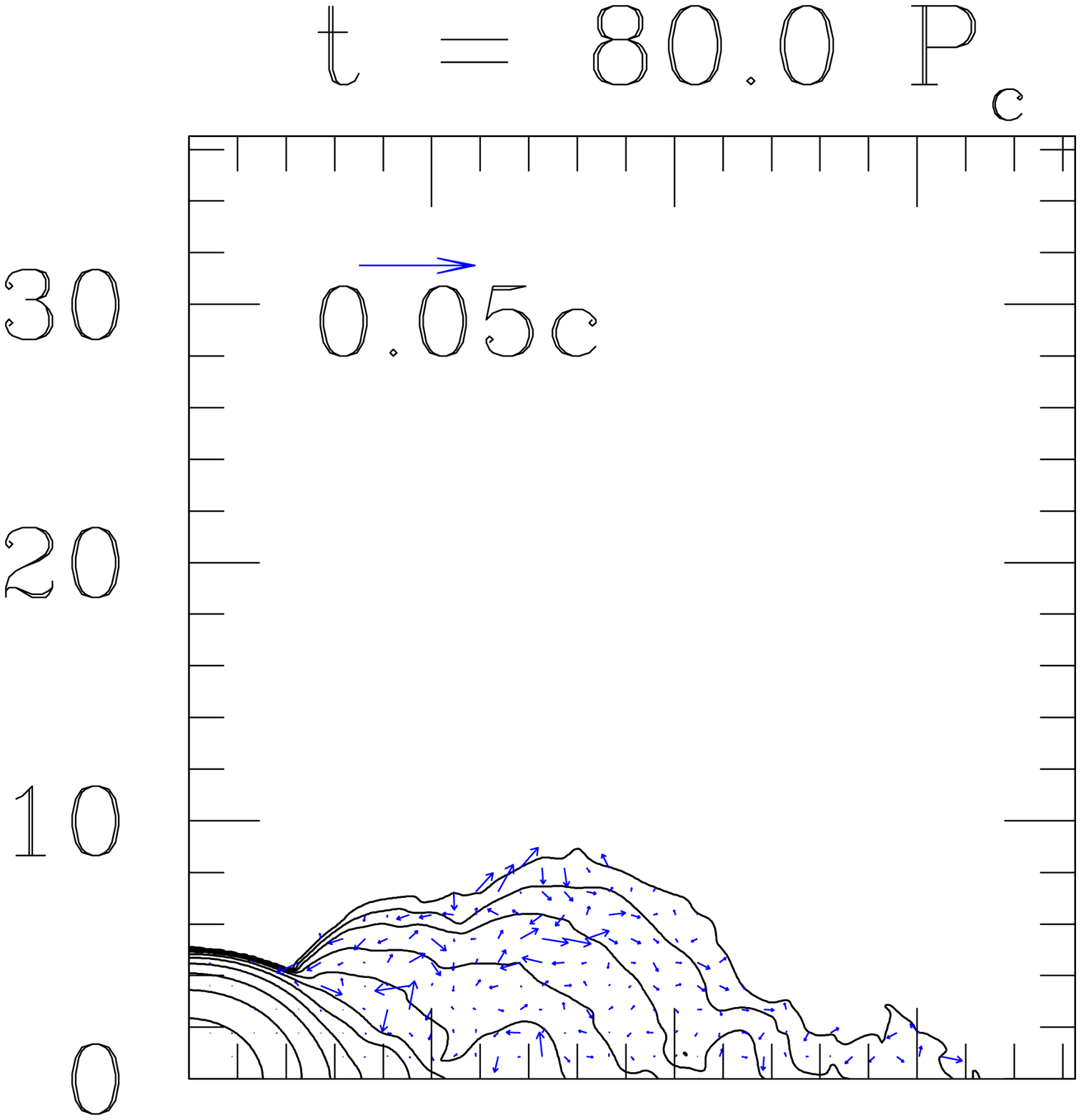}
\epsfxsize=1.8in
\leavevmode
\hspace{-0.5cm}\epsffile{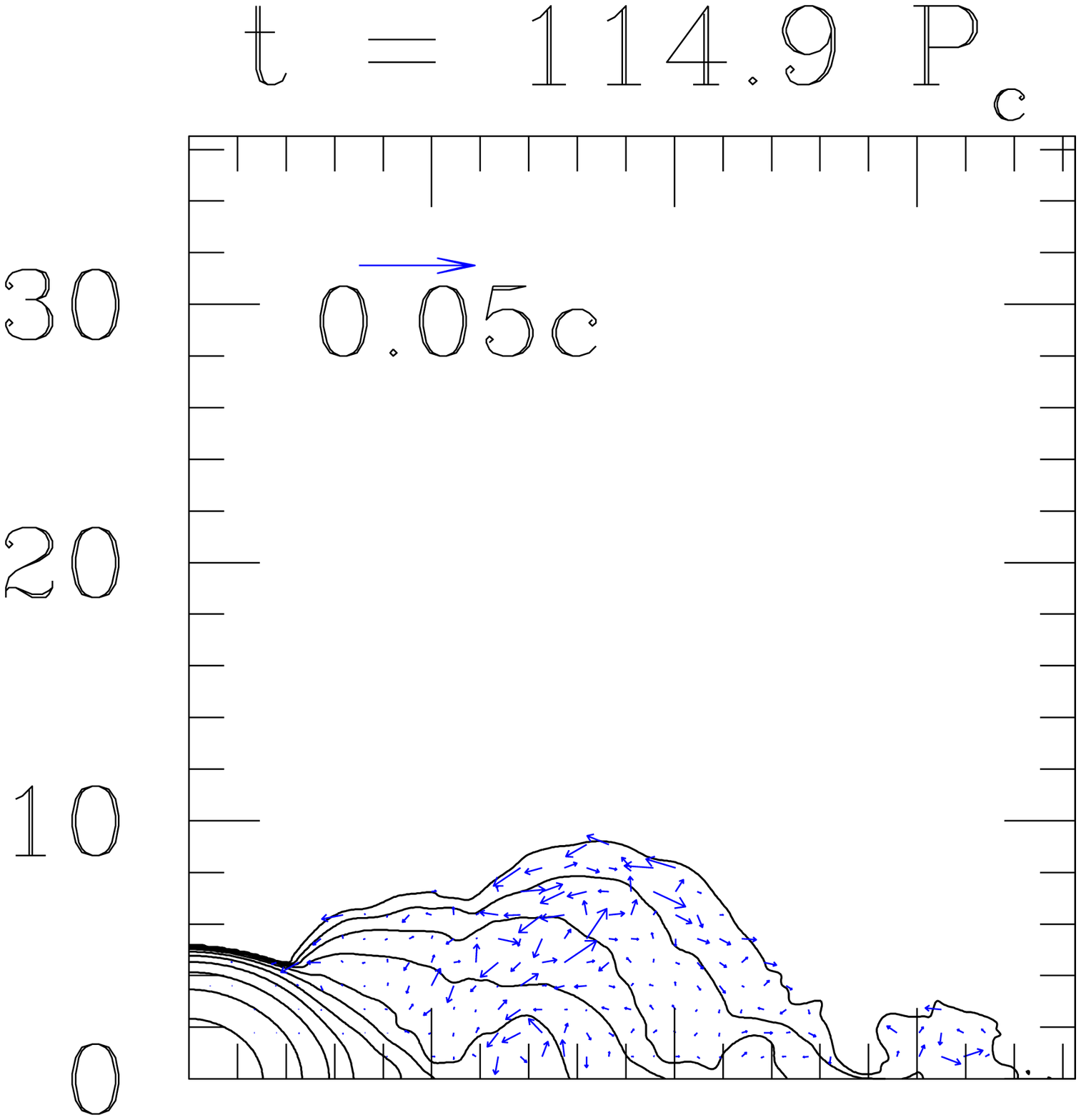} \\
\vspace{-0.5cm}
\epsfxsize=1.8in
\leavevmode
\hspace{-0.7cm}\epsffile{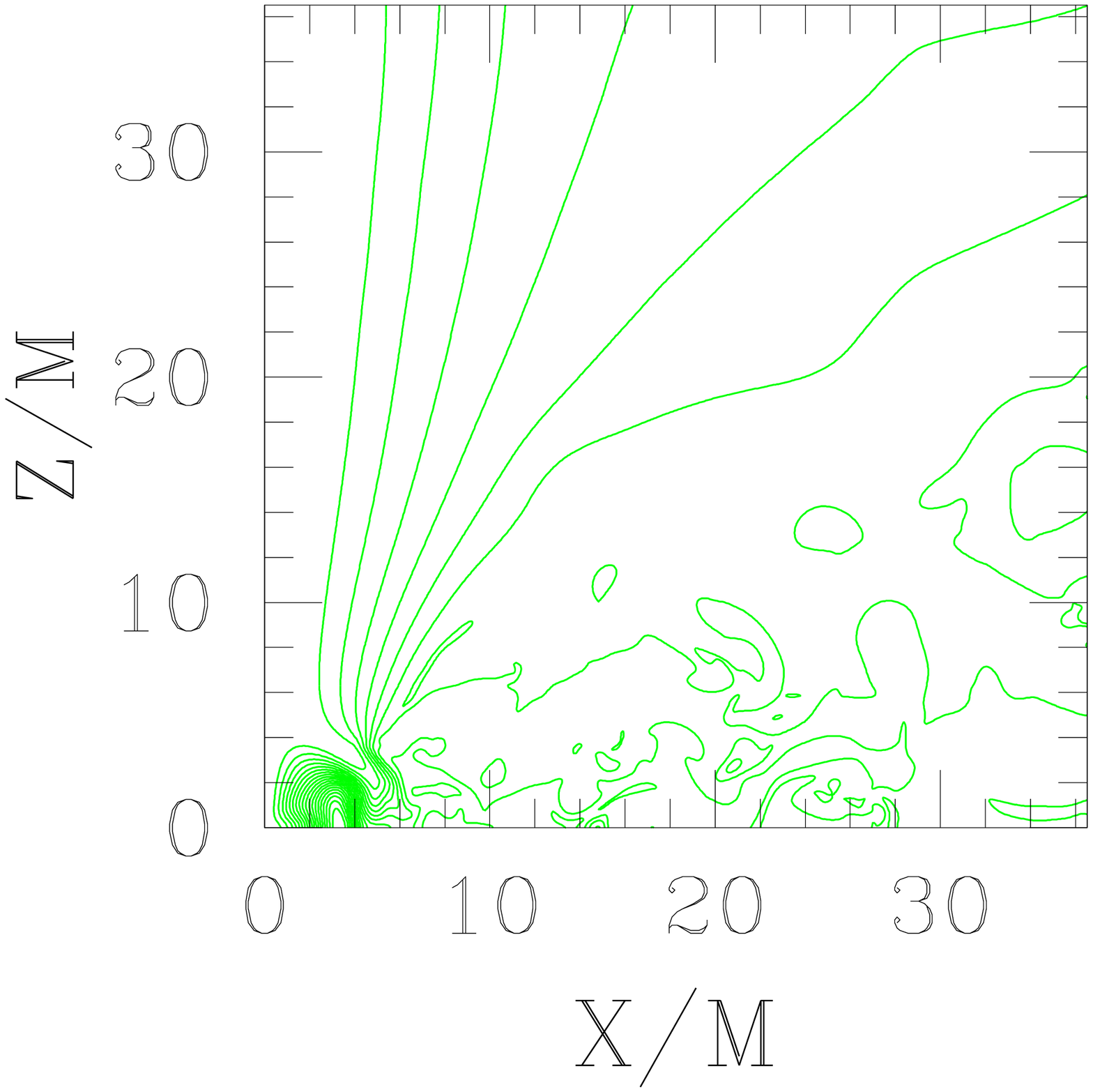}
\epsfxsize=1.8in
\leavevmode
\hspace{-0.5cm}\epsffile{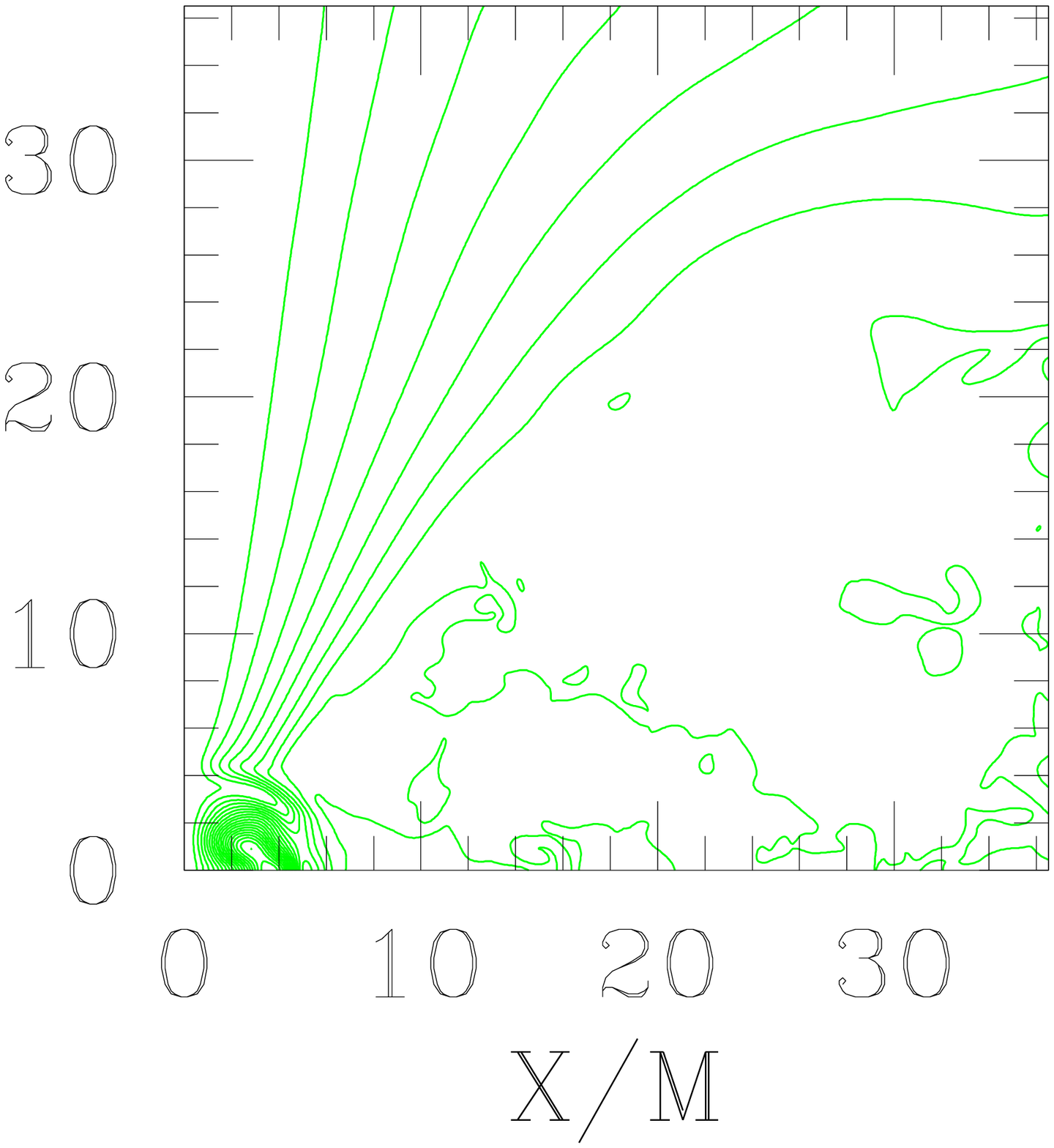}
\epsfxsize=1.8in
\leavevmode
\hspace{-0.5cm}\epsffile{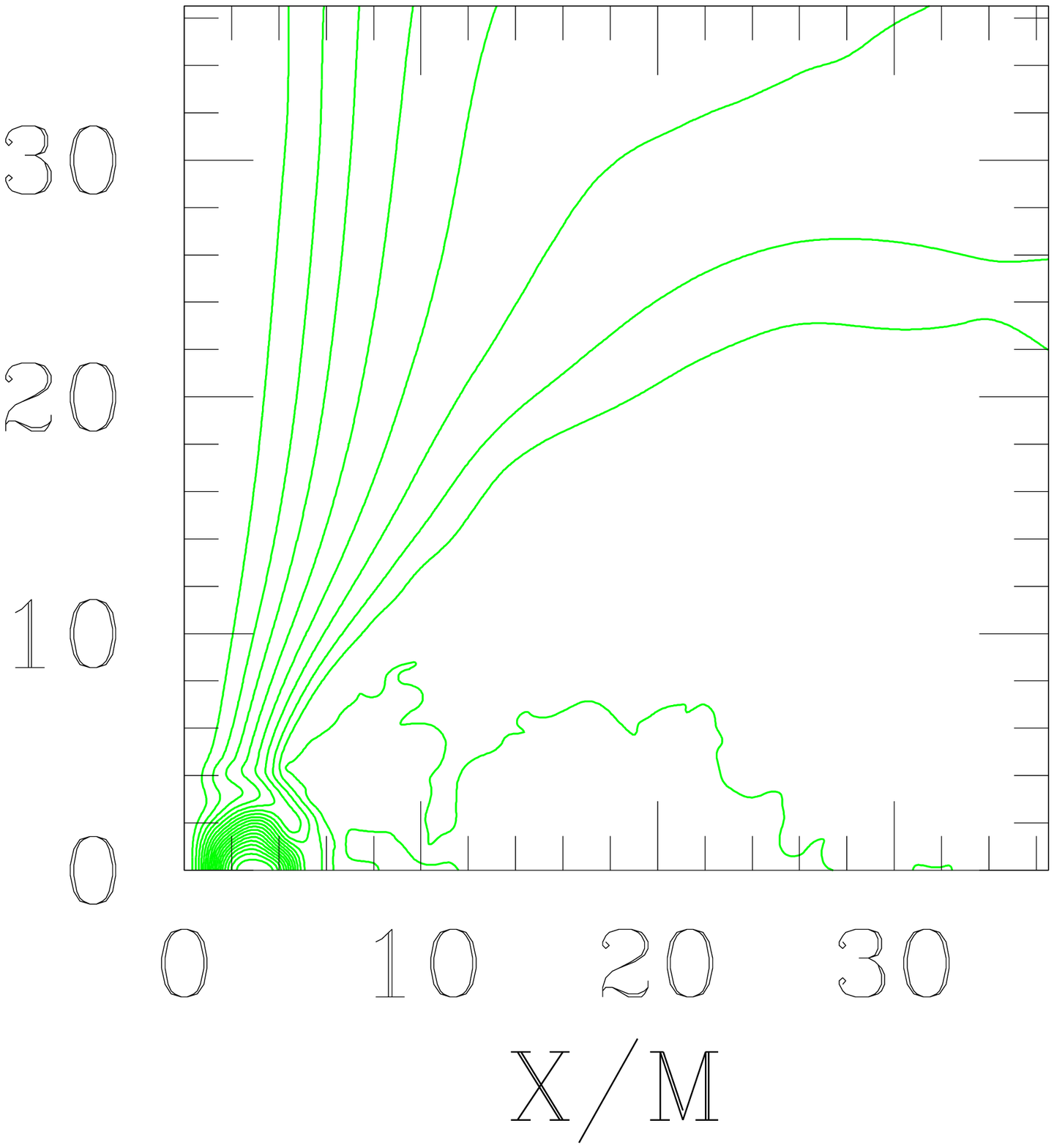}
\epsfxsize=1.8in
\leavevmode
\hspace{-0.5cm}\epsffile{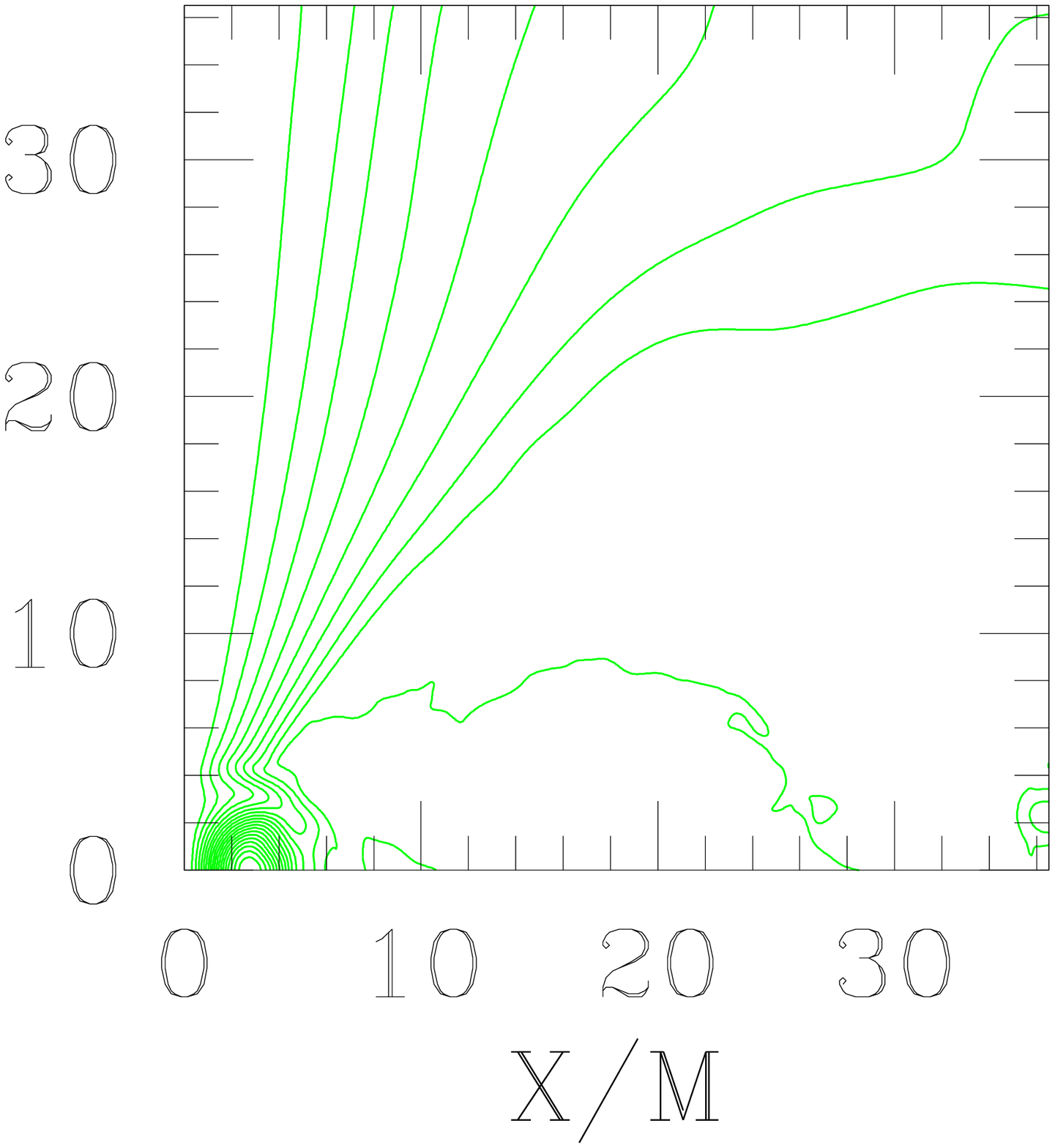}
\caption{Snapshots of density contours and poloidal magnetic field lines for 
star~B1. The first and third rows show snapshots of the rest-mass density
contours and velocity vectors on the meridional plane. The second and fourth
rows show the field lines (lines of constant $A_{\varphi}$)
for the poloidal magnetic field at the same times
as the first and third rows.
The density contours are drawn for $\rho_0/\rho_{\rm max}(0)=
10^{-0.36 i - 0.09}~(i=0$--10).
The field lines are drawn for $A_{\varphi} = A_{\varphi,\rm min}
+ (A_{\varphi,\rm max} - A_{\varphi,\rm min}) i/20~(i=1$--19),
where $A_{\varphi,\rm max}$ and $A_{\varphi,\rm min}$ are the maximum
and minimum values of $A_{\varphi}$ respectively at the given time.
Note that the field lines and the density contours show
little change for $t \gtrsim 35P_c$, indicating that the
star has settled down to an equilibrium state. 
\label{sBmerid}}
\end{center}
\end{figure*}

\begin{figure*}
\begin{center}
\epsfxsize=1.9in
\leavevmode
\epsffile{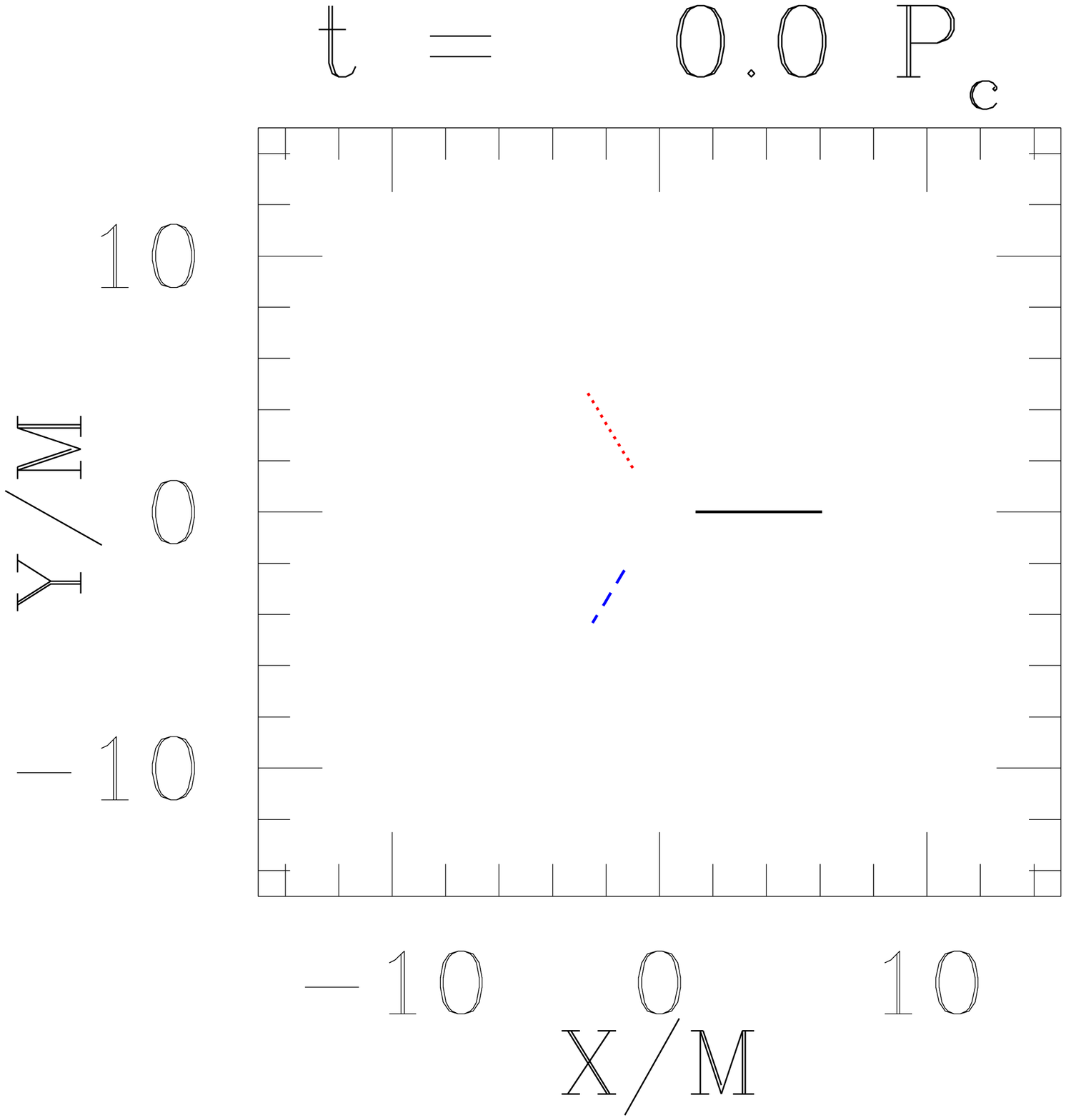}
\epsfxsize=1.9in
\leavevmode
\epsffile{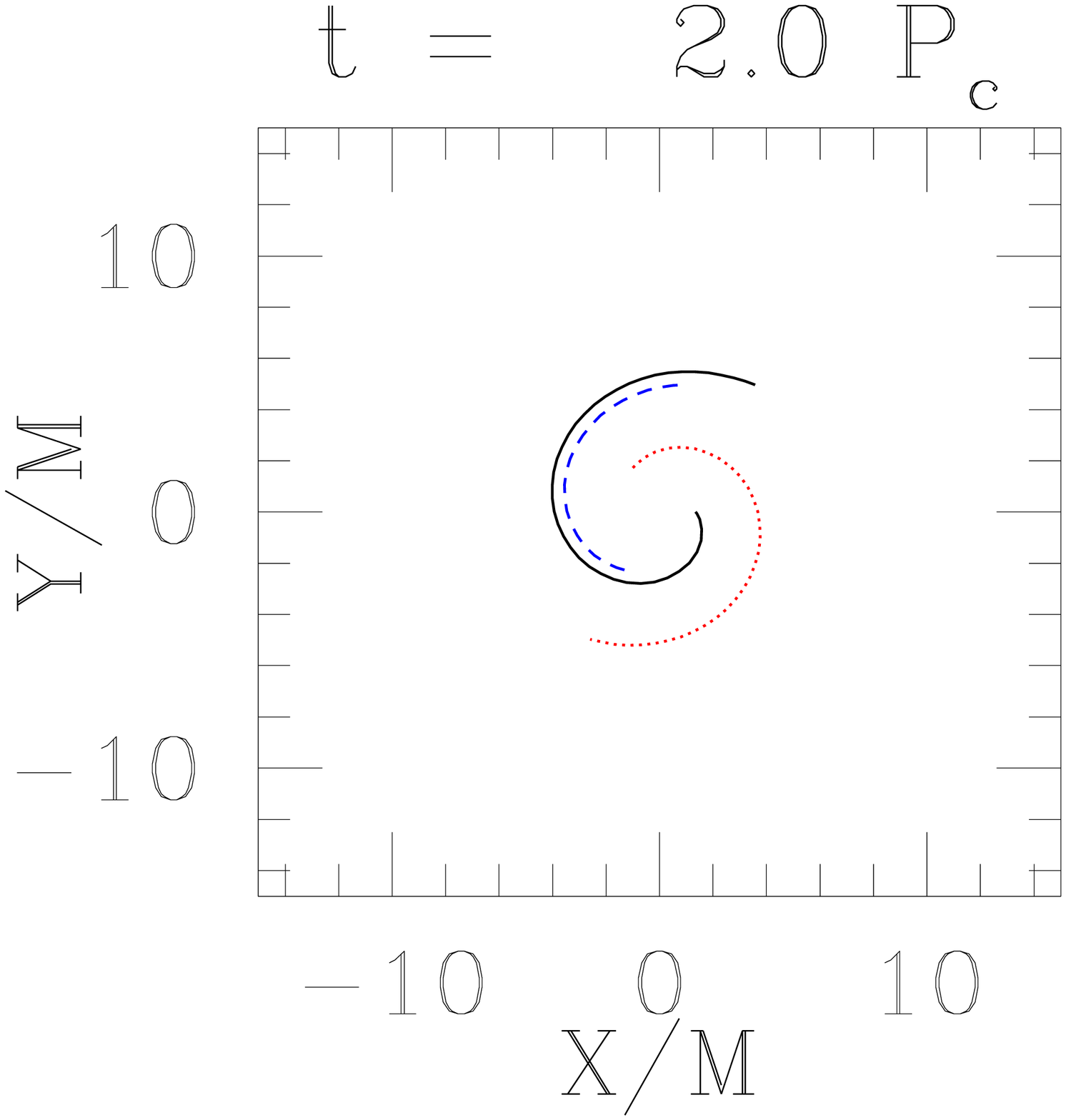}
\epsfxsize=1.9in
\leavevmode
\epsffile{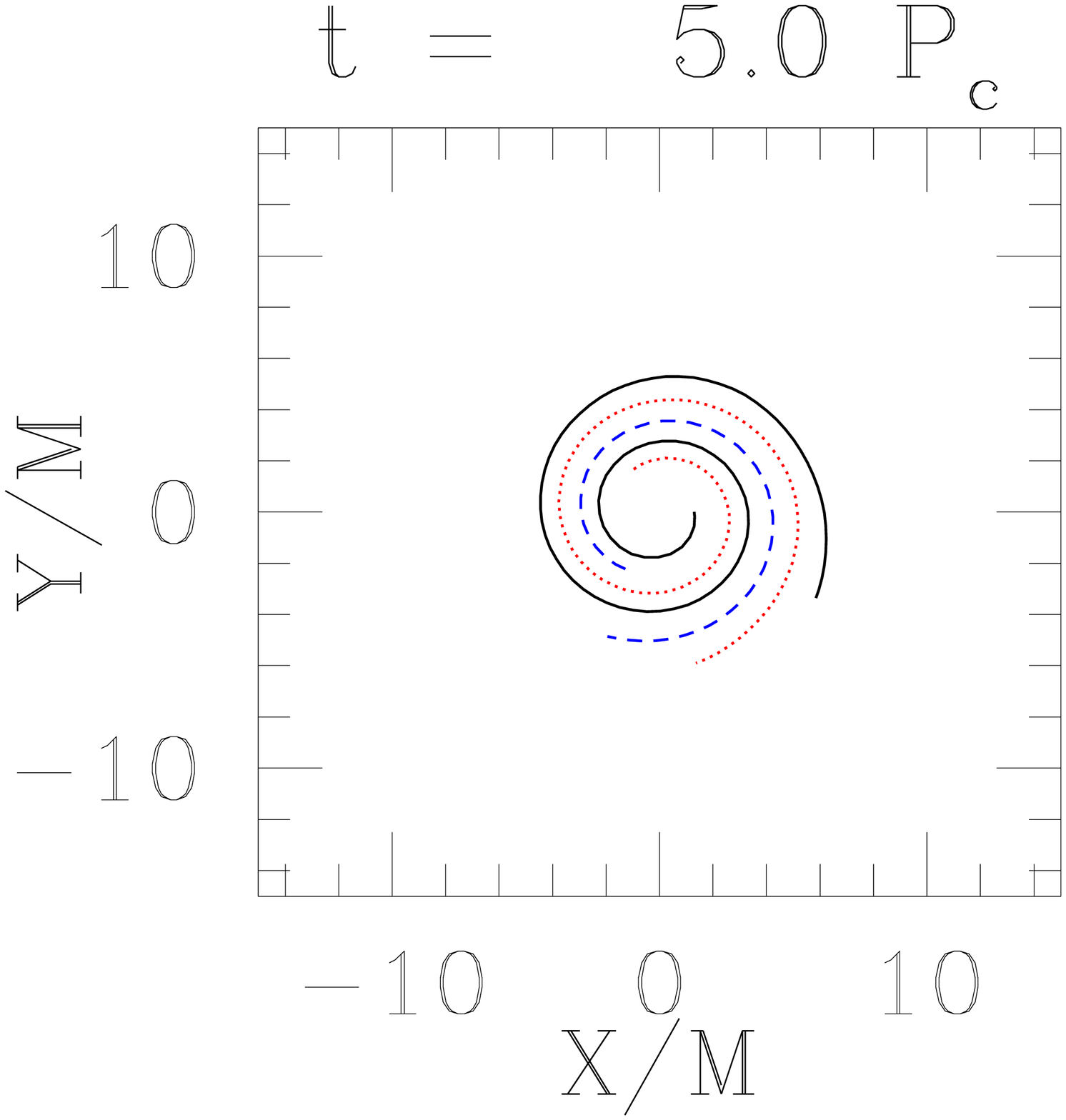}
\end{center}

\begin{center}
\epsfxsize=1.9in
\leavevmode
\epsffile{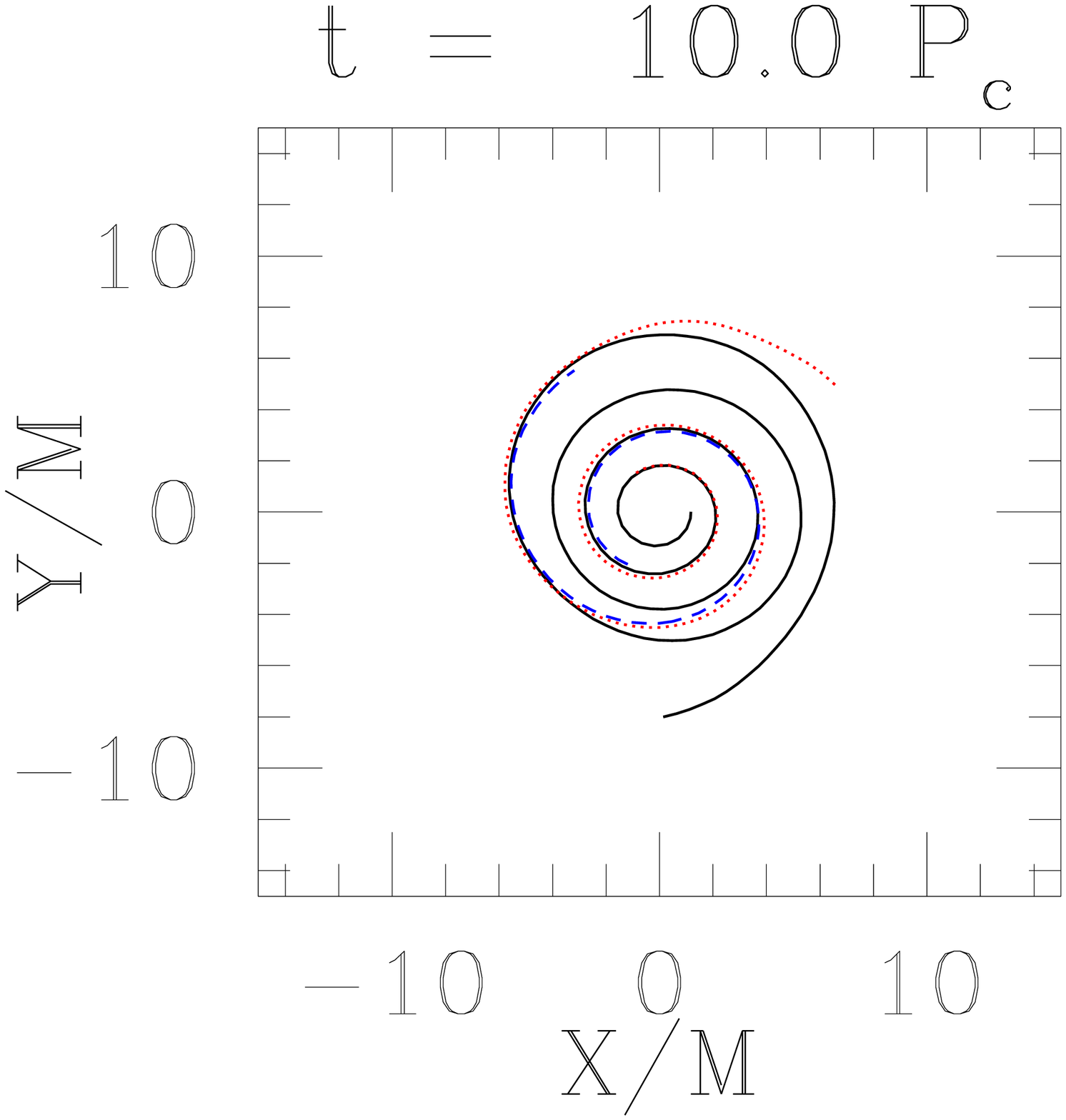}
\epsfxsize=1.9in
\leavevmode
\epsffile{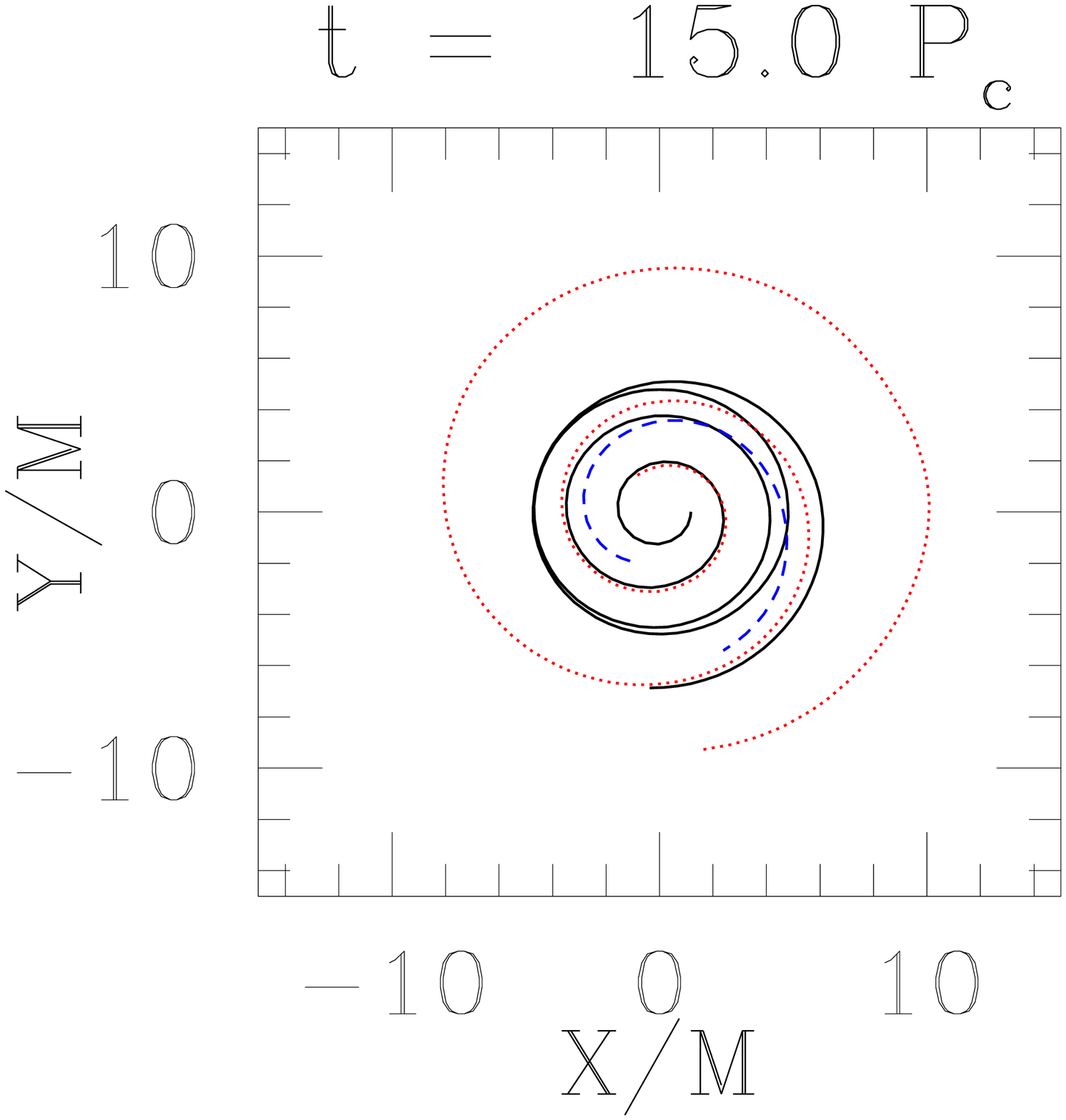}
\epsfxsize=1.9in
\leavevmode
\epsffile{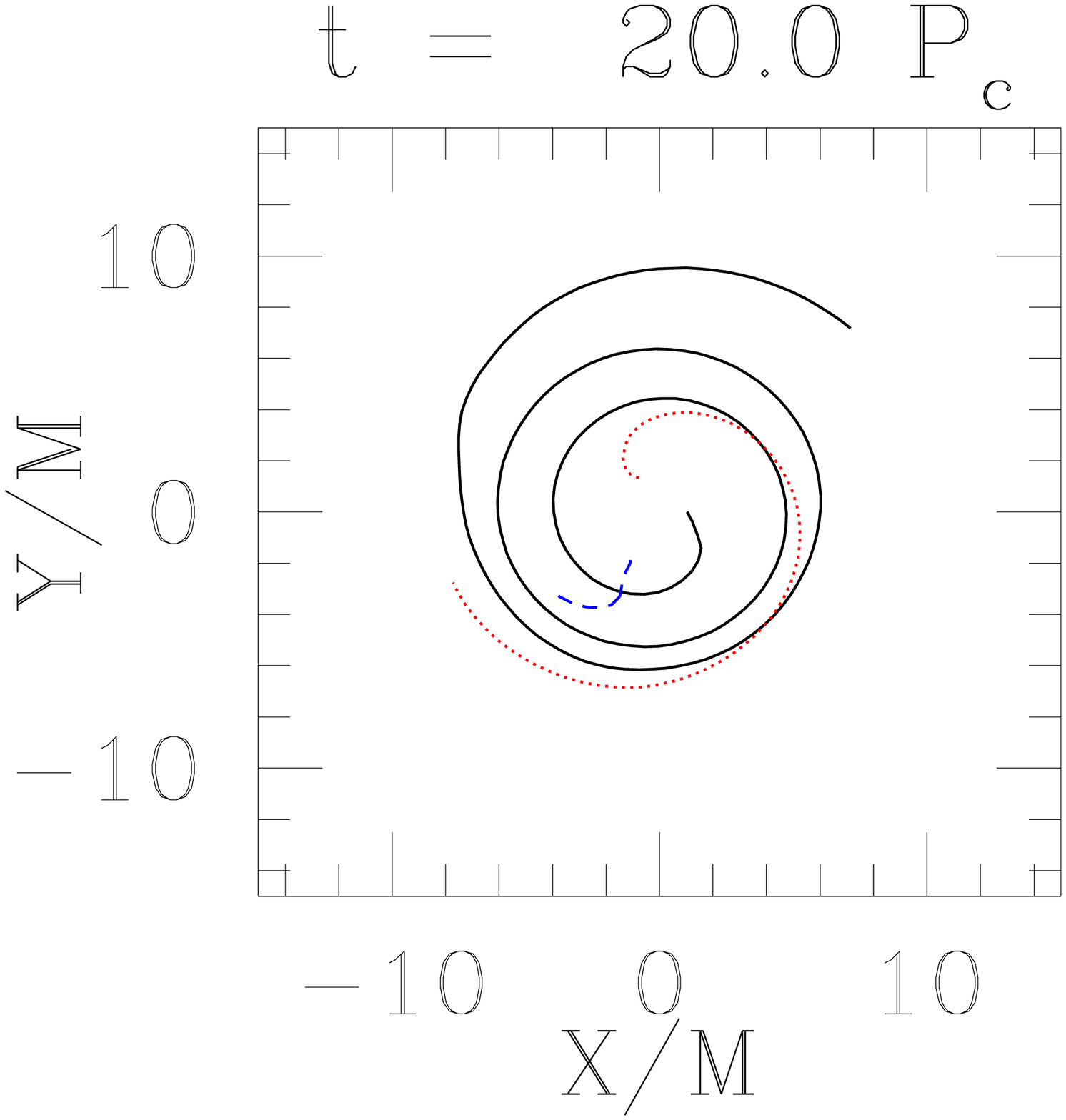}
\end{center}

\begin{center}
\epsfxsize=1.9in
\leavevmode
\epsffile{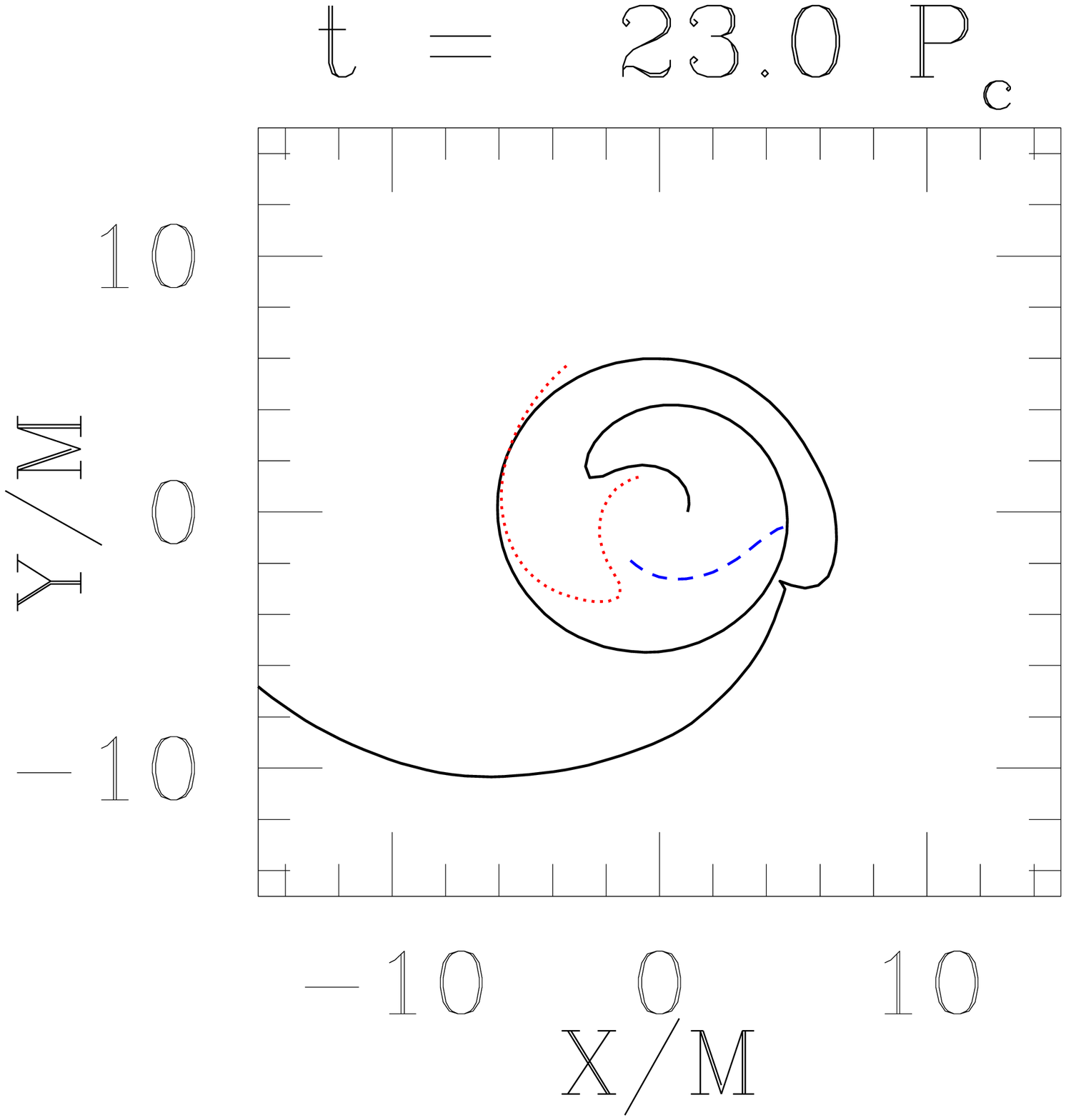}
\epsfxsize=1.9in
\leavevmode
\epsffile{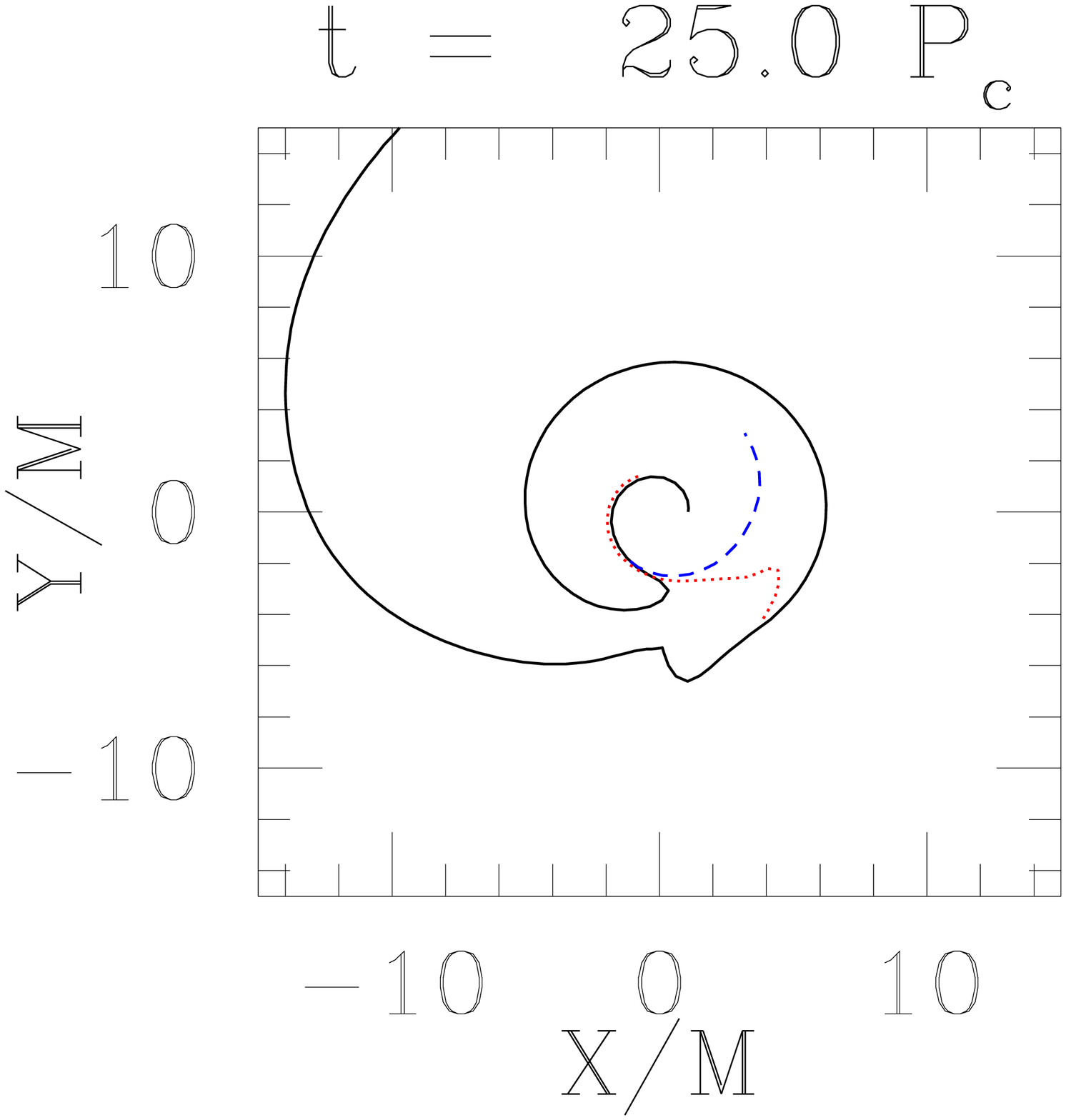}
\epsfxsize=1.9in
\leavevmode
\epsffile{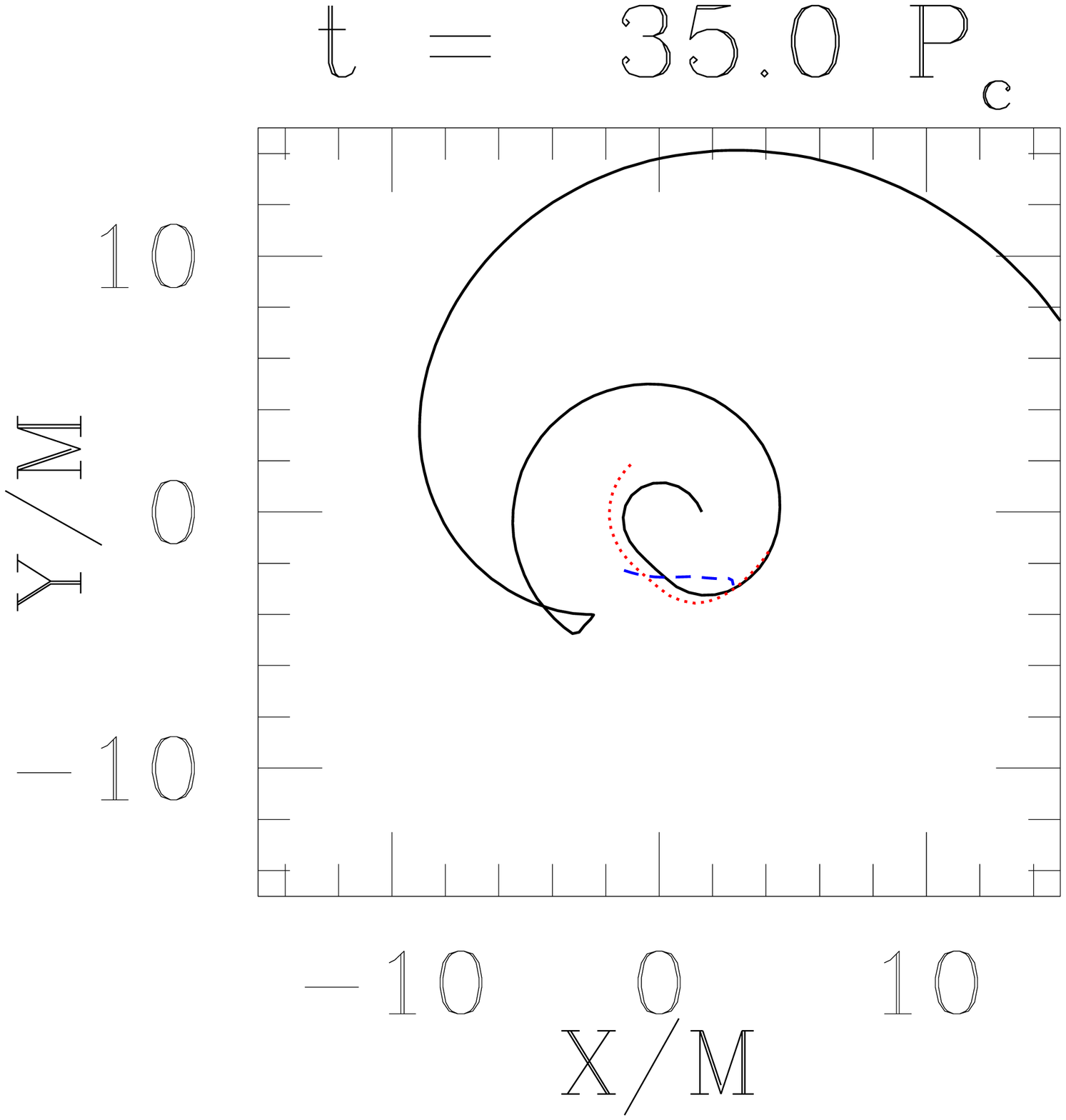}
\end{center}

\begin{center}
\epsfxsize=1.9in
\leavevmode
\epsffile{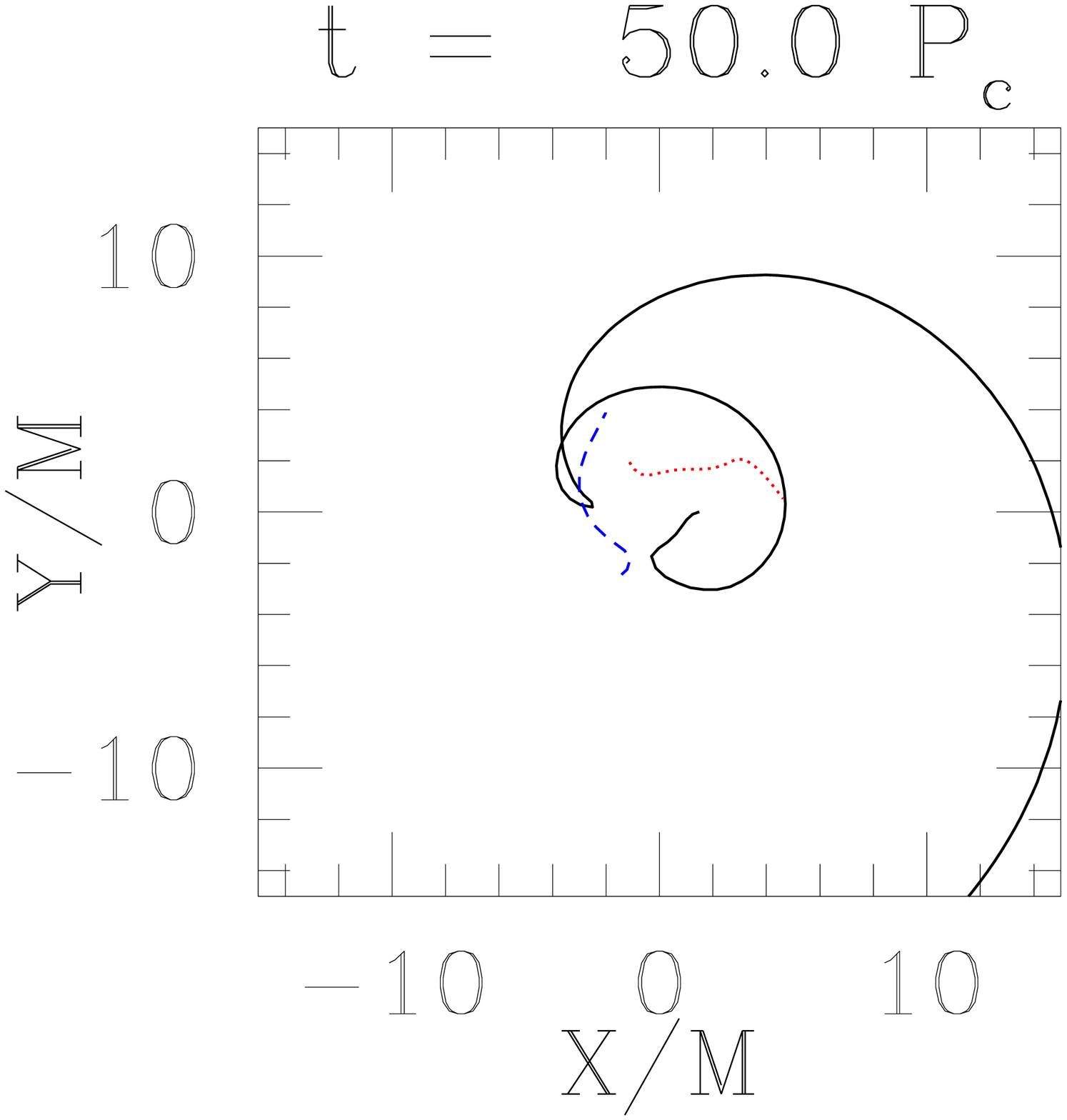}
\epsfxsize=1.9in
\leavevmode
\epsffile{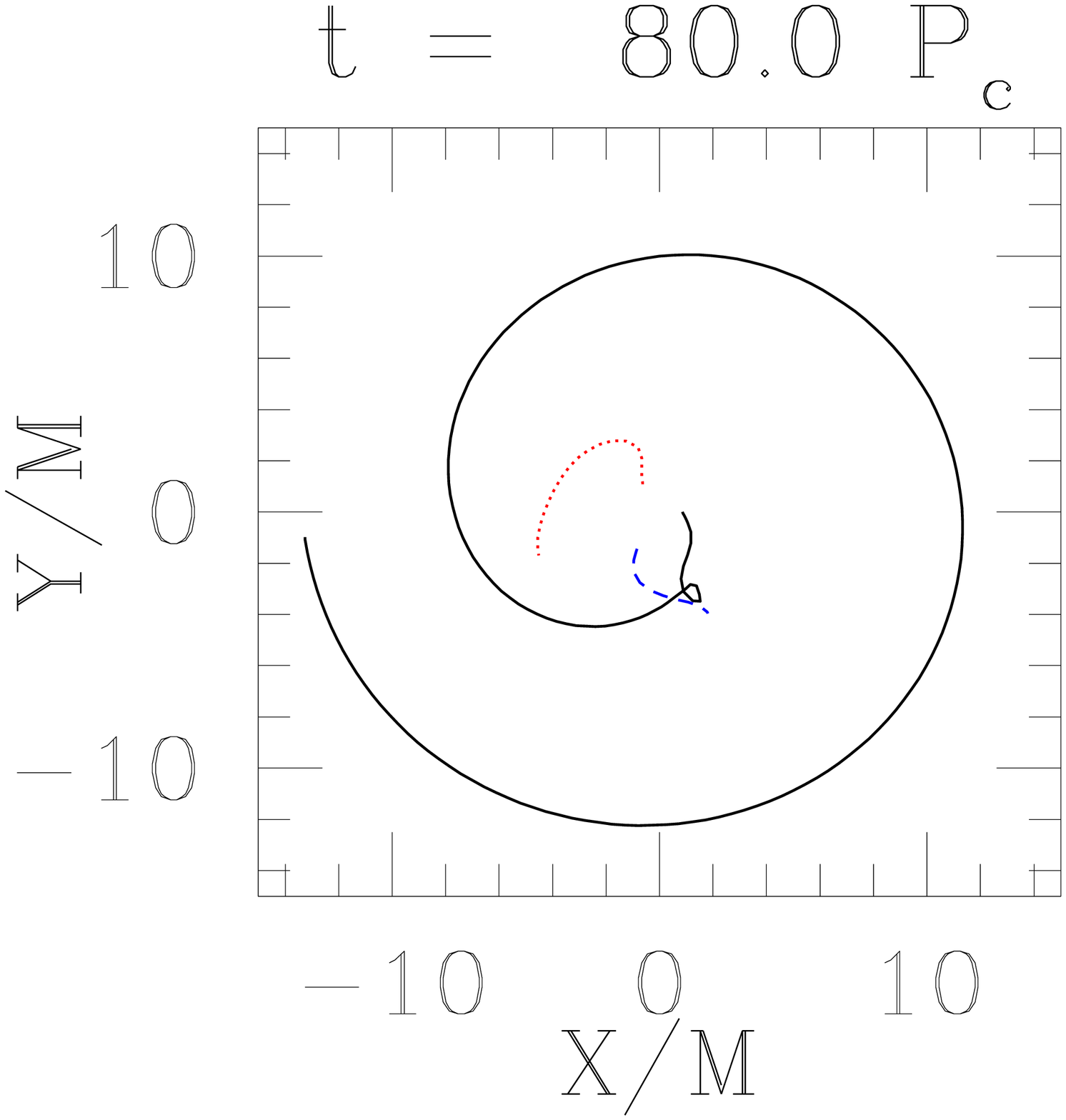}
\epsfxsize=1.9in
\leavevmode
\epsffile{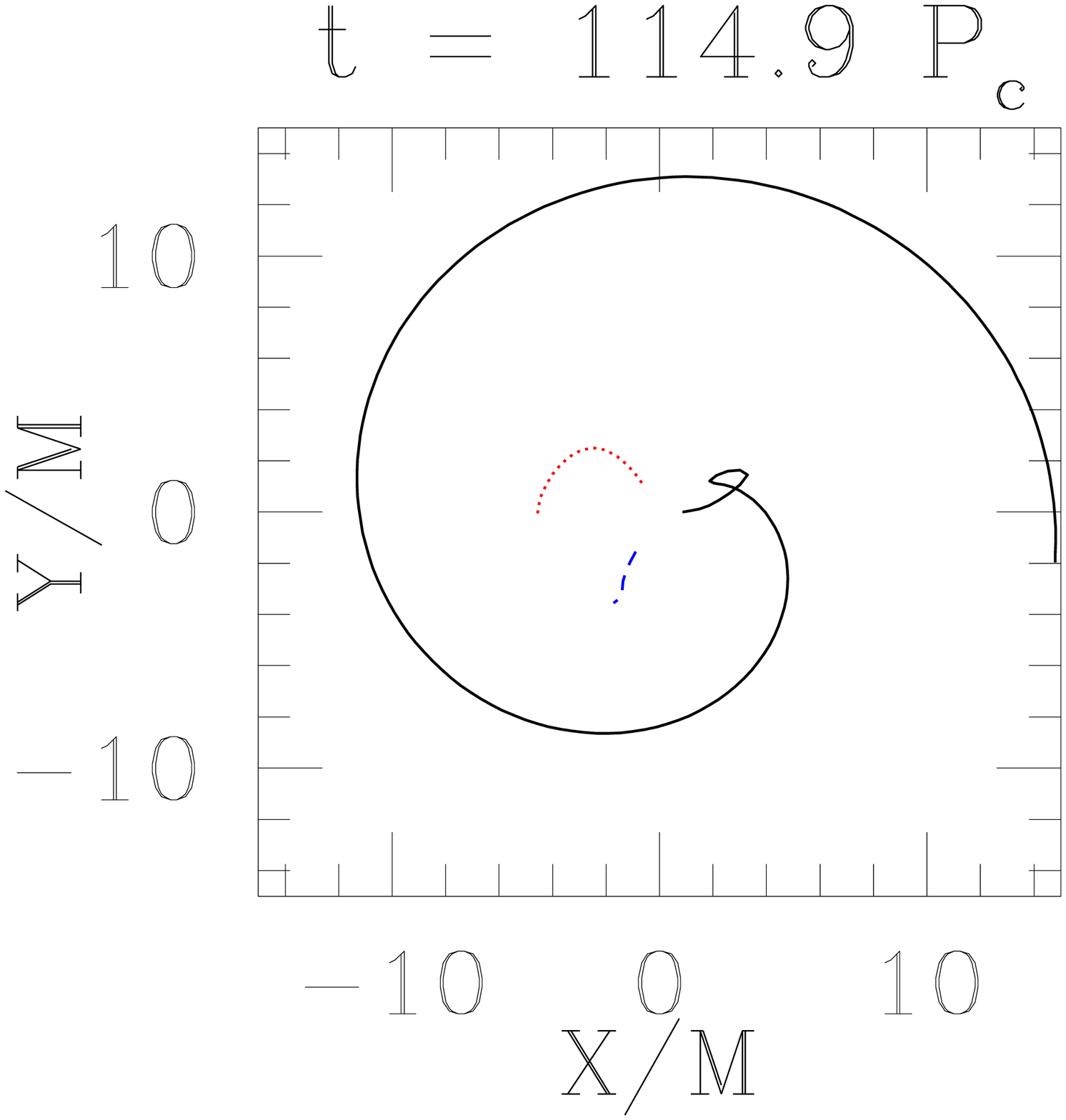}
\end{center}
\caption{Snapshots of the projected 3D magnetic field lines for star~B1.
\label{sBequat} }
\end{figure*}

Snapshots of the evolution in the $x$-$z$ plane are shown in 
Fig.~\ref{sBmerid}. The density contours for times $t=0$ through
$25.0 P_c$ show that angular momentum redistribution leads to the formation 
of a more compact star surrounded by a torus.  At $t=10 P_c$, the distortions 
of the magnetic field lines due to the MRI are clearly visible.  As the disk 
expands, magnetic field lines attached to this low density material open 
outward, eventually leading to the field structure seen in the last four 
times shown in Fig.~\ref{sBmerid}, for which some field lines are still 
confined inside the star while others have become somewhat collimated 
along the $z$-axis.   For $t \agt 35 P_c$, the density contours and poloidal
magnetic field lines change very little, indicating that the system has
reached an equilibrium state which is quite different from the initial state.
The effects of magnetic braking in this case are demonstrated by 
the series of snapshots in Fig.~\ref{sBequat}, which is analogous to 
Fig.~\ref{fig:StarA_Blines}.  The field lines become very tightly wound for 
$t \sim 10 P_c$ and relax at later times.  However, a significant toroidal 
field persists at late times when the system has essentially settled down to 
a final state.

\begin{figure}
\begin{center}
\epsfxsize=3.in
\leavevmode
\epsffile{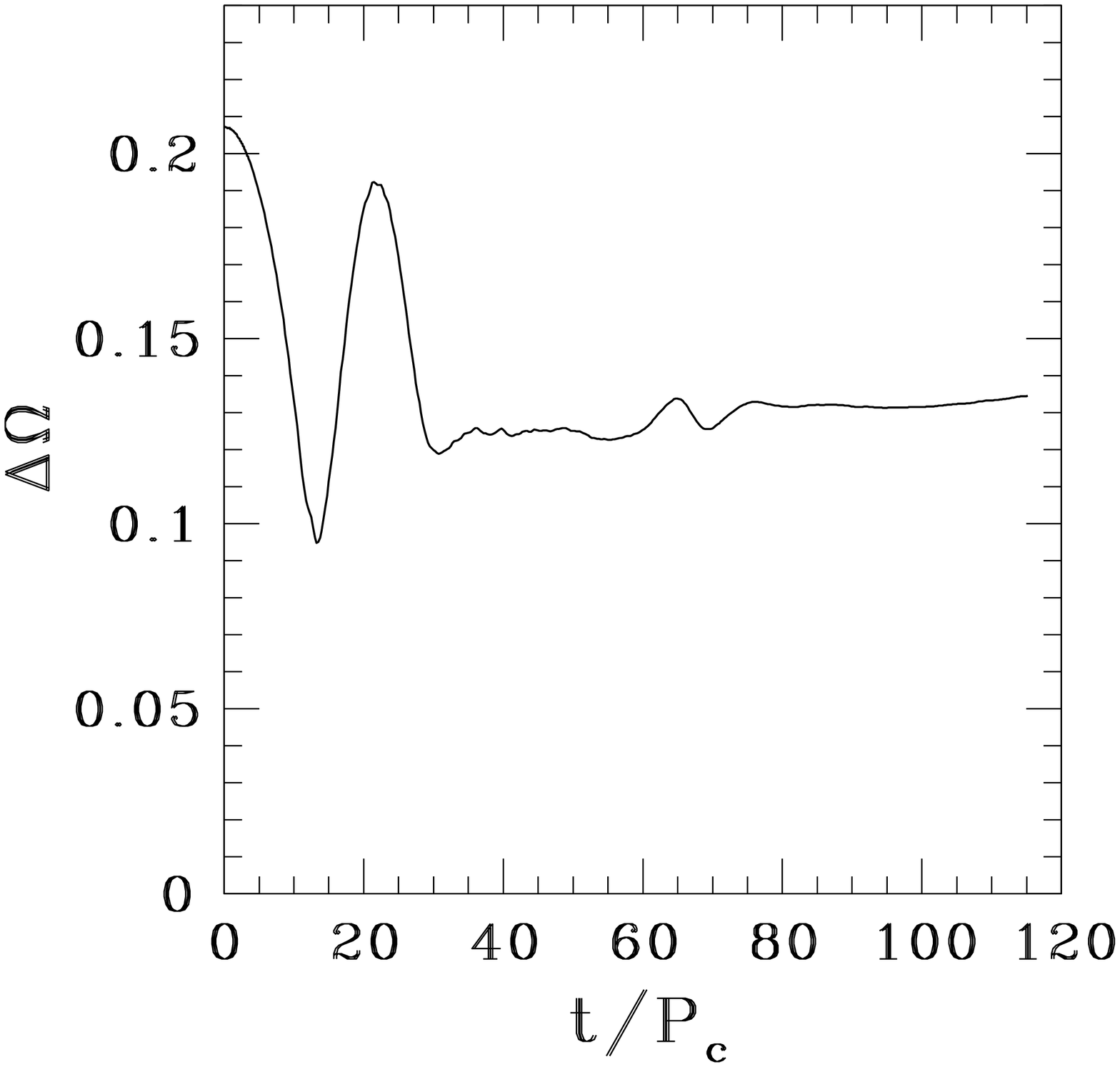}
\caption{Evolution of the degree of differential rotation $\Delta \Omega$ 
for star~B1. At late times, $\Delta \Omega$ approaches a non-zero constant 
value. This shows that the final equilibrium state of star~B1 is still 
differentially rotating.\label{deltaOmega}}
\end{center}
\end{figure}

\begin{figure}
\begin{center}
\epsfxsize=3.in
\epsffile{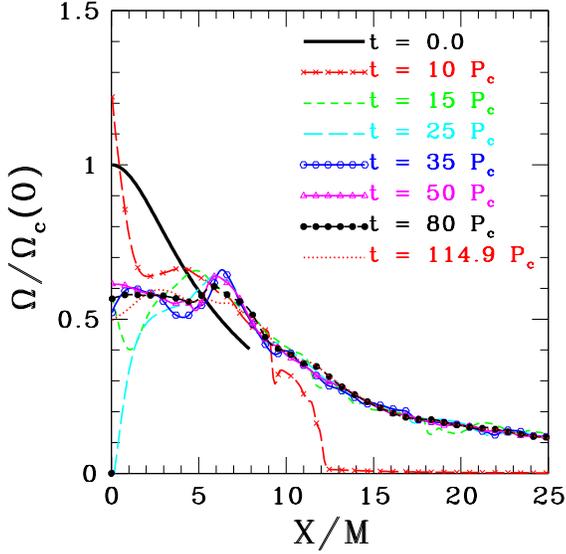}
\caption{Angular velocity profiles at selected times (corresponding
to the times in Fig.~\ref{sBmerid}) for star~B1.  
\label{omprof2}}
\end{center}
\end{figure}

\begin{figure}
\begin{center}
\epsfxsize=3.in
\leavevmode
\epsffile{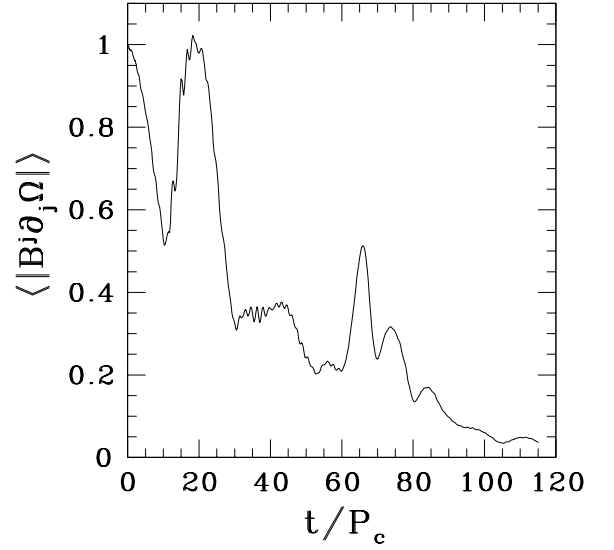}
\caption{Evolution of $\langle |B^j \partial_j \Omega| \rangle$ (normalized 
to unity at $t=0$). Note that this quantity drops toward zero at late time, 
indicating that the star is driven to a differentially rotating equilibrium 
state in which 
$\Omega$ is constant along the magnetic field lines. \label{bdotgradOm}}
\end{center}
\end{figure}

In order to understand the behavior of this case, we plot in 
Fig.~\ref{deltaOmega} the degree of differential rotation $\Delta \Omega$, 
defined as follows: 
\beq
\Delta\Omega = \frac{\sqrt{\langle\Omega^2\rangle - \langle\Omega\rangle^2}}
{\langle\Omega\rangle_0} \ ,
\eeq
where the angular brackets refer to density weighted averages 
($\langle f \rangle \equiv \int d^3 x \rho_* f / M_0$) and 
$\langle\Omega\rangle_0$ is the average angular velocity at $t=0$.  
Rather than approaching zero at late times, this quantity approaches a 
roughly constant value.  Thus, the equilibrium final state still has 
significant differential rotation.  
The evolution of the angular velocity profile for star~B1 is shown
in Fig.~\ref{omprof2} for the equatorial plane.  
Figures.~\ref{sBmerid} and~\ref{omprof2} suggest that the final state 
consists of a fairly uniformly rotating 
core surrounded by a differentially rotating torus.
However, this differential 
rotation no longer winds up the magnetic field lines (i.e., the 
toroidal field strength does not grow).  This is because the rotation 
profile has adjusted so that $\Omega$ is approximately constant along 
magnetic field lines.  This is demonstrated in Fig.~\ref{bdotgradOm}, 
which shows that 
\beq
\langle |B^j \partial_j \Omega| \rangle \to 0  
\eeq
at late times.  Since the rotation 
profile is adapted to the magnetic field structure, a stationary final 
state is reached which allows differential rotation and a nonzero 
toroidal field.  

Since the final state is still differentially rotating and is threaded
with magnetic fields, this configuration must be checked for the presence 
of the MRI.  From the linear (and local) analysis discussed in 
Section~\ref{overview}, we found that the predicted wavelength for the 
fastest growing mode is $\sim 2-3 M$ at late times, whereas the radius of 
the final star is $\sim 6 M$.  (Since the local analysis does not 
take into account gradients in the vertical direction, it is qualitative
at best in this regime.  However, this does suggest that 
$\lambda_{\rm max} \sim R$.)  Thus, the magnetic field is no longer weak,
and the MRI is likely suppressed.  This is corroborated by the fact that
we do not see any rapid magnetic field growth at late times.

\begin{figure}
\begin{center}
\epsfxsize=3.in
\leavevmode
\epsffile{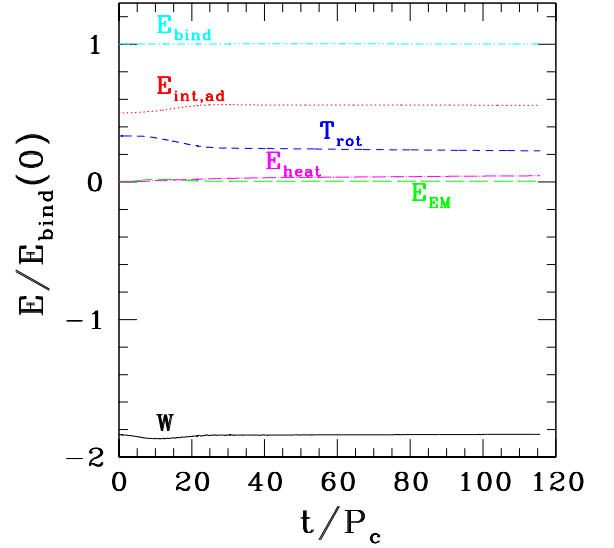}
\caption{Components of the energy vs.\ time for star~B1. 
All energies are normalized to the binding energy at $t=0$.
In the evolution, $E_{\rm bind}$ should be nearly conserved.
\label{energies2}}
\end{center}
\end{figure}

\begin{figure}
\begin{center}
\epsfxsize=3.in
\leavevmode
\epsffile{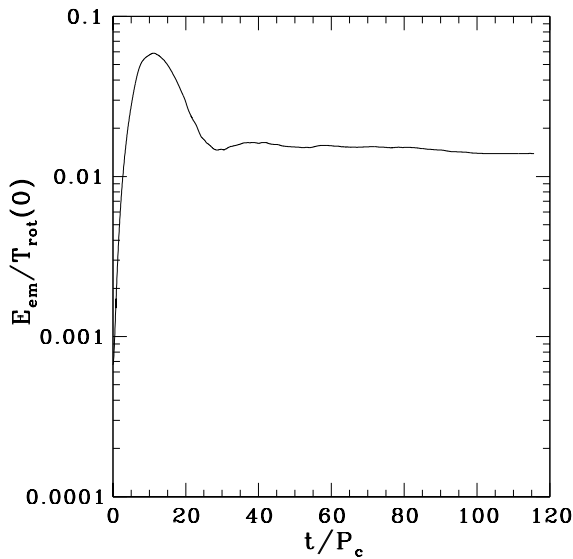}
\caption{Evolution of magnetic energy $E_{\rm EM}$ for star~B1. The energy 
is normalized by the initial rotational kinetic energy of the star, 
$T_{\rm rot}(0)$.
\label{E_mag2}}
\end{center}
\end{figure}

Figure~\ref{energies2} shows the evolution of various energies. As in 
the case of star~A, the magnetic energy $E_{\rm EM}$ shows a much 
smaller change in amplitude than $T_{\rm rot}$, $E_{\rm int,ad}$ and $W$. 
These results are very different from those found in~\cite{Shapiro,zys06} 
and from those of the star~B2 evolution (see the next subsection), where 
the change in $E_{\rm EM}$ is comparable to the change in $T_{\rm rot}$. 
This is probably because star~B1 is ultraspinning, 
in contrast to the ``normal'' models 
in~\cite{Shapiro,zys06} and star~B2. 
Here the seed magnetic field in star~B1 causes a substantial 
change (on a secular Alfv\'en timescale) in the structure of the star. 
The energies 
$E_{\rm int,ad}$, $T_{\rm rot}$ and $W$ readjust to the values of the new 
configuration, which is significantly different from the initial state. 
On the other hand, star~B2 and the models studied in~\cite{Shapiro,zys06} show 
little or no change in the density profile. As a result, a decrease in 
$T_{\rm rot}$ results in an increase in $E_{\rm EM}$. 

Figure~\ref{E_mag2} shows the evolution of the magnetic energy $E_{\rm EM}$, 
normalized to the initial rotational kinetic energy of the star, 
$T_{\rm rot}(0)$. The value of $E_{\rm EM}/T_{\rm rot}(0)$ rises from 
its initial value of $6.7\times 10^{-4}$ to a peak of $\sim 0.06$ 
(the corresponding field strength is about 90 times the initial
field strength)  mainly due to 
magnetic braking. Then it gradually decreases to the equilibrium value of 
0.014. The final magnetic field strength $|B_{\rm final}|$ is about 4.5 times 
the initial value. In cgs units, we find that for the initial field 
considered here,
\begin{equation}
  |B_{\rm final}| \sim 10^{17} \left(\frac{2 M_{\odot}}{M}\right)~{\rm G} \ .
\end{equation}
This field is comparable to the field strength of a magnetar. 
Since the strength 
of the initial seed magnetic field is much smaller than the strength when 
it saturates, it is possible that the final equilibrium state will be the same 
even if the initial seed field is much smaller than the present value. 
If this is true, a new-born neutron star with mass and angular momentum 
distribution similar to star~B1 is likely to end up as a magnetar due 
to MHD processes.

\subsection{Star B2}
\label{sec:starB2}

\begin{figure}
\vspace{-4mm}
\begin{center}
\epsfxsize=3.in
\leavevmode
\epsffile{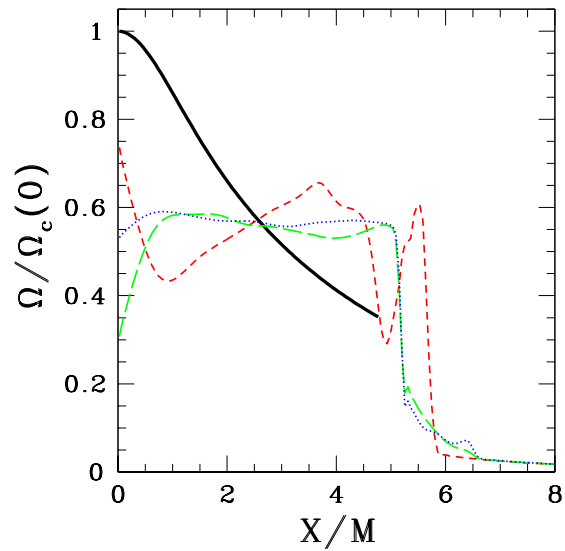}
\caption{Angular velocity profiles in the equatorial plane for star~B2 
at times $t=0$ 
[thick solid (black) line], $t=8.1P_c \approx 1t_A$ [dashed (red) line], 
$t=32.5P_c \approx 4 t_A$ [long dashed (green) line], and 
$t=46.5P_c \approx 5.8 t_A$ [dotted (blue) line]. At late time 
($t \gtrsim 30P_c \approx 4 t_A$), the bulk 
of the star is nearly uniformly rotating.
\label{omprof_b2}}
\end{center}
\end{figure}

\begin{figure}
\vspace{-4mm}
\begin{center}
\epsfxsize=1.8in
\leavevmode
\hspace{-0.7cm}\epsffile{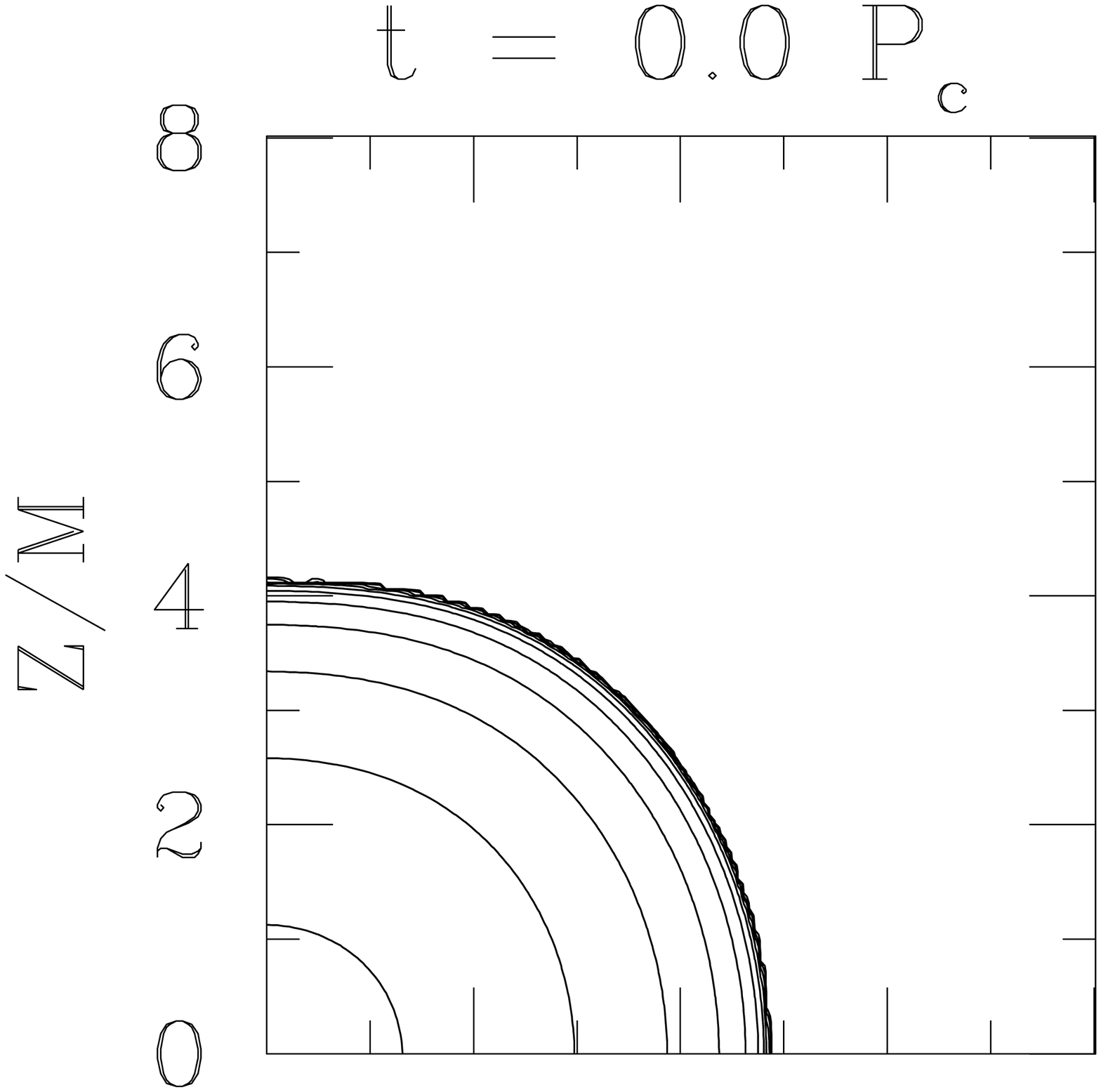}
\epsfxsize=1.8in
\leavevmode
\hspace{-0.5cm}\epsffile{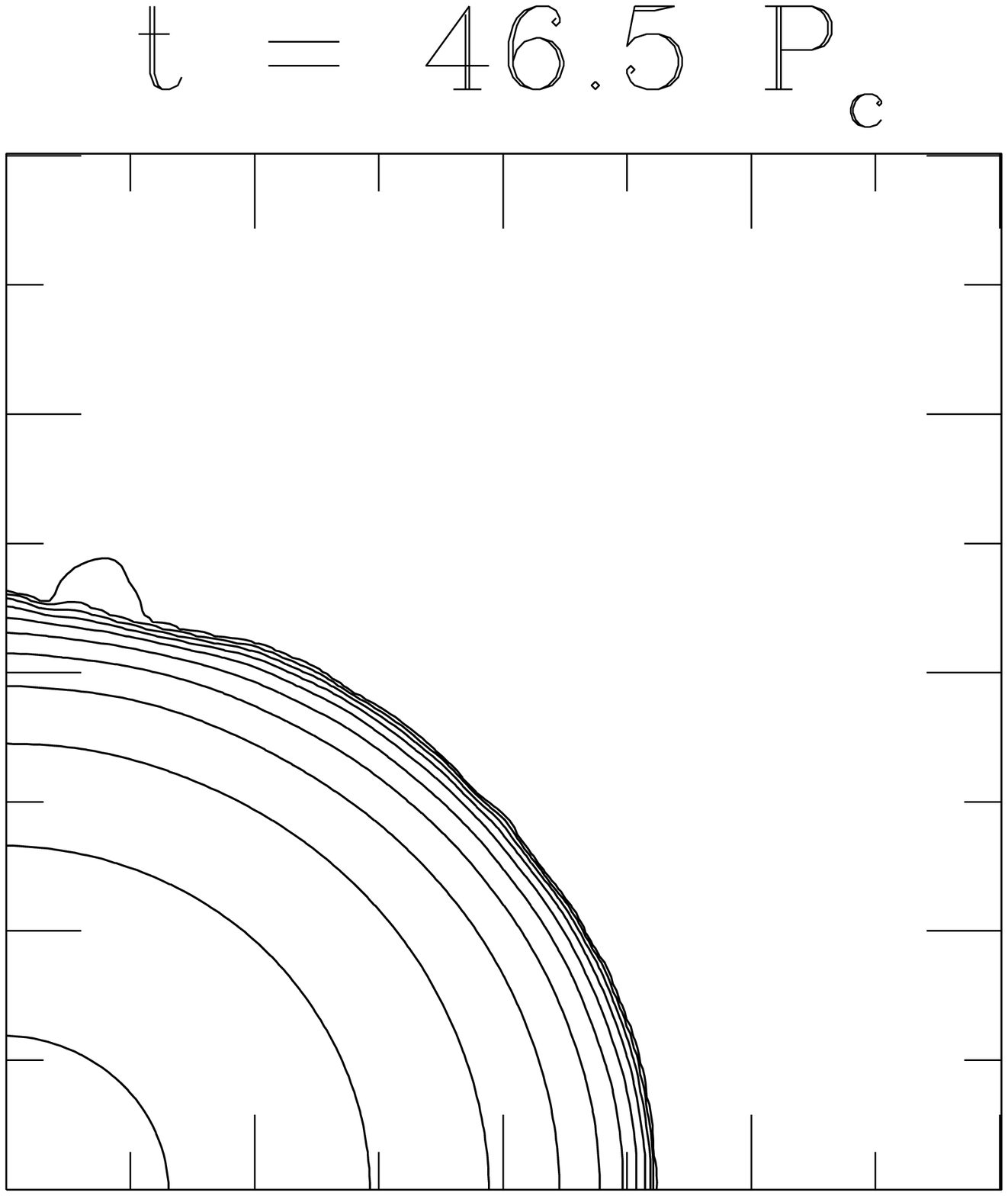} \\
\vspace{-0.3cm}
\epsfxsize=1.8in
\leavevmode
\hspace{-0.7cm}\epsffile{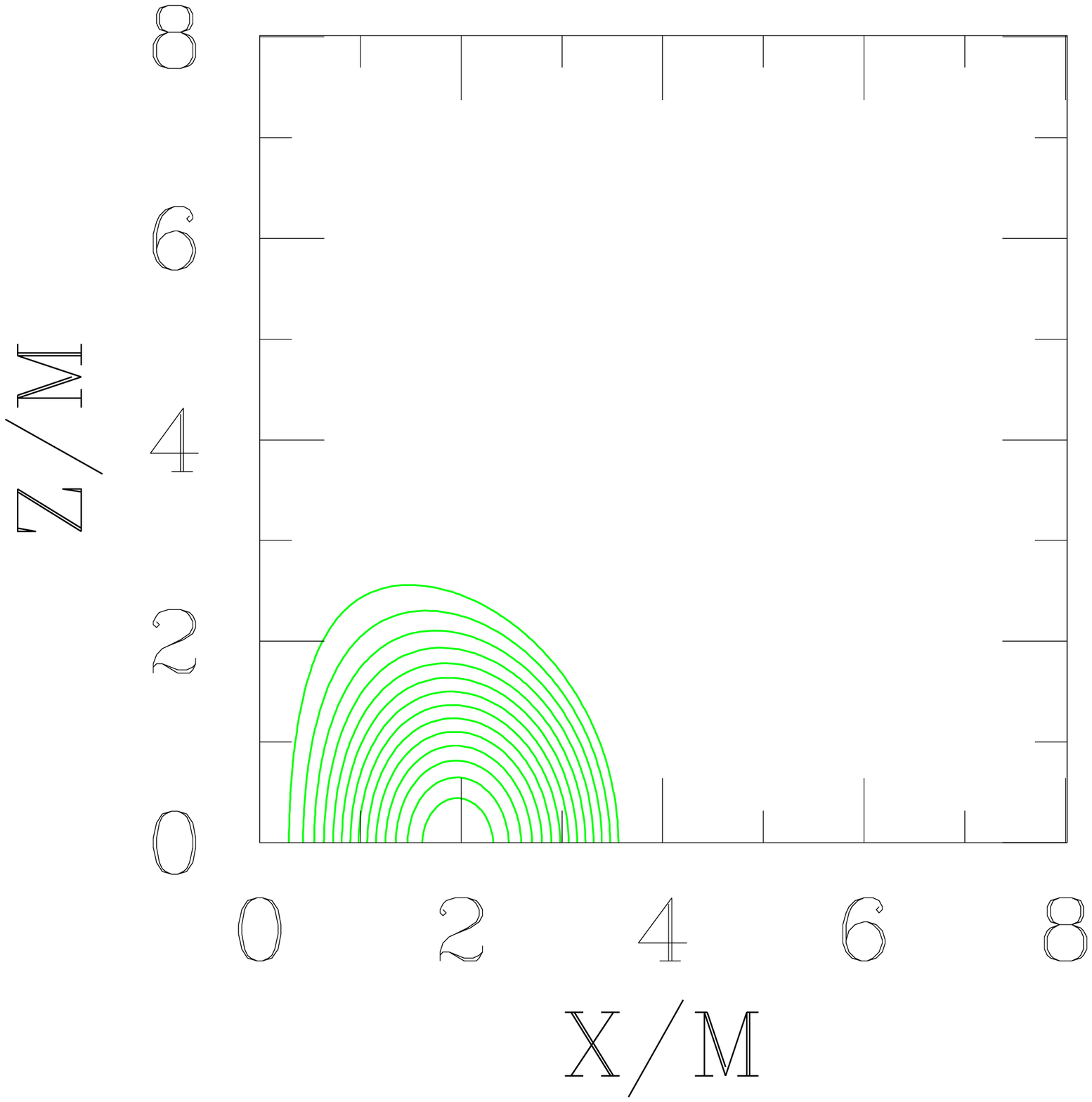}
\epsfxsize=1.8in
\leavevmode
\hspace{-0.5cm}\epsffile{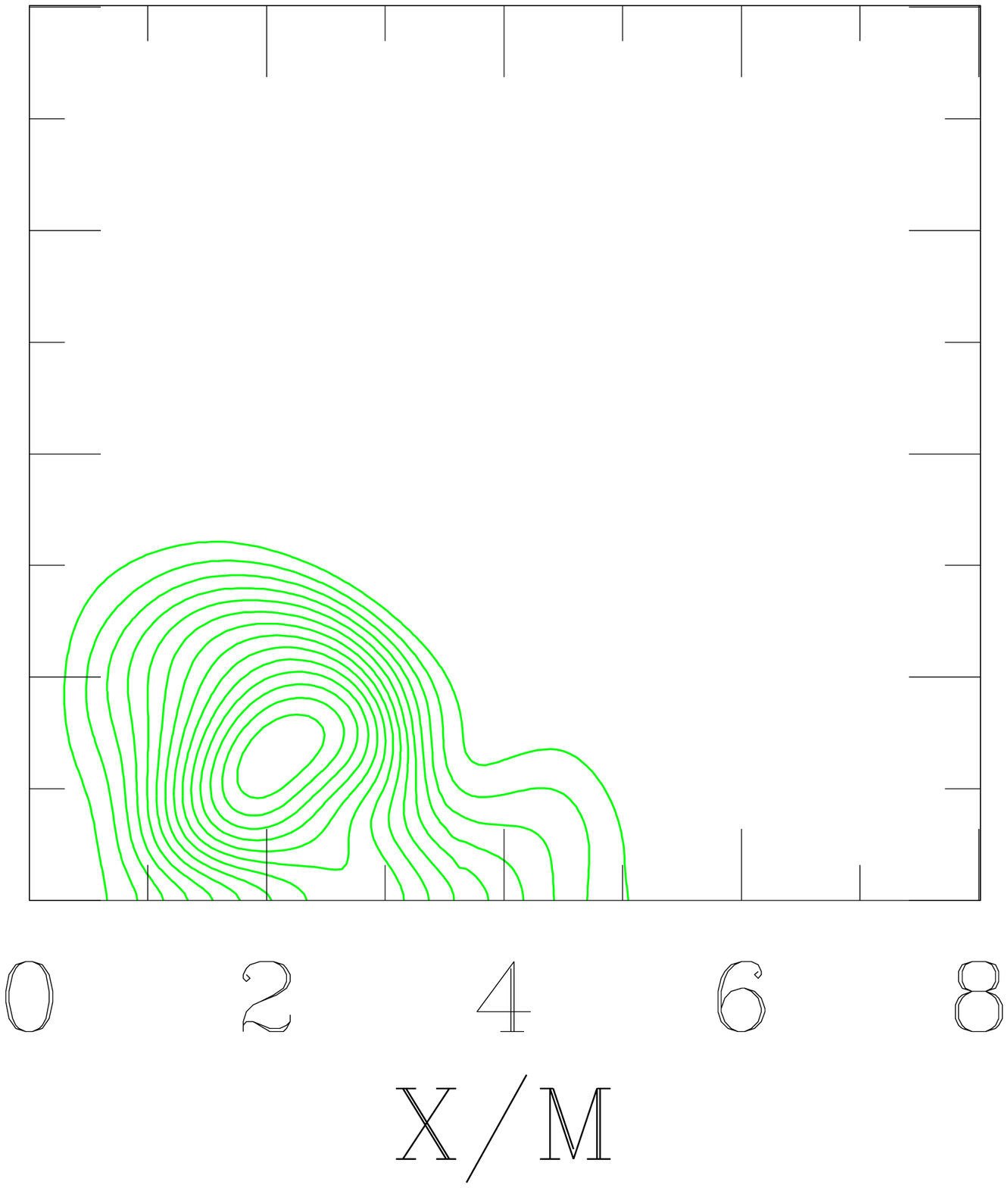}
\caption{Snapshots of the rest-mass density contours and poloidal
magnetic field lines for star~B2 at times $t=0$ and $t=46.5P_c$. 
The first row shows snapshots of the rest-mass density
contours on the meridional plane. The second 
row shows the corresponding field lines 
for the poloidal magnetic field at the same times.
The density contours are drawn for $\rho_0/\rho_{\rm max}(0)=
10^{-0.36 i - 0.09}~(i=0$--10), where $\rho_{\rm max}(0)$ is the
maximum rest-mass density at $t=0$.
The field lines are drawn for $A_{\varphi} = A_{\varphi,\rm min}
+ (A_{\varphi,\rm max} - A_{\varphi,\rm min}) i/15~(i=1$--14),
where $A_{\varphi,\rm max}$ and $A_{\varphi,\rm min}$ are the maximum
and minimum values of $A_{\varphi}$, respectively, at the given time. 
The meridional components of the velocity (which are zero initially) 
at $t=46.5P_c$ are very small and so are not shown here.
\label{fig:cont_b2}}
\end{center}
\end{figure}

\begin{figure}
\vspace{-4mm}
\begin{center}
\epsfxsize=3.in
\leavevmode
\epsffile{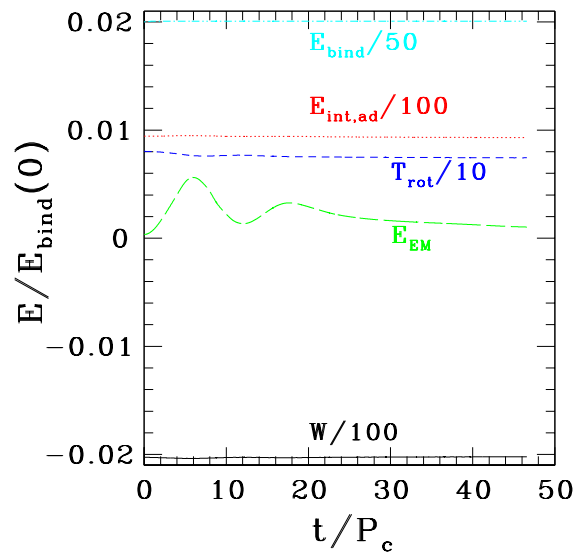}
\caption{Components of the energy vs.\ time for star~B2.
All energies are normalized to the binding energy at $t=0$.  
Some quantities are normalized by an additional numerical factor
(as indicated) to ease visualization.
\label{fig:energy_b2}}
\end{center}
\end{figure}

\begin{figure}
\vspace{-4mm}
\begin{center}
\epsfxsize=3.in
\leavevmode
\epsffile{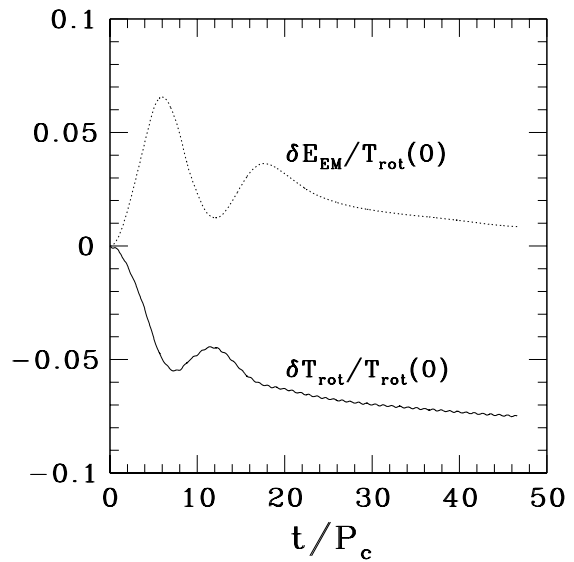}
\caption{Change of $T_{\rm rot}$ and $E_{\rm mag}$ vs.\ time. 
Here, $\delta T_{\rm rot} \equiv T_{\rm rot}-T_{\rm rot}(0)$ and 
$\delta E_{\rm mag} \equiv E_{\rm mag}-E_{\rm mag}(0)$. 
\label{fig:Trot_Emag_b2}}
\end{center}
\end{figure}

Both stars~B1 and B2 are nonhypermassive. However, star~B1 is 
ultraspinning, whereas B2 is normal. 
We evolve this star with a seed magnetic field strength $C=2.5\times 
10^{-3}$. Our simulation shows that this star evolves 
to a uniformly rotating configuration with little structural 
change (see Figs.~\ref{omprof_b2} and~\ref{fig:cont_b2}). 

Figure~\ref{fig:cont_b2} shows the density contours and poloidal magnetic 
field at the initial time ($t=0$) and at $t=46.5P_c\approx 5.8t_A$. 
We see that the density profile of the star does not change 
appreciably. This is not surprising since the main effect of 
the MHD processes is to redistribute the angular momentum 
inside the star. However, the rotational kinetic energy of 
star~B2 is not very large (the initial $T/|W|=0.040$).
Hence, the change of the centrifugal force inside the star 
as a result of angular momentum transport does not 
disturb the initial equilibrium significantly, unlike the cases of 
stars~A, B1 and C (see the next section).

Figure~\ref{fig:energy_b2} shows the evolution of various energy 
components. Unlike stars~A, B1 and C (see Fig.~\ref{energiesC}), 
the magnetic energy $E_{\rm EM}$ and rotational kinetic energy 
$T_{\rm rot}$ show the largest fractional variations. The adiabatic internal 
energy $E_{\rm int,ad}$ and gravitational potential energy $W$ have 
very small fractional changes. This is because the configuration of the 
star does not deviate significantly from the initial equilibrium 
(see Fig.~\ref{fig:cont_b2}). A large fraction of the growth of magnetic 
energy comes from the rotational kinetic energy (see 
Fig.~\ref{fig:Trot_Emag_b2}). This is similar to the results reported 
in~\cite{Shapiro,zys06}. 

\subsection{Star C}\label{StarC}

\begin{figure*}[t]
\begin{center}
\epsfxsize=2.1in
\leavevmode
\epsffile{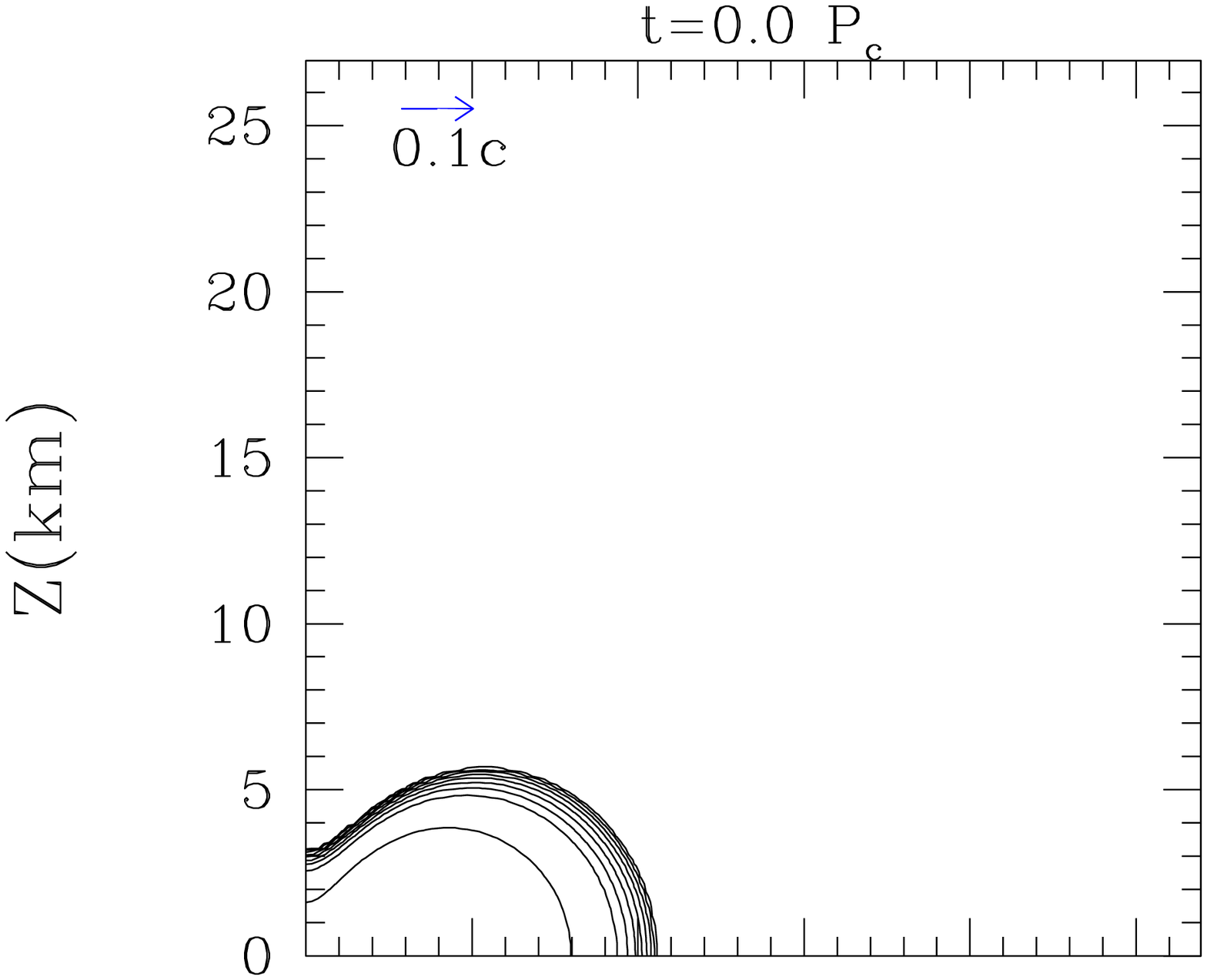}
\epsfxsize=2.1in
\leavevmode
\hspace{-1.65cm}\epsffile{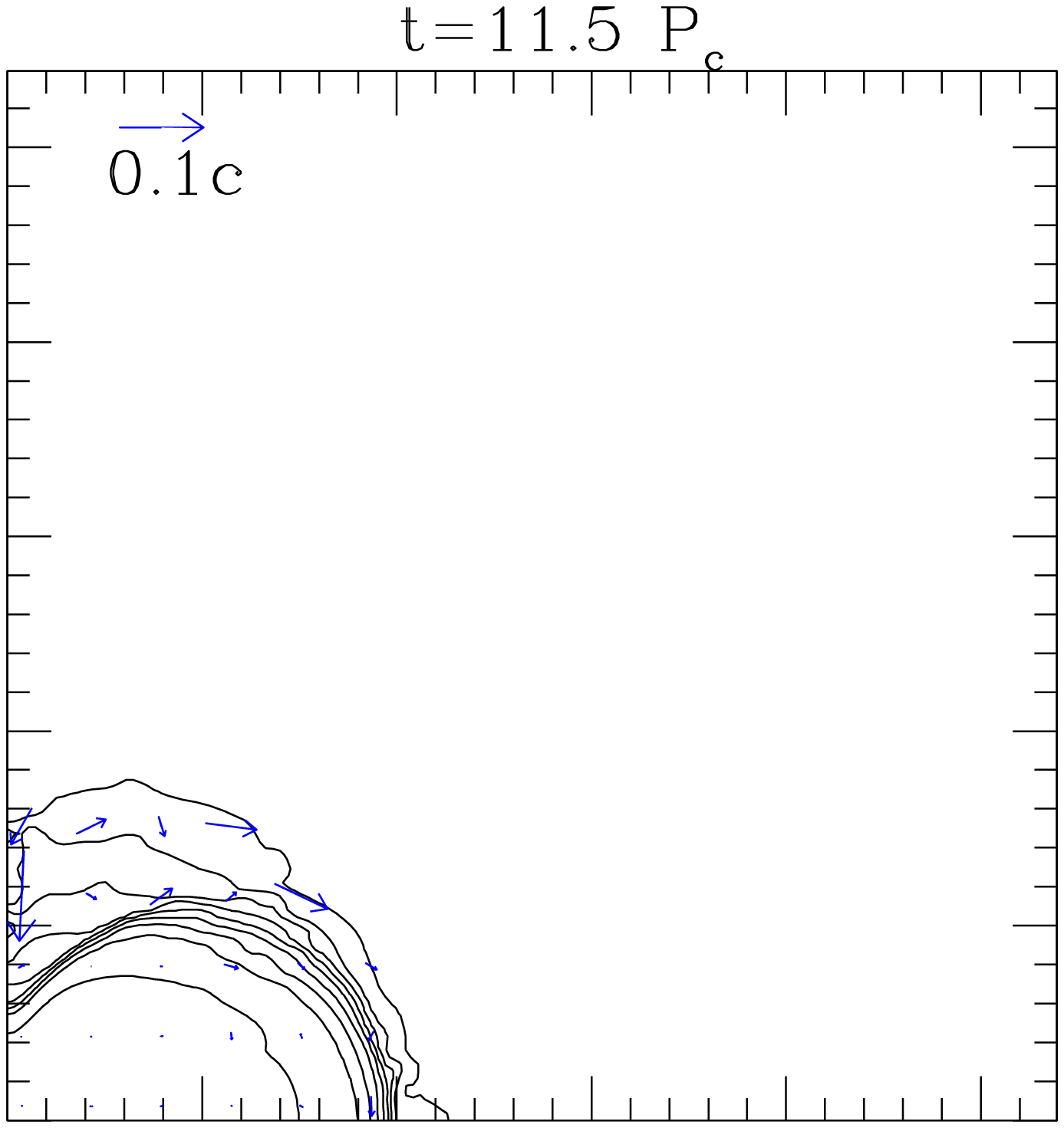}
\epsfxsize=2.1in
\leavevmode
\hspace{-1.65cm}\epsffile{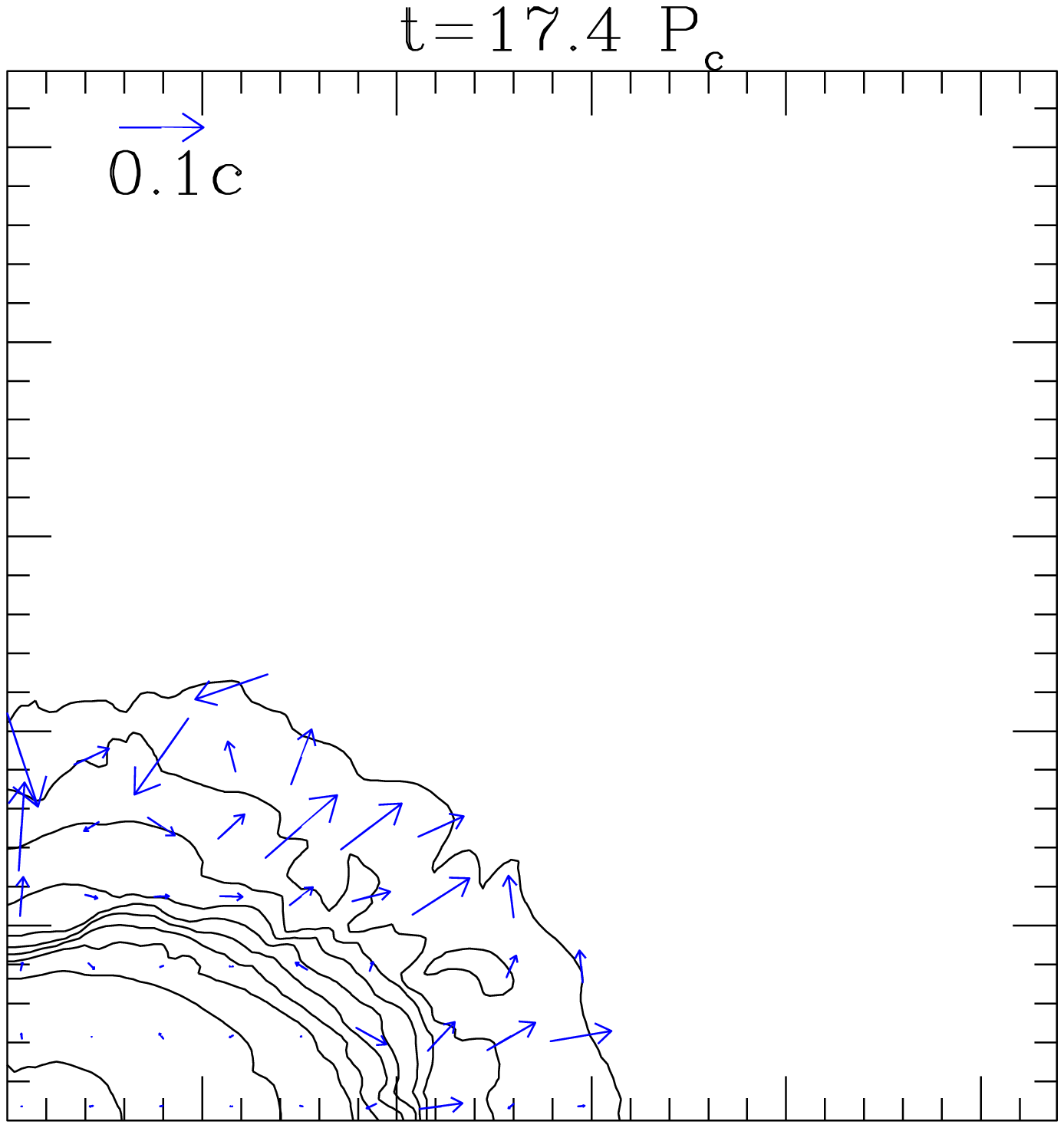}
\epsfxsize=2.1in
\leavevmode
\hspace{-1.65cm}\epsffile{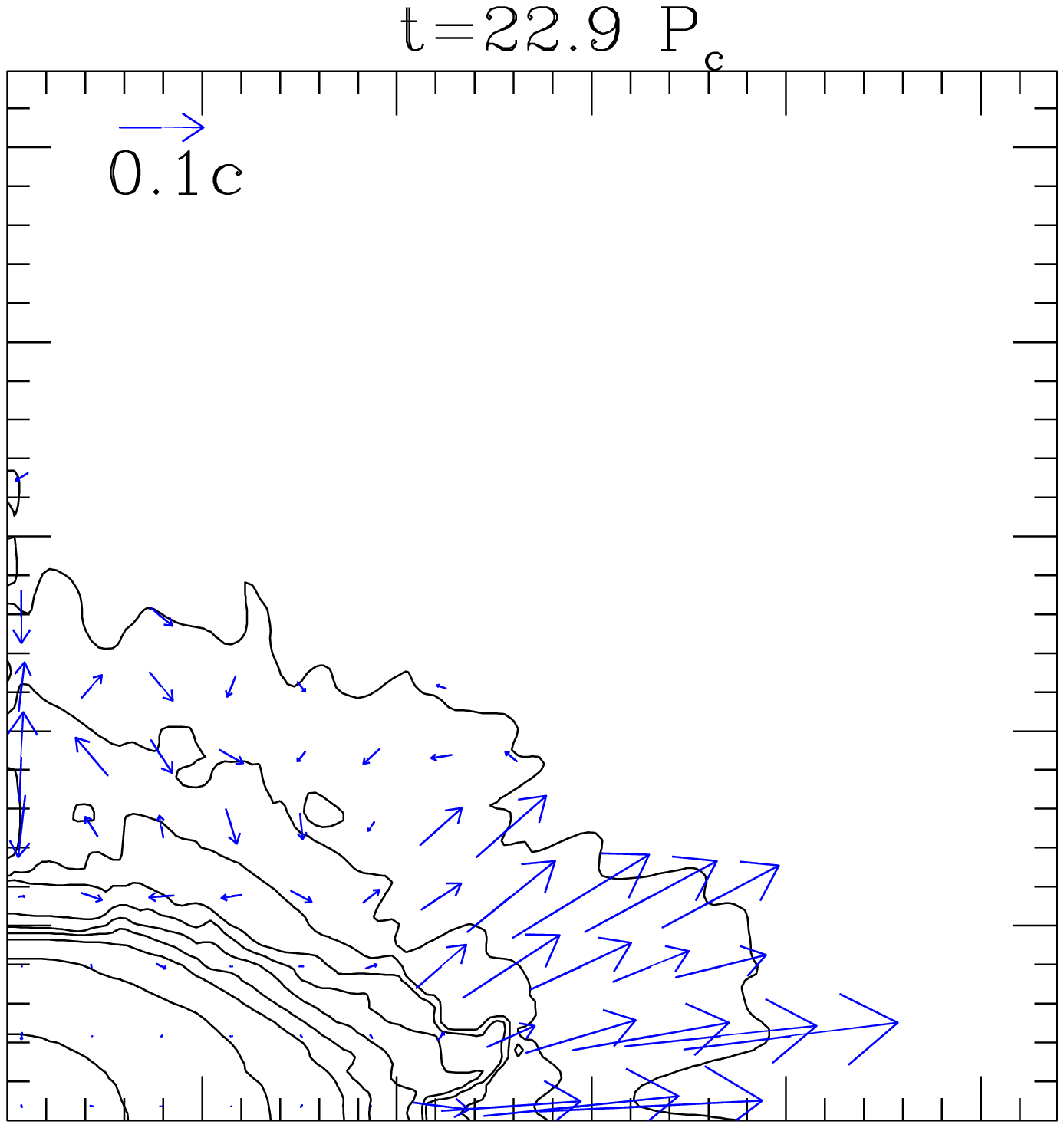}\\
\vspace{-1.7cm}
\epsfxsize=2.1in
\leavevmode
\epsffile{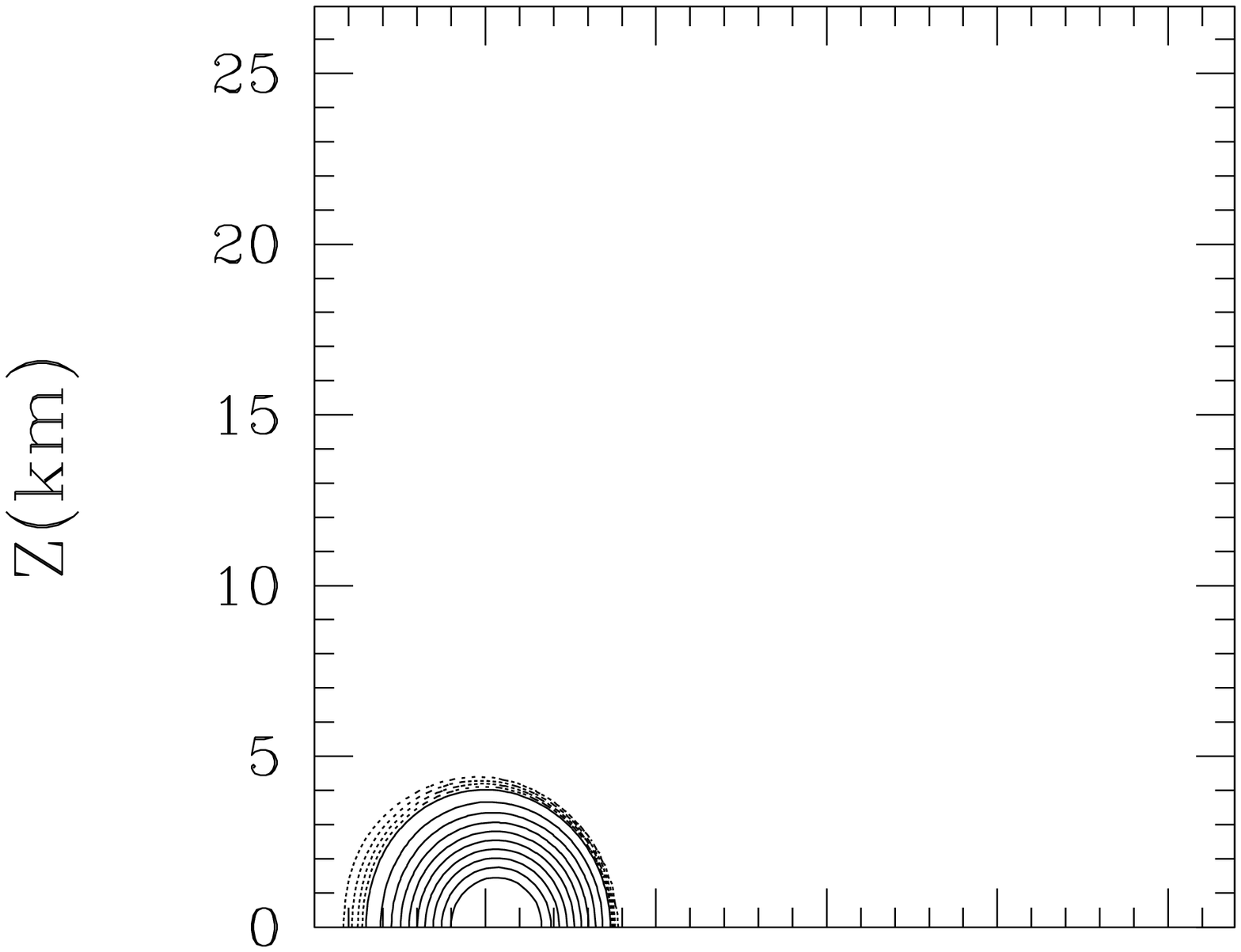}
\epsfxsize=2.1in
\leavevmode
\hspace{-1.65cm}\epsffile{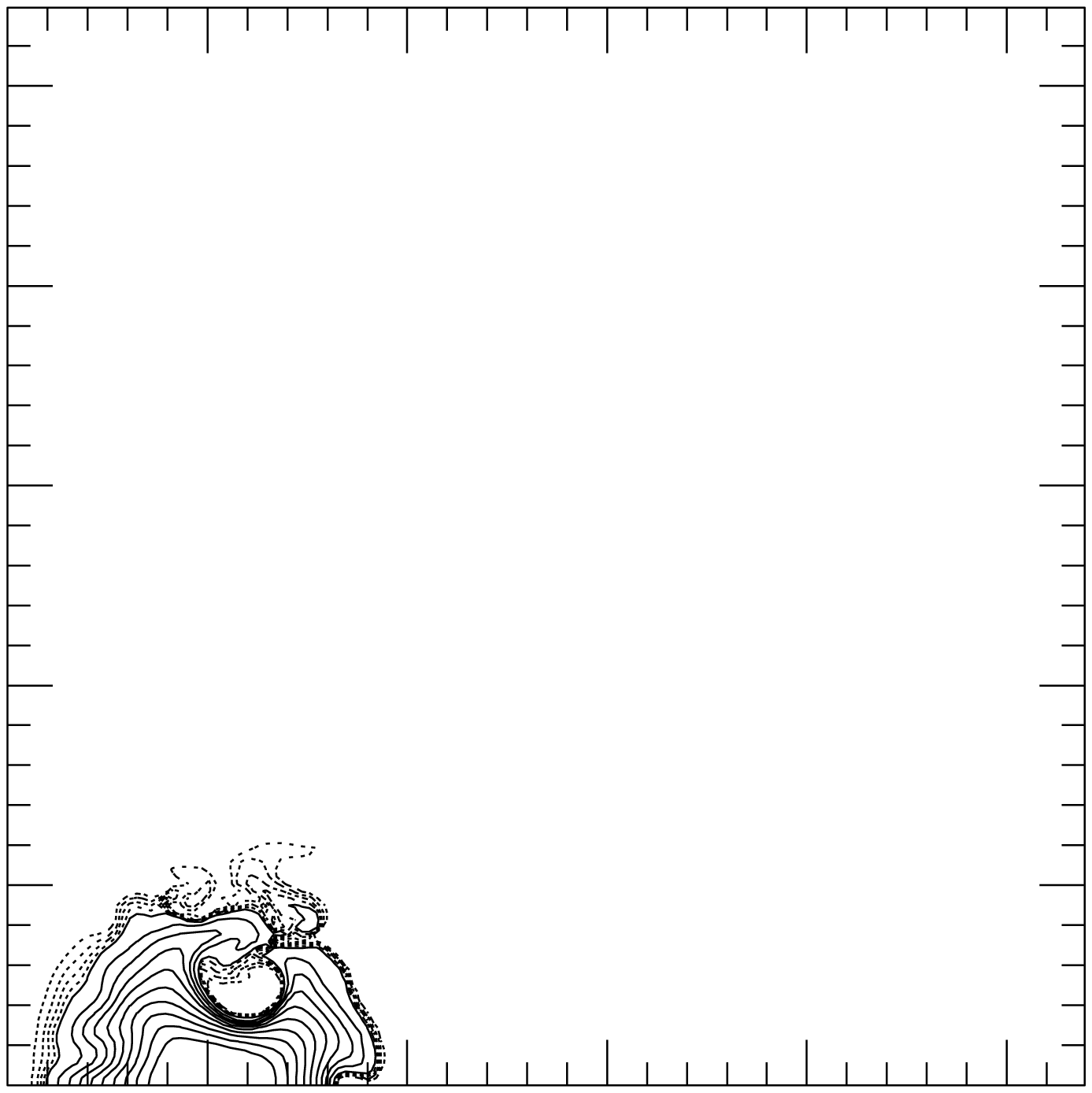}
\epsfxsize=2.1in
\leavevmode
\hspace{-1.65cm}\epsffile{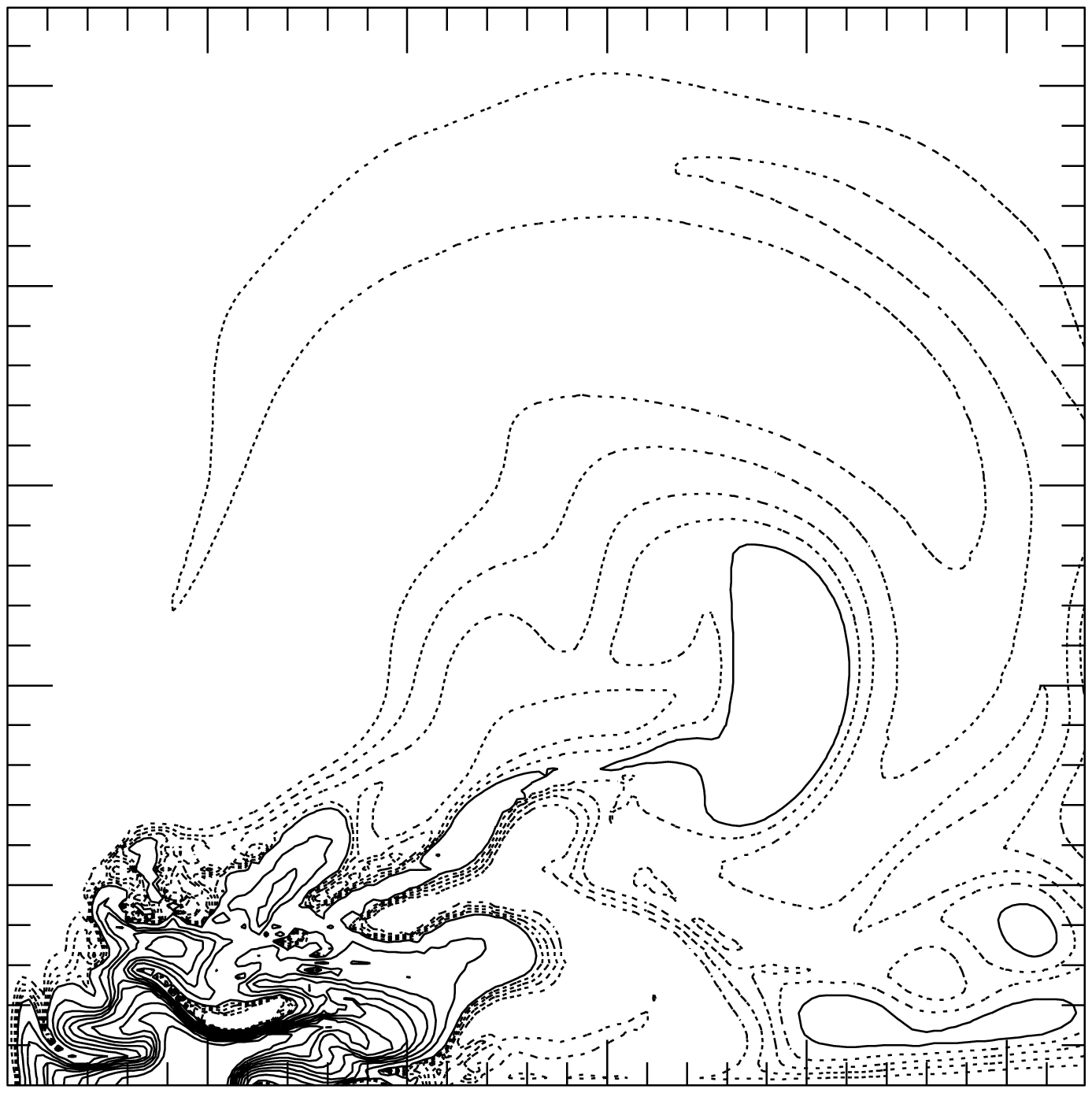}
\epsfxsize=2.1in
\leavevmode
\hspace{-1.65cm}\epsffile{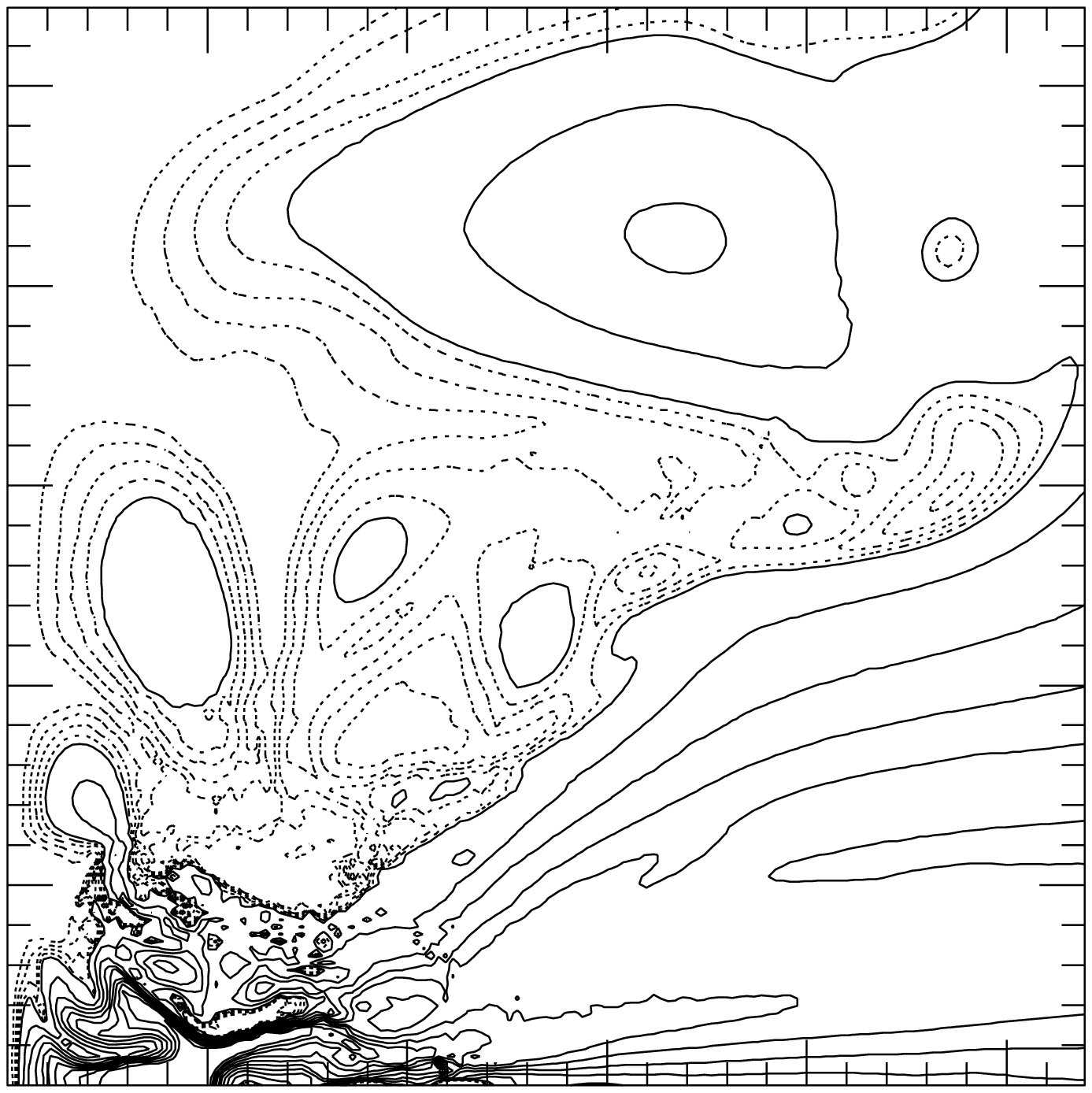}\\
\vspace{-1.cm}
\epsfxsize=2.1in
\leavevmode
\epsffile{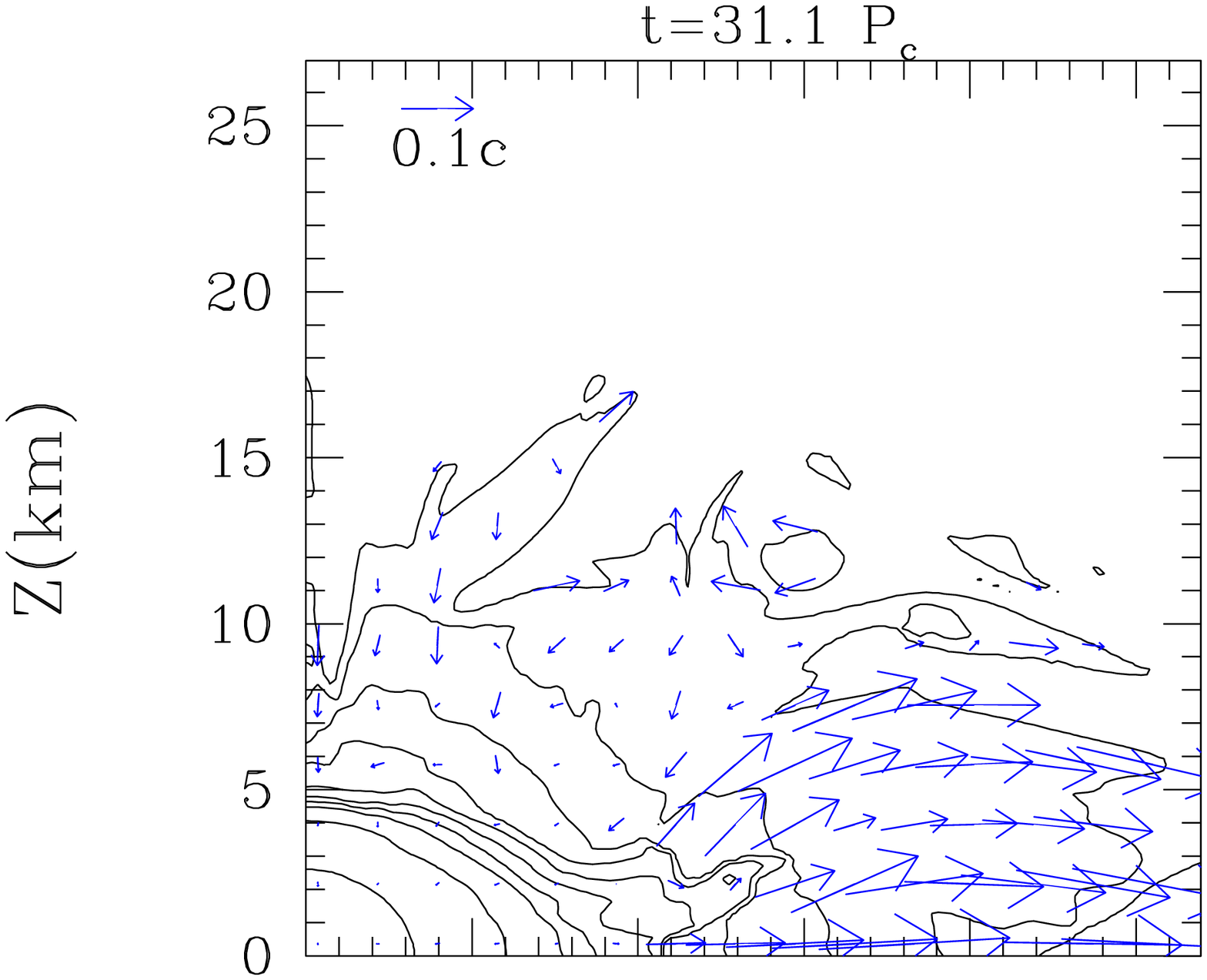}
\epsfxsize=2.1in
\leavevmode
\hspace{-1.65cm}\epsffile{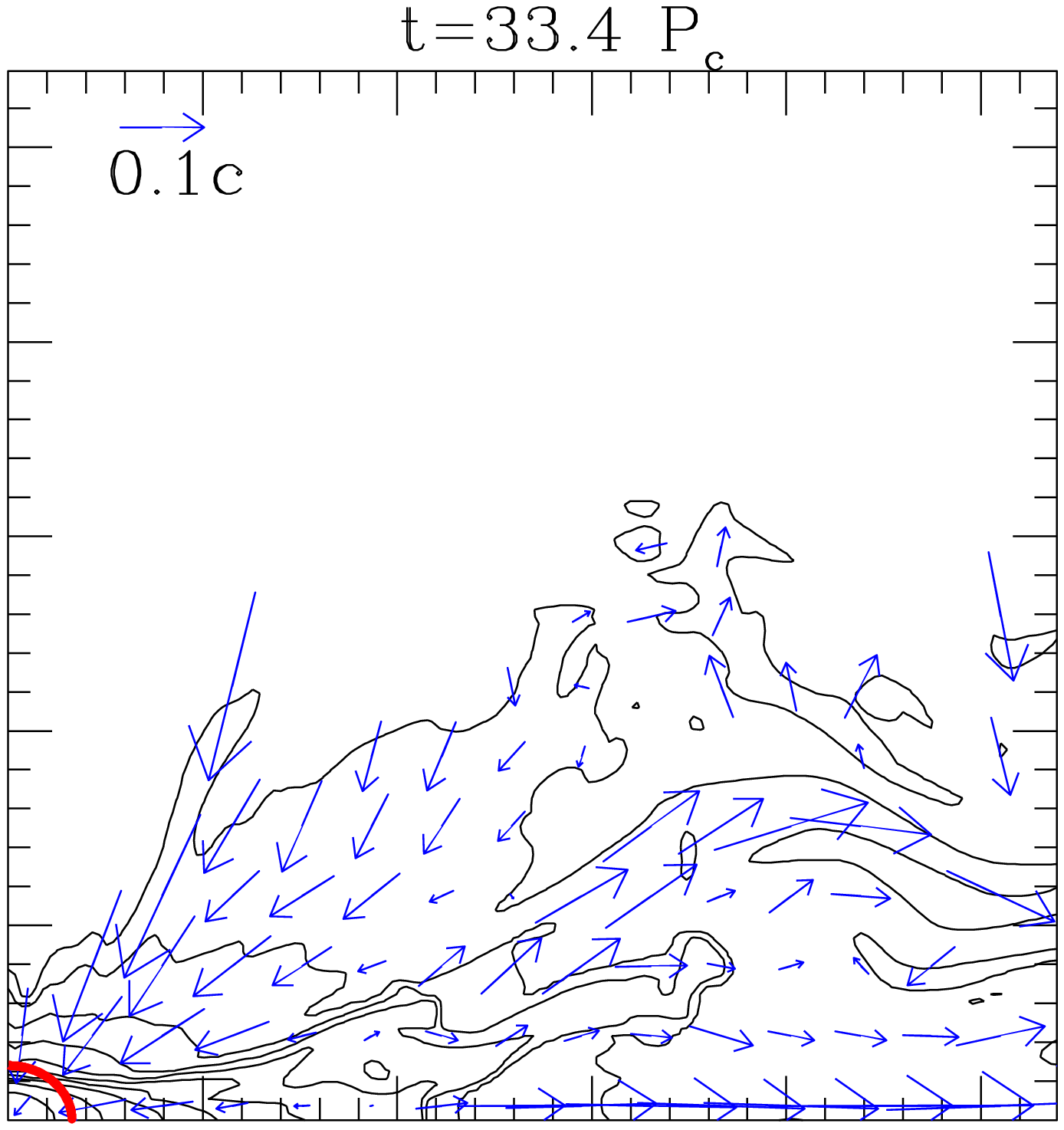}
\epsfxsize=2.1in
\leavevmode
\hspace{-1.65cm}\epsffile{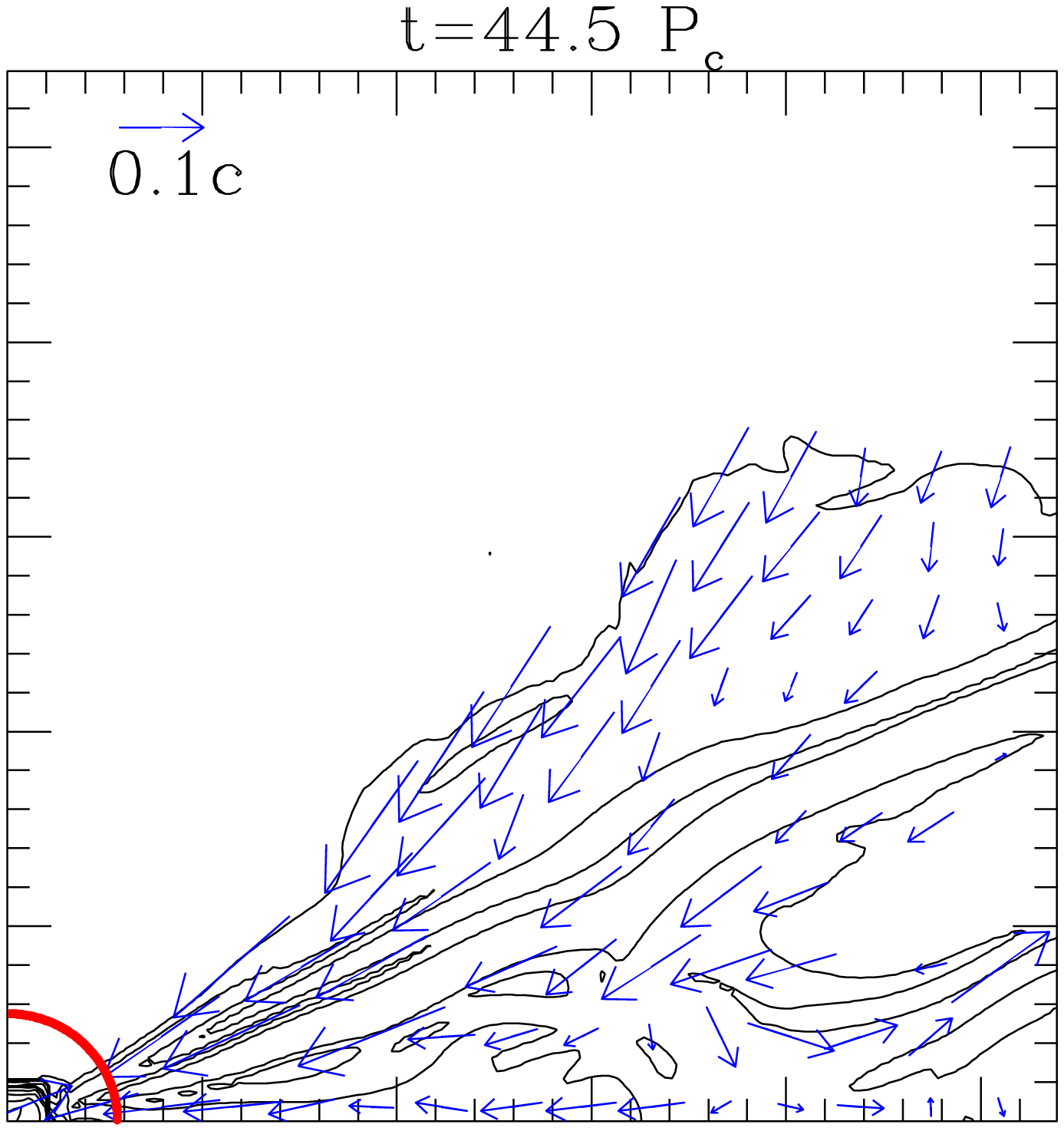}
\epsfxsize=2.1in
\leavevmode
\hspace{-1.65cm}\epsffile{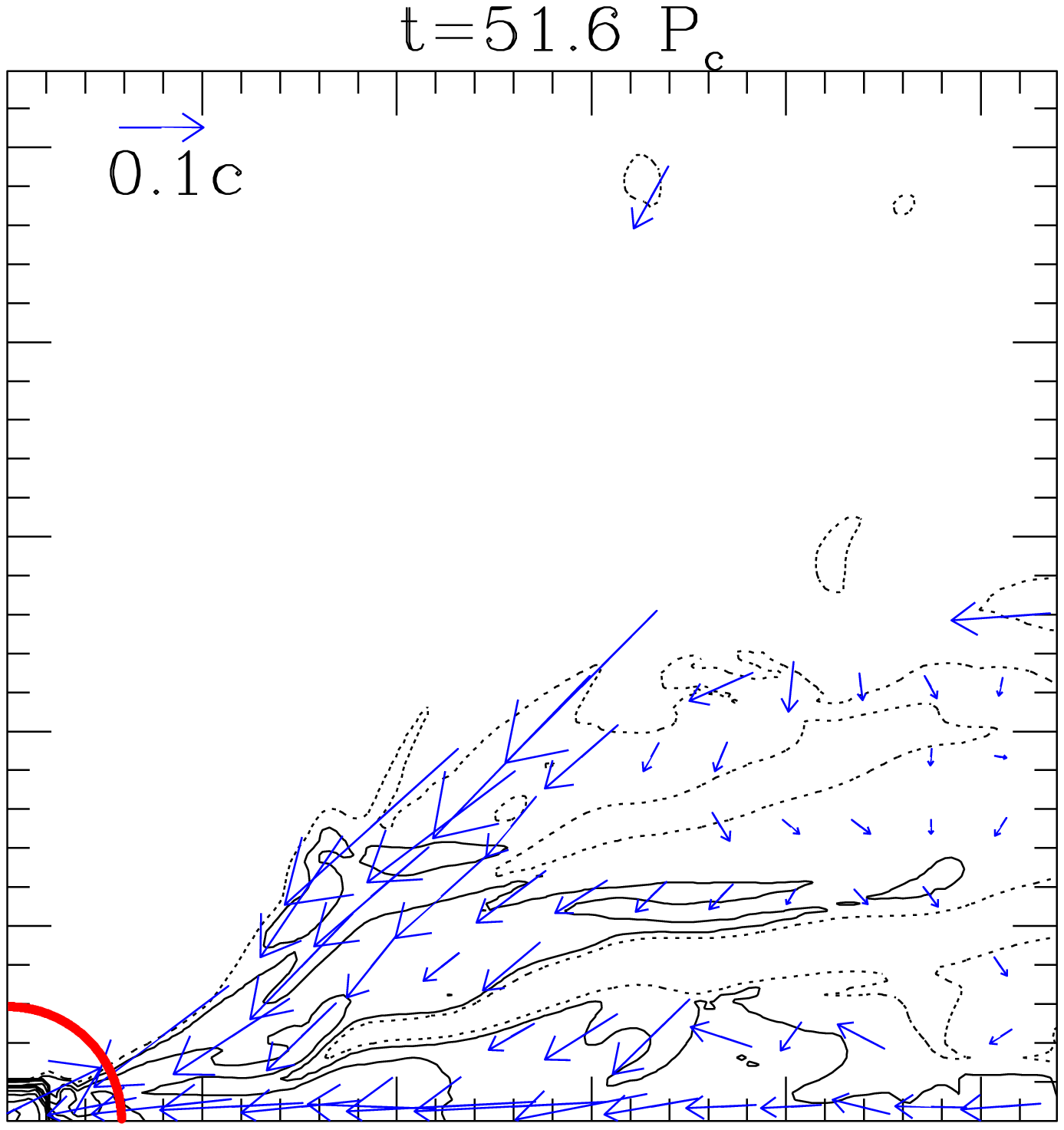}\\
\vspace{-1.7cm}
\epsfxsize=2.1in
\leavevmode
\epsffile{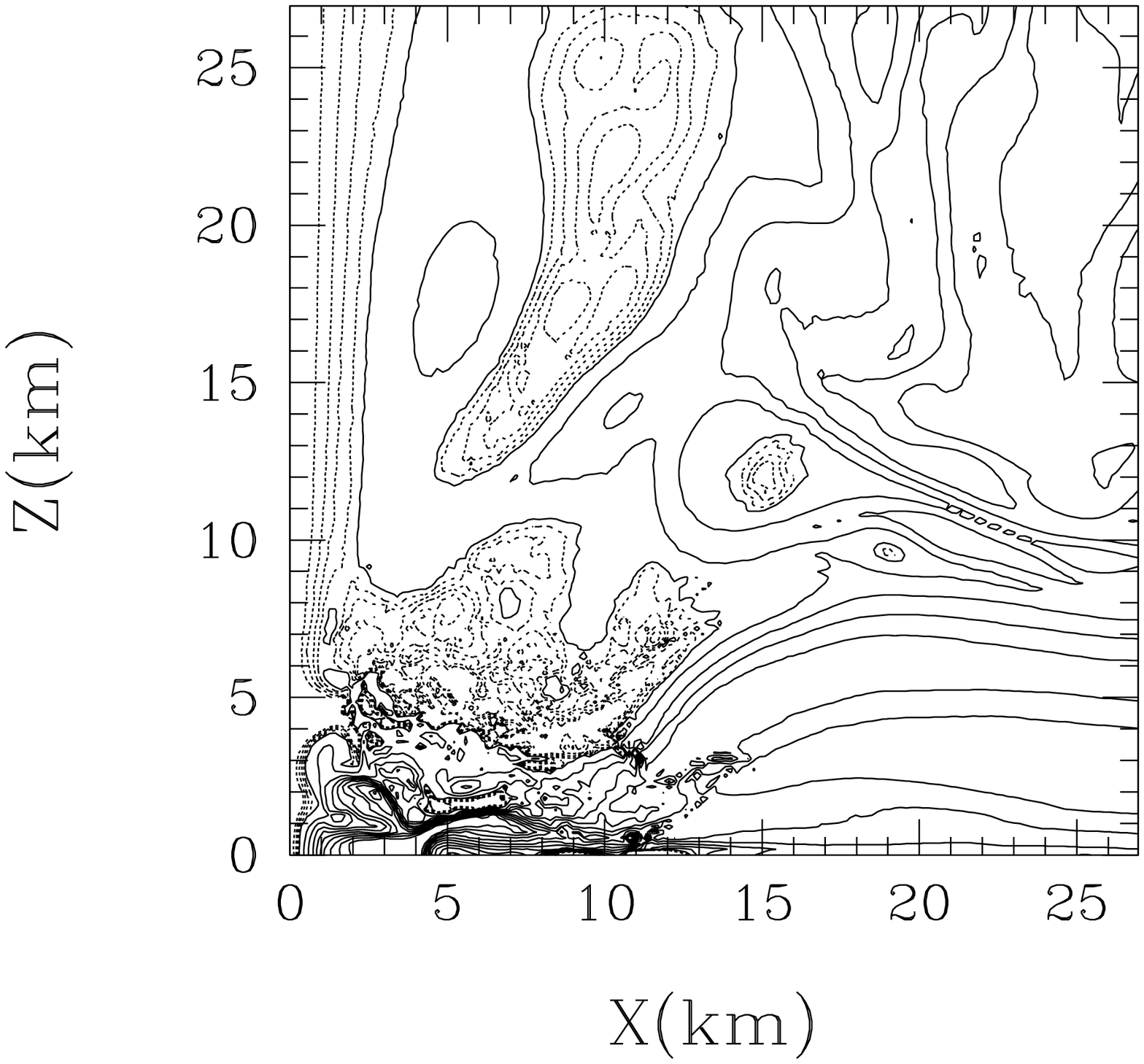}
\epsfxsize=2.1in
\leavevmode
\hspace{-1.65cm}\epsffile{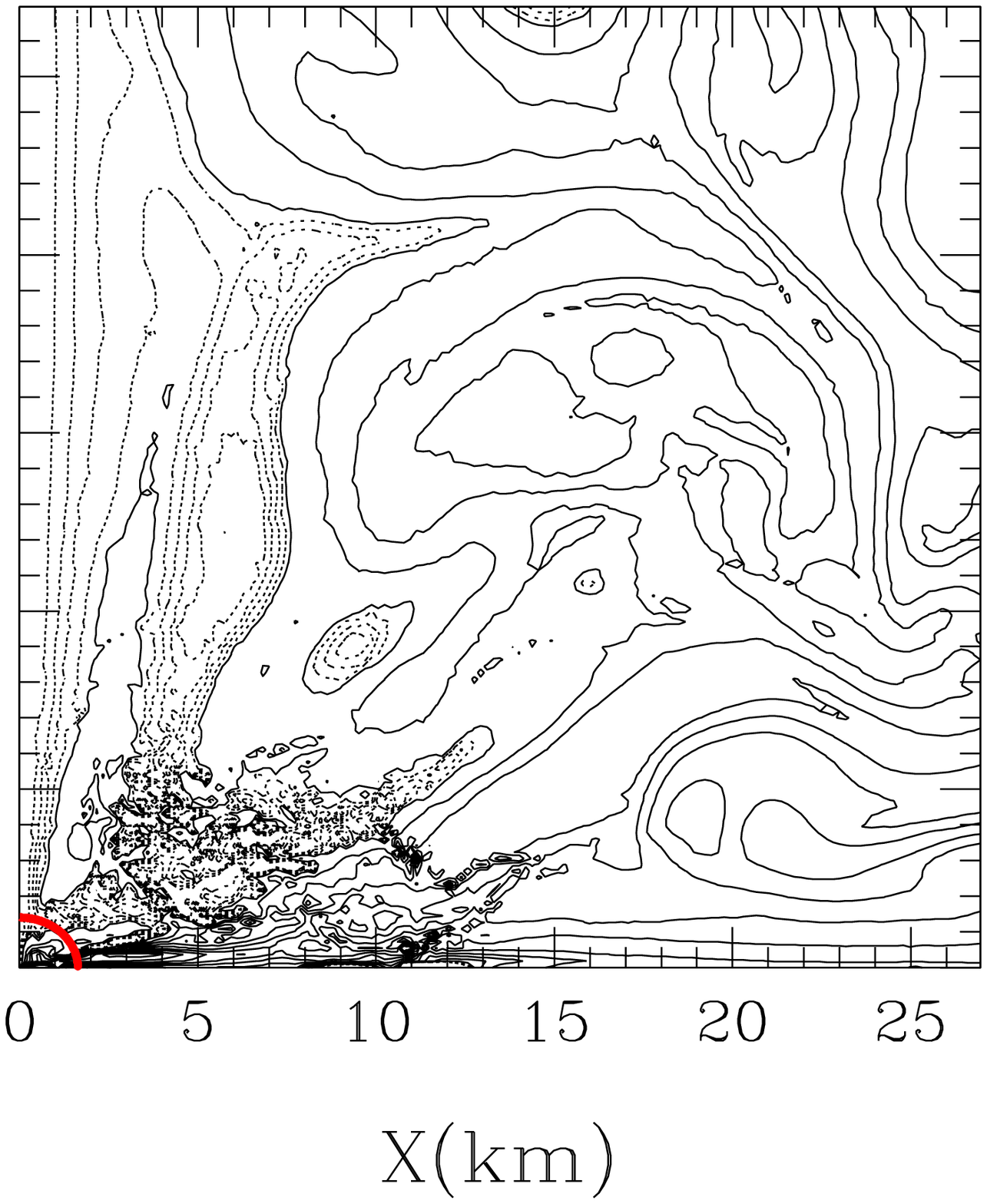}
\epsfxsize=2.1in
\leavevmode
\hspace{-1.65cm}\epsffile{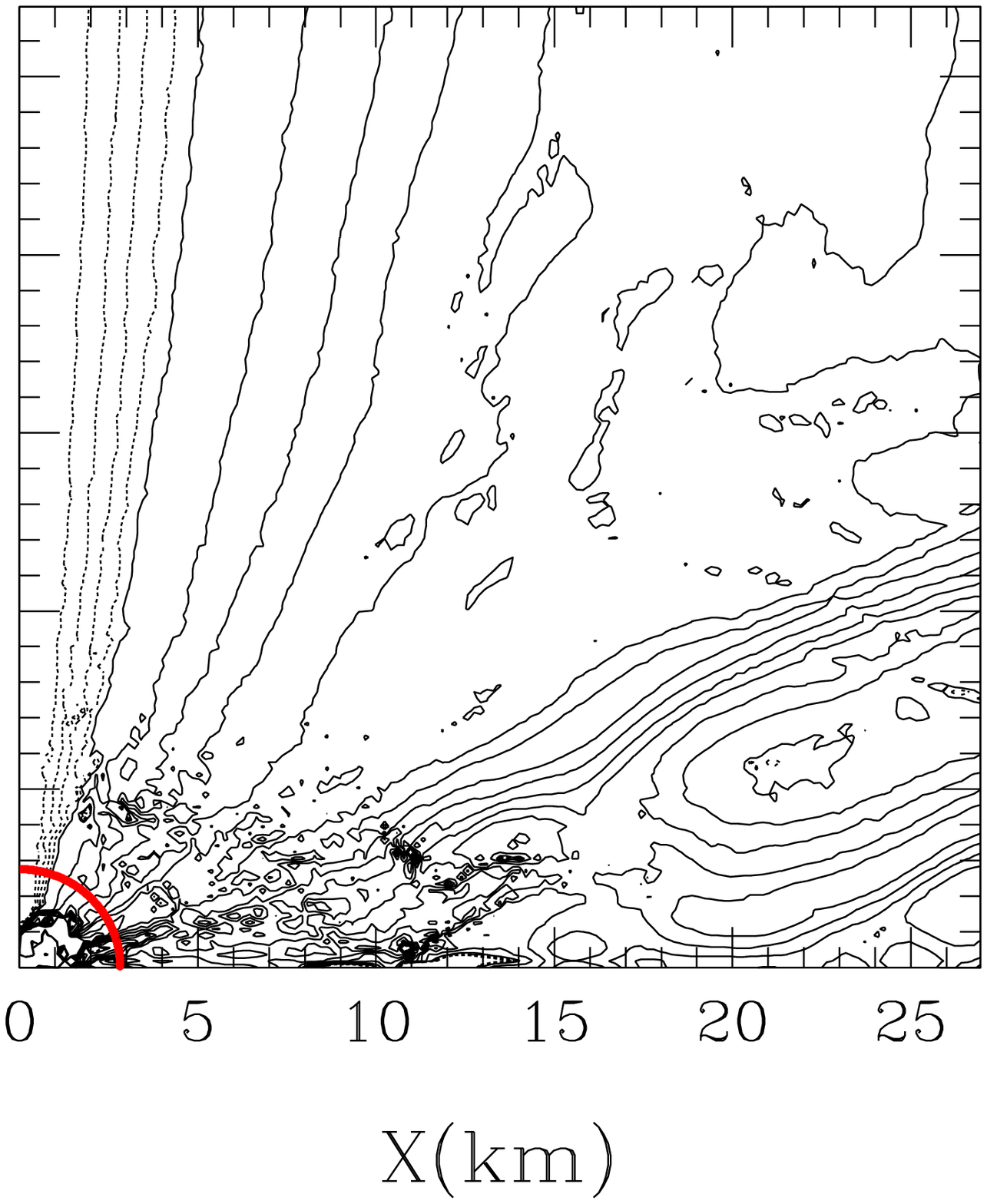}
\epsfxsize=2.1in
\leavevmode
\hspace{-1.65cm}\epsffile{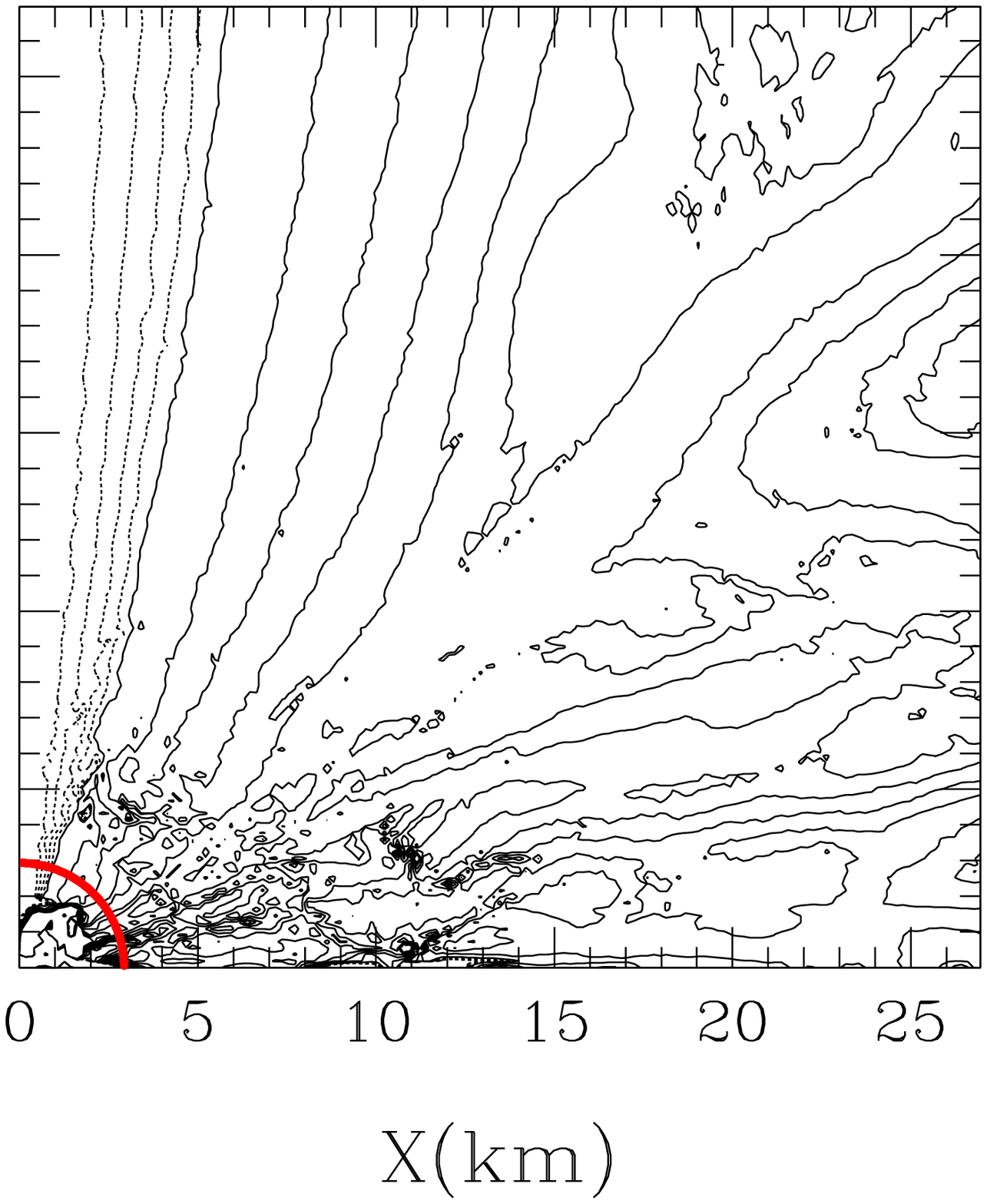}
\vspace{-5mm}
\caption{The same as Fig.~\ref{fig:StarA_contours} but for star C. The
contours for the first and third rows are drawn for
$\rho_0=10^{15-0.4i}~{\rm g/cm^3}~(i=0$--9).  In the last two panels,
curves with $\rho_0=10^{11}~{\rm g/cm^3}$ (solid curves) and with
$\rho_0=5 \times 10^{10}~{\rm g/cm^3}$ (dotted curves) are also drawn.
The circular arc near the bottom-left corner in last three panels denotes 
an apparent horizon.  The second and fourth rows show the poloidal magnetic 
field lines at the corresponding times. The solid contour curves are
drawn for $A_{\varphi}= 0.8(1-0.1i)A_{\varphi,{\rm max,0}}~(i=0$--9)
and the dotted curves are for $A_{\varphi}=0.08(1-0.2i)
A_{\varphi,{\rm max,0}}~(i=1$--4).  Here, $A_{\varphi}=A_{\varphi,{\rm
max,0}}$ is the maximum value of $A_{\varphi}$ at $t=0$. Note that the
outer computational boundary in this simulation is located at $x
\approx 54$ km and $z \approx 54$ km, and that $P_c \approx 0.2$ ms.
The results with $N=601$ are shown here. 
\label{sCmerid}}
\end{center}
\end{figure*}

We next demonstrate that the same qualitative features of the
MHD-induced hypermassive collapse discussed in Section~\ref{starA} are
also present with a more realistic EOS.  To do this, we evolve star~C,
which was constructed using the hybrid EOS described in
Section~\ref{models}.  The ADM mass of this star is $2.65M_{\odot}$,
which is 17\% larger than the mass limit of a rigidly rotating neutron
star for the adopted hybrid EOS. We choose an initial magnetic field
with $C=7.1 \times 10^{-3}$ as the fiducial model. In this case,
the maximum strength of the magnetic field is $\sim 5 \times 10^{16}$
G.  The computational domain is $[0,L]$ in the $x$- and
$z$-directions, with $L=5R_{\rm eq} \approx 54$km. We performed the
same evolution with resolutions $N=501$, $601$, and $751$ to check  
convergence.

\begin{figure}
\epsfxsize=3.4in \epsffile{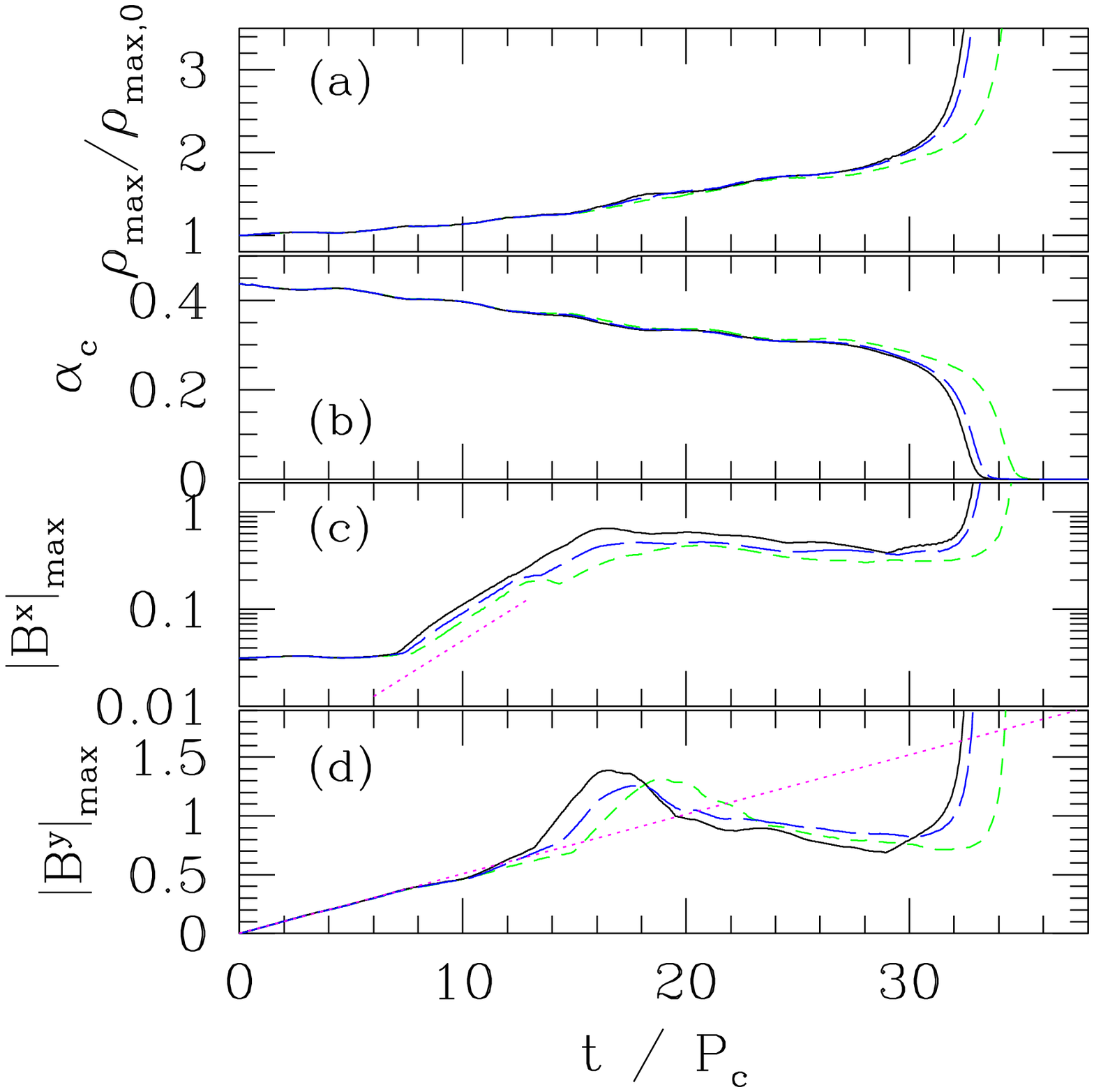}
\vspace{-8mm}
\caption{Evolution of the maximum rest-mass density $\rho_{\rm max}$,
central lapse $\alpha_c$, and maximum values of $|B^x|$ and $|B^y|$
for star~C.  $|B^x|_{\rm max}$ and $|B^y|_{\rm max}$ are plotted in
units of $\sqrt{\rho_{\max}(0)}$.  The dashed (green), long-dashed
(blue), and solid (black) curves denote the results with resolutions
of $N=501$, $601$, and $751$ respectively.  The dotted lines
in (c) and (d) correspond to an exponential growth rate of
$1/3P_c$ and predicted linear growth of $|B^y|_{\rm max}$ at early
times from Eq.~(\ref{eqn:dtBphi}), respectively. 
\label{fig:jkl}}
\end{figure}

\begin{figure*}
\begin{center}
\epsfxsize=5in
\epsffile{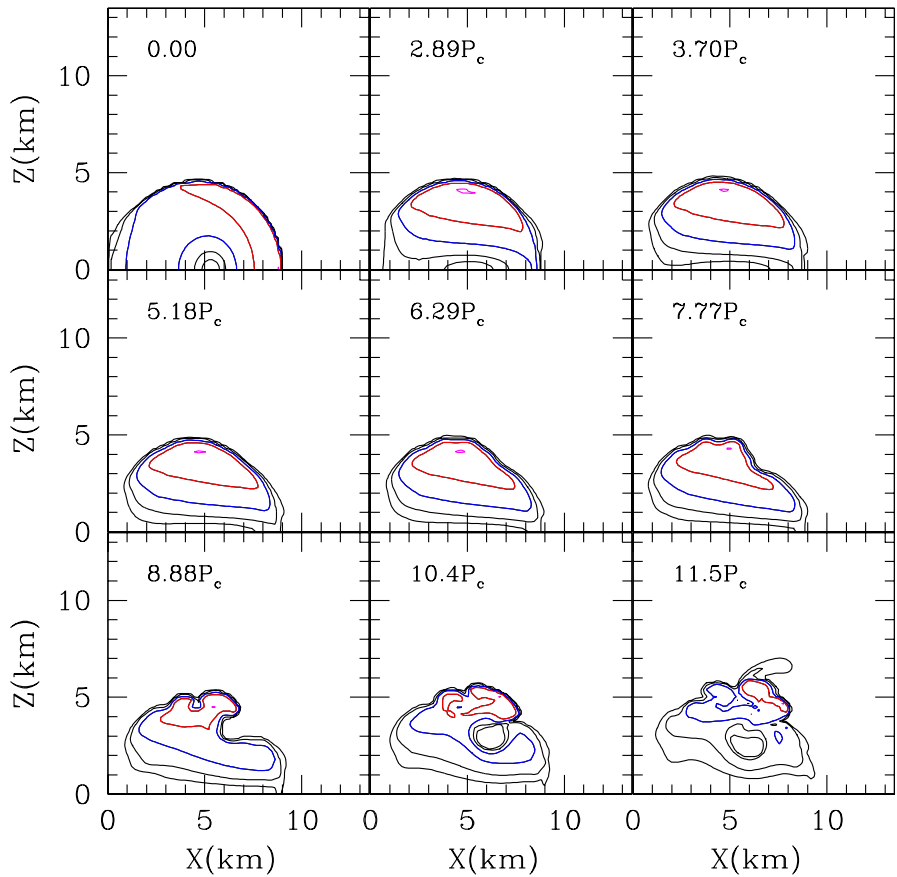}
\caption{Contour curves for the ratio of the magnetic pressure
$P_{\rm mag}$ to the gas pressure $P$
for $t \leq 11.5 P_c$. The contour curves
are drawn for $P_{\rm mag}/P=(P_{\rm mag}/P)_{\rm max}
\times 10^{-0.01}$ (magenta), $10^{-0.5}$ (red), $10^{-1}$ (blue), and
$10^{-0.5i}~(i=3,4)$ (black). The time $t$ is indicated for each 
snapshot.
\label{fig:pressureC}}
\end{center}
\end{figure*}

\begin{figure}
\epsfxsize=3.2in
\epsffile{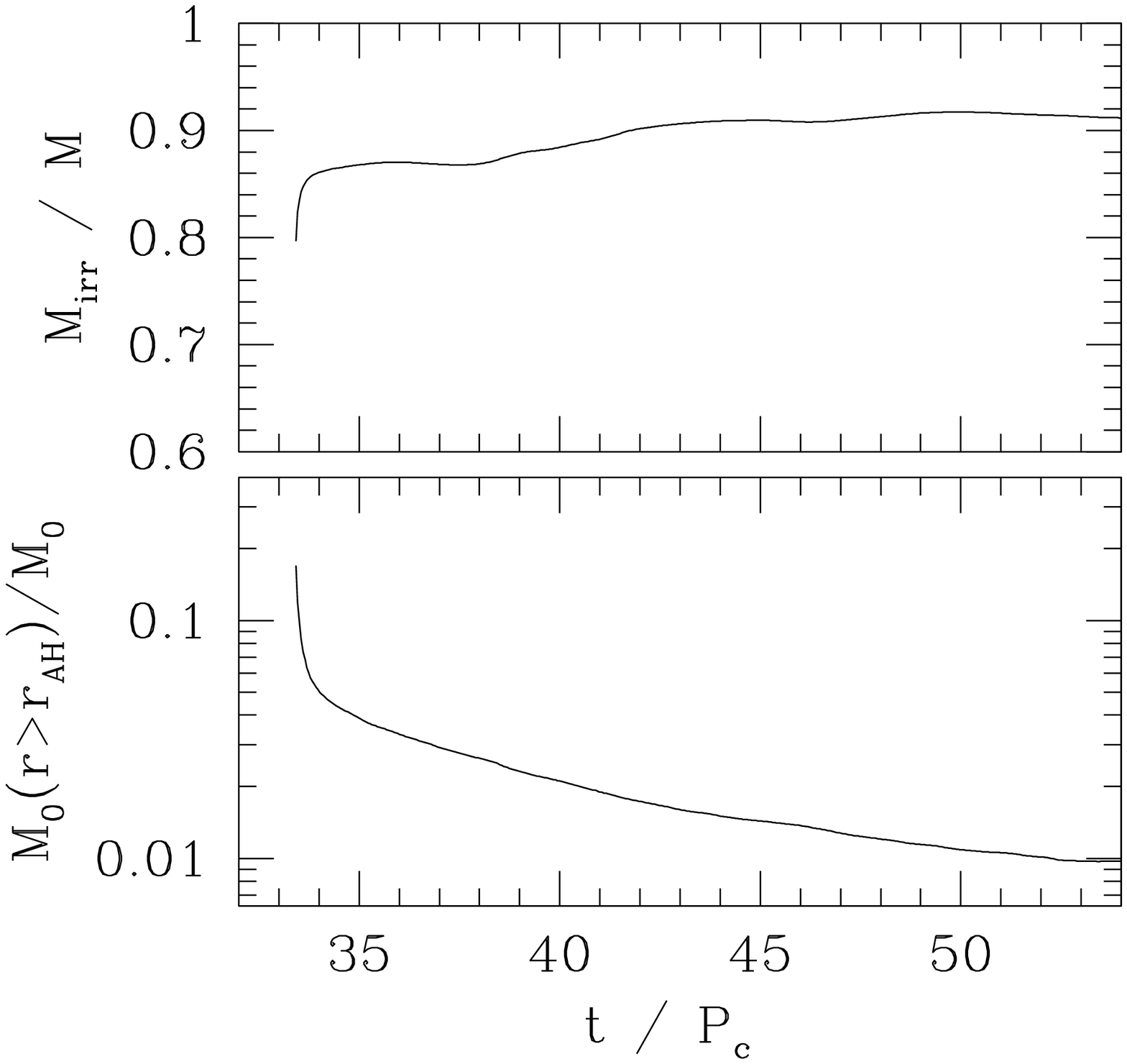}
\vspace{-8mm}
\caption{Post-excision evolution of star~C. The irreducible mass $M_{\rm irr}$ 
of the black hole and the rest mass of the torus surrounding the black 
hole settle down to their quasiequilibrium values at late times.
The results with $N=601$ are shown. 
\label{fig:jkl1}}
\end{figure}

Snapshots of the evolution at eight selected times are shown in
Fig.~\ref{sCmerid}.  Figure~\ref{fig:jkl} shows the evolution of the
maximum density, central lapse, maximum values of $|B^x|$ and $|B^y|$
for the three values of $N$, indicating approximate convergence.  The
maximum values of $|B^x|$ and $|B^y|$ increase as the value of $N$ is
increased. This is a natural consequence of the fact that the profile 
of the magnetic field is better resolved with increasing $N$.

As in the case of star~A, the early phase of the evolution ($t
\alt 13 P_c$) is dominated by magnetic winding.  The linear growth
then saturates and the subsequent evolution is dominated by the MRI
(see the snapshots at $t=11.5 P_c$ in Fig.~\ref{sCmerid} in which a
clear distortion of the poloidal magnetic field lines is seen for
$1~{\rm km} \alt x \alt 4~{\rm km}$ and $z \alt 3$ km).  Soon after
the onset of the MRI, the outer layers of the stellar envelope are 
blown off (see the snapshots for $t/P_c=11.5$--31.1). This explosion 
causes an expansion and redistribution of the magnetic field lines.  
Eventually, the removal of angular momentum from the central regions 
by the MRI results in collapse and black hole formation at 
$t \simeq 33 P_c$.

The winding up of the toroidal magnetic field leads to strong, 
inhomogeneous magnetic pressure.  The toroidal field is
primarily generated in regions where the initial poloidal magnetic 
field has a significant radial component. Thus, material at high 
latitudes gains a high magnetic pressure at early times in the 
evolution.  In Fig.~\ref{fig:pressureC}, contour curves for the ratio 
of the magnetic pressure $P_{\rm mag}=b^2/2$ to the gas pressure $P$ are shown 
for $t \leq 11.5P_c$.  It is seen that the region around $x \sim 5$ km 
and $z \sim 4$ km has the maximum ratio $P_{\rm mag}/P$, and 
the initial seed magnetic field is roughly radial in this region 
(see the first snapshots of Fig. \ref{sCmerid}).  This region of
strong magnetic pressure beneath the surface of the star is subject to 
the effects of magnetic buoyancy~\cite{Kulsrud,Parker}, and 
toroidal magnetic field lines suddenly emerge from inside the HMNS,
propelling material outward in the explosion (see the snapshots of 
Fig. \ref{fig:pressureC} for $t/P_c \agt 8$).  This behavior may
be due to the interchange instability~\cite{Kulsrud}. A similar 
magnetic buoyancy phenomenon is also observed in star~A. However, 
unlike star~C, the magnetic buoyancy does not cause an explosion 
in star~A's outer layers.

The time scale for the rearrangement of the field and the fluid 
due to buoyancy is approximately the same as that of the convection 
instability, and hence, of order 
$\tau_{\rm buoy} \sim (R^2 H /GM)^{1/2}(\rho/\Delta
\rho)^{1/2}$ or $\sim (R^2 H /GM)^{1/2}(P/\Delta P_{\rm
mag})^{1/2}$ \cite{Parker} where $R$ and $H$ are the equatorial radius
and scale height of the inhomogeneity of magnetic pressure and $\Delta
\rho/\rho$ is the degree of inhomogeneity of the density due to the
inhomogeneity of the magnetic pressure $\Delta P_{\rm mag}$. For the outer
layers of star C, we find $H \sim 2$ km, and $\Delta P_{\rm mag}$ is 
approximately equal to the magnetic pressure $P_{\rm mag}$. Since 
$(P/P_{\rm mag})^{1/2} \sim c_s /V_A$, we have
\beqn 
\tau_{\rm buoy} & \sim & (R^2H/GM)^{1/2}c_s/V_A \sim t_A c_s/(GM/H)^{1/2} \nonumber \\
& & \sim 0.4 t_A \sim 0.9~{\rm ms} \ , 
\eeqn
which is comparable to the Alf\'ven time scale. Indeed, this churning of 
field lines and fluid due to magnetic buoyancy seems to begin as soon 
as the toroidal field is wound up to a significant strength.

The formation of the black hole is accompanied by the formation of a 
torus (see the last three snapshots of Fig. \ref{sCmerid}). To follow 
the growth of the black hole due to accretion, the subsequent evolution 
of the system is computed with excision.  Since the torus is magnetized, 
turbulent motion is induced which transports angular momentum outward 
in the accretion torus and encourages the accretion of matter onto the 
black hole.

In the accretion torus, the magnetic fields have a strong radial
component (see, e.g., the snapshots at $t=44.5P_c$). This is because,
during the formation of the black hole, some material in the envelope 
of the HMNS is ejected in the radial direction (see the snapshot at
$t/P_c=31.1$), enhancing the radial magnetic field.  The ejected matter 
soon falls onto the accretion torus, and compresses the magnetic fields. 
This process leads to a strong magnetic field in the accretion torus.  
(The typical value of $P_{\rm mag}/P$ is $10^3$--$10^4$ near the surface of 
the accretion torus.)  As a result, material from high latitudes (which 
is originally blown away from the HMNS as a wind during the collapse)
does not fall toward the equatorial plane, but collides with the surface
of the torus, and then falls into the black hole along the surface of
the torus (see the vector fields at $t=44.5P_c$). Hence, accretion 
occurs along high latitudes as well as along the equatorial plane. This 
scenario for black hole accretion is slightly different from those 
presented, e.g., in~\cite{mg04,dvhkh05}. We also note that the last 
three panels of Fig.~\ref{sCmerid} show that the density of the accretion 
torus gradually decreases, indicating that accretion is quite rapid 
in the first $\sim 15P_c$ after the formation of the black hole.

For $t \agt 45P_c$, the accretion relaxes to a steady rate 
$\dot M \sim 5 \times 10^{-4} M_0/P_c \sim 5 M_{\odot}/$s. The final 
state consists of a rotating black hole surrounded by a hot torus
undergoing quasistationary accretion.  At $t=50P_c$, the irreducible
mass of the black hole is $M_{\rm irr}\approx 0.9M$, while the torus
consists of $\sim 1\% $ of the original rest mass and $\sim
4\%$ of the original angular momentum of the system (see
Fig.~\ref{fig:jkl1}). During the simulation, $\sim 1\%$ of the total
rest mass and $\sim 5\%$ of the total angular momentum escape from 
the computational domain through outflows. Following the same calculations 
as in Sec.~\ref{sec:SA-excision}, we estimate the mass and spin 
parameter of the black hole at $t \approx 50 P_c$ to be 
$M_{\rm hole} \approx 0.98 M$ and $J_{\rm hole}/M_{\rm hole}^2
\approx 0.75$.

\begin{figure}
\vspace{-4mm}
\begin{center}
\epsfxsize=3.2in
\leavevmode
\epsffile{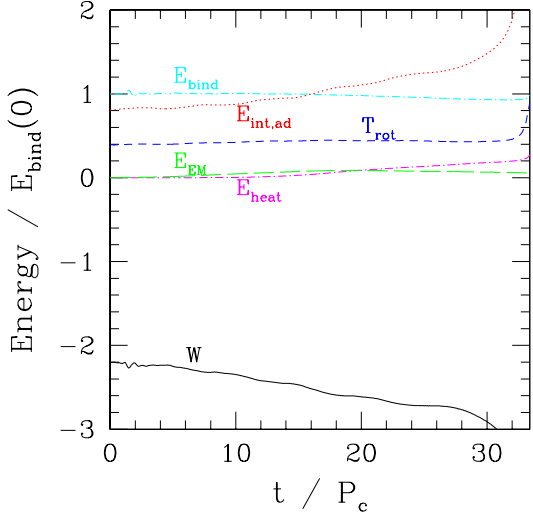}
\vspace{-8mm}
\caption{Components of the energy vs.\ time for star~C.  
All energies are normalized to the binding energy at
$t=0$, where the binding energy is defined as $E_{\rm bind} \equiv M_0 - M$.
In the evolution, $E_{\rm bind}$ should be nearly conserved.
\label{energiesC}}
\end{center}
\end{figure}

Figure~\ref{energiesC} shows the evolution of the various energies 
defined in Sec.~\ref{sec:diagnostics}. The magnetic energy $E_{\rm EM}$ 
reaches a value of at most 8\% of the binding energy ($E_{\rm bind}$) 
throughout the entire evolution, even though the magnetic field drives the 
secular evolution. The gravitational potential energy $W$ and the adiabatic 
part of the internal energy $E_{\rm int,ad}$ change the most, which results 
from the drastic contraction of the stellar core. The fraction of $E_{\rm
int,ad}$ is 60--70\% larger than that for star A. This is simply due
to the fact that star~C is more compact. The rotational kinetic energy
is nearly constant. A substantial amount of heat ($E_{\rm heat}$) is
generated by shocks. When the apparent horizon first appears, this heat 
is $\sim 1\%$ of the rest-mass energy (i.e., $\sim 5 \times 10^{52}$
ergs). Most of the heat is swallowed by the black hole, but a substantial 
fraction remains in the accretion torus (see below). Finally, the binding 
energy decreases by $\approx 7\%$ by the end of the pre-excision evolution.  
This is mainly due to the violation of approximate conservation of the ADM 
mass by $\approx 0.5\%$ and to the escape of $\sim 1\%$ of the mass from 
outer boundaries.

The internal energy in the torus corresponds to a typical
thermal energy per nucleon of approximately $10^2$~MeV
\cite{GRB2}, giving an equivalent temperature $T \approx $1--2
$\times 10^{11}$ K for the density $\sim 10^{10}$--$10^{12}~{\rm g/cm^3}$
if the assumed components are free nucleons, ultra-relativistic electrons, 
positrons, neutrinos, and thermal radiation \cite{MPN}. The opacity to 
neutrinos inside the torus (considering only neutrino absorption and 
scattering interactions with nucleons) is \cite{MPN}
\beq
\kappa \sim 7\times 10^{-17} \biggl({T \over 10^{11}~{\rm K}}\biggr)^2
~{\rm cm}^2~{\rm g}^{-1}. 
\eeq
Because of its high temperature and density, the
torus is optically thick to neutrinos. Thus, the neutrino
luminosity is estimated~\cite{ST} as $L_{\nu} \sim \pi R^2 F$, 
where $R$ is the typical radius of the emission zone and $F$ is the flux 
from the neutrinosphere. In the diffusion limit, $F$ is approximated by
\beq
F \sim {7N_{\nu} \over 3}{\sigma T^4 \over \kappa \Sigma}
\eeq
where $\sigma$ is the Stefan-Boltzmann constant, $N_{\nu}$ is the 
number of thermal neutrino species, taken as 3, and $\Sigma$ is the 
surface density of the torus $\sim 10^{17}$--$10^{18}~{\rm g/cm^2}$. 
We then obtain
\beqn
L_{\nu} &\sim & 2 \times 10^{52}~{\rm ergs/s}
\biggl({R \over 10~{\rm km}}\biggr)^2 \nonumber \\ 
&~&~~ \times \biggl( {T \over 10^{11}~{\rm K}} \biggr)^2
\biggl({\Sigma \over 10^{17}~{\rm g/cm^2}}\biggr)^{-1}. 
\eeqn
This luminosity will be present for the total duration of the accretion,
$\sim 10$ ms. Since the torus has a geometrically thick structure, a
substantial fraction of neutrinos are emitted toward the rotation 
axis, leading to enhanced neutrino-antineutrino pair annihilation along
the axis. The pair annihilation could produce a relativistic fireball 
since the baryon density near the rotation axis is much lower than that 
in the torus. Furthermore, the luminosity is expected to have a strong 
time-variability because of the turbulent nature of the torus. Therefore, 
this massive and hot torus has many favorable properties which may explain 
a short GRB of energy $\sim 10^{48}$--$10^{49}$ ergs \cite{MPN}. This 
possibility was explored by Shibata et al.\ in~\cite{GRB2}.

\begin{figure}
\epsfxsize=3.2in \epsffile{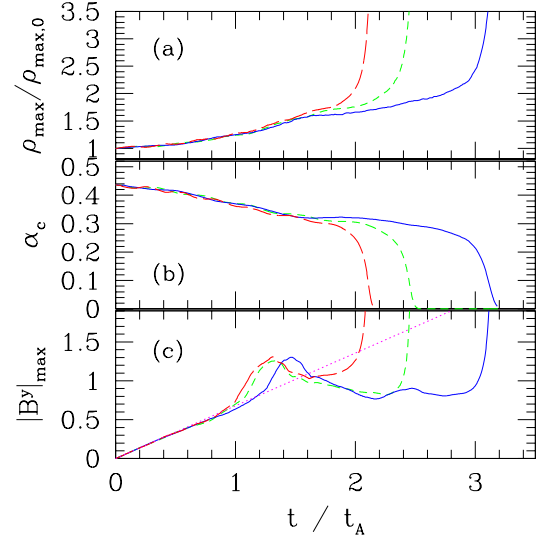}
\vspace{-6mm}
\caption{Evolution of the maximum rest-mass density $\rho_{\rm max}$,
central lapse $\alpha_c$, and maximum values of
$|B^y|/\sqrt{\rho_{\max}(0)}$ as a function of $t/t_A$
for star~C with three values of $C$.
The solid (blue), dashed (green), and long-dashed (red) curves correspond 
to results with $C=1.55 \times 10^{-2}$, $7.1 \times 10^{-3}$, and
$3.8\times 10^{-3}$, respectively. The grid size is $N=601$ for all
cases. The dotted line in (c) corresponds to the predicted linear growth 
of $|B^y|_{\rm max}$ at early times from Eq.~(\ref{eqn:dtBphi}).
\label{fig:jkl2}}
\end{figure}

\begin{figure}
\epsfxsize=3.2in \epsffile{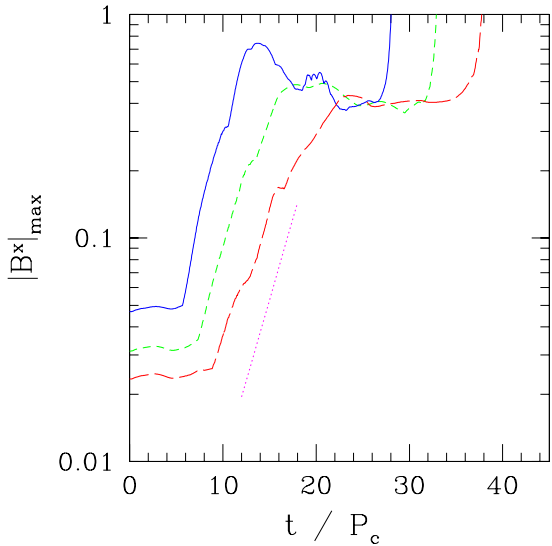}
\vspace{-8mm}
\caption{Evolution of the maximum values of
$|B^x|/\sqrt{\rho_{\max}(0)}$ as a function of $t/P_c$ for star~C with
three values of $C$. The solid (blue), dashed (green), and
long-dashed (red) curves denote the results with $C=1.55 \times 10^{-2}$,
$7.1 \times 10^{-3}$, and $3.8\times 10^{-3}$, respectively. The grid
size is $N=601$ for all cases. The dotted line segment corresponds
to an exponential growth rate of $1/3P_c$.
\label{fig:jkl3}}
\end{figure}

Two other simulations are performed (with $N=601$) for different values 
of the initial magnetic field strength: $C=3.8 \times 10^{-3}$ and 
$1.55 \times 10^{-2}$. In Fig.~\ref{fig:jkl2}, we show the evolution of 
the maximum density, central lapse, and maximum value of $|B^y|$ as a 
function of $t/t_A$. For star~C, we find 
$t_A/P_c=8.94$, 13.4, and 17.9 for $C=1.55 \times 10^{-2}$, 
$7.1 \times 10^{-3}$, and $3.8 \times 10^{-3}$, respectively. 
Figure~\ref{fig:jkl2} shows that the scaling relationship holds for
$t/t_A \alt 1$ as in Fig.~\ref{scaling}.  The scaling breaks down when 
$t/t_A \gtrsim 1$, indicating that the MRI and other effects such 
as magnetic buoyancy determine the evolution of the system.

In Fig.~\ref{fig:jkl3}, we show the evolution of the maximum value of
$|B^x|_{\rm max}$ as a function of $t/P_c$ for three values of $C$. As
in Fig.~\ref{bxmax}, the sudden exponential growth signals the onset
of the MRI, and the approximate agreement of the growth rate for
different values of $C$ indicates that the exponential growth rate of
the MRI does not depend on the initial magnetic field strength. After
the exponential growth, the magnitude of $|B^x|_{\rm max}$ remains
roughly constant until the dynamical collapse occurs. During this
phase before collapse, the angular momentum is transported outward 
gradually by the turbulence. The duration for this angular momentum 
transport is $\sim 15P_c$ irrespective of the value of $C$ as long as 
$C \agt 3.8 \times 10^{-3}$.  This indicates that the angular momentum 
is transported by a mechanism independent of the initial magnetic field 
strength (probably the turbulent transport associated with the MRI).

We note that the collapse time of the HMNS reported here depends slightly
on the parameters of the atmosphere, although the timescales for growth of
the magnetic field due to winding and the MRI do not. This is inevitable 
since, just before the collapse, the HMNS is only marginally stable 
against a quasiradial instability, and thus, a slight increase in the 
atmospheric mass-energy sensitively shortens the collapse time.

We have also studied models with masses slightly different from that
of star~C presented here. We find that the mass of the resulting torus
varies significantly. For more massive stars, the torus mass is
smaller. This is probably due to the fact that the star collapses
sooner and hence there is less time for outward angular momentum
transport. For a sufficiently large mass, the resulting torus mass is 
smaller than 0.1\% of the total mass, which is probably too small for 
the system to trigger a short GRB. On the other hand, less massive 
stellar models result in larger torus masses. This result is interesting 
since it might explain the variety of short GRBs. The details of this 
study will be reported in a future paper.

\section{Discussion and Conclusions}
\label{sec:conclusions}
We have discussed in detail the evolution of magnetized HMNSs as
first reported in~\cite{DLSSS1,GRB2}.  In addition, we have performed 
simulations of two differentially rotating, but nonhypermassive, neutron 
star models with the same initial magnetic field geometry.  These 
simulations have revealed a rich variety of behavior with possible 
implications for astrophysically interesting systems such as binary 
neutron star remnants, nascent neutron stars, and GRBs, 
where magnetic fields and strong gravity both play important roles.

The two hypermassive models considered in this study, stars~A and C,
both collapse to BHs due to the influence of the initially poloidal,
seed magnetic field.  The early phase of evolution for both 
models is dominated by magnetic winding.  As the strength of the toroidal
magnetic field grows, the resulting magnetic stress begins to transport
angular momentum from rapidly moving fluid elements in the 
inner region to the more slowly moving fluid elements in the outer layers.
During this magnetic braking phase, the inner regions of the stars undergo
quasistationary contraction, while the outer layers expand and begin to 
form a low-density torus.  

The winding of the magnetic field proceeds until the back-reaction
on the fluid becomes strong enough that the growth of the toroidal field 
ceases.  This happens after roughly one Alfv\'en time.  After several 
rotation periods, we also see the effects of the MRI.  Plots of the poloidal 
magnetic field lines display perturbations with wavelengths similar to 
$\lambda_{\rm max}$ (the wavelength of the fastest growing mode estimated 
from the linear analysis).  These perturbations first appear in the outer 
layers of the star, which is consistent with the linear analysis. 
In order to diagnose the sudden local growth of the 
poloidal magnetic field due to the MRI, we track the maximum value of $|B^x|$ 
on the grid.  We found that $|B^x|_{\rm max}$ grows exponentially at a rate 
which does not depend on the strength of the initial magnetic field, in accord 
with the properties of the MRI. However, the growth rate observed in our 
numerical simulations differs significantly from that predicted by the 
linear analysis. This is due probably to the fact that (a) the linear analysis 
is based on Newtonian gravity, but the models we study here are highly 
relativistic, and/or (b) the small MRI wavelength assumption in 
the analysis might not be applicable to our magnetic field configuration.

The nonlinear outcome of the MRI is turbulence, and this turbulence leads to 
further angular momentum transport.  
Eventually, the inner cores of stars~A and C become unstable and
collapse to BHs.  Surrounding the BHs, significant amounts of material 
remain in  magnetized tori which have been heated considerably by 
shocks resulting from the turbulent motions of the fluid.  This final 
state consisting of a BH surrounded by a massive, hot accretion disk may
be capable of producing highly relativistic outflows and 
a fireball (either through
$\nu-\bar{\nu}$ annihilation or MHD processes) and is hence a promising
candidate for the central engine of short-hard GRBs.  This model predicts 
that such GRBs should accompany a burst of gravitational radiation and neutrino 
emission from the HMNS delayed collapse. 

The behavior of the nonhypermassive, ultraspinning star~B1 under the 
influence of a 
seed magnetic field is quite different.  Magnetic braking and the MRI 
operate in this model as well, leading to a mild contraction of the 
inner core and the expansion of the outer layers into a high angular
momentum torus-like structure.  
The final state consists of a fairly uniformly rotating core surrounded
by a differentially rotating torus.  The remaining differential rotation 
does not shear the magnetic field lines (i.e.\ 
$\langle |B^j \partial_j \Omega| \rangle$ approaches zero in the final 
state), so that the toroidal field settles down.  We find
that this configuration is not subject to the MRI, probably because it
is suppressed by the strong magnetic field (the MRI wavelength 
is comparable to the size of the star, and the standard local linear 
analysis breaks down in this regime).  The rotation state of the 
final configuration naturally depends on the geometry of the initial 
magnetic field. On the other hand, the normal star B2 
simply evolves to a uniformly rotating configuration. 

Two issues in particular warrant further study.  The first is the 
scaling behavior of our solutions.  We begin our simulations with a 
seed magnetic field which, though far too weak to be dynamically 
important, may be significantly larger than magnetic fields present 
in HMNSs formed through stellar collapse or a binary neutron star merger.
We have demonstrated that, by varying the strength of the initial
magnetic field through a factor of $\sim 3$ (See Fig.~\ref{scaling}), 
our evolution obeys the expected scaling during the magnetic winding 
phase, and the qualitative outcome of the simulations remains the same.  
However, since the MRI grows on a timescale $\sim \mbox{few}\times P_c$ 
regardless of the initial magnetic field strength, it is possible that, 
for very weak initial fields, the effects of the MRI could dominate the 
evolution long before the effects of magnetic braking become important.
In this case, the scaling of our numerical results with the Alfv\'en
time (relevant for magnetic winding) may break down.  The 
relative importance of magnetic winding and the MRI for different
seed field strengths deserves further study.  Unfortunately, the 
wavelength of the fastest growing MRI mode becomes very difficult
to resolve numerically as the strength of the initial magnetic
field decreases.  However, our results seem to indicate that 
magnetic braking and MRI-induced turbulence have similar effects
in magnetized HMNSs.  Thus, the qualitative features of the 
evolutions described here may also be present for HMNSs with much 
weaker initial seed fields.

Another issue which warrants further study concerns the effects
on our evolutions of relaxing the axisymmetry assumption.  Rapidly
and differentially rotating neutron stars may be subject to bar
and/or one-armed spiral mode instabilities which could affect the 
dynamics (though star~A was shown in~\cite{BSS,DLSS} to be stable 
against such instabilities, at least on dynamical timescales).  
Additionally, the development of the MRI in 2D differs from the 
3D case~\cite{hgb95}.  Turbulence arises and persists more readily 
in 3D due to the lack of symmetry.  More specifically, according to 
the axisymmetric anti-dynamo theorem~\cite{moffatt78}, sustained 
growth of the magnetic field energy is not possible through axisymmetric
turbulence. This 
phenomenon has been demonstrated by numerical simulations~\cite{hb92}.
However, McKinney and Gammie~\cite{mg04} 
have performed axisymmetric simulations of magnetized tori accreting 
onto Kerr BHs and have found good quantitative agreement with the 3D 
results of De Villiers and Hawley~\cite{dvhkh05} for the global 
quantities $\dot{E}/\dot{M}_0$ and $\dot{J}/\dot{M}_0$~\cite{fn4}, 
which are the 
rates of total energy and angular momentum falling into the horizon, 
normalized by the accretion rate.  Though simulations in full 3D will 
eventually be necessary to capture the full behavior of magnetized 
HMNSs, the 2D results presented here likely provide (at least) 
a good qualitative picture.  

\acknowledgments
It is a pleasure to thank C.~Gammie for useful suggestions and discussions.
Numerical computations were performed at the National Center for
Supercomputing Applications at the University
of Illinois at Urbana-Champaign (UIUC), and
on the FACOM xVPP5000 machine at
the data analysis center of NAOJ and the NEC SX6 machine in ISAS,
JAXA. This work was in part supported by NSF Grants PHY-0205155
and PHY-0345151, NASA Grants NNG04GK54G and NNG046N90H
at UIUC, and
Japanese Monbukagakusho Grants (Nos.\ 17030004 and 17540232).

\vskip 1cm
\appendix
\section{Drawing magnetic field lines}
\label{app:field_lines}

The vector potential $A^i$ is related to the magnetic field $B^i$ by 
$B^i = n_{\mu} \epsilon^{\mu ijk} \partial_j A_k$, where $\epsilon^{\alpha 
\beta \gamma \delta}$ is the Levi-Civita tensor. It is easy to show that 
in axisymmetry, the poloidal components of a magnetic field ($B^{\varpi}$ 
and $B^z$) are determined by $A_{\varphi}$ alone as follows: 
\begin{eqnarray}
  \varpi \sqrt{\gamma}\, B^{\varpi} &=& -\partial_z A_{\varphi} \ , 
\label{eq:Bvarphi}\\
  \varpi \sqrt{\gamma}\, B^z &=& \partial_{\varpi} A_{\varphi} \ .
\label{eq:Bz}
\end{eqnarray}
Poloidal magnetic field lines are two-dimensional curves on 
which $d\varpi/dz = B^{\varpi}/B^z=-\partial_z A_{\varphi}/\partial_{\varpi} 
A_{\varphi}$. Hence we have 
\begin{equation} 
  d A_{\varphi} = (\partial_{\varpi} A_{\varphi}) d\varpi + (\partial_z 
A_{\varphi}) dz = 0
\end{equation} 
on the curves. This means that contours of constant $A_{\varphi}$ are the 
poloidal magnetic field lines. All the poloidal field lines shown in 
this paper are drawn by the contours of $A_{\varphi}$. There are two ways 
of calculating $A_{\varphi}$. One method is to integrate Eqs.~(\ref{eq:Bvarphi}) 
and~(\ref{eq:Bz}). The other method is to evolve $A_{\varphi}$ according to 
the equation
\beq
  \partial_t A_{\varphi}=\varpi \sqrt{\gamma} (v^z B^{\varpi}-v^{\varpi} B^z) \ .
\eeq
This equation can be derived from Eqs.~(35) and~(42) of~\cite{bs03}.

In order to show the toroidal component of the magnetic field, we 
draw field lines projected onto the equatorial plane. To do this, 
we first choose
three points $\varpi_{(j)}$ ($j=$1, 2, 3) in the equatorial plane 
so that at the given time,
\begin{equation}
  A_{\varphi}(\varpi_{(j)},z_{\rm min}) = A_{\varphi,{\rm min}} +
(A_{\varphi,{\rm max}}  - A_{\varphi,{\rm min}})j/4 \ ,
\label{eq:initial_points}
\end{equation}
where $A_{\varphi,{\rm max}}$ and $A_{\varphi,{\rm min}}$ are the maximum
and minimum values of $A_{\varphi}$ at the given time~\cite{fn5}, and 
$z_{\rm min}=0$ when there is no apparent horizon in the
time slice. If there is an apparent horizon, we set $z_{\rm min} =0$ if
$\varpi_{(j)}>0.5r_{AH}$ and $z_{\rm min}=\sqrt{0.25r_{AH}^2-\varpi_{(j)}^2}$
if $\varpi_{(j)}<0.5r_{AH}$. Here $r_{AH}$ is the coordinate radius
of the apparent horizon. Next we integrate
the equations
\begin{eqnarray}
\frac{dx_{(j)}}{d\lambda} &=& B^x(x_{(j)},y_{(j)},z_{(j)}) \ , \\
\frac{dy_{(j)}}{d\lambda} &=& B^y(x_{(j)},y_{(j)},z_{(j)}) \ , \\
\frac{dz_{(j)}}{d\lambda} &=& B^z(x_{(j)},y_{(j)},z_{(j)}) \ ,
\end{eqnarray}
with the initial locations $(x_{j},y_{j},z_{j})$ given by:
\beqn
x_{(j)}|_{\lambda = 0} & = & \varpi_{(j)}\cos\left[\frac{2(j-1)\pi}{3}\right], \nonumber \\
y_{(j)}|_{\lambda = 0} & = & \varpi_{(j)}\sin\left[\frac{2(j-1)\pi}{3}\right], \nonumber \\ 
z_{(j)}|_{\lambda = 0} & = & z_{\rm min}  \ \ \ \ \ \ \ \  (j = 1,2,3) \ .
\eeqn
Here $\lambda$ serves as a parameter of the 3D curves. The integration
is terminated when the curve goes beyond the boundary of the grid.
The projected field lines are the trajectories
$(x_{(j)}(\lambda),y_{(j)}(\lambda))$ traced out by $\lambda$. On the 
other hand, the poloidal field lines determined by the contours of 
$A_{\varphi}$ 
are equivalent to the trajectories of $(\sqrt{x^2 + y^2},z)$.


\begin{thebibliography}{99}
\bibitem{fn3} Note that this may be due to the artifact of the axisymmetry 
assumption. As explained in~\cite{hb92} and \cite{MRIrev}, this 
artifact is related to 
the axisymmetric anti-dynamo theorem~\cite{moffatt78}.
\bibitem{fn4} The expression for $\dot{E}$, $\dot{M}_0$ and $\dot{J}$ 
are as follows: $\dot{E}=2\pi \int_0^{\pi} d\theta \sqrt{-g} T^r{}_t$, 
$\dot{M}_0 = 2\pi \int d\theta \sqrt{-g} \rho_0 u^r$, and 
$\dot{J}=2\pi \int_0^{\pi} d\theta \sqrt{-g} T^r{}_{\varphi}$. 
All the integrands are evaulated at a constant spherical radius 
inside the black hole horizon.
\bibitem{fn5} Note that there may be more than one value of
$\varpi_{(j)}$ for each $j$ that satisfies Eq.~(\ref{eq:initial_points}).
We choose the minimum value in this case. In the last panel
in Fig.~\ref{fig:StarA_Blines} ($t=74.6P_c$), Eq.~(\ref{eq:initial_points})
has no solution for all $j$. We instead choose $\varpi_{(j)}$ so that 
$A_{\varphi}(\varpi_{(j)},z_{\rm min}) = A_{\varphi,{\rm min}} 
+ (A_{\varphi,{\rm max}} - A_{\varphi,{\rm min}})j/10$ ($j$=1, 2, 3).
\end{thebibliography}
\end{document}